\documentclass[11pt]{report}

\usepackage{amsmath}
\usepackage{amssymb}
\usepackage{graphicx}
\usepackage{subfigure}
\usepackage{setspace}
\usepackage{bm}

\newcommand{\Cal}[1]{\ensuremath{\mathcal{#1}}}

\newcommand{\eqn}[1]{Eqn. \eqref{#1}}
\newcommand{\eqns}[1]{Eqns. \eqref{#1}}
\newcommand{\Cite}[1]{Ref. \cite{#1}}
\newcommand{\Cites}[1]{Refs. \cite{#1}}
\newcommand{\Sec}[1]{Sec. \ref{#1}}
\newcommand{\fig}[1]{Fig. \ref{#1}}
\newcommand{\figs}[1]{Figs. \ref{#1}}
\newcommand{\ph}[1]{\phantom{#1}}
\newcommand{\tab}[1]{Table \ref{#1}}

\newcommand{\G}{\ensuremath{\Gamma}}
\newcommand{\GF}{\ensuremath{\Gamma_F}}
\newcommand{\dG}{\ensuremath{\delta\Gamma}}
\newcommand{\dg}{\ensuremath{\delta\Gamma^{(1)}}}
\newcommand{\Gb}{\ensuremath{\widetilde{\Gamma}}}
\newcommand{\dGb}{\ensuremath{\widetilde{\delta\Gamma}}}
\newcommand{\dgb}{\ensuremath{\widetilde{\delta\Gamma}^{(1)}}}

\newcommand{\aD}{\ensuremath{a_{\Cal{D}}}}
\newcommand{\aDdot}{\ensuremath{\dot a_\calD}}
\newcommand{\aDddot}{\ensuremath{\ddot a_\calD}}
\newcommand{\calD}{\ensuremath{{\Cal{D}}}}
\newcommand{\calI}{\ensuremath{\mathcal{I}}}
\newcommand{\calP}{\ensuremath{\mathcal{P}}}
\newcommand{\eplog}{\ensuremath{\epsilon\ln\epsilon}}
\newcommand{\eps}{\ensuremath{\epsilon}}

\newcommand{\W}[2]{\ensuremath{\Cal{W}{}^{#1^\prime}_{#2}}}
\newcommand{\Winv}[2]{\ensuremath{\Cal{W}{}^{#1}_{#2^\prime}}}
\newcommand{\Wi}{\ensuremath{W^{-1}}}
\newcommand{\dW}{\ensuremath{\delta W}}
\newcommand{\dw}{\ensuremath{\delta W^{(1)}}}

\newcommand{\p}{\ensuremath{\partial}}
\newcommand{\pp}{\ensuremath{\partial^\prime}}
\newcommand{\nab}{\ensuremath{\nabla}}

\newcommand{\avg}[1]{\ensuremath{\langle \,#1\, \rangle}}
\newcommand{\stavg}[1]{\ensuremath{\langle #1\rangle_{ST}}}
\newcommand{\avgD}[1]{\ensuremath{\langle \,#1\, \rangle_{\Cal{D}}}} 
\newcommand{\eavg}[1]{\ensuremath{[ \,#1\, ]}_{ens}}

\newcommand{\be}{\begin{equation}}
\newcommand{\ee}{\end{equation}}
\newcommand{\la}[1]{\label{#1}}

\newcommand{\ti}[1]{{\tilde #1}}
\newcommand{\wti}[1]{{\widetilde #1}}
\newcommand{\vphi}{\varphi}
\newcommand{\om}{\omega}
\newcommand{\com}{{\hat\omega}}
\newcommand{\cchi}{{\hat\chi}}
\newcommand{\cdelta}{{\hat\delta}}
\newcommand{\ctheta}{{\hat\Theta}}
\newcommand{\cv}{{\hat V}}

\newcommand{\adot}{{\dot a}}
\newcommand{\addot}{{\ddot a}}
\newcommand{\uD}[1]{\ensuremath{#1_\Cal{D}}}

\newcommand{\B}{Buchert}
\newcommand{\Z}{Zalaletdinov}

\newcommand{\Om}[2]{{\bm{\Omega}{}^{#1}_{\ph{#1}#2}}} 
\newcommand{\bom}{\bm{\omega}} 
\newcommand{\barOm}[2]{\bar{\bm{\Omega}}{}^{#1}_{\ph{#1}#2}} 
\newcommand{\rmb}[1]{{\bf #1}} 
\newcommand{\bZ}[4]{{\rmb{Z}{}^{#1\ph{i}#3}_{\ph{i}#2\ph{i}#4}}} 
\newcommand{\ext}{{\rmb{d}}} 
\newcommand{\dx}[1]{\ext x^{#1}} 

\newcommand{\dbil}{{\ext\negthickspace\negmedspace^{-}}}

\newcommand{\pbil}{{\p\negthickspace\negthickspace\textbf{--}}}

\newcommand{\Dbil}[1]{{\rmb{D}_{#1}\negthickspace\negthickspace
    \negthickspace\negthickspace\negthickspace
    \text{--\thickspace\thickspace\thickspace\,\,\,}}}       

\newcommand{\bil}[1]{\widetilde{#1}}

\newcommand{\Wxx}[2]{\ensuremath{\Cal{W}{}^{#1^\prime}_{#2}(x^\prime,x)}}
\newcommand{\Wxxinv}[2]{\ensuremath{\Cal{W}{}^{#1}_{#2^\prime}(x,x^\prime)}}

\newcommand{\Mbar}{\ensuremath{\bar{\Cal{M}}}}

\newcommand{\pap}[1]{Paper #1}

\newcommand{\Linh}{\ensuremath{L_{\rm inhom}}}
\newcommand{\Lfrw}{\ensuremath{L_{\rm FLRW}}}
\newcommand{\Lhub}{\ensuremath{L_{\rm Hubble}}}
\newcommand{\bt}{\ensuremath{\bar t}}
\newcommand{\ba}{\ensuremath{\bar a}}

\begin{document}

\begin{titlepage}
\begin{center}
{\LARGE {\bf The Averaging Problem in Cosmology}}\\
\vspace{1.2in}
{\Large A Thesis}\\
\vspace{1.2in}
{\Large Submitted to the}\vskip 0.1in
{\Large Tata Institute of Fundamental Research, Mumbai}\vskip 0.1in
{\Large for the degree of Doctor of Philosophy}\vskip 0.1in
{\Large in Physics}\\
\vspace{0.8in}
{\Large by}\\
\vspace{0.5in}
{\Large Aseem Paranjape}\\
\vspace{1.5in}
{\Large Dept of Astronomy \& Astrophysics}\vskip 0.1in
{\Large Tata Institute of Fundamental Research}\vskip 0.1in
{\Large Mumbai}\vskip 0.1in
{\Large May, 2009}
\end{center}
\end{titlepage}
~\thispagestyle{empty}\newpage
\begin{center}
{\Large {\bf Acknowledgments}}
\end{center}
This thesis, and the academic life which surrounds it in both time and
space, would not have been possible without the
constant, unfailing support and encouragement of my parents Madhu
and Sudhir Paranjape. They will always have my gratitude for
introducing me to the joys of science. 
\vskip 0.1in
\noindent
I am indebted to my advisor T. P. Singh for his calm and patient
guidance throughout the period over which this work was done, and for
granting me the freedom of choosing which avenue to go down while
tackling the problem at hand. It has been a pleasure working with
him. It is a pleasure also to acknowledge the
guidance and support of T. Padmanabhan (my ``other advisor''), who
has generously hosted me at his parent institute IUCAA, Pune, several 
times during my PhD lifetime. These periodic breaks working on other
problems while recharging my batteries, have played an important role
in the timely completion of this work. I am grateful also to both
institutes, TIFR and IUCAA, for making this arrangement work
smoothly. 
\vskip 0.1in
\noindent
I have benefited immensely from detailed correspondence and
conversations with Thomas Buchert, Roustam Zalaletdinov, Syksy
R\"as\"anen and David Wiltshire. I am grateful to them for sharing
with me their opinions and ideas, which have influenced several parts
of this work. 
\vskip 0.1in
\noindent
It is a pleasure to thank my friend and colleague Karel Van Acoleyen
for many collaborative discussions which have had a direct bearing on
parts of this work, especially in chapter 5. We may yet write a paper
together someday!
\vskip 0.1in
\noindent
I have intermittently bored many of my friends at TIFR with long and 
winding descriptions of my work. Noteworthy among these bravehearts
are Rakesh, Sarang, Aditya, Satej and Tridib. I thank them for
their patience! I also thank the entire HEP journal club gang for
many many weeks of fun physics, and Sourendu Gupta for inviting me to
be part of the group. They can be found at 
{\tt http://theory.tifr.res.in/$\sim$sgupta/hepjc.php}
\vskip 0.1in
\noindent
It is because of the following people, old friends and new, that
leaving TIFR will be like leaving home : Aditya, Ajesh, Alok, Argha,
Basu, Benny, Bhargava, Ganesh, Manjusha, Naren, Partha, Prerna,
Rakesh, Ronnie, Sanchari, Sarang, Satej, Sayantani, Shamayita,
Somnath, Sonal, Subhabrata, Sudarshan, Suresh, Swapnil, Tridib,
Upasana, Vaibhav, Vandna, among several others.  
\vskip 0.1in
\noindent
And finally, a BIG thank you to Samuel (the SAM) Morris and his
nightingales Anagha, Aparna, Garima, Sanchari, Sonal, Sonia and
Upasana; and his other not-so-fair associates Arkarup, Deepak, DJ,
Naren, Rajeev, Richard, Sujaan and Sudarsan; for sharing with me the
musical experience of a lifetime. May the sounds of silence stay long
with you, despite what everybody said!

~\thispagestyle{empty}\newpage
~\thispagestyle{empty} \newpage
\pagenumbering{roman} \setcounter{page}{1}
\tableofcontents
\newpage
\chapter*{Synopsis}

\begin{onehalfspacing}
\section*{Introduction}
A central assumption in modern cosmology is that the universe on large
scales is homogeneous and isotropic \cite{weinberg}. This assumption
leads to tremendous simplifications in the application of general
relativity (GR) to cosmology, since it reduces the ten independent
components of the metric of spacetime $g_{ab}(t,\vec{x})$ to
essentially a single function of time $a(t)$ known as the scale
factor. In the early days of modern cosmology, beginning with
stalwarts such as Einstein and deSitter, the assumption of homogeneity
and isotropy was largely motivated on grounds of simplicity and
aesthetic appeal. In recent times however, it has become possible to
confront this assumption with observations, which remarkably appears
to be justified to a large extent (based on observations of the cosmic
microwave background (CMB) radiation \cite{wmap}, and on analyses of
galaxy surveys \cite{2df,sdss}). This
indicates that a model based on essentially a single function of time
might in fact go a long way in furthering our understanding of the
behaviour of the universe.  

Of course the real universe is not homogeneous; we see a rich variety
of structure around us from stellar systems to galaxies to clusters of
galaxies and even larger structures. The study of the large scale
structure (LSS) in the universe has a long history going 
back several decades \cite{lss-rev}. Perhaps one of the biggest
successes of cosmological theory based on GR, has been the explanation
of how  statistical properties of the LSS arise \cite{peebles}. The
relevant calculations are largely based on linear perturbation theory,
in which one describes inhomogeneities in the universe as
perturbations around the smooth solution characterised by the scale
factor $a(t)$ and expands the Einstein equations as a series in these
small perturbations. While such a treatment has met with great success
in the description of the statistical properties of the tiny
fluctuations (anisotropies) in the temperature of the CMB, there are
two causes for concern. 

The first is a purely theoretical issue, and is the basis of this
work. The idea that the large scale universe is homogeneous and
isotropic necessarily entails an implicit notion of averaging on these
large scales. In other words, what one is really saying is, ``When the 
spatially fluctuating parts of the solution of GR describing our
universe are averaged out, what is left is the homogeneous and
isotropic solution of Einstein's equations''\footnote{This solution is
known as the Friedmann-Lema\^itre-Robertson-Walker (FLRW) solution
after those who first studied it.}. The immediately obvious
problem with this statement is that the details of the averaging
operation are not at all clear, and indeed are usually never
specified. A bigger problem  is one noted by Ellis \cite{ellis}, and 
can be stated in the following symbolic way. If $g$ denotes the
metric, $\Gamma$ the Christoffel connection and $E[g]$ the Einstein
tensor for the metric $g$, then we have the relations
\be
\Gamma \sim \partial g ~~;~~ E[g] \sim \partial\Gamma + \Gamma^2\,, 
\label{syn-eq1}
\ee 
with $\partial$ denoting spacetime derivatives. The Einstein equations
are therefore
\be
E[g] = T\,,
\label{syn-eq2}
\ee
with $T$ denoting the energy-momentum tensor of the matter
components. Now, irrespective of any details of the averaging
operation, one notes that 
\be
E[\avg{g}] - \avg{E[g]}\sim \avg{\Gamma}^2 - \avg{\Gamma^2} \neq0\,,
\label{syn-eq3}
\ee
with the angular brackets denoting the averaging. The FLRW solution
would amount to solving the equations $E[\avg{g}]=\avg{T}$. In general 
therefore, it is \emph{not} true that averaging out the fluctuating
inhomogeneities leaves behind the FLRW solution, since what we are
actually left with is
\be
E[\avg{g}] = \avg{T} - \Cal{C} ~~;~~ \Cal{C} \sim \avg{\Gamma^2} -
\avg{\Gamma}^2 \,,
\label{syn-eq4}
\ee
and the homogeneous solution that we are looking for will depend on
the details of the correction terms \Cal{C}.

The second cause for concern comes from observations. It has now been
established beyond a reasonable doubt, that the FLRW metric confronted
with observations indicates an accelerating scale factor
\cite{accln-evid}. Conventional sources of energy such as radiation
and nonrelativistic matter cannot explain the acceleration, and it is
now common to attribute this effect to a hitherto unknown ``Dark
Energy'', which in its simplest form is a cosmological constant. The
true nature of this additional component in the cosmological
equations, is perhaps the most challenging puzzle facing both
theorists and observers today. A huge amount of research has gone into
(a) explaining the value that a cosmological constant term must take
to explain data or (b) assuming a zero cosmological constant,
constructing models of a dynamical dark energy which explains the
observed acceleration \cite{sami}. It is fair to say however that
there is no theoretical consensus on what the origin of Dark Energy
is. 

Since we have seen above that the effects of averaging lead to some
extra, as yet unknown terms in the equations, it is natural to ask
whether these two issues are connected. Could the acceleration of the
universe be explained by the effects of averaging inhomogeneities
(``backreaction'') in the universe? Regardless of the answer to this
question, what is the nature and magnitude of this backreaction? The
purpose of this thesis is to answer these questions as rigorously as
possible. 

\section*{The conventional wisdom, and loopholes}
One should note that the conventional wisdom on the issue of
backreaction in the sense described in the previous section, is that
this effect can never be significant. It is of importance therefore to
understand this argument and its shortcomings. The argument goes as
follows \cite{wald}. One starts by assuming that inhomogeneities in the
universe can be described in the Newtonian approximation of GR, by the
gravitational potential $\Phi(t,\vec{x})$ with $|\Phi|\ll1$, which
satisfies the Poisson equation $\nab^2\Phi=4\pi G a^2\delta\rho$ where
$\delta\rho(t,\vec{x})$ is the fluctuation of matter density about the
mean homogeneous value $\bar\rho(t)$, and can in general have a large
value. (E.g., in clusters of galaxies one finds
$\delta\equiv\delta\rho/\bar\rho\sim10^2$, and the ratio increases on
smaller length scales). One then argues that the universe we observe 
\emph{does} seem to be very well-described by the above model, and
effects of averaging this model can only arise at second order in
$\Phi$ and should hence be extremely small. 

There is a loophole in this argument though. The catch is that the
background expansion $a(t)$ is defined completely ignoring
the backreaction, which is an integrated effect with contributions
from a large range of length scales. This means that the following
possibility cannot be \emph{a priori} ruled out : Initial conditions
are specified as a perturbation around a specific FLRW solution, but
the integrated effect of the backreaction grows (with time) in such a
manner as to effectively yield a late time solution which is a
perturbation around a \emph{different} FLRW model. Indeed, there are
calculations in the literature that do indicate such a possibility
being realised \cite{martineau,kolb-2005,notari}. We therefore see the
need to actually perform a rigorous calculation that will describe the
time evolution of the backreaction, and thereby either confirm or
overthrow the conventional wisdom.  

\section*{Averaging schemes}
A major hurdle in computing the effects of averaging has been the lack
of reliable averaging procedures which can be used in GR, mainly
because defining and physically interpreting averaging operations
suitable for tensors, is a challenging prospect. A number of authors
have attempted to solve this problem, both in the specific context of
cosmology and also as a more general problem of the mathematics of GR
(see, e.g. \Cites{noonan,futamase,kasai-avg,boersma,avg}). To date,
the most promising averaging schemes have been 
the spatial averaging of scalars due to Buchert
\cite{buchert1a,buchert1b}, and the fully covariant tensor averaging
due to Zalaletdinov \cite{zala1}.  

Buchert's scalar averaging deals with a chosen $3+1$ splitting of
spacetime, and only averages two of the Einstein equations. This
averaging scheme is simple to implement and intuitively easy to
grasp, however it is ultimately difficult to interpret its physical
significance. Zalaletdinov's scheme on the other hand, is technically
challenging to handle, since its averaging operation can deal with
full-fledged tensors at the cost of introducing some new mathematical
structures into the problem. The appeal of this scheme lies in the
fact that ultimately one has in hand an object which can legitimately
be called the ``averaged metric'' on an ``averaged manifold''.

In this thesis we use both these schemes to address certain specific
questions concerning the backreaction problem. As we discuss below
however, ultimately we rely upon Zalaletdinov's scheme to make
realistic statements regarding the nature of cosmological
backreaction. 

\subsection*{Can backreaction ever be large? \\{\small A Paranjape and 
    T P Singh, Class. Quant. Grav. 23, 6955 (2006).}}
One question to ask in the context of the conventional wisdom
presented above, is whether it is technically possible within GR to
have a situation in which the backreaction dominates the averaged
expansion. We answer this in the affirmative by studying a toy
model. We use the exact, spherically symmetric Lema\^itre-Tolman-Bondi
(LTB) solution of GR, to construct a parametrized toy model for a
\emph{curvature dominated universe}, i.e. -- a spacetime in which the
spatial curvature of the 3-dimensional slices dominates over the
contribution of the (nonrelativistic) matter. Using Buchert's
averaging scheme we find that the effective scale factor obtained in
this toy model, does in fact accelerate for a wide class of parameter
values. Now one needs to ask whether this can happen in the real
universe, for which we turn to Zalaletdinov's approach.

\subsection*{Simplifying Zalaletdinov's framework \\{\small A Paranjape
    and T P Singh, Phys. Rev. D76, 044006 (2007).}}
As it stands, Zalaletdinov's averaging framework deals with averaging
an arbitrary spacetime, and due to its generality it is technically
challenging and difficult to work with. By restricting its application
to cosmology and requiring consistency with basic cosmological
assumptions, we find that we can simplify this framework and bring it 
to a form which can be readily applied to perform calculations. Doing
this also clarifies the nature of the backreaction as being a
physically relevant quantity on the same footing as the scale factor
of the FLRW spacetime. An important point is that for cosmology one
must necessarily consider the \emph{spatial averaging limit} of
Zalaletdinov's 4-dimensional spacetime averaging. In this limit we
further highlight the similarities and differences between
Zalaletdinov's and Buchert's approaches, and the fact that
the structure of the correction terms in both approaches is very
similar, being essentially the same as expected from the heuristic
arguments of Ellis discussed earlier.

There is a significant difference between the original philosophy of
the averaging formalism, common to both the Buchert and Zalaletdinov
schemes, and the manner in which we employ Zalaletdinov's
averaging. The original idea as developed by these authors was to
construct a framework which would independently describe a suitably
defined averaged dynamics, with no reference to the inhomogeneous
spacetime whose average leads to this dynamics. So, for example,
Zalaletdinov formulates a new \emph{theory} of gravity (named
Macroscopic Gravity or MG) which attempts to describe the dynamics of
an averaged manifold, with no recourse to the underlying manifold
which is described by the usual Einstein equations. The backreaction
in this approach is actually a new field in the problem which
satisfies its own equations and whose dynamics must be solved for
simultaneously with that of other fields such as the averaged metric
and the averaged energy-momentum tensor for matter.

Our approach to the backreaction issue is different : We consider it
central to be able to \emph{self-consistently} describe both the
inhomogeneous geometry as well as its averaged counterpart. We find
this necessary since modern cosmology crucially relies on observations
of inhomogeneities around us, and ignoring the evolution of
inhomogeneities when solving for the averaged dynamics does not appear
to be satisfactory. Put another way, when faced with a solution of the
averaged dynamics, we find it essential to answer the question ``which
(if any) inhomogeneous solution could lead to this averaged
homogeneous solution?'' All our calculations therefore focus on
solving for the averaged dynamics of specific inhomogeneities, which
we attempt to keep as realistic as possible.

\section*{Backreaction in (linear) perturbation theory \\{\small A 
    Paranjape, Phys. Rev. D78, 063522 (2008).}}
We apply the simplified version of Zalaletdinov's scheme to the
problem of calculating the backreaction in the perturbative framework
and determining its evolution. We discuss the issue of a possible
\emph{gauge dependence} of the backreaction, which is essentially the
problem that in the perturbative framework one must be careful to
distinguish physical effects from artifacts of choosing a specific
coordinate system. We show how the backreaction can be calculated in a
gauge independent manner, although one is forced to make certain
choices concerning the averaging operation itself, which are not fixed
by Zalaletdinov's formalism. Once the formalism is developed, we are
left with expressions for the backreaction that are valid whenever the
\emph{metric} of spacetime can be described as a perturbation around
its FLRW form, regardless of the magnitude of matter density
fluctuations.  

To deal with the issue of self-consistency, we propose an iterative
procedure to compute the backreaction. Since order-of-magnitude
estimates of the backreaction indicate that the effect is expected to
be small in the early universe (around the epoch of last scattering
say), we begin with a ``zeroth iteration'' in which the backreaction
is assumed to vanish entirely. This of course is simply the setup for
the standard treatment of cosmology, in which the evolution of
inhomogeneities can be numerically evaluated. We do this and
consequently obtain a first estimate for the backreaction using the
formalism developed earlier. We find that the magnitude of the
backreaction in this first estimate remains negligible ($\sim10^{-4}$
at present epoch in dimensionless units) compared to the
background contribution at all times, and its evolution indicates that
continuing to further iterations would not lead to any instability;
the final answer is expected to converge to a form very close to the
original ``zeroth order'' choice for the background. 

\section*{The nonlinear regime}
The preceding arguments however are valid in the \emph{linear regime}
of perturbation theory where \emph{matter fluctuations} are
small. However, the structure of the integrands of the backreaction
functions indicate that the contribution from length scales where
matter fluctuations have become nonlinear, should in fact also remain
negligible. Yet one would like to see this in an actual nonlinear
calculation rather than relying on heuristic arguments. Specifically,
in the nonlinear regime when matter fluctuations are large, one needs
needs to address two issues : (a) Is a perturbative expansion in the 
\emph{metric} still valid? (b) If so, then is the contribution to the
backreaction from nonlinear scales in fact negligible?

\subsection*{A toy model for structure formation \\{\small A Paranjape   
    and T P Singh, JCAP03(2008)023;}}
It has been claimed in the literature \cite{rasanen} that perturbation 
theory does not give correct insight into the problem of structure
formation in the late universe, and that when one studies simple but
nonperturbative examples of structure formation, the contribution of
backreaction to the averaged dynamics is in fact large. These claims
are clearly contradictory to the conventional wisdom and to the
arguments presented earlier. We attempt to sort out this debate by
studying a toy model of structure formation, using the spherically
symmetric LTB solution.

The matter source in the LTB solution is pressureless ``dust'', which
is sufficient for our purposes since we wish to enquire whether a
universe dominated by nonrelativistic matter can have a large
backreaction component in realistic situations. We assume
initial conditions to be a perturbation around an FLRW model without
dark energy. Our model contains an inner spherical overdense 
ball surrounded by an underdense shell, outside which we take the
matter density to be homogeneous. The overdense ball initially
expands but eventually turns around and begins collapsing, mimicking
for example the infalling region outside clusters of galaxies. This
happens because this inner region satisfies equations which are
identical to those of a ``closed'' FLRW model in which the universe
eventually recollapses. Naively one would expect that the underdense
region behaves like the ``open'' FLRW models which expand forever,
however the situation is more involved. It turns out that imposing
appropriate matching conditions at the boundary of the over- and
underdense regions, implies that a section of the underdense 
region immediately surrounding the overdense ball, will in fact
eventually collapse.

This result has interesting consequences. By simply ignoring this part
of the underdense shell that eventually turns around, we can show
using our model that arguments such as in \Cite{rasanen} which ignore 
matching conditions across boundaries can lead to an accelerating
effective scale factor. However, as soon as this region is correctly
accounted for, the acceleration disappears.

Consistently with this result, we show that one can rewrite the
nonperturbative LTB solution as a perturbation around the same FLRW
model we started with, \emph{provided} the peculiar velocity of the
dust remains nonrelativistic\footnote{The peculiar velocity is defined
as the difference between the physical velocity and the Hubble flow of
the dust element.}. This is exactly what one expects from standard
textbook results concerning cosmology in the Newtonian limit of
GR. The small parameter governing the linear perturbation theory valid
in the early universe is the magnitude of the matter density contrast
$\delta$, while the small parameter relevant for the late time
Newtonian limit is the nonrelativistic peculiar velocity $v$. Our
calculation is an explicit demonstration in a physically clear and
simple setting, of how the perturbation theory in $\delta$ becomes a
perturbation theory in $v$.

\subsection*{Backreaction in the nonlinear regime \\{\small A
    Paranjape and T P Singh, Phys. Rev. Lett. 101, 181101 (2008).}} 
As a final step, for completeness we compute the backreaction in our
model of structure formation described above. The formalism described
in the preceding sections can be applied to this model since we can
bring the metric of this model to the perturbed FLRW form. There are
some subtleties regarding the numerical calculations, since the
coordinates that are natural to the model are not natural to the
backreaction formalism, but these can be handled in a straightforward
manner. As expected, we find that the backreaction for such a model of
structure formation is in fact negligible. The significance of this
calculation is that it is the first one in which the backreaction has
been calculated as a physically meaningful quantity even in the late
time nonlinear phase of the cosmological evolution.

\section*{Conclusions and Outlook}
The question of whether backreaction from averaging of inhomogeneities
can lead to significant effects, has generated a heated debate in the
literature. There have been conflicting results on issues such as the
stability of perturbation theory in the presence of these
corrections. Our calculations using Zalaletdinov's covariant averaging
scheme applied to both linear perturbation theory and toy models of
nonlinear structure formation, form the first systematic demonstration
that perturbation theory \emph{is} in fact stable against corrections
due to backreaction, and that backreaction \emph{cannot} explain the
late time acceleration of the universe. In principle such calculations
can be extended to numerical simulations of structure formation,
however that is beyond the scope of this work. 

Given that cosmological data is rapidly increasing in quantity and
improving in quality, it will soon become possible to determine
cosmological parameters with percent level accuracy \cite{future-obs},
and even perform tests of fundamental assumptions such as the
Copernican principle \cite{copernican-test}. In this context, it
becomes interesting to ask whether the contribution of the
backreaction, while very small compared to the background, could be
tested or used to improve parameter estimation. This remains a subject
for future work. 

\end{onehalfspacing}

\newpage
~
\vskip 0.75in
\noindent
{\Huge \bf List of publications}
\vskip 0.5in
\begin{itemize}
\item Publications contributing to this thesis
\begin{description}
\item{\bf I.}  {\bf ``The Possibility of Cosmic Acceleration via
  Spatial Averaging in Lemaitre-Tolman-Bondi Models''}, Aseem
  Paranjape and T. P. Singh, {\it Class. Quant. Grav.} {\bf 23}, 6955
  (2006) [arXiv:astro-ph/0605195]. 
\item{\bf II.} {\bf ``The Spatial Averaging Limit of Covariant
  Macroscopic Gravity -- Scalar Corrections to the Cosmological
  Equations''}, Aseem Paranjape and T. P. Singh, {\it Phys. Rev.} {\bf
  D76}, 044006 (2007) [arXiv:gr-qc/0703106].
\item{\bf III.} {\bf ``Backreaction of Cosmological Perturbations in
  Covariant Macroscopic Gravity''}, Aseem Paranjape, {\it Phys. Rev.}
  {\bf D78}, 063522 (2008) [arXiv:0806.2755]. 
\item{\bf IV.} {\bf ``Structure Formation, Backreaction and Weak
  Gravitational Fields''}, Aseem Paranjape and T. P. Singh, {\it JCAP}
  {\bf 03}(2008)023 [arXiv:0801.1546]. 
\item{\bf V.} {\bf ``Cosmic Inhomogeneities and the Average
  Cosmological Dynamics''}, Aseem Paranjape and T. P. Singh, {\it
  Phys. Rev. Lett.} {\bf 101}, 181101 (2008) [arXiv:0806.3497].
\end{description}
\item Other publications
\begin{enumerate}
\item {\bf ``Nonlinear Structure Formation, Backreaction and Weak
  Gravitational Fields''}, Aseem Paranjape, arXiv:0811.2619 (2008),
  talk presented at {\bf CRAL-IPNL Conference on Dark Energy and Dark
    Matter, Observations, Experiments and Theories}, July, 2008, Lyon
  Center for Astrophysics Research (CRAL), Lyon, France; to appear in
  the Conference Proceedings.
\item {\bf ``A Covariant Road to Spatial Averaging in Cosmology -- 
  Scalar Corrections to the Cosmological Equations''}, Aseem
  Paranjape, {\it Int. J. Mod. Phys.} {\bf D17}, 597 (2008)
  [arXiv:0705.2380]. (This essay received an Honourable Mention in the
  Gravity Research Foundation's Essay Competition, 2007.) 
\item  {\bf ``Entropy of Null Surfaces and Dynamics of Spacetime''},
  T. Padmanabhan and  Aseem Paranjape, {\it Phys. Rev.} {\bf D75},
  064004 (2007) [arXiv:gr-qc/0701003]. 
\item  {\bf ``Explicit Cosmological Coarse Graining via Spatial 
  Averaging''}, Aseem Paranjape and  T. P. Singh,
  {\it Gen. Rel. Grav.} {\bf 40}, 139 (2008) [arXiv:astro-ph/0609481].   
\item  {\bf ``Thermodynamic route to field equations in 
  Lanczos-Lovelock gravity''}, Aseem Paranjape, Sudipta Sarkar and
  T. Padmanabhan, {\it Phys. Rev.} {\bf D74}, 104015 (2006)
  [arXiv:hep-th/0607240]. 
\item  {\bf ``Embedding diagrams for the Reissner-Nordstrom 
  space-time''}, Aseem Paranjape and Naresh Dadhich, {\it
  Gen. Rel. Grav.} {\bf 36}, 1189 (2004) [arXiv:gr-qc/0307056]. (This
  paper was the result of part of the work done under the JNCASR
  Summer Research Fellowship, April -- June 2003.)  
\end{enumerate}
\end{itemize}

\newpage



\listoffigures
\newpage
\begin{onehalfspacing}

~
\cleardoublepage
\pagenumbering{arabic} \setcounter{page}{1}
\chapter{Introduction}
Our understanding of the universe has undergone dramatic changes in
the last century. Edwin Hubble's discovery in 1924 that stars known
as Cepheid variables could be found in the Andromeda nebula and
appeared fainter than Cepheids in the Milky Way, established that such
nebulae were not part of the Milky Way but were in fact distant
galaxies themselves. And his demonstration 5 years later that these 
galaxies appear to be receding from the Milky Way at speeds
proportional to their distances, has found its way into popular
consciousness as the maxim ``the universe is expanding''
\cite{hubble-1929}. On the theoretical front, Einstein's general
relativity had at this time quickly gained acceptance as a fundamental
theory of gravity, and its application to cosmology was being studied
by several workers including Einstein himself, deSitter, Lema\^itre,
Friedmann among others. The simple models of a universe described by
the homogeneous and isotropic geometries characterised by the
Friedmann-Lema\^itre-Robertson-Walker (FLRW) metric, were very
successful at describing the then limited amount of cosmological
observations. The decades since have seen the emergence of the highly 
successful Big Bang model of cosmology, which posits that the universe
went through a very hot dense phase at early times and cooled as it
expanded, with tiny fluctuating inhomogeneities in the past that have
grown to form structures such as galaxies today \cite{history}.

A central assumption in this (widely accepted) model of cosmology is
that the universe on large scales is homogeneous and isotropic
\cite{weinberg}. This assumption leads to tremendous simplifications
in the application of general relativity (GR) to cosmology, since it
reduces the ten independent components of the metric of spacetime
$g_{ab}(t,\vec{x})$ to essentially a single function of time $a(t)$
known as the scale factor. In the early days of $20^{\rm th}$ century
cosmology, the assumption of homogeneity  and isotropy was largely
motivated on grounds of simplicity and aesthetic appeal. In recent
times however, it has become possible to confront this assumption with
observations, which remarkably appears to be justified to a large
extent (based on observations of the CMB radiation \cite{COBE,wmap},
and on analyses of galaxy surveys \cite{2df,sdss,homog}, although see 
\Cite{sylos}). This indicates that a model based on essentially a
single function of time might in fact go a long way in furthering our
understanding of the behaviour of the universe.   

Of course the real universe is not homogeneous; we see a rich variety
of structure around us from stellar systems to galaxies to clusters of
galaxies and even larger structures \cite{lss-rev}. Perhaps one of the 
biggest successes of cosmological theory based on GR, has been the
explanation of how  statistical properties of the large scale
structure arise \cite{peebles}. The relevant calculations are largely
based on linear perturbation theory (i.e. linearizing Einstein's
equations around the smooth FLRW solution) which is valid at all
length scales of interest at early times and on large scales at late
times \cite{dodelson,mukhanov}. Dynamics on small scales at late times 
involves nonlinear theory, and is dealt with using
approximation schemes such as the Press-Schechter formalism and its
extensions \cite{press-schechter}, ``Newtonian'' nonlinear
perturbation analyses \cite{nonlinearPT} and numerical simulations
\cite{simulations}. While such treatments have met with great success
in the description of the statistical properties of the anisotropies
in the temperature of the CMB, as well as of the inhomogeneous
distribution of galaxies, there are two causes for concern.  

The first is a purely theoretical issue, and is the basis of this
work. The idea that the large scale universe is homogeneous and
isotropic necessarily entails an implicit notion of averaging on these
large scales. In other words, what one is really saying is, ``When the 
spatially fluctuating parts of the solution of GR describing our
universe are averaged out, what is left is the homogeneous and
isotropic FLRW solution of Einstein's equations''. The immediately
obvious problem with this statement is that the details of the
averaging operation are not at all clear, and indeed are usually never 
specified. A bigger problem is one noted by Ellis \cite{ellis}, and 
can be stated in the following symbolic way. If $g$ denotes the
metric, $\Gamma$ the Christoffel connection and $E[g]$ the Einstein
tensor for the metric $g$, then we have the relations
\be
\G \sim \partial g ~~;~~ E[g] \sim \p\G + \G^2\,, 
\label{1eq1}
\ee 
with $\p$ denoting spacetime derivatives. The Einstein equations
are therefore
\be
E[g] = T\,,
\label{1eq2}
\ee
with $T$ denoting the energy-momentum tensor of the matter
components. Now, irrespective of any details of the averaging
operation, one notes that 
\be
E[\avg{g}] - \avg{E[g]}\sim \avg{\G}^2 - \avg{\G^2} \neq0\,,
\label{1eq3}
\ee
with the angular brackets denoting the averaging. The FLRW solution
would amount to solving the equations $E[\avg{g}]=\avg{T}$. In general 
therefore, it is \emph{not} true that averaging out the fluctuating
inhomogeneities leaves behind the FLRW solution, since what we are
actually left with is
\be
E[\avg{g}] = \avg{T} - \Cal{C} ~~;~~ \Cal{C} \sim \avg{\G^2} -
\avg{\G}^2 \,,
\label{1eq4}
\ee
and the homogeneous solution that we are looking for will depend on
the details of the correction terms \Cal{C}.

The second cause for concern comes from observations. It has now been
established beyond a reasonable doubt, that the FLRW metric confronted
with observations indicates an accelerating scale factor
\cite{accln-evid}. Conventional sources of energy such as radiation
and nonrelativistic matter cannot explain the acceleration, and it is
now common to attribute this effect to a hitherto unknown ``dark
energy'', which in its simplest form is a cosmological constant. The
true nature of this additional component in the cosmological
equations, is perhaps the most challenging puzzle facing both
theorists and observers today. A huge amount of research has gone into
(a) explaining the value that a cosmological constant term must take
to explain data or (b) assuming a zero cosmological constant,
constructing models of a dynamical dark energy which explains the
observed acceleration \cite{sami}. It is fair to say however that
there is no theoretical consensus on what the origin of dark energy
is. Since we have seen above that the effects of averaging lead to
some extra, as yet unknown terms in the equations, it is natural to
ask whether these two issues are connected. Could the acceleration of
the universe be explained by the effects of averaging inhomogeneities 
(``backreaction'') in the universe? Regardless of the answer to this
question, what is the nature and magnitude of this backreaction? The
purpose of this thesis is to answer these questions as rigorously as
possible. 

\section{History of the averaging problem}
The problem of averaging in general relativity has a history going back
even further than Ellis' work of 1984. In the context of gravitational
radiation, the problem of second order effects of gravity waves on the
large scale background metric of spacetime was studied by Isaacson in
the 1960's \cite{isaacson} in the ``short-wavelength''
approximation. Isaacson used an averaging operation which he called
the ``BH assumption'' after Brill and Hartle \cite{brill-hartle},
which was suited to studying the effects of perturbative gravity waves
in a spacetime region encompassing many wavelengths. An attempt to
generalize Isaacson's results was made by Noonan \cite{noonan}, who
introduced a different averaging procedure which was also constructed
for situations where inhomogeneities were perturbative in
nature. Interest in the cosmological consequences of such an averaging
picked up only after Ellis very clearly laid down the problems and
possibilities that open up when the idea of averaging in general
relativity is taken seriously. An example is the work of Futamase
\cite{futamase}, who introduced a spatial averaging procedure after
performing a $3+1$ splitting of spacetime, and computed backreaction
terms arising from averaging second order perturbations, finding them
to be negligibly small (see also \Cite{kasai-avg}). Another
example is the work of Boersma \cite{boersma}, who attempted to
construct a gauge-invariant (i.e. coordinate independent) averaging
procedure in perturbation theory, and also estimated that backreaction
effects remain negligibly small at the present epoch. (For other work
on the averaging problem, see \Cite{avg}.)

It may seem intuitively obvious that perturbatively small
inhomogeneities can only lead to negligibly small backreaction
effects. Indeed, this has been the conventional wisdom on this
subject, and has recently been spelt out by Ishibashi and Wald
\cite{wald}.  One starts by assuming that inhomogeneities in the
universe can be described in the Newtonian approximation of GR, by the
gravitational potential $\Phi(t,\vec{x})$ with $|\Phi|\ll1$, which
satisfies the Poisson equation $\nab^2\Phi=4\pi G a^2\delta\rho$ where
$\delta\rho(t,\vec{x})$ is the fluctuation of matter density about the
mean homogeneous value $\bar\rho(t)$, and can in general have a large
value. (E.g., in clusters of galaxies one finds
$\delta\equiv\delta\rho/\bar\rho\sim10^2$, and the ratio increases on 
smaller length scales). One then argues that the universe we observe 
\emph{does} seem to be very well-described by the above model, and
effects of averaging this model can only arise at second order in
$\Phi$ and should hence be extremely small. 

There is a loophole in this argument though. The catch is that the
background expansion $a(t)$ is defined completely ignoring
the backreaction, which is an integrated effect with contributions
from a large range of length scales. This means that the following
possibility cannot be \emph{a priori} ruled out : Initial conditions
are specified as a perturbation around a specific FLRW solution, but
the integrated effect of the backreaction grows (with time) in such a
manner as to effectively yield a late time solution which is a
perturbation around a \emph{different} FLRW model.  Indeed, there are
calculations in the literature that do indicate that this may
happen. For example, Martineau and Brandenberger \cite{martineau}
showed in a toy model that long wavelength fluctuations can give rise
to a backreaction contribution which has a late-time effective
equation of state similar to a cosmological constant. Their
calculations were based on the averaging procedure developed by Abramo
et al. \cite{abramo} in the context of backreaction in inflationary
cosmology. Other claims to solving the dark energy problem using
backreaction from long wavelength fluctuations were made by Barausse
et al. \cite{barausse} and Kolb et al. \cite{kolb-2005}. It is fair to
say however, that such claims have been controversial.  A number of
authors have argued that when effects of long wavelength fluctuations
are suitably ``renormalized'' and the background suitably redefined,
the backreaction cannot lead to acceleration of the scale factor
\cite{nosuperhorznbackrxn}. Nevertheless, what is definitely true   
is that the idea of backreaction of cosmological fluctuations has
generated a lively debate in the community
\cite{wald,misc-backrxn,notari,karel}.

In this thesis we will not deal with the effects of long wavelength
fluctuations, although we will see that certain assumptions need to be
made in order to define a self-consistent perturbation theory in the
presence of an averaging operation. A separate and equally interesting
question, which will be the main focus of this work, is whether
cosmological perturbation theory is \emph{stable} in the presence of
the backreaction contribution. There are results in the literature
which indicate that this might not be the case, and that the
backreaction can grow with time in such a manner that at late times
(when matter fluctuations have become nonlinear) perturbation theory
\emph{in the metric} also no longer holds \cite{notari} (see also
\Cite{schwarz-1}). In the same vein, there are arguments using
\emph{nonperturbative} toy models of gravitational collapse and
nonlinear structure formation, which suggest that perturbation theory
may not give correct insight into gravitational dynamics at late times
in cosmology \cite{rasanen}. If these results are relevant for the
real world, then it not only means that the conventional wisdom is
badly failing, but in fact implies that all of late-time cosmology
must be reworked from scratch (see, e.g. \Cite{wiltshire}; also see
however \Cite{kwan}). On the other hand, if these results are for some
reason or other not realistic, then it is important ask what is wrong
with such arguments, and further what the correct approach to the
problem is.

Clearly, to make any headway in this problem, it is first essential to
have a reliable averaging scheme at hand. Since the questions one is
asking deal with the stability of cosmological perturbation theory,
this averaging scheme needs to be inherently \emph{nonperturbative},
i.e. the validity of perturbation theory should not be a prerequisite
to defining the averaging prescription. This thesis will deal with 
two averaging schemes present in the literature : the spatial
averaging of scalars defined by Buchert
\cite{buchert1a,buchert1b,buchert2}, and the fully covariant tensor
averaging defined by Zalaletdinov \cite{zala1,mars,zala2}.  Some very 
interesting early work on possible nonperturbative effects of
averaging was by Buchert and Ehlers \cite{buchert-ehlers}, followed up
by \Cite{kerscher}, in the context of spatial averaging in Newtonian
cosmology. Buchert's averaging operation in general relativity has
since been used by several authors to explore the effects of
backreaction in various situations
\cite{wiltshire,buchertavg-misc,behrend,buch-lar-alimi,marra-phd},
including the perturbative contexts mentioned
above \cite{notari,schwarz-1}, and has also been compared against
observations \cite{buchert-obs}. As we shall see later, this averaging
scheme has an appealing simplicity of  implementation, which could be
a reason for the amount of attention it has received. In contrast,
Zalaletdinov's averaging scheme (which was developed earlier than
Buchert's work) is technically rather challenging to handle and
involves a fair amount of complicated algebra. Its strength however
lies in the fact that it is a fully covariant prescription which, at
the end of the day, yields an object which can be legitimately called
the ``averaged metric'' on an ``averaged manifold''. This ultimately
allows us to make physically  clear statements regarding the
backreaction, which is difficult to do in Buchert's scheme as it
stands. While Zalaletdinov's scheme has not received the same amount
of attention as Buchert's, there \emph{has} been a series of very
interesting results derived in this framework by Zalaletdinov and
coworkers \cite{zala-results}.   

\subsection{The ``Special Observer'' assumption}
The idea of using inhomogeneities to explain the dark energy problem
has generated a flurry of research in the backreaction problem in
recent years, as we saw above (see also \Cite{backrxn-DE}). It is
important to also mention another approach which has gained popularity
in this context, namely that of ascribing the dark energy phenomenon
to light propagation effects in an inhomogeneous universe
\cite{inhom-lightprop-early}. The central idea here is that light
propagation through an inhomogeneous underdensity or ``void'' can be
significantly different from that in a homogeneous space. In fact, it
is possible to show that luminosity distance data from supernovae can
always be fit by modelling ourselves as observers in a void with a
suitable density profile. Typically however, the (usually spherical)
voids invoked for this purpose are very large (in the range of
$\sim200h^{-1}$Mpc to $\sim1h^{-1}$Gpc in diameter), and are difficult
to reconcile with the typical sizes of voids seen in galaxy surveys,
which are in the range of $30$-$50h^{-1}$Mpc, with some ``supervoids''
reaching $\sim100h^{-1}$Mpc \cite{voids-obs}. Nevertheless, this idea
has been rather vigorously investigated in the last several
years. Since this thesis will not directly deal with this approach to
the dark energy problem, we will simply point the reader to a list of
references \cite{voids-DE} dealing with the study of light propagation
in an inhomogeneous universe and of supernovae data and the CMB
from the point of view of void-based observers. Unlike the
backreaction issue which requires mainly theoretical work, a detailed
description of the inhomogeneous universe belongs squarely in the
regime of observational cosmology \cite{obs-cosmo}. Due to the obvious
observational difficulties involved in such a program (for example due
to the lack of homogeneous samples of galaxy data), this approach at
present is largely restricted to being an exercise in building toy
models of the local large scale structure \cite{voids-DE}. As a final
comment on this topic, we note that this ``non-Copernican'' approach
(even at the level of building toy models) is amenable to
observational verification or disproof in the coming generation of
surveys, as pointed out by \Cite{copernican-test}.  

\vskip 0.25in
\noindent
With this introduction, our main results (the thesis of the thesis!)
are :
\begin{itemize}
\item {\it Although technically possible, in the
  real world backreaction does not significantly affect the expansion
  history of the universe.}
\item {\it Cosmological perturbation theory is stable
  against backreaction effects, well into the nonlinear regime.}
\item {\it Dark energy cannot therefore be an effect of the
  backreaction of inhomogeneities. } 
\end{itemize}
The outline of this thesis is as follows : \vskip 0.1in

Chapters 2 and 3 will deal respectively with Buchert's and
Zalaletdinov's averaging schemes. We will first describe Buchert's
scheme in chapter 2 and show using a toy 
model of a spherically symmetric inhomogeneous spacetime (the
Lema\^itre-Tolman-Bondi or LTB solution \cite{LTB}), how an
averaged effective description of the spacetime can have an
accelerating scale factor even when the underlying exact solution has
no exotic elements. This calculation will be based on Paranjape and
Singh, CQG (2006) (henceforth \pap{1}). We will then turn to
Zalaletdinov's formalism in chapter 3 and give a pedagogical
introduction to his 4-dimensional covariant averaging scheme and the
derivation of the equations in his effective theory of Macroscopic
Gravity (MG). We will then specialize this formalism for use in
cosmology and emphasize the need for a \emph{spatial averaging limit}
of this averaging. By explicitly writing out the backreaction terms in
a $3+1$-splitting of spacetime, we will also be in a position to give
a detailed comparison between Zalaletdinov's formalism and Buchert's
spatial averaging, and to demonstrate the physical relevance of the
backreaction terms. This part will be based on Paranjape and Singh,
PRD (2007) (henceforth \pap{2}). 

In chapter 4 we will use the spatial averaging limit of
Zalaletdinov's scheme in the setting of cosmological perturbation
theory. We will show how the leading order contribution to the
backreaction can be calculated in a gauge invariant manner, and derive
expressions for the backreaction which are valid whenever the
\emph{metric} of spacetime can be written as a perturbation around the
FLRW form. We will also discuss the issue of self-consistency of the
backreaction calculation, and propose an iterative scheme to calculate
the backreaction. As concrete examples we will perform the first
iteration in such a process, for some well-studied models of
perturbative inhomogeneities in the \emph{linear} regime. This set of
calculations will be based on Paranjape, PRD (2008) (henceforth
\pap{3}). 

Chapter 5 will deal with the regime when \emph{matter}
fluctuations have become large, so that linear perturbation theory no
longer holds. We will use the LTB solution once more to describe a
semi-realistic situation of spherical collapse, which we will follow
well into the nonlinear regime. Initially working in the Buchert
formalism, we will emphasize the importance of correctly accounting
for boundary conditions when building models, by showing how spurious
results can be obtained for the averaged expansion, by ignoring
boundary conditions. We will also show how the late time
nonperturbative behaviour in our LTB model can be recast in the
perturbed FLRW form, by a straightforward coordinate
transformation, \emph{provided} matter velocities remain
non-relativistic. This will demonstrate in a clear and unambiguous
manner, how a perturbation theory in the density contrast $\delta$ at
early times becomes a perturbation theory in peculiar velocity $v$ at
late times. These results will be based on Paranjape and Singh, JCAP
(2008) (henceforth \pap{4}). Finally, to complete the picture, we will
apply the formalism developed in chapter 3, to the toy model described
above, and explicitly show that the backreaction in the nonlinear
regime of structure formation does remain small in this model. This
calculation will be based on Paranjape and Singh, PRL (2008)
(henceforth \pap{5}).

We will conclude in chapter 6 with a summary and a brief comparison
with other work in the literature. Additionally, in the Appendices we
have collected some results which will be
used in the main text. Appendix A describes the homogeneous and
isotropic FLRW cosmology and serves to fix notation. Appendix B
describes the LTB metric which is a solution of the Einstein equations
sourced by a spherically symmetric pressureless fluid.  Appendix C
contains proofs of some results quoted in chapter 2. Throughout this
work we shall set the speed of light $c$ to unity, and use the metric
signature $(-+++)$. Lowercase Latin indices $a,b,... i,j,...$ will
take values $0,1,2,3$, while uppercase indices $A,B,... I,J,...$ will
take spatial values $1,2,3$. The Hubble constant $H_0$ is parametrized
when necessary in usual astronomers' units as $H_0=100h\,{\rm km
s^{-1}Mpc^{-1}}$.

\label{intro}

\chapter{Averaging schemes : \B's spatial averaging}
This chapter and the next describe the averaging schemes we will use
for the backreaction calculations. In this chapter we deal with \B's
spatial averaging scheme and show in a toy example how a
large, acceleration-inducing backreaction can arise even in the absence of
any exotic dark energy. 

\section{\B's spatial averaging of scalars}
\label{buch-spatavg}
The most straightforward and intuitively clear application of
\B's spatial averaging is in the case when the matter source is a 
pressureless ``dust'' with an energy-momentum tensor $T^{ab}=\rho
u^au^b$, with $u^a$ the dust 4-velocity which satisfies
$u_au^a=-1$. Assuming further that the dust is irrotational, the
4-velocity will be orthogonal to 3-dimensional spatial sections and
the metric can be written in ``synchronous and comoving'' coordinates
(in which $u^a=(1,\vec{0})$) 
\cite{landau-tof} as, 
\begin{equation}
ds^2=\,-dt^2+h_{AB}(t,\vec{x})dx^Adx^B\,.
\label{avgbuch1}
\end{equation}
The expansion tensor $\Theta^A_B$ is given by $\Theta^A_B\equiv
(1/2)h^{AC}\dot h_{CB}$ where the dot refers to a derivative with
respect to  time $t$. The traceless symmetric shear tensor is defined
as  $\sigma^A_B\equiv \Theta^A_B-(\Theta/3) \delta^A_B$ where $\Theta
=\Theta^A_A$  is the expansion scalar. The Einstein equations can be
split into a set of scalar equations and a set of vector
and traceless tensor equations. The scalar equations are the
Hamiltonian constraint \eqref{avgbuch2a} and the evolution equation for
$\Theta$ \eqref{avgbuch2b}, 
\begin{subequations}
\begin{align}
{}^{(3)}\mathcal{R}+\frac{2}{3}\Theta^2-2\sigma^2&=16\pi G\rho\,,
\label{avgbuch2a}\\
{}^{(3)}\mathcal{R}+\dot\Theta+\Theta^2&=12\pi G\rho\,,
\label{avgbuch2b}
\end{align}
\label{avgbuch2}
\end{subequations}
where $^{(3)}\mathcal{R}$ is the Ricci scalar of the 3-dimensional
hypersurface of  constant $t$ and $\sigma^2$ is the rate of shear
defined by $\sigma^2\equiv(1/2)\sigma^A_B\sigma^B_A$. \eqns{avgbuch2a}
and \eqref{avgbuch2b} can be combined to give Raychaudhuri's equation   
\begin{equation}
\dot\Theta+\frac{1}{3}\Theta^2+2\sigma^2+4\pi G\rho=0\,.
\label{avgbuch3}
\end{equation}
The continuity equation $\dot\rho=-\Theta\rho$ which gives the
evolution of $\rho$, is consistent with Eqns. \eqref{avgbuch2a}, 
\eqref{avgbuch2b}. We only consider the scalar equations, since the
spatial average of a scalar quantity can be defined in a gauge
covariant manner within a given  foliation of spacetime. For the
spacetime described by \eqref{avgbuch1}, the spatial average of a
scalar $\Psi(t,\vec{x})$ over a {\em comoving}  domain \calD\ at time
$t$ is defined by 
\begin{equation}
\avgD{\Psi}=\frac{1}{V_\calD}\int_\calD{d^3x\sqrt{h}\,\Psi}\,,
\label{avgbuch4}
\end{equation}
where $h$ is the determinant of the 3-metric $h_{AB}$ and $V_\calD$ is 
the volume of the comoving domain given by
$V_\calD=\int_\calD{d^3x\sqrt{h}}$. The following commutation relation
then holds \cite{buchert1a} 
\begin{equation}
\avgD{\Psi}^\cdot-\avgD{\dot\Psi}=
\avgD{\Psi\Theta}-\avgD{\Psi}\avgD{\Theta}\,,
\label{avgbuch5}
\end{equation}
which yields for the expansion scalar $\Theta$
\begin{equation}
\avgD{\Theta}^\cdot-\avgD{\dot\Theta}=
\avgD{\Theta^2}-\avgD{\Theta}^2\,.
\label{avgbuch6}
\end{equation}
Introducing the dimensionless scale factor
$\aD\equiv\left(V_\calD/V_{\calD in}\right)^{1/3}$ normalized by the
volume of the domain \calD\ at some initial time $t_{in}$, we can
average the scalar Einstein equations \eqref{avgbuch2a},
\eqref{avgbuch2b} and the continuity  equation to obtain
\cite{buchert1a} 
\begin{subequations}
\begin{align}
\left(\frac{\aDdot}{\aD}\right)^2&=\frac{8\pi G}{3}\avgD{\rho}
-\frac{1}{6}\left(\Cal{Q}_\calD + \avgD{\Cal{R}}\right)\,,
\label{avgbuch7a}\\
\left(\frac{\aDddot}{\aD}\right)&= -\frac{4\pi G}{3} \avgD{\rho} + 
\frac{1}{3} \Cal{Q}_\calD\,, 
\label{avgbuch7b}\\
\avgD{\rho}^\cdot&=\,-\avgD{\Theta}\avgD{\rho}=\,-
3\frac{\aDdot}{\aD}\avgD{\rho}\,.   
\label{avgbuch7c}
\end{align}
\label{avgbuch7}
\end{subequations}
Here $\avgD{\Cal{R}}$, the average of the spatial Ricci scalar 
$^{(3)}\Cal{R}$, is a domain dependent spatial constant. The
`backreaction' $\Cal{Q}_\calD$ is given by  
\begin{equation}
\Cal{Q}_\calD\equiv\frac{2}{3}\left(\avgD{\Theta^2}-
\avgD{\Theta}^2\right)-2\avgD{\sigma^2}\,,
\label{avgbuch8}
\end{equation}
and is also a spatial constant. The last equation \eqref{avgbuch7c} simply
reflects the fact that the mass contained in a comoving domain is
constant by construction : since $\Theta=\p_t\ln\sqrt{h}$, the local
continuity equation $\dot\rho=-\Theta\rho$ can be solved to give
$\rho\sqrt{h}=\rho_0\sqrt{h_0}$ where the subscript $0$ refers to some
arbitrary reference time $t_0$. The mass $M_\calD$ contained in a
comoving domain \calD\ is then $M_\calD=\int_\calD{\rho\sqrt{h}d^3x}
=\int_\calD{\rho_0\sqrt{h_0}d^3x}=\,$constant. Hence 
\begin{equation}
\avgD{\rho}=\frac{M_\calD}{V_{\calD in} \aD^3}
\label{avgbuch9}
\end{equation}
which is precisely what is implied by \eqn{avgbuch7c}.
Equations \eqref{avgbuch7a}, \eqref{avgbuch7b} can be thought of as ``modified
Friedmann equations'' (compare \eqns{flrw3}), with the modifications
arising due to the presence of the backreaction $\Cal{Q}_\calD$ and
the fact that the averaged Ricci curvature in general need not evolve
like $\sim\aD^{-2}$ as in the FLRW case. 

A necessary condition for \eqref{avgbuch7b} to integrate to
\eqref{avgbuch7a} takes the form of the following differential equation
involving $\Cal{Q}_\calD$ and $\avgD{\Cal{R}}$
\begin{equation}
\Cal{\dot Q}_\calD+6\frac{\aDdot}{\aD}\Cal{Q}_\calD
+\avgD{\Cal{R}}^{\cdot}+2\frac{\aDdot}{\aD}\avgD{\Cal{R}}=0\,,
\label{avgbuch12}
\end{equation}
which is a very interesting equation because it shows that the
evolution of the backreaction is intimately tied to that of the
average spatial curvature. Scaling solutions for this equation have
been explored by \B, Larena and Alimi \cite{buch-lar-alimi}, a
simple example being $\avgD{\Cal{R}}\propto\aD^{-2}$,
$\Cal{Q}_\calD\propto\aD^{-6}$. Clearly the FLRW solution with
$\Cal{Q}_\calD=0$ is a special case. In this thesis we will mainly be
concerned with the behaviour of the backreaction arising from
explicitly averaging an inhomogeneous spacetime, and will therefore
not discuss these scaling solutions which make no reference to the
underlying inhomogeneous geometry. We note that the criterion to be
met in order for the effective scale factor $\aD$ to accelerate, is
\begin{equation}
\Cal{Q}_\calD>4\pi G\avgD{\rho}\,.
\label{avgbuch13}
\end{equation}
Equations \eqref{avgbuch7} and \eqref{avgbuch12} describe the essence
of \B's averaging formalism, for the simplest case of
irrotational dust. We note that the remaining eight Einstein equations
for the inhomogeneous geometry, which are not scalar equations, are
not averaged. These are the five evolution equations for the
trace-free part of the shear,  
\begin{equation}
\partial_t\left(\sigma{}^A_B\right) = - \Theta
\sigma{}^A_B -  \Cal{R}{}^A_B +\frac{2}{3}\delta{}^A_B \left(\sigma^2
- \frac{1}{3}\Theta^2 + 8\pi G\rho \right) \,.
\label{avgbuch14}
\end{equation}
and the three equations relating the spatial variation of the shear
and the expansion,
\begin{equation}
\sigma{}^A_{B || A} = \frac{2}{3} \Theta_{|| B}\,.
\label{avgbuch15}
\end{equation}
Here, $\Cal{R}{}^A_B$ is the spatial Ricci tensor and a $||$ denotes
covariant derivative with respect to the $3$-metric. Later, when we
apply \Z's formalism to cosmology, we will see that
accounting for the full set of Einstein's equations when performing
the averaging, leads to additional constraints on the inhomogeneous
geometry, which are not present in \B's formalism.

For completeness, and to enable a fuller comparison with \Z's
formalism, we also display the results of applying
\B's averaging to the case of a perfect fluid with energy-momentum
tensor $T^{ab}=(\rho+p)u^au^b + pg^{ab}$ \cite{buchert1b}. In
this case comoving coordinates are not in general synchronous, and for
an irrotational perfect fluid in comoving coordinates, the metric
takes the form
\begin{equation}
ds^2 = -N^2 dt^2 + h_{AB}dx^Adx^B \,.
\label{avgbuch16}
\end{equation}
The averaged scalar Einstein equations for the scale factor \aD\ are 
\begin{align}
3\frac{\partial_t^2 \aD}{\aD} + 
4\pi G \avgD{N^2\left(\rho+3p\right)}&= \uD{\bar{\Cal{Q}}} +
\uD{\bar{\Cal{P}}} \,,
\label{avgbuch17}\\
6 \uD{H}^2 - 16\pi G \avgD{N^2\rho}&=
-\uD{\bar{\Cal{Q}}} - \avgD{N^2\Cal{R}}~~;~~
\uD{H}=\frac{\partial_t\aD}{\aD}\,,   
\label{avgbuch18}
\end{align}
where the kinematical backreaction $\uD{\bar{\Cal{Q}}}$ is given by
\begin{equation}
\uD{\bar{\Cal{Q}}} = \frac{2}{3}\left( \avgD{\left(N\Theta\right)^2} -
\avgD{N\Theta}^2\right) 
 - 2\avgD{N^2\sigma^2}\,,
\label{avgbuch19}
\end{equation}
and the dynamical backreaction $\uD{\bar{\Cal{P}}}$ is given by 
\begin{equation}
\uD{\bar{\Cal{P}}} = \avgD{N^2\Cal{A}} + \avgD{\Theta\partial_tN}
\,,
\label{avgbuch20}
\end{equation}
where $\Cal{A}=\nabla_j(u^i\nabla_iu^j)$ is the $4$-divergence of the
$4$-acceleration of the fluid. \eqn{avgbuch18} follows as an integral
from \eqn{avgbuch17} if and only if the relation 
\begin{align}
\partial_t \uD{\Cal{Q}} &+ 6\uD{H}\uD{\Cal{Q}} + 
\partial_t \avgD{N^2\Cal{R}} + 2\uD{H}\avgD{N^2\Cal{R}} +
4\uD{H}\uD{\bar{\Cal{P}}} \nonumber\\
&- 16\pi G \left[\partial_t \avgD{N^2\rho} +
3\uD{H}\avgD{N^2\left(\rho+p\right)} \right]  =0\,, 
\label{avgbuch21}
\end{align}
is satisfied. There are also the unaveraged equations (which we do not  
display here) for the shear, analogous to the shear equations
\eqref{avgbuch14} and \eqref{avgbuch15} for dust.  

\section{Acceleration from averaging}
In this section we use the spherically symmetric dust solution of
Einstein's equations (the Lema\^itre-Tolman-Bondi or LTB solution
described in Appendix B) to construct an explicit example where \B's
effective scale factor accelerates even though the underlying model
has no cosmological constant or dark energy. Our model contains a
single underdense and ``curvature-dominated'' region (in a sense to be 
described below), whose evolution we study at late times. 

\subsection{Late time and curvature dominated unbound models}
Consider the LTB metric \eqref{LTB1}
\begin{equation}
ds^2=\,-dt^2+\frac{R^{\prime 2}(t,r)}{1-k(r)r^2}dr^2+
R^2(t,r)\left(d\theta^2 + \sin^2\theta d\phi^2\right)\,,
\label{avgLTB1}
\end{equation}
which satisfies
\begin{equation}
\dot R^2(t,r)=\frac{2GM(r)}{R(t,r)}-k(r)r^2\,,
\label{avgLTB2}
\end{equation}
for the specific case $k(r)<0$ in the region of interest, so that the
solution is \eqref{LTB4a}
\begin{equation}
R=\frac{GM(r)}{-k(r)r^2}\left(\cosh\eta -1\right)~~~
;~~~t-t_0(r)=\frac{GM(r)}{\left(-k(r)r^2\right)^{3/2}}\left(\sinh\eta 
- \eta\right)~,~0\leq\eta<\infty~,~~~\text{for}~k(r)<0\,.
\label{avgLTB4}
\end{equation}
Although it is straightforward to numerically evaluate $R(t,r)$ for
any given choice of the free functions, we would like to try and
analytically simplify these expressions as far as possible.
Since $R(t,r)$ in the unbound case is an increasing function of time
for arbitrary $k(r)<0$, \eqn{avgLTB2} shows that the late time
solution (after neglecting the $1/R$ term in the equation) can be
expressed as 
\begin{equation}
R\simeq ~r\sqrt{-k(r)}\left(t-t_0(r)\right)\,,
\label{avgLTB5}
\end{equation}
[This leading order solution can also be derived from an asymptotic
  expansion of the solution \eqref{avgLTB4} for large $\eta$, see 
  below.] The function $t_0(r)$ can be obtained using the closed form
expression for $t(R)$ obtained by integrating \eqn{avgLTB4} \cite{tps}
and the scaling $R_{in}(r)=r$, as
\begin{equation}
t_0(r)=t_{in}-r\left(\frac{r}{g}\right)^{1/2}
F\left(\frac{r}{g}\right)
~~~;~~~F(x)\equiv\frac{1}{x}(1+x)^{1/2}-\frac{1}{x^{3/2}}\sinh^{-1}
\left(x^{1/2}\right)\,,
\label{avgLTB6}
\end{equation}
where we have defined
\be
g(r) \equiv \frac{2GM(r)}{-k(r)r^2}\,.
\la{avgLTB6a}
\ee
If we further assume that the matter
contribution encoded in $M(r)$ is negligible compared to that of the
spatial curvature as encoded in $k(r)$, i.e. if
$|g(r)/r|\ll1$, then the expression for $t_0(r)$ also
simplifies at the leading order to\footnote{We are assuming that
  $g(r)/r$ remains finite at all $r$. In particular as
  $r\to0$ this implies $k(r\to0)\sim r^\mu$; $\mu\leq0$ (see
  \eqn{reg2}).} 
\be
t_0(r)\simeq ~t_{in}-\frac{1}{\sqrt{-k(r)}}\,.
\la{avgLTB7}
\ee
The leading order unbound LTB solution in this late time, negligible
matter limit is then given by
\begin{equation}
R(t,r)=r\left[\lambda_r + \sqrt{-k(r)}\left(t-\lambda_tt_{in}\right) 
  \right] \,,
\label{avgLTB8}
\end{equation}
where we have introduced two placeholders $\lambda_r$ and $\lambda_t$
(ultimately set to unity) which will remind us that we are working
with a late time solution with large $t$. Note that \eqn{avgLTB8} (with
$\lambda_r=\lambda_t=1$), is the exact solution \eqn{avgLTB2} in the
special case $M(r)\to0$ for all $r$. This actually corresponds to 
Minkowski spacetime, with the corresponding Riemann tensor being
exactly zero. The constant time $3$-spaces are hypersurfaces of 
negative curvature, with the $3$-curvature being determined by the
function $k(r)$. The `FLRW' limit of this solution is in fact the
Milne universe; the solution \eqref{avgLTB8} could hence be thought of
as the `LTB' type generalization of the Milne universe. Although we
will use this form of the solution to draw conclusions regarding
acceleration of \B's \aD, we will later argue that these conclusions
are not altered by the presence of a nonzero but small amount of
matter. Although it may appear at this stage that requiring $M\to0$
renders the late time approximation redundant, we will see below that
additionally imposing the late time approximation allows us to write
down a fairly straightforward sufficient condition for acceleration
of \aD, which would not be possible with only the $M\to0$ condition.

For the metric \eqref{avgLTB1}, the volume of a spherical comoving
domain of radius $r_\calD$ is
\begin{equation}
V_\calD=4\pi\int_0^{r_\calD}{\frac{R^\prime
R^2}{\sqrt{1-k(r)r^2}}~dr}\,.
\label{avgLTB9}
\end{equation}
Substituting the solution \eqref{avgLTB8} in this expression, we find  
\begin{equation}
V_\calD = (t-\lambda_tt_{in})^3\calI_k
+ \lambda_r(t-\lambda_tt_{in})^2\calI_{kr}  
+\lambda_r^2(t-\lambda_tt_{in})\calI_{kr^2}+\lambda_r^3\calI_{r^2}\,,
\label{avgLTB10}
\end{equation}
where we have defined the domain dependent integrals 
\begin{align}
\calI_k=2\pi\int_0^{r_\calD}{ \frac{\sqrt{-kr^2}(-kr^2)^\prime}{\sqrt{1-kr^2}} 
~dr} 
~~~~&;~~~~
\calI_{kr}=4\pi\int_0^{r_\calD}{\frac{\left(-kr^3\right)^\prime}
{\sqrt{1-kr^2}}~dr}\,,\nonumber\\
\calI_{kr^2}=4\pi\int_0^{r_\calD}{ \frac{\left(r^3\cdot\sqrt{-k}\right)^\prime} 
{\sqrt{1-kr^2}}~dr}~~~~&;~~~~
\calI_{r^2}=4\pi\int_0^{r_\calD}{\frac{r^2}{\sqrt{1-kr^2}}~dr}\,.
\label{avgLTB11} 
\end{align}
The sum of the exponents of $\lambda_r$ and $\lambda_t$ in each 
term in \eqref{avgLTB10} indicates the relative order of that term
with respect to the leading $t^3$ term. This approach of treating some  
terms as small compared to others is valid since the various integrals 
which multiply the powers of $t$, are all finite and non-zero. 
Expanding $V_\calD$ in powers of $\lambda_t,\lambda_r$, we find for the
effective scale factor,
\begin{equation} 
3\frac{\aDddot}{\aD}=\frac{\ddot
V_\calD}{V_\calD}-\frac{2}{3}\left(\frac{\dot
V_\calD}{V_\calD}\right)^2
=\frac{2\lambda_r^2}{\calI_kt^4}\left(\calI_{kr^2}-\frac{1}{3\calI_k}
\left(\calI_{kr}\right)^2\right) +\Cal{O}\left(3\right) \,,
\label{avgLTB12} 
\end{equation} 
where $\Cal{O}\left(3\right)$ represents terms involving
$\lambda_r^m\lambda_t^n$ (i.e. containing $(1/t^{m+n})$) with
$m+n\geq3$.   

We see that the generic late time (i.e. $t\to\infty$) behaviour   
of the unbound models under consideration is $\aDddot\to 0$, and that
deviations from zero are small, being a second order effect. Whether
the approach to $\aDddot=0$ is via an accelerating or decelerating
phase, depends upon the relative magnitudes of the domain integrals 
involved. A sufficient condition for an unbound model with negligible
matter to accelerate at late times, is 
\begin{equation}
\calP\equiv\calI_{kr^2}-\frac{1}{3\calI_k}\left(\calI_{kr}
\right)^2>0\,.  
\label{avgLTB13}
\end{equation}
\begin{figure}[t]
\subfigure[$\calP/\calI_k$]{\includegraphics[width=.45\textwidth]
{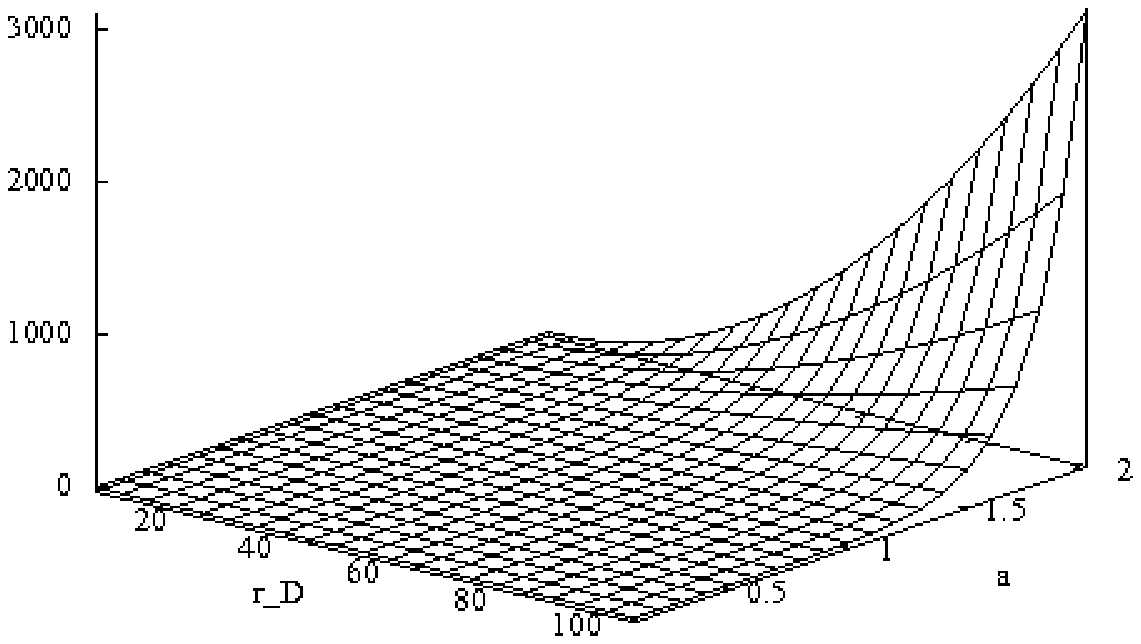}  \label{fig1a}}  
\hspace{.05\textwidth} 
\subfigure[$\calP/\calI_k$]{\includegraphics[width=.45\textwidth]
{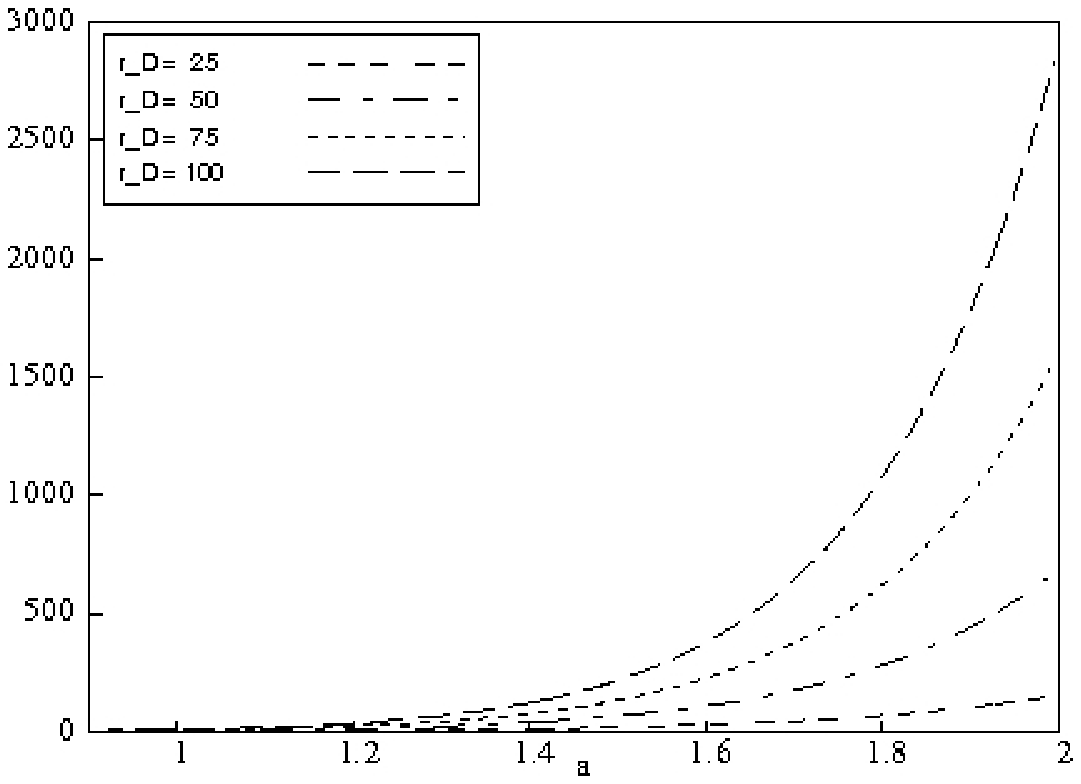}  \label{fig1b}}  
\caption{\small{The models described by $k(r)=-1/(1+r^a)$. (a) The
scaled function $\calP/\calI_k$. (b) $\calP/\calI_k$ plotted against
$a$ for specific values of $r_\calD$.}} 
\label{fig1}
\end{figure}
As an explicit example, consider a model with $k(r)$ given by
\begin{equation}
k(r)=-\frac{1}{1+r^a}~~~;~~~0<a<2\,,
\label{avgLTB14}
\end{equation}
in arbitrary units. The condition $0<a<2$ ensures that the regularity 
conditions of Appendix B are satisfied.
The function $\calP/\calI_k$ for these models, which controls the
magnitude of the late time acceleration (see \eqn{avgLTB12}) is shown
in \fig{fig1}, against $r_\calD$ and $a$. For clarity, in the
second panel we have shown $\calP/\calI_k$ against $a$ for specific
values of $r_\calD$. We find that $\calP/\calI_k$ is positive
everywhere in the region shown. To explicitly demonstrate
acceleration, we plot the evolution of the dimensionless quantity
$q_\calD$ defined by    
\begin{equation}
q_\calD\equiv-\frac{\aDddot\aD}{\aDdot^2}=
2-3\frac{\ddot V_\calD V_\calD}{(\dot V_\calD)^2}\,,
\label{avgLTB15}
\end{equation}
for various fixed values of $a$ and $r_\calD$, using the full
expression for $V_\calD$ in \eqref{avgLTB10}. The results are shown
in \fig{fig2}. We have used units in which $t_{in}=1$, and have
displayed the evolution for times $t>100\,t_{in}$. 

Even though the acceleration condition \eqref{avgLTB13} is strictly
derived for the case $M(r)=0$, one can easily see that it remains
valid \emph{at late enough times} even in the presence of a small
amount of matter. Introducing another place holder
$\epsilon$ to keep track of the smallness of the function $g(r)$
defined in \eqn{avgLTB6a}, one sees that by waiting long enough, the
solution for $R(t,r)$ will be approximately given by \eqn{avgLTB5},
with $t_0(r)$ given by \eqn{avgLTB6}. The expression for the volume
$V_\calD$ will have the same form as \eqn{avgLTB10}, but the integrals
involved will be different due to the presence of terms involving
$\epsilon\neq0$. A condition similar to \eqref{avgLTB13}, say
$\Cal{P}(\epsilon)>0$, will then be obtained for late time
acceleration, with different integrals involved in the definition of
the functional $\Cal{P}(\epsilon)$. The point to note is that
$\Cal{P}(\epsilon)=\Cal{P}+\,$ terms containing $\epsilon$
with \Cal{P}\ defined as in \eqn{avgLTB13}, and \emph{for
small enough} $\epsilon$, $\Cal{P}(\epsilon)$ will be positive
whenever $\Cal{P}$ is positive\footnote{More precisely, we will have
$\Cal{P}(\epsilon) = \Cal{P}+\Cal{O}(\eplog)$.}. Hence the
acceleration condition is robust against adding a small but nonzero
amount of matter. 
\begin{figure}
\subfigure[Domain range
$r_\calD=250$]{\includegraphics[width=.45\textwidth]{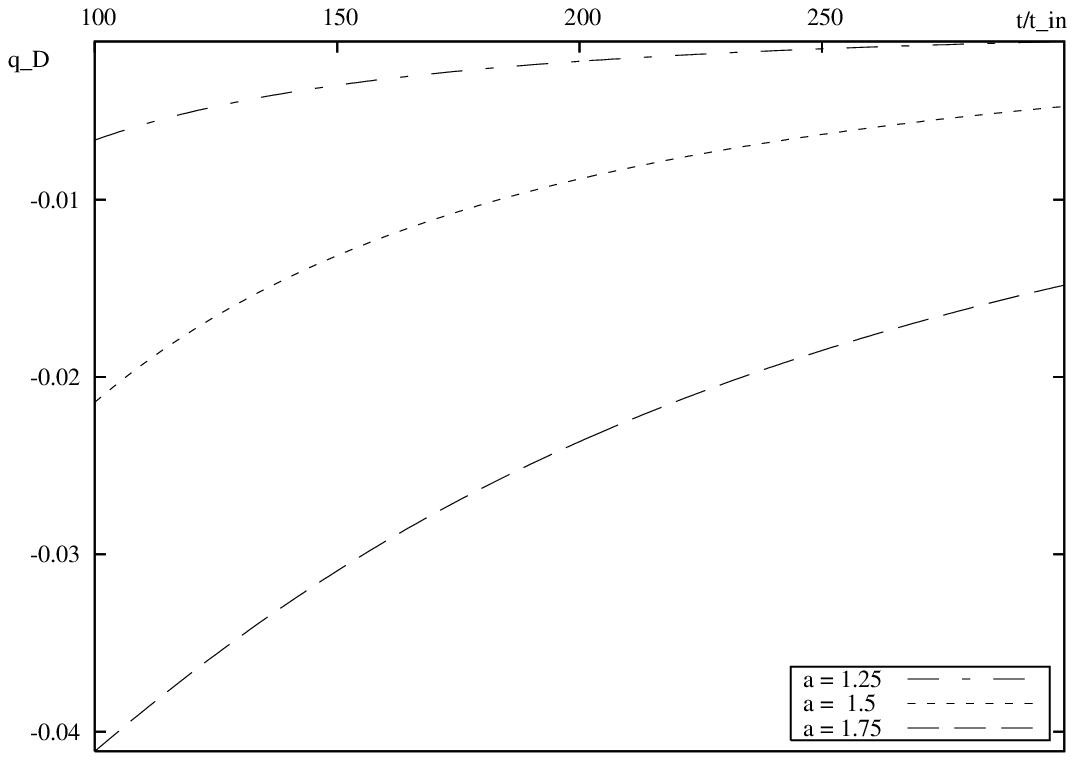}
\label{fig2a}}
\hspace{.05\textwidth}  
\subfigure[Exponent
$a=1$]{\includegraphics[width=.45\textwidth]{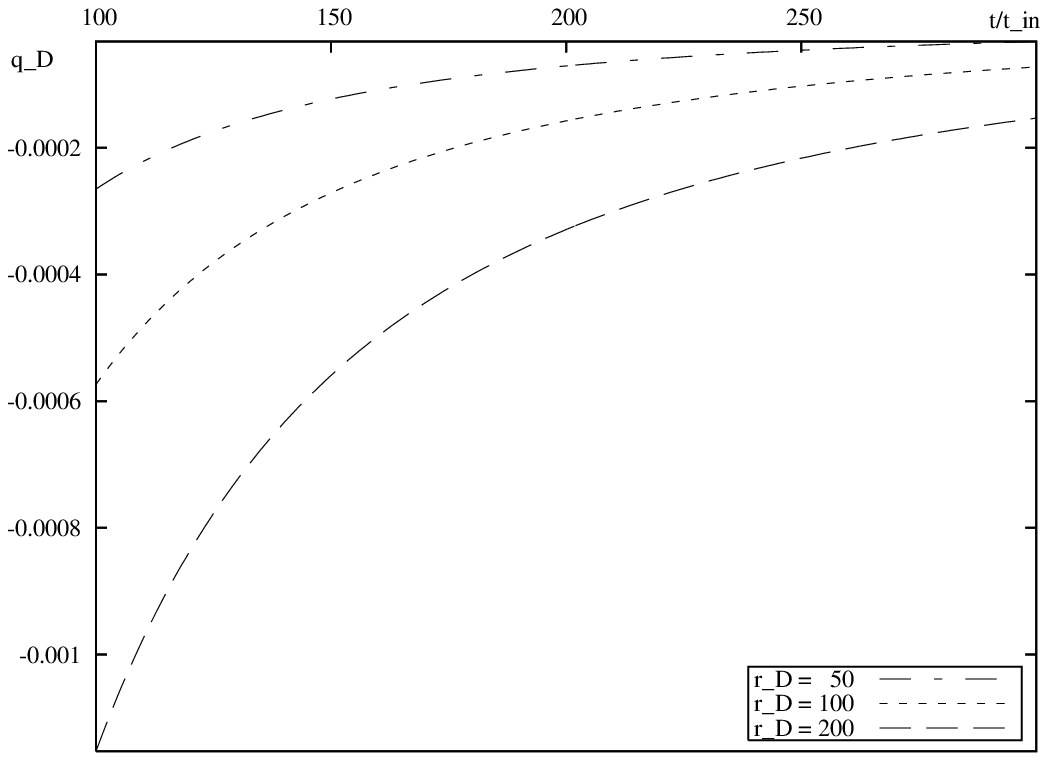}  
\label{fig2b}}  
\caption{\small{Evolution of $q_\calD$ in the models with
$k(r)=-1/(1+r^a)$, plotted against $t/t_{in}$ for (a) three values of
$a$ with $r_\calD=250$, and (b) three values of $r_\calD$ with
$a=1$.}}   
\label{fig2}
\end{figure}

\vskip 0.25in
\noindent
{\large\bf Chapter summary and discussion:}\vskip 0.1in
\noindent
This chapter dealt with details of Buchert's scheme for spatially
averaging scalar quantities, and the effective cosmological equations
it leads to. Using exactly solvable toy models of inhomogeneities,
which were explicitly averaged using this scheme, we saw how an
accelerating effective scale factor can arise even in situations where
there is no exotic matter component.

However, the fact that Buchert's scheme deals with only two of the ten
Einstein equations, makes it difficult to relate the effective scale
factor $\aD$ with observations. In particular, it is not clear whether
or not $\aD$ should replace the usual scale factor in the FLRW
metric. We see this difficulty of interpretation as arising from the
inherent non-covariant structure of Buchert's averaging scheme. To get
around this problem we will study a different averaging scheme in the
next chapter, namely \Z's fully covariant Macroscopic Gravity. This
scheme will allow us to deal with objects which are structurally
similar to Buchert's scale factor and backreaction, while being easier
to interpret in a physically clear manner.

\chapter{Averaging schemes : \Z's covariant Macroscopic Gravity}
\label{zala-avg}
In this chapter we turn to the averaging defined by \Z\ \cite{zala1},
which is a $4$-dimensional generally covariant procedure. This
averaging is used on the Einstein equations and, together with some
additional assumptions, leads to what \Z\ has called Macroscopic
Gravity (henceforth MG). After introducing MG, we will describe its
spatial averaging limit as discussed in \pap{2}.

\section{A covariant averaging scheme}
The starting point in any covariant averaging scheme has to be the
question : ``How does one average tensors while retaining their
transformation properties under coordinate changes?'' If the averaging
operation is to involve an integral over a spacetime region, then
clearly only scalar objects can be averaged, since they change only
trivially under coordinate transformations. To define a scheme for
general tensors then, it is essential to introduce some additional
structure in the formalism. The most convenient option is to introduce
a \emph{bivector} \Wxx{a}{b}\ which transforms as a vector at event
$x^\prime$ and as a co-vector at event $x$. In \Z's scheme one
postulates the existence of such a bivector, requires it to have
certain desirable properties, and then explicitly \emph{constructs} an
object which has all these properties. We will see how this is done in
what follows. [Throughout this chapter primed indices
(e.g. $P^{a^\prime}$) will refer to ``primed events'' ($x^\prime$) and 
unprimed indices to unprimed events.]

To begin with, we require that this bivector be idempotent
(i.e. ``square to itself'')
\be
\Cal{W}^{a^\prime}_{c^{\prime\prime}}(x^\prime,x^{\prime\prime}) 
\Cal{W}^{c^{\prime\prime}}_j(x^{\prime\prime},x) =\Wxx{a}{j}\,,
\la{avgZala1}
\ee
and have the coincidence limit 
\be
\lim_{x^\prime\to x}\Wxx{a}{j} =
\delta{}^a_j\,,
\la{avgZala2}
\ee
This ensures that \Wxx{a}{j}\ has the inverse operator
\Wxxinv{a}{j} (which is easily seen by taking the $x\to
x^\prime$ limit in \eqn{avgZala1} and using the condition
\eqref{avgZala2}). The bivector is then used to define the ``bilocal 
extension'' of a general tensorial object (denoted by an overtilde) :
for a vector $P{}^a(x)$ this takes the form 
\begin{equation}
\bil{P}{}^a(x^\prime,x) =
\Wxxinv{a}{a}P{}^{a^\prime}(x^\prime) \,,
\label{avgZala3}
\end{equation}
with the obvious generalisation to higher rank objects.
Notice that the bilocal extension as defined above transforms like the
original tensor at the event $x$, but as a \emph{scalar} at  the event
$x^\prime$. This allows us to define the ``average'' of 
$P{}^a(x)$ over a 4-dimensional spacetime region $\bm{\Sigma}$ with
a supporting point $x$, as
\begin{equation}
\bar P{}^a(x) = \stavg{\bil{P}{}^a} = \frac{1}{V_\Sigma}
\int_\Sigma{d^4x^\prime\sqrt{-g^\prime}\bil{P}{}^a(x^\prime,x)}
 ~~;~~ V_\Sigma =
\int_\Sigma{d^4x^\prime\sqrt{-g^\prime}} \,,
\label{avgZala4}
\end{equation}
the subscript $ST$ standing for `spacetime'. While this averaged
tensor has the correct transformation properties at the event $x$, in
order to be a local \emph{function} of its argument, one needs to
ensure that $\bar P{}^a(x)$ has appropriate differential
properties. Since the $x$ dependence of this object arises not only
from the explicit appearance of \Wxx{a}{j}\ but also through the
dependence of the domain $\bm{\Sigma}$ on the support point, in order
to correctly calculate the derivative of $\bar P{}^a(x)$ we need to
specify how neighbouring domains are related to each other.

This is done as follows : The same bivector \W{a}{j}\ is used to
specify a \emph{Lie dragging} of the averaging region $\bm{\Sigma}$, 
ensuring that the volumes of the averaging regions constructed at
nearby supporting points are coordinated in a well defined
manner (which motivates the terminology ``coordination bivector'' for
\W{a}{j}, which we will follow henceforth). Suppose $x^a$ and
$x^a+\xi^a\Delta\lambda$ are the coordinates of two support points,
where $\Delta\lambda$ is a small change in the parameter along the
integral curve of a given vector field $\xi^a$. Symbolically denote
the two points as $x$ and $x+\xi\Delta\lambda$. Then the averaging
region at $x+\xi\Delta\lambda$ is defined in terms of the averaging
region $\bm{\Sigma}(x)$ at $x$, by transporting every point
$x^\prime\in\bm{\Sigma}(x)$ around $x$ along the appropriate integral 
curve of a \emph{new} bilocal vector field $S^{a^\prime}$ defined as 
$S^{a^\prime}(x^\prime,x) = \Wxx{a}{j}\xi^j(x)$, thereby constructing
the averaging region $\bm{\Sigma}(x,\Delta\lambda)$ with support point 
$x+\xi\Delta\lambda$.

We can now evaluate the Lie derivative of $\bar P{}^a(x)$ along the
vector field $\xi^a$, by first noting that the Lie derivative of the
volume $V_{\Sigma(x)}$ is
\be
\frac{d}{d\lambda}V_{\Sigma(x)} = \xi^a(x)
\stavg{\Cal{W}{}^{j^\prime}_{a;j^\prime}} V_{\Sigma(x)}\,,
\la{avgZala5}
\ee
where the semicolon denotes a covariant derivative. An easy way to see
this is to note that since 
\be
\bm{\Sigma}(x,\Delta\lambda) =
\left\{y^{a^\prime}\left|\right. y^{a^\prime}=x^{a^\prime}+S^{a^\prime}(x^\prime, 
x)\Delta\lambda~;~ x^\prime\in\bm{\Sigma}(x) \right\} \,,
\la{avgZala6}
\ee
or symbolically $y^\prime=x^\prime + S\Delta\lambda$, we will have
\be
\sqrt{-g(y^\prime)} = \sqrt{-g(x^\prime)}\left( 1 + \Delta\lambda
S^{a^\prime} \p_{a^\prime}\ln\sqrt{-g(x^\prime)} \right) ~~;~~
d^4y^\prime = d^4x^\prime
\left(1+\Delta\lambda\p_{a^\prime}S^{a^\prime} \right)\,, 
\la{avgZala7}
\ee
and the result \eqref{avgZala5} follows from writing
\be
\frac{d}{d\lambda}V_{\Sigma(x)} = \lim_{\Delta\lambda\to0}
\frac{1}{\Delta\lambda} \left(
 V_{\Sigma(x,\Delta\lambda)} - V_{\Sigma(x)}
\right)\,, 
\la{avgZala8}
\ee
and using $S^{a^\prime} = \W{a}{b}\xi^b$. For the derivative of $\bar
P{}^a(x)$, we need $\bar P{}^a(y=x+\xi\Delta\lambda)$. Explicitly we
have 
\be
\bar P{}^a(x+\xi\Delta\lambda) = \frac{1}{V_{\Sigma(x,\Delta\lambda)}} 
  \int_{\Sigma(x,\Delta\lambda)}{d^4y^\prime\sqrt{-g(y^\prime)}\,
    \bil{P}^a(y^\prime, x+\xi\Delta\lambda)}\,,
\la{avgZala9}
\ee
which after writing $y^\prime=x^\prime+S\Delta\lambda$ with
$x^\prime\in\bm{\Sigma}(x)$ and using \eqref{avgZala7}, finally gives
\begin{align}
\frac{d}{d\lambda}\bar P{}^a &= \lim_{\Delta\lambda\to0}
\frac{1}{\Delta\lambda} \left( 
\bar P{}^a(x+\xi\Delta\lambda) - \bar P{}^a(x)
\right) \nonumber\\
&\nonumber\\
&= \xi^b(x) \left[ \stavg{\pbil_b\bil{P}^a} 
+ \stavg{\Cal{W}{}^{j^\prime}_{b;j^\prime}\bil{P}^a} -
\stavg{\Cal{W}{}^{j^\prime}_{b;j^\prime}}\bar P{}^a  \right] \,,
\la{avgZala10}
\end{align}
where we have defined the ``bilocal partial derivative''
\be
\pbil_b \equiv \p_b + \W{j}{b}\p_{j^\prime}\,.
\la{avgZala11}
\ee
Since the vector field $\xi^a$ is arbitrary, \eqn{avgZala10} gives us
an expression for the partial derivative $\p_b\bar P{}^a$ (recalling
that $d\bar P{}^a/d\lambda = \xi^b\p_b\bar P{}^a -
\bar P{}^b\p_b\xi^a$). Requiring that partial derivatives of $\bar 
P{}^a(x)$ commute, leads to the following condition
\be
\p_{[b}\p_{c]}\bar P{}^a = \stavg{\pbil_{[b}\pbil_{c]}\bil{P}^a} +
\stavg{\bil{P}^a\pbil_{[b}\Cal{W}{}^{k^\prime}_{c];k^\prime} } -
\stavg{\pbil_{[b}\Cal{W}{}^{k^\prime}_{c];k^\prime} }\bar P^a = 0\,.
\la{avgZala12}
\ee
Straightforward algebra shows that
\begin{align}
\pbil_{[b}\pbil_{c]}\bil{P}^a &=
(\pbil_{[b}\Cal{W}{}^{k^\prime}_{c]})\,\p_{k^\prime}\bil{P}^a \,, 
\la{avgZala13} \\
\pbil_{[b}\Cal{W}{}^{k^\prime}_{c];k^\prime} &=
(\pbil_{[b}\Cal{W}{}^{k^\prime}_{c]})_{;k^\prime} \,,
\la{avgZala14}
\end{align}
and hence the necessary and sufficient condition for \eqref{avgZala12}
to hold, is
\be
\pbil_{[b}\Cal{W}{}^{k^\prime}_{c]}  = \W{k}{[b,\,c]} +
\W{k}{[b,\,\underline{j}^\prime}\W{j}{c]} = 0 \,,
\la{avgZala15}
\ee
where underlined indices are not antisymmetrized.

At this stage it is convenient to re-express results in the language
of differential forms, since it will make the algebra that follows
concise and readable. The original papers by \Z\ introduce a
full-fledged bilocal exterior calculus, involving $(k,l^\prime)$-forms
which are $k$-forms at $x$ and $l$-forms at $x^\prime$. In what
follows however, we will almost exclusively only need to deal with
$(k,0^\prime)$-forms that are differential forms at a \emph{single}
event $x$, although they may be bilocal in their \emph{functional}
dependence on events $x$ and $x^\prime$. This is because we will only
deal with \emph{bilocal extensions} of local $k$-forms, which are
defined to be $(k,0^\prime)$-forms. For example, the bilocal extension
of a $2$-form $\bm{\alpha}$ is defined as
\begin{equation}
\bil{\bm{\alpha}}(x^\prime,x) = \frac{1}{2!}\alpha_{j^\prime
  k^\prime}(x^\prime)\Wxx{j}{a}\Wxx{k}{b} \dx{a}\wedge\dx{b}\,,
\label{avgZala16}
\end{equation}
which is a $(2,0^\prime)$-form. We define the bilocal exterior
derivative as  
\begin{equation}
\dbil = \ext + \ext^\prime_{\Cal{W}}\,,
\label{avgZala17}
\end{equation}
where \ext\ is the usual exterior derivative which ``differentiates
and antisymmetrizes at $x$'', and the ``shifted''
exterior derivative $\ext^\prime_{\Cal{W}}$ ``differentiates at
$x^\prime$ but antisymmetrizes at $x$'' so that, say
for the $(2,0^\prime)$-form $\bil{\bm{\alpha}}$, we have
\begin{equation}
\ext^\prime_{\Cal{W}}\bil{\bm{\alpha}} = \frac{1}{2!}
\bil{\alpha}_{ab,j^\prime}(x^\prime,x)
\Wxx{j}{c} \dx{c} \wedge \dx{a} \wedge \dx{b} \,.
\label{avgZala18}
\end{equation}
For a bilocal function $f(x,x^\prime)$ we will have $\dbil f =
\pbil_af\dx{a}$. In this language, the preceding results for
differential conditions on the coordination bivector can be
generalised to arbitrary tensor-valued $k$-forms. In particular for a
vector-valued $k$-form $\rmb{p}^a(x)$ whose bilocal extension is the 
$(k,0^\prime)$-form $\bil{\rmb{p}}^a(x^\prime,x)$, we have
\begin{equation}
\ext \bar{\rmb{p}}{}^a = \stavg{\dbil\bil{\rmb{p}}{}^a} +
\stavg{{\rm div}_\epsilon \rmb{W}\wedge\bil{\rmb{p}}{}^a} -
\stavg{{\rm div}_\epsilon \rmb{W}}\wedge\bar{\rmb{p}}{}^a\,, 
\label{avgZala19}
\end{equation}
where we have defined,
\begin{align}
{\rm div}_\epsilon \rmb{W} &=
 \Cal{W}{}^{j^\prime}_{a;j^\prime}\dx{a}\nonumber\\
 &= \left(\Cal{W}{}^{j^\prime}_{a,j^\prime} +
 \W{j}{a}\partial_{j^\prime}\ln\sqrt{-g^\prime} \right)\dx{a} \,. 
\label{avgZala20}
\end{align}
The condition \eqref{avgZala15} for the averaged object to be a local
function of its argument, reduces to
\begin{equation}
\dbil\rmb{W}^{j^\prime} = 0 ~~;~~ \rmb{W}^{j^\prime} =
\W{j}{a}\dx{a}\,, 
\label{avgZala21}
\end{equation}
\eqn{avgZala19} shows that it is desirable to choose a coordination
bivector which satisfies
\begin{equation}
{\rm div}_\epsilon \rmb{W}=0\,,
\label{avgZala22}
\end{equation}
since firstly, this allows us to commute the exterior derivative with
the averaging according to
\begin{equation}
\ext \bar{\rmb{p}}{}^a = \stavg{\dbil\bil{\rmb{p}}{}^a}\,,
\label{avgZala23}
\end{equation}
and secondly, it implies that the volume of the averaging region is
held constant during the coordination (see \eqn{avgZala5}), and is
therefore a free parameter in the formalism.

Mars and \Z\ \cite{mars} show in their Theorems 1, 3 and 4, that 
firstly, the general solution of \eqn{avgZala21} for an idempotent 
coordination bivector is given by 
\begin{equation}
\Wxx{a}{j} = f^{a^\prime}_m(x^\prime) f^{-1\,m}_{\ph{-1}j}(x) \,, 
\label{avgZala24}
\end{equation}
\noindent
where $f^a_m(x)\bm{\partial}_a = \rmb{f}_m$ is any vector basis
satisfying the commutation relations
\begin{equation}
\left[\rmb{f}_i,\rmb{f}_j\right] = C{}^k_{ij}\rmb{f}_k ~~;~~
C{}^k_{ij} = {\rm constant} \,,
\label{avgZala25}
\end{equation}
\noindent
and secondly, that \eqn{avgZala22} with the coordination bivector
given by \eqn{avgZala24} is always integrable on a differentiable
manifold with a given volume $n$-form. The proofs given in \Cite{mars}
are very clear, although somewhat lengthy, and we will hence omit them
here. Further, these authors also show that for the special class of
bivectors for which $C{}^k_{ij} = 0$, the vectors $\{\rmb{f}_k\}$
form a coordinate basis, with `proper' coordinate functions
$\phi^m(x)$ say, so that 
\begin{equation}
f^{a}_m(x(\phi^n)) = \frac{\partial x^a}{\partial\phi^m} ~~;~~
f^{-1\,m}_{\ph{-1}j}(\phi(x^k)) = \frac{\partial \phi^m}{\partial x^j}
\,,
\label{avgZala26}
\end{equation} 
\noindent
and satisfying \eqn{avgZala22} makes this proper coordinate system
volume preserving, with $g(\phi^m)=\,$constant. When expressed in terms
of such a volume preserving coordinate (VPC) system, the coordination
bivector takes its most simple form, namely 
\begin{equation}
\Wxx{a}{j}\mid_{\rm proper} = \delta{}^{a^\prime}_j\,.
\label{avgZala27}
\end{equation}
Volume preserving coordinates in fact form a large class in
themselves, generalizing the Cartesian coordinate system of Minkowski
spacetime. For a discussion on the properties of VPCs and the
associated bivectors \W{j}{a}, see Sec. 8 of \Cite{mars}. The problem
of defining an averaging operator has now been reduced to the far
simpler problem of choosing a specific VPC system which then fixes the
coordination bivector. We emphasize that the averaging is still fully
covariant; choosing a coordination bivector is distinct from choosing
a coordinate system to perform calculations in. This freedom in
defining the coordination bivector leads to a lack of uniqueness of
the average in the formalism as it stands. We will return to this
issue when we study the backreaction arising from perturbative
inhomogeneities in chapter 4.

\section{The averaged manifold}
With the averaging operation in place, we can turn to the description
of an \emph{averaged geometry}. Let us begin by recalling some
standard results from differential geometry. Given a differentiable 
manifold \Cal{M}\ endowed with a metric $g_{ab}$ of Lorentzian
signature $(-+++)$, the connection $1$-forms $\bom^a_{\ph{i}b}$ are
defined by the action of the exterior derivative on the basis vectors
$\rmb{e}_a$ \cite{mtw} : 
\begin{equation}
\ext \rmb{e}_a = \rmb{e}_a\bom^a_{\ph{i}b} =
\rmb{e}_a\left(\Gamma{}^a_{bc}\right)\dx{c} \Rightarrow \left(
\dx{a} , \ext\rmb{e}_b \right) =  \bom^a_{\ph{i}b}\,, 
\label{avgZala28}
\end{equation}
where $\Gamma{}^a_{bc}$ are the Christoffel symbols and the
parentheses represent the inner product with $(\dx{a},\rmb{e}_b
)=\delta{}^a_b$. We define the ``exterior covariant derivative''
$\rmb{D}_\omega$ associated with the connection $\bom^a_{\ph{a}b}$, as 
follows : for a $k$-form $\rmb{p}{}^{a}_{b}$,
\begin{equation}
\rmb{D}_\omega\rmb{p}{}^{a}_{b} =
\ext\rmb{p}{}^{a}_{b} -
\bom^k_{\ph{k}b}\wedge\rmb{p}{}^{a}_{k} +
\bom^a_{\ph{a}k}\wedge\rmb{p}{}^{k}_{b} \,,
\label{avgZala29}
\end{equation}
with an obvious generalisation to higher rank objects. The
compatibility between the metric and the connection on \Cal{M}\ is
expressed by the condition 
\begin{equation}
\rmb{D}_\omega g_{ab} = \ext g_{ab} - g_{ak}\bom^k_{\ph{k}b} -
g_{bk}\bom^k_{\ph{k}a} = 0 \,,
\label{avgZala30}
\end{equation}
with a similar condition for the inverse metric
$g^{ab}$. The Cartan structure equations are given by
\begin{subequations}
\begin{align}
\bom^a_{\ph{a}b}\wedge\dx{b}&=0 \,, \label{avgZala31a}\\
\ext\bom^a_{\ph{a}b} + \bom^a_{\ph{a}c}\wedge\bom^c_{\ph{c}b} &= 
\rmb{r}^a_{\ph{a}b} \,, \label{avgZala31b}
\end{align}
\label{avgZala31}
\end{subequations}
where \eqn{avgZala31a} expresses the symmetry of the connection
$\bom^a_{\ph{a}b}$ and $\rmb{r}^a_{\ph{a}b}$ is the curvature $2$-form
on \Cal{M}\ which defines the Riemann curvature tensor via
$\rmb{r}^a_{\ph{a}b} = (1/2!)r^a_{\ph{a}bcd}\dx{c}\wedge\dx{d}$.
Finally, the structure equations \eqref{avgZala31} and the metric
compatibility condition \eqref{avgZala30} are supplemented by their
respective integrability conditions, given by equations
\eqref{avgZala32a}, \eqref{avgZala32b} and \eqref{avgZala32c}
\begin{subequations}
\begin{align}
\rmb{r}^a_{\ph{a}b}\wedge\dx{b}&=0 \,, \label{avgZala32a}\\
\ext\rmb{r}^a_{\ph{a}b} - \bom^c_{\ph{c}b}\wedge\rmb{r}^a_{\ph{a}c} 
+ \bom^a_{\ph{a}c}\wedge\rmb{r}^c_{\ph{c}b} &=
0\,, \label{avgZala32b}\\ 
g_{ak}\rmb{r}^k_{\ph{k}b} + g_{bk}\rmb{r}^k_{\ph{k}a} &= 0 \,.
\label{avgZala32c}
\end{align}
\label{avgZala32}
\end{subequations}
Note that \eqn{avgZala32a} corresponds to the cyclic identity
$r^a_{\ph{a}[bcd]}=0$ and \eqn{avgZala32b} to the Bianchi identity 
$r^a_{\ph{a}b[ij;k]}=0$.

Consider now the bilocal version of the basis vector $\rmb{e}_a$,
given by $\rmb{W}_a = \W{j}{a}\rmb{e}_{j^\prime}$, which is a scalar
at $x$ and a vector at $x^\prime$, and whose inverse is the
$(0,1^\prime)$-form $\rmb{W}^{-1b} = \Winv{b}{j}\dx{j^\prime}$. The
construction of an \emph{averaged manifold} begins with the
observation that the $(1,0^\prime)$-form defined by
\be
\Om{a}{b}(x,x^\prime) \equiv \left( \rmb{W}^{-1a}, \dbil \rmb{W}_b
\right) \,,
\la{avgZala33}
\ee
transforms like a connection under coordinate transformations at $x$
and, equally importantly, as a \emph{scalar} at $x^\prime$, and
further has the coincidence limit
\be
\lim_{x^\prime\to x} \Om{a}{b}(x,x^\prime) = \bom^a_{\ph{a}b}(x) \,.
\la{avgZala34}
\ee
Explicitly, we have 
\begin{align}
\Om{a}{b}(x,x^\prime) &= \left[ \Wxxinv{a}{i}\Wxx{j}{b}\Wxx{k}{c}
\Gamma{}^{i^\prime}_{j^\prime k^\prime}(x^\prime) + \Wxxinv{a}{j}
\pbil_b \Wxx{j}{c} \right]\dx{c} \nonumber\\
&\equiv \bil{\Gamma}{}^a_{bc}(x,x^\prime)\dx{c}\,,
\label{avgZala35}
\end{align}
where $\bil{\Gamma}{}^a_{bc}$ is symmetric in $(bc)$ due to the
condition \eqref{avgZala21} or \eqref{avgZala15}. This explicit form
can be used to directly check the transformation properties at $x$ and
$x^\prime$, and $\Om{a}{b}(x,x^\prime)$ can therefore be considered as
the bilocal extension of the connection $\bom^a_{\ph{a}b}(x)$. 

The transformation properties of $\Om{a}{b}$ imply that its
\emph{average} $\barOm{a}{b}(x)$,
\be
\barOm{a}{b} \equiv \avg{\Om{a}{b}}\,,
\label{MG-avgcond}
\ee
has the transformation properties of
a local connection, and the coincidence limit \eqref{avgZala34}
implies the limit $\lim_{V_\Sigma\to0}\barOm{a}{b}(x) =
\bom^a_{\ph{a}b}(x)$. This leads to the key idea of MG, which is that
$\barOm{a}{b}$, is defined as the connection $1$-form on a new,
averaged manifold \Mbar. As a set, \Mbar\ is identical to \Cal{M};
however for consistency in the definition of averaged quantities, the
guiding principle one adopts is that each \emph{averaging domain}
$\bm{\Sigma}$ is effectively treated as a single point on \Mbar. This
ensures, for example, that the averaging operation is idempotent
\cite{mars}. 

The goal now is to average out the bilocal extensions of the structure
equations \eqref{avgZala31} and the compatibility condition
\eqref{avgZala30} and their integrability conditions
\eqref{avgZala32}, and to express them in terms of appropriate
differential forms defined on \Mbar. The bilocal extensions of
Eqns. \eqref{avgZala31} and \eqref{avgZala30}, are respectively given
by  
\begin{subequations}
\begin{align}
\Om{a}{b}\wedge\dx{b}&=0 \,,\label{avgZala36a}\\
\dbil\Om{a}{b} + \Om{a}{c}\wedge\Om{c}{b} &= \bil{\rmb{r}}^a_{\ph{a}b} 
\,,\label{avgZala36b}\\
\Dbil{\Omega} \bil{g}_{ab} = \dbil \bil{g}_{ab} -
\bil{g}_{ak}\Om{k}{b} - \bil{g}_{bk}\Om{k}{a} &= 0 \,,
\label{avgZala36c}
\end{align}
\label{avgZala36}
\end{subequations}
\noindent
where, in the last equation, $\Dbil{\Omega}$ is the bilocal covariant
exterior derivative associated with the bilocal connection
$\Om{a}{b}$. The integrability conditions of \eqns{avgZala36} are given
by the bilocal extensions of \eqns{avgZala32},  
\begin{subequations}
\begin{align}
\bil{\rmb{r}}^a_{\ph{a}b}\wedge\dx{b}&=0 \,,
\label{avgZala37a}\\
\Dbil{\Omega}\bil{\rmb{r}}^a_{\ph{a}b}= \dbil\bil{\rmb{r}}^a_{\ph{a}b}
- \Om{c}{b}\wedge\bil{\rmb{r}}^a_{\ph{a}c} +
\Om{a}{c}\wedge\bil{\rmb{r}}^c_{\ph{c}b} &= 0 \,,  
\label{avgZala37b}\\
\bil{g}_{ak}\bil{\rmb{r}}^k_{\ph{k}b} +
\bil{g}_{bk}\bil{\rmb{r}}^k_{\ph{k}a} &= 0 \,.
\label{avgZala37c}
\end{align}
\label{avgZala37}
\end{subequations}
In the above equations, the $(2,0^\prime)$-form
$\bil{\rmb{r}}^a_{\ph{a}b}$  is the bilocal extension of the
curvature $2$-form constructed according to the rules set out in
\eqns{avgZala16} and \eqref{avgZala3}. Similarly the
$(0,0^\prime)$-form  $\bil{g}_{ab}$ is the bilocal extension of the
metric. To proceed with the averaging, a correlation $2$-form is
defined  
\begin{equation}
\bZ{a}{b}{i}{j} = \stavg{\Om{a}{b}\wedge\Om{i}{j}} -
\barOm{a}{b}\wedge\barOm{i}{j}\,.
\label{avgZala38}
\end{equation} 
\noindent
The average of the curvature $2$-form $\rmb{r}^a_{\ph{a}b}$ on
\Cal{M}\ is denoted
$\rmb{R}^a_{\ph{a}b}\equiv\stavg{\,\bil{\rmb{r}}^a_{\ph{a}b}}$ , and the
curvature $2$-form on the averaged manifold \Mbar\ is denoted
$\rmb{M}^a_{\ph{a}b}$, 
\begin{equation}
\rmb{M}^a_{\ph{a}b} = \ext\barOm{a}{b} +
\barOm{a}{k}\wedge\barOm{k}{b}\,.  
\label{avgZala39}
\end{equation}
\noindent
Equations \eqref{avgZala36a} and \eqref{avgZala36b} then average out
to give 
\begin{subequations}
\begin{align}
\barOm{a}{c}\wedge\dx{c} &= 0\,,
\label{avgZala40a}\\
\rmb{M}^a_{\ph{a}b} &= \rmb{R}^a_{\ph{a}b} - \bZ{a}{c}{c}{b}\,.
\label{avgZala40b}
\end{align}
\label{avgZala40}
\end{subequations}
The averages of \eqn{avgZala37a} and the identity
$\bil{\rmb{r}}^a_{\ph{a}a}=0$ give 
us 
\be
\rmb{R}^a_{\ph{a}b}\wedge\dx{b} = 0 ~~;~~ \rmb{R}^a_{\ph{a}a} = 0\,,
\la{avgZala41}
\ee
and the symmetry of $\Om{a}{b}$ and hence of $\barOm{a}{b}$ give us 
\be
\bZ{a}{c}{c}{b}\wedge\dx{b} = 0 ~~;~~ \bZ{a}{c}{c}{a} = 0\,.
\la{avgZala42}
\ee
This ensures that the curvature $2$-form $\rmb{M}^a_{\ph{a}b}$
satisfies the correct algebraic identities
\be
\rmb{M}^a_{\ph{a}b}\wedge\dx{b} = 0 ~~;~~ \rmb{M}^a_{\ph{a}a} = 0\,,
\la{avgZala43}
\ee
The formalism at this stage becomes somewhat complicated. The reason
is that there is no simple way of averaging out equations
\eqref{avgZala36c}, \eqref{avgZala37b} and \eqref{avgZala37c}, which
become 
\begin{subequations}
\begin{align}
\ext \bar{g}_{ab} - \stavg{\bil{g}_{ak}\Om{k}{b}} -
\stavg{\bil{g}_{bk}\Om{k}{a}} &= 0 \,, 
\label{avgZala44a} \\
\ext{\rmb{R}}^a_{\ph{a}b} -
\stavg{\Om{c}{b}\wedge\bil{\rmb{r}}^a_{\ph{a}c}} +
\stavg{\Om{a}{c}\wedge\bil{\rmb{r}}^c_{\ph{c}b}} &= 0 \,,  
\label{avgZala44b}\\
\stavg{\bil{g}_{ak}\bil{\rmb{r}}^k_{\ph{k}b}} +
\stavg{\bil{g}_{bk}\bil{\rmb{r}}^k_{\ph{k}a}} &= 0 \,,
\label{avgZala44c}
\end{align}
\la{avgZala44}
\end{subequations}
where $\bar g_{ab} = \stavg{\bil{g}_{ab}}$. What we need are
``splitting rules'' for the various products appearing inside the
averager in these equations. Of these, \eqn{avgZala44b} can be split
by noting that the exterior covariant derivative of the correlation
$2$-form becomes 
\be
\rmb{D}_{\bar\Omega}\bZ{a}{b}{i}{j}=
-2\mathbb{P}\,\rmb{Y}{}^{a\ph{i}m\ph{i}i}_{\ph{i}m\ph{i}b\ph{i}j} +
2\mathbb{P}\, \left( \stavg{\bil{\rmb{r}}^a_{\ph{a}b} \wedge
  \Om{i}{j}} - \rmb{R}^a_{\ph{a}b} \wedge \barOm{i}{j}\right) \,, 
\la{avgZala45}
\ee
where the symbol $\mathbb{P}$ permutes the free indices in, say
$\rmb{K}^{a\ph{i}i\ph{i}m}_{\ph{i}b\ph{i}j\ph{i}n}$ pairwise according 
to $\mathbb{P}(\rmb{K}^{a\ph{i}i\ph{i}m}_{\ph{i}b\ph{i}j\ph{i}n}) = 
(1/3!)(\rmb{K}^{a\ph{i}i\ph{i}m}_{\ph{i}b\ph{i}j\ph{i}n} -
\rmb{K}^{i\ph{i}a\ph{i}m}_{\ph{i}j\ph{i}b\ph{i}n} +
\rmb{K}^{i\ph{i}m\ph{i}a}_{\ph{i}j\ph{i}n\ph{i}b})$, and any summed
indices are ignored, and the correlation \emph{3-form} is defined as 
\be
\rmb{Y}{}^{a\ph{i}i\ph{i}m}_{\ph{i}b\ph{i}j\ph{i}n} = \stavg{\Om{a}{b}
\wedge \Om{i}{j} \wedge \Om{m}{n}} - 3\mathbb{P}(\bZ{a}{b}{i}{j}
\wedge \barOm{m}{n}) - \barOm{a}{b} \wedge \barOm{i}{j} \wedge
\barOm{m}{n} \,.
\la{avgZala46}
\ee
Tracing \eqn{avgZala45} on the indices $b$ and $i$ kills the term
involving the correlation $3$-form due to the presence of the
permutation symbol, leaving behind
\be
- \stavg{\bil{\rmb{r}}^a_{\ph{a}i} \wedge \Om{i}{b}} +
\stavg{\bil{\rmb{r}}^i_{\ph{a}b} \wedge \Om{a}{i}}  =
- \rmb{R}^a_{\ph{a}i} \wedge \barOm{i}{b} + \rmb{R}^i_{\ph{a}b} \wedge 
\barOm{a}{i} - \rmb{D}_{\bar\Omega}\bZ{a}{i}{i}{b}\,,
\la{avgZala47}
\ee
which averages out \eqn{avgZala44b} to give the Bianchi identity for
the curvature $2$-form $\rmb{M}^a_{\ph{a}b}$,
\be
\rmb{D}_{\bar\Omega}\rmb{M}^a_{\ph{a}b} = \ext\rmb{M}^a_{\ph{a}b} -
\barOm{k}{b}\wedge\rmb{M}^a_{\ph{a}k} +
\barOm{a}{k}\wedge\rmb{M}^k_{\ph{k}b} = 0\,.
\la{avgZala48}
\ee
This was achieved at the cost of introducing a new object, the
correlation $3$-form
$\rmb{Y}{}^{a\ph{i}i\ph{i}m}_{\ph{i}b\ph{i}j\ph{i}n}$, which fixes the
differential properties of the correlation $2$-form $\bZ{a}{b}{i}{j}$
and hence of the $2$-form $\rmb{R}^a_{\ph{a}b}$. The differential
properties of this $3$-form are in turn fixed by introducing a
correlation \emph{4-form} in an analogous manner. Due to the
$4$-dimensionality of spacetime there are no higher correlation
$p$-forms that need to be defined.

In practice, it is cumbersome to keep track of the correlation
$3$-form and $4$-form, and furthermore it is only the correlation
$2$-form which will appear in the averaged Einstein equations. We will
ultimately be interested in explicit calculations of the correlation
objects in a \emph{perturbative} setting at leading order, and will
hence ignore the $3$-form and $4$-form. Remarkably, it is also
possible to self-consistently ignore these forms in the
\emph{nonperturbative} setting \cite{zala1} as follows : We set the
$3$-form and $4$-form to zero and impose the conditions 
\begin{equation}
\rmb{D}_{\bar\Omega} \bZ{a}{b}{i}{j} = 0 = \rmb{D}_{\bar\Omega}
\rmb{R}^a_{\ph{a}b}\,, 
\label{avgZala49}
\end{equation}
with the second equality required since \eqn{avgZala48} holds. It can
be shown \cite{zala1} that requiring the $4$-form to vanish also
imposes the condition
\begin{equation}
\mathbb{P}\left(\bZ{a}{b}{c}{d} \wedge \bZ{d}{i}{j}{k} \right) = 0
\,.
\label{avgZala50}
\end{equation}
The integrability condition for $\rmb{D}_{\bar\Omega} \bZ{a}{b}{i}{j}
= 0$ is
\begin{equation}
\mathbb{P}\left(\rmb{R}^a_{\ph{a}c}\wedge\bZ{c}{b}{i}{j} -
\bZ{a}{b}{i}{k}\wedge\rmb{R}^k_{\ph{k}j} \right) = 0\,,
\label{avgZala51}
\end{equation}
which also contains the integrability condition for
$\rmb{D}_{\bar\Omega} \rmb{R}^a_{\ph{a}b} = 0$. 

So far we have only managed to average out \eqn{avgZala44b}. To make
progress with \eqns{avgZala44a} and \eqref{avgZala44c} we need
additional assumptions. \Z\ argues \cite{zala1} that for a class of 
slowly varying  tensor fields (tensor-valued $k$-forms)
$\rmb{c}{}^{m\cdots}_{n\cdots}$ on \Cal{M}\ such as the metric and 
other covariantly constant tensors, and Killing tensors, etc., the
following assumptions may be reasonable
\begin{subequations}
\begin{equation}
\stavg{\Om{a}{b}\wedge\bil{\rmb{c}}{\,}^{m\cdots}_{n\cdots}} =
\barOm{a}{b}\wedge\bar{\rmb{c}}{\,}^{m\cdots}_{n\cdots} \,,
\label{avgZala52a}
\end{equation}
\begin{equation}
\stavg{\Om{a}{b}\wedge\Om{i}{j}\wedge
  \bil{\rmb{c}}{\,}^{m\cdots}_{n\cdots}} =
  \stavg{\Om{a}{b}\wedge\Om{i}{j}}\wedge
  \bar{\rmb{c}}{\,}^{m\cdots}_{n\cdots} \,.
\label{avgZala52b}
\end{equation}
\label{avgZala52}
\end{subequations}
Then \eqn{avgZala44a} and its analogue for $g^{ab}$ average out to
give 
\begin{equation}
\rmb{D}_{\bar\Omega} \bar g_{ab} = 0 ~~;~~ \rmb{D}_{\bar\Omega} \bar
g^{ab} = 0 \,.
\label{avgZala53}
\end{equation}
Further, for a general slowly varying object
$\rmb{c}{}^{m\cdots}_{n\cdots}$, the following identity holds
\begin{align}
\stavg{\bil{\rmb{r}}{}^a_{\ph{a}b} \wedge
  \bil{\rmb{c}}{}^{m\cdots}_{n\cdots}} - \rmb{R}^a_{\ph{a}b} \wedge
\bar{\rmb{c}}{}^{m\cdots}_{n\cdots} &- \stavg{\Om{a}{b} \wedge
  \Dbil{\Omega}\bil{\rmb{c}}{}^{m\cdots}_{n\cdots}} + \barOm{a}{b}
  \wedge \rmb{D}_{\bar\Omega} \bar {\rmb{c}}{}^{m\cdots}_{n\cdots}
  \nonumber\\ 
 &=  -\bZ{a}{b}{m}{j} \wedge \bar{\rmb{c}}{}^{j\cdots}_{n\cdots}
  -\ldots +  \bZ{a}{b}{j}{n} \wedge  \rmb{c}{}^{m\cdots}_{j\cdots} 
  +\ldots~~\,, 
\label{avgZala54}
\end{align}
which follows from differentiating \eqn{avgZala52a}, and which
averages out \eqn{avgZala44c} (and its analogue for $g^{ab}$) to 
give 
\begin{equation}
\bar g_{ak}\rmb{M}^k_{\ph{k}b} + \bar g_{kb}\rmb{M}^k_{\ph{k}a} = 0
~~;~~ \rmb{M}^a_{\ph{a}k}\bar g^{kb} + \rmb{M}^a_{\ph{a}k}\bar g^{kb}
= 0\,.
\label{avgZala55}
\end{equation}
\eqn{avgZala53} allows one to choose $G_{ab} = \bar g_{ab}$, where
$G_{ab}$ is the metric on the averaged manifold \Mbar. In general
however, we have $G^{ab}\neq\bar g^{ab}$, and one defines the tensor
$U^{ab}\equiv \bar g^{ab}-G^{ab}$ to keep track of this
difference. However, we shall see
later that when the averaged manifold is highly symmetric, as in the
case of a manifold with homogeneous and isotropic spatial sections
which we will consider, one finds that $U^{ab}=0$. 

\section{Averaging Einstein's equations}
In the general
case, it turns out that \eqn{avgZala54} is all that is needed to
average out the Einstein equations 
\begin{equation}
g^{ak}r_{kb} - \frac{1}{2}\delta{}^a_b g^{ij}r_{ij} = -\kappa
t{}^{a({\rm mic})}_b\,,
\label{avgZala56}
\end{equation}
where $\kappa = 8\pi G_N$, $t{}^{a({\rm mic})}_b$ is the microscopic
energy momentum tensor of the matter distribution, and the Ricci
tensor $r_{ab}$ on \Cal{M}\ is defined according to the sign
convention $r_{ab} = r^j_{\ph{j}abj}$. The averaging leads to the
equations  
\begin{equation}
G^{ak}M_{kb} - \frac{1}{2}\delta{}^a_b G^{ij}M_{ij} = -\kappa
\stavg{\bil{t}{\,}^{a({\rm mic})}_b} + \left(Z^a_{\ph{a}ijb} -
\frac{1}{2}\delta{}^a_b Z^k_{\ph{k}ijk}\right)\bar g^{ij} -
\left(U^{ak}M_{kb} - \frac{1}{2}\delta{}^a_bU^{ij}M_{ij}\right)\,,  
\label{avgZala57}
\end{equation}
\noindent
where $M_{ab} = M^j_{\ph{j}abj}$ is the Ricci tensor on \Mbar\ and we
have defined
\begin{equation}
Z^a_{\ph{a}ijb} = 2 Z^{a\ph{ik}k}_{\ph{a}ik\ph{k}jb} ~~;~~
  \bZ{a}{b}{i}{j} = Z^{a\ph{bm}i}_{\ph{a}bm\ph{i}jn}\dx{m}\wedge\dx{n} 
  \,. 
\label{avgZala58}
\end{equation}
\noindent
The averaged equations \eqref{avgZala57} differ from the usual Einstein
equations by the correlation tensor which we define as
\begin{equation}
C{}^a_b = \left(Z^a_{\ph{a}ijb} - \frac{1}{2}\delta{}^a_b
Z^m_{\ph{a}ijm} \right)\bar g^{ij} - \left(U^{ak}M_{kb} -
\frac{1}{2}\delta{}^a_bU^{ij}M_{ij}\right) \,. 
\label{avgZala59}
\end{equation}
\noindent
Hence, denoting the Einstein tensor on \Mbar\ as $E{}^a_b$, and
defining the tensor $T{}^a_b$ via
\begin{equation}
T{}^a_b = \stavg{\bil{t}{\,}^{a({\rm mic})}_b}\,,
\label{avgZala60}
\end{equation}
\noindent
the averaged Einstein equations read
\begin{equation}
E{}^a_b = -\kappa T{}^a_b + C{}^a_b\,.
\label{avgZala61}
\end{equation}
\noindent
Since the left hand side of \eqn{avgZala61} is covariantly conserved
by construction ($E{}^a_{b;a} = 0$), where the semicolon denotes
covariant differentiation with respect to the connection on \Mbar, in
general one has
\begin{equation}
\left(-\kappa T{}^a_b + C{}^a_b\right)_{;a} = 0\,,
\label{avgZala62}
\end{equation}
\noindent
with no condition on $T{}^a_b$ and $C{}^a_b$ separately. If, however,
we assume that $\rmb{D}_{\bar\Omega} \bZ{a}{b}{i}{j} = 0$, it follows
that 
\begin{equation}
C{}^a_{b;a} = 0\,,
\label{avgZala63}
\end{equation}
\noindent
which implies that the averaged energy-momentum tensor $T{}^a_b$ is
also covariantly conserved. In our explicit calculations in the
perturbative setting in chapter 4, we will see that this condition
does not hold in general.

It can also be shown that in 4 dimensions, the 720 \emph{a priori}
independent components of $Z^{a\ph{bm}i}_{\ph{a}bm\ph{i}jn}$ are
subject to 680 constraints arising from \eqns{avgZala51} and
\eqref{avgZala50}. This leaves 40 independent components which
combine to give the 10 independent components of the correlation
tensor $C{}^a_b$. The conditions in \eqns{avgZala51} and
\eqref{avgZala50} do not constrain the components of $C{}^a_b$, which
follows from considering the structure of those equations.

\section{A $3+1$ spacetime splitting and the spatial averaging
  limit} 
\noindent
We are now in a position to apply the MG formalism to the problem of
cosmology. The main idea we wish to emphasize is that in the
cosmological context, it is essential to consider a spatial averaging
limit of the covariant averaging used in MG. The simplest way to see
this is to note that the homogeneous and isotropic FLRW
spacetime\footnote{Appendix A describes the main features of the FLRW
  spacetime.} \emph{must} be left invariant under the averaging
operation, and this is only possible if the averaging is tuned to the
uniquely defined spatial slices of constant curvature in the FLRW
spacetime. We will elaborate on this below. Our main motivation here
is to spell out all the assumptions usually made in the standard
approach to cosmology, and ask what they imply in the context of the
averaging paradigm. It has been suggested that the observationally
relevant averaging must necessarily be performed on the light
cone \cite{coley-lightcone}. This would not preserve the symmetries of
the FLRW spacetime. Our point of view is that one needs
a \emph{theoretically} self-consistent mathematical framework in which
to study cosmological expansion, structure formation, etc. We will be
conservative in our approach and only make the minimum assumptions
necessary in order to make standard cosmology compatible with the
averaging paradigm.

We start with the assumption that Einstein's equations are
to be imposed on length scales \Linh where stars are pointlike
objects, and that there exists a length scale \Lfrw\ such that
averaging on this length scale yields a geometry which has homogeneous 
and isotropic spatial sections. We expect $\Lfrw\gtrsim100h^{-1}$Mpc
and we assume $\Linh\ll\Lfrw\ll\Lhub$ where \Lhub\ is the length scale
of the observable universe. In other words, we will assume that
the averaged manifold \Mbar\ admits a preferred,
hypersurface-orthogonal unit timelike vector field $\bar v^a$, which
defines $3$-dimensional spacelike hypersurfaces of constant curvature,
and that $\bar v^a$ is tangent to the trajectories of observers who
see an isotropic cosmic background radiation. For simplicity we will
work with the special case where these spatial sections on
\Mbar\ are flat. One can then choose coordinates $(t,x^A)$, $A=1,2,3$,
on \Mbar\ such that the spatial line element takes the form  
\begin{equation}
^{(\Mbar)}ds^2_{\rm spatial} = a^2(t)\delta_{AB}dx^Adx^B\,,
\label{spatlim1}
\end{equation}
where $\delta_{AB}=1$ for $A=B$, and $0$ otherwise, and we have
$\bar v^a=(\bar v^t,0,0,0)$ so that the spatial coordinates are
comoving with the preferred observers. The vector field $\bar v^a$
also defines a proper time (the cosmic time) $\tau$ such that
$\partial_\tau = \bar v^a\partial_a = \bar v^t\partial_t$. We will
further assume that the averaged energy-momentum tensor $T{}^a_b$ can
be written in the form of a perfect fluid, as   
\begin{equation}
T{}^a_b = \rho \bar v^a \bar v_b + p\pi^a_b\,, 
\label{spatlim2}
\end{equation} 
where the projection operator $\pi^a_b$ is defined as
\begin{equation}
\pi^a_b = \delta{}^a_b + \bar v^a\bar v_b \,,
\label{spatlim3}
\end{equation} 
and satisfies $\pi^a_b \bar v^b = 0 =\, \pi^a_b \bar v_a$,
$\pi^a_c\pi^c_b = \pi^a_b$, and $\rho$ and $p$ are the
homogeneous energy density and pressure respectively, as
measured by observers moving on trajectories (in \Mbar) with the
tangent vector field $\bar v^a$, 
\begin{equation}
\rho\equiv T{}^a_b\bar v^b\bar v_a ~~;~~ p\equiv\frac{1}{3}\pi^b_a
T{}^a_b\,.
\label{spatlim-T-ab1}
\end{equation}
An important consequence of the above assumptions is that the
correlation tensor $C{}^a_b$, when expressed in terms of the natural
coordinates adapted to the spatial sections defined by the vector
field $\bar v^a$, is \emph{homogeneous} and depends only on the time
coordinate. This is clear when the modified Einstein equations
\eqref{avgZala61} are written in these natural coordinates. 

Using the vector field $\bar v^a$, the (FLRW) Einstein tensor
$E{}^a_b$ can be written as 
\begin{align}
E{}^a_b = j_1(x) \bar v^a\bar v_b &+ j_2(x) \pi^a_b\,, \nonumber\\ 
j_1(x) \equiv E{}^a_b\bar v^b\bar v_a ~~&;~~ j_2(x) \equiv
\frac{1}{3}\left( \pi^b_a E{}^a_b\right)\,, 
\label{spatlim-FLRW1}
\end{align}
where $j_1(x)$ and $j_2(x)$ are scalar functions whose form depends
upon the coordinates used. The remaining components given by
$\pi^b_k E{}^a_b\bar v_a$ and the traceless part of
$\pi^i_a\pi^b_k E{}^a_b$, vanish identically. Since the
energy-momentum tensor $T{}^a_b$ in \eqn{spatlim2} also has an
identical structure, this structure is therefore \emph{also imposed}
on the correlation tensor $C{}^a_b$. Namely, only the components
$C{}^a_b\bar v^b\bar v_a$ and the trace $\pi^b_a C{}^a_b$ are 
relevant to the dynamics of the averaged metric. The remaining
components, namely $\pi^b_k C{}^a_b\bar v_a$ and the traceless part 
of $\pi^i_a\pi^b_k C{}^a_b$, \emph{must vanish}. This is a 
condition on the underlying inhomogeneous geometry, irrespective of
the coordinates used to describe the geometry on either \Cal{M}\ or
\Mbar, and is clearly a consequence of demanding that the averaged
geometry have the symmetries of the FLRW spacetime.

This leads us to the crucial question of the choice of \emph{gauge}
for the underlying geometry : namely, what choice of spatial sections
for the \emph{inhomogeneous} geometry, will lead to the spatial
sections of the FLRW metric in the comoving coordinates defined in
\eqn{spatlim1}? Since the matter distribution at scale 
\Linh\ need not be pressure-free (or, indeed, even of the perfect
fluid form), there is clearly no natural choice of gauge available,
although locally, a synchronous reference can always be chosen. We
note that there must be \emph{at least one} choice of gauge in which
the averaged metric has spatial sections in the form \eqref{spatlim1}
-- this is simply a refinement of the Cosmological Principle, and of
the Weyl postulate, according to which the universe is homogeneous and
isotropic on large scales, and individual galaxies are considered as
the ``observers'' travelling on trajectories with tangent $\bar
v^a$. In the averaging approach, it makes more sense to replace
``individual galaxies'' with the \emph{averaging domains} considered
as physically infinitesimal cells -- the ``points'' of the averaged
manifold \Mbar. This is physically reasonable since we know after all,
that individual galaxies exhibit peculiar motions, undergo mergers and
so on. This idea is also more in keeping with the notion that the
universe is homogeneous and isotropic \emph{only on the largest
  scales}, which are much larger than the scale of individual
galaxies.

Consider any $3+1$ spacetime splitting in the form of a lapse function
${N}(t,x^J)$, a shift vector ${N}^A(t,x^J)$, and a
metric for the $3$-geometry ${h}_{AB}(t,x^J)$, so that the line
element on \Cal{M}\ can be written as 
\begin{equation}
^{(\Cal{M})}ds^2 = -\left({N}^2 -
  {N}_{\!A}{N}^A\right)dt^2 + 2{N}_{\!B}dx^Bdt + 
  {h}_{AB}dx^Adx^B\,,  
\label{spatlim4} 
\end{equation}
where ${N}_{\!A} = {h}_{\!AB}{N}^B$. At first
sight, it might seem reasonable to leave the choice of gauge
arbitrary. One could then formally consider a coordination bivector
given by the \eqns{avgZala24} and \eqref{avgZala26}, with $x^i$
denoting the coordinates in the chosen gauge and $\phi^m$ the VPCs;
and demand for example, that the metric \eqref{spatlim4} (with say 
${N}^A=0$) average out to the FLRW form (with a nonsynchronous
time coordinate in general). This would imply
\begin{equation}
G_{00} = \stavg{\bil{g}_{00}} = -f^2(t) ~~;~~ G_{0A} =
\stavg{\bil{g}_{0A}} = 0 ~~;~~ G_{AB} = \stavg{\bil{g}_{AB}} =
a^2(t)\delta_{AB} \,. 
\label{spatlim-gauge1}
\end{equation} 
Note that the condition on the bilocal extension
$\bil{g}_{0A}(x^\prime,x)$ is in general nontrivial even when the
components $g_{0A}(x)$ are chosen to be zero. In the Appendix
\eqref{app-gbar} we show that with the above assumptions, for a
general lapse function ${N}$, the conditions
$\rmb{D}_{\bar\Omega}\bar g^{ab}=0$ (\eqn{avgZala53}) also allow us to   
choose 
\begin{equation}
U^{ij}\equiv\bar g^{ij}-G^{ij}=0\,.
\label{spatlim-gauge2}
\end{equation}
However, it turns out that if we make the assumption that the
spatial sections on \Cal{M}\ leading to the spatial metric
\eqref{spatlim1} on \Mbar, are spatial sections \emph{in a volume
  preserving gauge}, then the correlation terms simplify
greatly. This is not surprising since the MG formalism is
nicely adapted to the choice of volume preserving
coordinates. Moreover, as we will see in chapter 4, at least in the
perturbative context a modified version of this ``VP gauge
assumption'' in fact becomes a necessity in order to consistently set
up the formalism. We will therefore introduce spatial averaging in MG
by making the VP gauge assumption, and will then
calculate the correlation terms and display the modified equations
resulting from this choice of gauge. Following that calculation we
will also show how the correlation terms can be generalized to the
case where the gauge in the inhomogeneous metric is (formally) left
unspecified.   

To begin our first calculation, we perform a coordinate transformation
and shift to the gauge wherein the new lapse function $N$ is given by
$N=1/\sqrt{h}$ where $h$ is the determinant of the new $3$-metric,
denoted $h_{AB}$. In general, one will now be left with a non-zero
shift vector $N^A$; however, the condition $N\sqrt{h}=1$ ensures that
the coordinates we are now using are volume preserving, since the
metric determinant is given by $g=-N^2h=-1={\rm constant}$. We denote
these volume preserving coordinates (VPCs) by $(\bt, \rmb{x}) = (\bt,
x^A) = (\bt, x, y, z)$, and will assume that the spatial coordinates
are noncompact. For simplicity, we make the added assumption that
$N^A=0$ in the inhomogenous geometry \footnote{We are making the
  assumption $N^A=0$ in the volume preserving gauge for algebraic
  convenience only -- in our case this assumption cannot be justified
  by the absence of vorticity. A more detailed (and complicated)
  analysis should retain an arbitrary $N^A$ in the inhomogeneous
  geometry, and make assumptions about its average -- such as
  $\avg{N_A}=0$ for example.}, so that $g_{\bt\,\bt}=-N^2=-1/h$ and
$g_{\bt A}=0$. The line element for the inhomogenous manifold
\Cal{M}\ becomes  
\begin{equation}
^{(\Cal{M})}ds^2=-\frac{d\bt^2}{h(\bt,\rmb{x})} +
  h_{AB}(\bt,\rmb{x})dx^Adx^B\,. 
\label{spatlim5}
\end{equation}
Note that in this gauge, the average takes on a particularly simple
form : for a tensor $p{}^i_j(x)$, with a spacetime averaging domain
given by the ``cuboid'' $\bm{\Sigma}$ defined by
\begin{equation}
\bm{\Sigma} =
\left\{({\bt}^\prime,x^{A\prime}) \mid \bt-T/2<{\bt}^\prime
< \bt+T/2,x^A-L/2 < x^{A\prime}<x^A+L/2; A=1,2,3\right\},    
\label{spatlim6}
\end{equation}
where $T$ and $L$ are averaging time and length scales respectively,
the average is given by     
\begin{equation}
\stavg{\bil{p}{\,}^i_j}(\bt,\rmb{x}) =
    \stavg{p{}^i_j}(\bt,\rmb{x}) =  \frac{1}{TL^3}
    \int_{\bt-T/2}^{\bt+T/2}{dt^\prime\int_{x-L/2}^{x+L/2}{    
    dx^\prime\int_{y-L/2}^{y+L/2}{dy^\prime\int_{z-L/2}^{z+L/2}{
	dz^\prime\bigg[
	  p{}^i_j(t^\prime,x^\prime,y^\prime,z^\prime)\bigg]}}}} \,.   
\label{spatlim7}
\end{equation}
We define the ``spatial averaging limit'' as the limit $T\to0$ (or
$T\ll\Lhub$) which is interpreted as providing a definition of the
average on a spatial domain corresponding to a ``thin'' time slice,
the averaging operation now being given by
\begin{equation}
\avg{p{}^i_j}(\bt,\rmb{x}) =  \frac{1}{L^3}
    \int_{x-L/2}^{x+L/2}{dx^\prime
    \int_{y-L/2}^{y+L/2}{dy^\prime
    \int_{z-L/2}^{z+L/2}{dz^\prime\bigg[ 
	  p{}^i_j(\bt,x^\prime,y^\prime,z^\prime)\bigg]}}} +
    \Cal{O}\left(TL_{\rm Hubble}^{-1} \right) \,.   
\label{spatlim8}
\end{equation}
(Note the time dependence of the integrand.) Henceforth, averaging
will refer to spatial averaging, and will be denoted by $\avg{...}$,
in contrast to the spacetime averaging considered thus far (denoted by 
$\stavg{...}$)\footnote{The choice of a cube with sides of length $L$ 
  as the spatial averaging domain was arbitrary, and is in fact not
  essential for any of the calculations to follow. In particular, all
  calculations can be performed with a spatial domain of arbitrary
  shape. We will only use the cube for definiteness and simplicity in
  displaying equations.}. The significance of introducing a spatial
averaging in this manner is that the construction of spatial averaging
is not isolated from spacetime averaging, but is a special limiting
case of the latter and is, in fact, still a fully covariant
operation. 


For the volume preserving gauge, the averaging assumption
\eqref{spatlim-gauge1} reduces to 
\begin{equation} 
G_{\bt\bt} = \avg{g_{\bt\,\bt}} = \avg{\frac{-1}{h}} = -f^2(\bt) ~~;~~ 
G_{AB} = \avg{h_{AB}}=\ba^2(\bt)\delta_{AB} \,,  
\label{spatlim9}
\end{equation}
where $\ba$ and $f$ are some functions of the time coordinate alone. A
few remarks are in order on this particular choice of assumptions. 
Apart from the fact that the spacetime averaging operation takes on
its simplest possible form \eqref{spatlim7} in this gauge and allows a
transparent definition of the spatial averaging limit, it can also be
shown that the assumptions in \eqn{spatlim9} are sufficient to
establish the following relations :  
\begin{equation}
f^2(\bt) = \avg{\frac{1}{h}} = \frac{1}{\avg{h}} = \frac{1}{\ba^6}\,. 
\label{spatlim10}
\end{equation}
Here the second equality arises from the condition $\bar
g^{ij}=G^{ij}$ which can be assumed whenever the averaged metric is of
the FLRW form (see Appendix \eqref{app-gbar}). The last equality
follows on considering the conditions $\avg{\bil{\Gamma}{}^a_{bc}}
=\,^{(\rm FLRW)}\Gamma{}^a_{bc}$ in obvious notation, (the basic
assumption of the MG averaging scheme), details of which can be found
in the Appendix \eqref{app-gamma}. \eqn{spatlim10} reduces the line
element on \Mbar\ to the form       
\begin{equation}
^{(\Mbar)}ds^2 = -\frac{d\bt^2}{\ba^6(\bt)} +
  \ba^2(\bt)\delta_{AB}dx^Adx^B \ .
\label{spatlim11}
\end{equation}
The line element in \eqn{spatlim11} clearly corresponds to the FLRW
metric in a volume preserving gauge which differs from the standard
synchronous and comoving gauge, only by a redefinition of the time
coordinate. The vector field $\bar v^a$ introduced at the beginning of
this section and which defines the FLRW spatial sections, is now given
by 
\begin{equation}
\bar v^a = \left(\ba^3,0,0,0\right) ~~;~~ \bar v_a = G_{ab}\bar v^b =
\left(-\frac{1}{\ba^3},0,0,0,\right) \,.
\label{spatlim12}
\end{equation}
Note that $\bar v^a$ is \emph{not} in general the average of the
vector field $u^a = (\sqrt{h},0,0,0)$ which defines the $3+1$
splitting on \Cal{M}, but (at least in the volume preserving gauge) is
related to it by 
\begin{equation}
\bar v^a = \frac{\ba^3}{\avg{\sqrt h}} \avg{u^a} \,.
\label{spatlim13}
\end{equation}
(A simple relation such as \eqref{spatlim13} cannot in general be
written for an arbitrary gauge.) 

As mentioned earlier, the spatial averaging limit of the covariant MG
averaging is important because we want the 
homogeneous and isotropic FLRW geometry to average to itself.
Since the FLRW geometry has a preferred set of spatial
sections, one therefore needs to average over these sections. Further,
since the FLRW metric adapted to its 
preferred spatial sections depends on the time coordinate, it is also
essential that the spacetime average involve a time range that
is short compared to the scale over which say the scale factor changes
significantly. Clearly then, averaging the FLRW metric (denoted
$^{(FLRW)}g_{ab}$) given in \eqn{spatlim11} will strictly yield the
same metric \emph{only} in the limit $T\to0$. Namely, for the cuboid
$\bm{\Sigma}$ defined in 
\eqn{spatlim6}   
\begin{equation}
\avg{^{(FLRW)}\bil{g}_{ab}} =
\lim_{T\to0}\frac{1}{TL^3}\int_\Sigma{dt^\prime d^3x^\prime\,
  ^{(FLRW)}g_{ab}(t^\prime,\rmb{x}^\prime)} =
\,^{(FLRW)}g_{ab}\,,  
\label{spatlim14}
\end{equation}
which should be clear from the definition of the metric. The result
$\avg{^{(FLRW)}\bil{g}_{ab}} = \,^{(FLRW)}g_{ab}$ in the spatial
averaging limit can also be shown to hold for the FLRW metric in
synchronous gauge, where the coordination bivector $\W{a}{j}$ can be
easily computed using the transformation from the VPCs $(\bt,x^A)$ to
the synchronous coordinates $(\tau,y^A)$ given by
\begin{equation}
\tau = \int^{\bt}{\frac{dt}{\ba^3(t)}} ~~;~~ y^A = x^A\,.
\label{spatlim15}
\end{equation}
The transformation \eqref{spatlim15} will also later allow us to write
the averaged equations in the synchronous gauge for the averaged
geometry.

\section{The correlation $2$-form and the averaged field equations}
\label{correln}
\subsection{Results for the Volume Preserving Gauge}
\noindent
In any gauge with $N^A=0$, the expansion tensor $\Theta{}^A_B$ is
given by 
\begin{equation}
\Theta{}^A_B\equiv \frac{1}{2N}h^{AC}\p_{\bt} h_{CB}\,.
\label{correln1}
\end{equation}
The notation is the same as used in chapter 2, and the expansion scalar
$\Theta$, shear tensor $\sigma{}^A_B$ and shear scalar $\sigma^2$ are
defined as before. Note that
\begin{equation}
\sigma^2 = \frac{1}{2}\Theta{}^A_B\Theta{}^B_A - \frac{1}{6}\Theta^2
~~;~~ \Theta^2 - \Theta{}^A_B\Theta{}^B_A = \frac{2}{3}\Theta^2 -
2\sigma^2 \,.
\label{correln4}
\end{equation}
The connection $1$-forms $\om^i_{\ph{i}j} = \Gamma{}^i_{jk}\dx{k}$ in
terms of the expansion tensor, are listed below for an arbitrary lapse
function $N$ : 
\begin{align}
\om^0_{\ph{i}0} = \partial_{\bt}(\ln N)\ext\bt +  \partial_A(\ln
N)\dx{A} ~~&;~~ \om^A_{\ph{i}0} = N^2h^{AC}\partial_C(\ln N)\ext\bt +
N\Theta{}^A_B\dx{B} \,,\nonumber\\
\om^0_{\ph{i}A} = \partial_A(\ln N)\ext\bt +
\frac{1}{N}\Theta_{AB}\dx{B} ~~&;~~ \om^A_{\ph{i}B} =
N\Theta{}^A_B\ext\bt + \,^{(3)}\Gamma{}^A_{BC}\dx{C}\,,
\label{correln5}
\end{align}
where $^{(3)}\Gamma{}^A_{BC}$ is the Christoffel symbol built from the
$3$-metric $h_{AB}$ and its inverse. Specializing now to the volume
preserving gauge ($N=h^{-1/2}$), the bilocal extensions $\Om{i}{j}$ of
the connection $1$-forms are trivial and are simply given by    
\begin{equation}
\Om{i}{j}(x^\prime,x) = \Gamma{}^i_{jk}(x^\prime)\dx{k}\,.
\label{correln6}
\end{equation}
Since $G_{ab} = \bar g_{ab}$, the connection $1$-forms $\barOm{i}{j}$
for the averaged manifold \Mbar\ are constructed using the FLRW metric
in volume preserving gauge given in \eqn{spatlim11}, and are given by
\begin{align}
\barOm{0}{0} = -3H\ext\bt ~~&;~~ \barOm{A}{0} = H\delta{}^A_B\dx{B}
\,,\nonumber\\  
\barOm{0}{A} = \ba^8H\delta_{AB}\dx{B} ~~&;~~ \barOm{A}{B} =
H\delta{}^A_B\ext\bt \,, 
\label{correln7}
\end{align}
where we have defined 
\begin{equation}
H\equiv \frac{1}{\ba}\frac{d\ba}{d\bt}\,.
\label{correln8}
\end{equation}
Using \eqns{correln5} with $N=h^{-1/2}$, \eqn{correln6} and
\eqns{correln7}, we can now easily construct the correlation $2$-form
$\bZ{a}{b}{i}{j}$ defined in \eqn{avgZala38}. For completeness, we
will display all the nontrivial components $\bZ{a}{b}{i}{j}$, although
not all of them will be relevant for the final equations. The
condition $N=h^{-1/2}$ has the effect that several of the Christoffel
symbols (which can be read off from \eqn{correln5}) become related to
each other. For example, we have
$\Gamma{}^0_{00}=-\partial_{\bt}(\ln\sqrt{h}) = -\Gamma{}^A_{0A} =
-(1/\sqrt{h})\Theta$, and so on. Denoting the spatial average of some
quantity $p{}^i_j$ as simply $\avg{p{}^i_j}$, we have 
{\small
\begin{subequations}
\begin{align}
\bZ{0}{0}{0}{A}&= - \bZ{0}{A}{0}{0} \nonumber\\ 
               &= \left[\avg{\Theta\Theta_{AJ}} +
               \avg{\partial_A(\ln\sqrt{h})\partial_J(\ln\sqrt{h})} -
               3\ba^8H^2\delta_{AJ}\right] \dx{J} \wedge \ext\bt
               \nonumber\\ 
               &\ph{\left[\avg{\Theta\Theta_{AJ}} +
               \avg{\partial_A(\ln\sqrt{h})\partial_J(\ln\sqrt{h})} -
               3\ba^8H^2\delta_{AJ}\right]}+
               \avg{\sqrt{h}\Theta_{AJ}\partial_K(\ln\sqrt{h})}\dx{J}
               \wedge \dx{K}    \,,
\label{correln9a} \\
\bZ{0}{0}{A}{0}&= - \bZ{A}{0}{0}{0} \nonumber\\
               &= \left[ \avg{\frac{1}{h}\Theta\Theta{}^A_J} +
               \avg{\frac{1}{h}h^{AK} \partial_K(\ln\sqrt{h})
               \partial_J(\ln\sqrt{h})} - 3H^2\delta{}^A_J
               \right]\dx{J} \wedge \ext\bt \nonumber\\
               &\ph{\left[ \avg{\frac{1}{h}\Theta\Theta{}^A_J} +
               \avg{\frac{1}{h}h^{AK} \partial_K(\ln\sqrt{h})
               \partial_J(\ln\sqrt{h})} - 3H^2\delta{}^A_J
               \right]}+  \avg{\frac{1}{\sqrt
               h}\Theta{}^A_J \partial_K(\ln\sqrt{h})} \dx{J} \wedge
               \dx{K} \,, 
\label{correln9b} \\
\bZ{0}{0}{A}{B}&= - \bZ{A}{B}{0}{0} \nonumber\\
               &= \left[
               \avg{\frac{1}{\sqrt{h}}\Theta\,^{(3)}\Gamma{}^A_{BJ}} -
               \avg{\frac{1}{\sqrt h} \Theta{}^A_B
               \partial_J(\ln\sqrt{h})} \right] \dx{J} \wedge \ext\bt
               + \avg{\,^{(3)}\Gamma{}^A_{BJ}\partial_K(\ln\sqrt{h})}
               \dx{J} \wedge \dx{K} \,,
\label{correln9c} \\
\bZ{0}{A}{0}{B}&= \left[2
  \avg{\sqrt{h}\partial_{\left[A\right.}(\ln\sqrt{h})
  \Theta_{\left.B\right]J}} \right] \dx{J}  \wedge \ext\bt + \left[
  \avg{h\Theta_{AJ}\Theta_{BK}} -  \ba^{16}H^2\delta_{AJ}\delta_{BK}
  \right] \dx{J} \wedge \dx{K} \,, 
\label{correln9d} \\
\bZ{0}{A}{B}{0}&= - \bZ{B}{0}{0}{A} \nonumber\\
               &= \left[ \avg{\frac{1}{\sqrt
               h}\partial_A(\ln\sqrt{h})\Theta{}^B_J} -
               \avg{\frac{1}{\sqrt  h}h^{BK}\partial_K(\ln\sqrt{h})
               \Theta_{AJ}} \right] \dx{J} \wedge \ext\bt\nonumber\\  
               &\ph{\left[ \avg{\frac{1}{\sqrt
               h}\partial_A(\ln\sqrt{h})\Theta{}^B_J} -
               \avg{\frac{1}{\sqrt  h}h^{BK}\partial_K(\ln\sqrt{h})
               \Theta_{AJ}} \right]}+ \left[
               \avg{\Theta_{AJ}\Theta{}^B_K} -
               \ba^8H^2\delta_{AJ}\delta{}^B_K\right]
               \dx{J}\wedge\dx{K}   \,,
\label{correln9e} \\
\bZ{0}{A}{B}{C}&= - \bZ{B}{C}{0}{A} \nonumber\\
               &= \left[
               \avg{\partial_A(\ln\sqrt{h})\,^{(3)}\Gamma{}^B_{CJ}} +
               \avg{\Theta_{AJ}\Theta{}^B_C} -
               \ba^8H^2\delta_{AJ}\delta{}^B_C \right] \dx{J} \wedge
               \ext\bt +
               \avg{\sqrt{h}\Theta_{AJ}\,^{(3)}\Gamma{}^B_{CK}} \dx{J}
               \wedge \dx{K} \,,
\label{correln9f} \\
\bZ{A}{0}{B}{0}&= \left[2
  \avg{\frac{1}{h^{3/2}}\partial_K(\ln\sqrt{h})h^{K\left[A\right.}
  \Theta{}^{\left.B\right]}_J} \right] \dx{J} \wedge \ext\bt + \left[
  \avg{\frac{1}{h}\Theta{}^A_J\Theta{}^B_K} -
  H^2\delta{}^A_J\delta{}^B_K \right] \dx{J} \wedge \dx{K} \,,
\label{correln9g} \\
\bZ{A}{0}{B}{C}&= - \bZ{B}{C}{A}{0} \nonumber\\
               &= \left[
               \avg{\frac{1}{h}h^{AK}\partial_K(\ln\sqrt{h})
               \,^{(3)}\Gamma{}^B_{CJ}} +
               \avg{\frac{1}{h}\Theta{}^A_J\Theta{}^B_C} - 
               H^2\delta{}^A_J\delta{}^B_C \right] \dx{J} \wedge
               \ext\bt \nonumber\\
               &\ph{\left[
               \avg{\frac{1}{h}h^{AK}\partial_K(\ln\sqrt{h})
               \,^{(3)}\Gamma{}^B_{CJ}} +
               \avg{\frac{1}{h}\Theta{}^A_J\Theta{}^B_C} - 
               H^2\delta{}^A_J\delta{}^B_C \right]}+
               \avg{\frac{1}{\sqrt
                   h}\Theta{}^A_J\,^{(3)}\Gamma{}^B_{CK}} \dx{J}
               \wedge \dx{K}  \,,
\label{correln9h} \\
\bZ{A}{B}{C}{D} &= \left[\avg{\frac{1}{\sqrt h} \Theta{}^C_D
    \,^{(3)}\Gamma{}^A_{BJ}} - \avg{\frac{1}{\sqrt h} \Theta{}^A_B
    \,^{(3)}\Gamma{}^C_{DJ}}\right] \dx{J} \wedge \ext\bt +
    \avg{\,^{(3)}\Gamma{}^A_{BJ}\,^{(3)}\Gamma{}^C_{DK}} \dx{J} \wedge
    \dx{K} \,, 
\label{correln9i}
\end{align}
\label{correln9}
\end{subequations}
}
\noindent where we have used the relation $\,^{(3)}\Gamma{}^J_{BJ} =
\partial_B(\ln\sqrt{h})$.

It is now straightforward to use the relations in \eqn{avgZala58}
(note the unconventional normalization of the $2$-form) to read off
the components $Z^{a\ph{bm}i}_{\ph{a}bm\ph{i}jn}$ and hence perform
the required summations to construct $Z^a_{\ph{a}ijb}$. This, together 
with the fact that $\bar g^{ab} = G^{ab}$ (see Appendix
\eqref{app-gbar}), allows us to construct the correlation tensor
$C{}^a_b$ defined in \eqn{avgZala59}
\begin{equation}
C{}^a_b = \left(Z^a_{\ph{a}ijb} - \frac{1}{2}\delta{}^a_b
Z^m_{\ph{a}ijm} \right)G^{ij}.
\label{correln10}
\end{equation}
Now, the components of the Einstein tensor $E{}^a_b$ for the averaged
spacetime with metric \eqref{spatlim11} are given by
\begin{align} 
E{}^{\bt}_{\bt} &= 3\ba^6H^2 ~~;~~ E{}^{\bt}_A = 0 = E{}^B_{\bt} \,,  
\nonumber\\ 
E{}^A_B &= \ba^6\delta{}^A_B\left[ 2\left(\frac{\ddot\ba}{\ba} +
  3H^2\right) + H^2\right] \,,
\label{correln11}
\end{align}
where the peculiar splitting of terms in the last equation is for
later convenience. Recall that the overdot denotes a derivative with
respect to the VPC time $\bt$, not synchronous time. In terms of the
coordinate independent objects introduced in \eqn{spatlim-FLRW1}, we
have 
\begin{equation}
j_1(x) = -3\ba^6H^2 ~~;~~ j_2(x) = \ba^6\left[
  2\left(\frac{\ddot\ba}{\ba} + 3H^2\right) + H^2\right] \,.
\label{correln-FLRW1}
\end{equation}
From the averaged Einstein equations in \eqref{avgZala61} we next 
construct the scalar equations which in the standard case
would correspond to the Friedmann equation and the Raychaudhuri
equation. These correspond to the Einstein tensor components, 
\begin{equation}
E{}^a_b\bar v^b\bar v_a  = j_1(x)  ~~;~~
\pi^b_aE{}^a_b + E{}^a_b\bar v^b\bar v_a = 3j_2(x) + j_1(x)\,,  
\label{correln12}
\end{equation}
and are given by
\begin{subequations}
\begin{align}
3\ba^6H^2&=\left(\kappa T{}^a_b - C{}^a_b\right)\bar v_a\bar v^b = 
\kappa\bar\rho - \frac{1}{2}\left[ \Cal{Q}^{(1)} +
  \Cal{S}^{(1)} \right]  \,,
\label{correln13a} \\
6\ba^6\left( \frac{\ddot\ba}{\ba} + 3H^2 \right) 
&= \left(-\kappa T{}^a_b + C{}^a_b\right) \left(\bar v_a\bar v^b + 
\pi^b_a \right) = -\kappa\left(\bar\rho+3\bar p\right) + 2\left[
  \Cal{Q}^{(1)} +  \Cal{Q}^{(2)} + \Cal{S}^{(2)} \right]\,.
\label{correln13b}
\end{align}
\label{correln13}
\end{subequations}
\noindent Here \eqn{correln13a} is the modified Friedmann equation and 
\eqn{correln13b} the modified Raychaudhuri equation (in the volume
preserving gauge on \Mbar). We have used \eqn{spatlim-T-ab1}, with the
overbar on $\rho$ and $p$ reminding us that they are expressed in
terms of the nonsynchronous time $\bt$, and we have defined the
correlation terms  
\begin{subequations}
\begin{align}
\Cal{Q}^{(1)} &= \ba^6\left[\frac{2}{3}\left(\avg{\frac{1}{h}\Theta^2} -
\frac{1}{\ba^6}(^{\rm F}\Theta^2)\right) -
2\avg{\frac{1}{h}\sigma^2}\right] 
~~;~~ \frac{1}{\ba^6}(^{\rm F}\Theta^2) = \left(3H\right)^2\,,
\label{correln14a} \\&\nonumber\\
\Cal{S}^{(1)} &= \frac{1}{\ba^2}\delta^{AB}\left[
  \avg{\,^{(3)}\Gamma{}^J_{AC}\,^{(3)}\Gamma{}^C_{BJ}} -
  \avg{\partial_A(\ln\sqrt{h})\partial_B(\ln\sqrt{h})}  
   \right] \,,   
\label{correln14b} \\&\nonumber\\
\Cal{Q}^{(2)} &= \ba^6\avg{\frac{1}{h}\Theta{}^A_B\Theta{}^B_A} -
\frac{1}{\ba^2}\delta^{AB}\avg{\Theta_{AJ}\Theta{}^J_B},
\label{correln14c} \\&\nonumber\\
\Cal{S}^{(2)} &= \ba^6\avg{\frac{1}{h}h^{AB}
  \partial_A(\ln\sqrt{h})\partial_B(\ln\sqrt{h})} -
  \frac{1}{\ba^2}\delta^{AB}
  \avg{\partial_A(\ln\sqrt{h})\partial_B(\ln\sqrt{h})}  \,.  
\label{correln14d}
\end{align}
\label{correln14}
\end{subequations}
We have used the second relation in \eqn{correln4} in defining
$\Cal{Q}^{(1)}$. $\Cal{Q}^{(1)}$ and $\Cal{Q}^{(2)}$ are correlations
of the extrinsic curvature, whereas $\Cal{S}^{(1)}$ and
$\Cal{S}^{(2)}$ are correlations restricted to the intrinsic
$3$-geometry of the spatial slices of \Cal{M}. Since the components of
$C{}^a_b$ are not explicitly constrained by \eqns{avgZala50} and
\eqref{avgZala51}, we can treat the combinations $(1/2)(\Cal{Q}^{(1)} 
+ \Cal{S}^{(1)})=-C{}^0_0$ and $2(\Cal{Q}^{(1)} + \Cal{Q}^{(2)} +
\Cal{S}^{(2)}) = (C{}^A_A-C{}^0_0)$ as independent, subject only to
the differential constraints \eqref{avgZala63} which follow if we
assume $\rmb{D}_{\Omega}\bZ{a}{b}{i}{j}=0$. We will return to these
below. 

As discussed earlier, the remaining
components of $C{}^a_b$ must be set to zero, giving constraints on the
underlying inhomogeneous geometry. In coordinate independent language,
these constraints read
\begin{equation}
\pi^b_k C{}^a_b\bar v_a = 0 =\,\pi^k_a C{}^a_b\bar v^b ~~;~~ 
\pi^i_a\pi^b_k C{}^a_b -
\frac{1}{3}\pi^i_k \left(\pi^b_a C{}^a_b\right) = 0\,.
\label{correln15}
\end{equation}
\eqns{correln15} reduce to the following for our specific choice of
volume preserving coordinates,  
\begin{equation}
C{}^0_A = 0 ~~;~~ C{}^A_0 = 0 ~~;~~ C{}^A_B -
\frac{1}{3}\delta{}^A_B(C{}^J_J) = 0 \,,
\label{correln16}
\end{equation}
or, in full detail,
\begin{subequations}
\begin{align}
&\left[\ba^6 \avg{\frac{h^{JK}}{h}\sqrt{h}
  \Theta{}_{JA}\partial_K(\ln\sqrt{h})} -
  \avg{h^{JK}}
  \avg{\sqrt{h}\Theta_{JA}\partial_K(\ln\sqrt{h})}  \right]
  \nonumber\\ 
&\ph{\ba^6[]\avg{\frac{1}{\sqrt h}
  \Theta{}^B_A\partial_B(\ln\sqrt{h})}} 
  +  \left[\avg{h^{JK}}
  \avg{\sqrt{h}\Theta_{JB}\,^{(3)}\Gamma{}^B_{AK}} -
  \ba^6 \avg{\frac{h^{JK}}{h}\sqrt{h}
  \Theta_{JK}\partial_A(\ln\sqrt{h})} \right] =  0 \,,      
\label{correln17a} \\
&\left[\ba^6 \avg{\frac{h^{JK}}{h}\frac{1}{\sqrt h}\Theta^A_J
    \partial_K(\ln\sqrt{h})} -
\avg{h^{JK}}\avg{\frac{1}{\sqrt 
      h}\Theta{}^A_J\partial_K(\ln\sqrt{h})} \right]
 \nonumber\\ 
&\ph{\ba^6[]\avg{\frac{1}{h^{3/2}}\Theta^{AB} \partial_B(\ln\sqrt{h})}}
+  \left[\avg{h^{JK}} \avg{\frac{1}{\sqrt h}
    \Theta{}^B_K\,^{(3)}\Gamma{}^A_{JB}}  - \ba^6  
  \avg{\frac{h^{JA}}{h}\frac{1}{\sqrt h}\Theta{}^K_K
    \,^{(3)}\Gamma{}^B_{JB}}  \right] = 0  \,,    
\label{correln17b}\\
&\left[\ba^6\avg{\frac{h^{JK}}{h}\Theta{}^A_J\Theta_{KB}} -
  \avg{h^{JK}}\avg{\Theta{}^A_J\Theta_{KB}}\right] + \ba^6\left[   
  \avg{\frac{1}{h}h^{AC} \partial_C(\ln\sqrt{h})
  \partial_B(\ln\sqrt{h})} \right. \nonumber\\
&\ph{\ba^6[]\avg{\frac{1}{h}\Theta{}^A_C\Theta{}^C_B} -
  \frac{1}{\ba^2}\delta^{JK}\left[
  \avg{\Theta{}^A_J\Theta_{KB}}\right]} 
 - \avg{h^{JK}}\left.
  \avg{\,^{(3)}\Gamma{}^A_{JC}\,^{(3)}\Gamma{}^C_{KB}} \right]
  =\frac{1}{3}\delta{}^A_B \left( \Cal{Q}^{(2)} - \Cal{S}^{(1)} +
  \Cal{S}^{(2)}  \right) \,,
\label{correln17c}
\end{align}
\label{correln17}
\end{subequations} 
\noindent 
where $\avg{h^{JK}} = G^{JK} = (1/\ba^2)\delta^{JK}$. \eqns{correln5} 
with the choice $g_{00} = -N^2 = -h^{-1}$ show that all the terms
paired within square brackets in \eqns{correln17} above, as also the
correlations $\Cal{Q}^{(2)}$ and $\Cal{S}^{(2)}$ defined in
\eqns{correln14c} and \eqref{correln14d}, are of the form 
\begin{equation}
\frac{1}{\avg{g_{00}}}\avg{g_{00}g^{AB}\Gamma{}^{a_1}_{b_1c_1}
  \Gamma{}^{i_1}_{j_1k_1}} - \avg{g^{AB}}\avg{\Gamma{}^{a_2}_{b_2c_2}
  \Gamma{}^{i_2}_{j_2k_2}}\,.   
\label{correln18}
\end{equation}
The assumption in \eqn{avgZala52b} shows that one can write
\begin{equation}
\avg{g_{00}g^{AB}\Gamma{}^a_{bc}\Gamma{}^i_{jk}} =
\avg{g_{00}g^{AB}}\avg{\Gamma{}^a_{bc}\Gamma{}^i_{jk}}  =
-\avg{\frac{h^{AB}}{h}}\avg{\Gamma{}^a_{bc}\Gamma{}^i_{jk}} \,.
\label{correln19}
\end{equation}
An interesting point is that the VPC assumption $N=h^{-1/2}$ also
allows us to assume $\avg{h^{AB}/h} = \avg{h^{AB}}\avg{1/h}$
consistently with the formalism (details in Appendix
\eqref{app-gamma}). Using \eqn{spatlim10} this gives us 
\begin{equation}
\avg{\frac{h^{AB}}{h}} = \frac{1}{\ba^6}\avg{h^{AB}}\,.
\label{correln20}
\end{equation}
This leads to some remarkable cancellations in \eqns{correln17}, and
also shows that the correlation terms $\Cal{Q}^{(2)}$ and
$\Cal{S}^{(2)}$ in fact vanish,
\begin{equation}
\Cal{Q}^{(2)} = 0 = \Cal{S}^{(2)}\,.
\label{correln21}
\end{equation}
\eqns{correln17} simplify to give
\begin{subequations}
\begin{align}
&\delta^{JK}\left[\avg{\sqrt{h}\Theta_{JB}\,^{(3)}\Gamma{}^B_{AK}} -
  \avg{\sqrt{h}\Theta_{JK}\,^{(3)}\Gamma{}^B_{AB}} \right] = 0\,,  
\label{correln22a}\\
&\delta^{JK}\avg{\frac{1}{\sqrt h}\Theta{}^B_K\,^{(3)}\Gamma{}^A_{JB}}
  - \delta^{AJ}\avg{\frac{1}{\sqrt
      h}\Theta{}^K_K\,^{(3)}\Gamma{}^B_{JB}} = 0\,,  
\label{correln22b}\\
&\delta^{JK}\avg{\,^{(3)}\Gamma{}^A_{JC}\,^{(3)}\Gamma{}^C_{KB}} -
\delta^{AJ}\avg{\,^{(3)}\Gamma{}^C_{JC}\,^{(3)}\Gamma{}^K_{BK}} = 
\frac{1}{3}\delta{}^A_B\left(\ba^2\Cal{S}^{(1)}\right) \,.
\label{correln22c}
\end{align}
\label{correln22}
\end{subequations}
These simplifications are solely a consequence of assuming that the
inhomogeneous metric in the volume preserving gauge averages out to
give the FLRW metric in standard form. In general, these
simplifications will not occur when the standard FLRW metric arises
from an arbitrary choice of gauge for the inhomogeneous metric.

In order to come as close as possible to the standard approach in 
cosmology, we will now rewrite the scalar equations \eqref{correln13}
(which are the cosmologically relevant ones) after performing the
transformation given in \eqn{spatlim15} in order to get the FLRW
metric to the form  
\begin{equation}
^{(\Mbar)}ds^2 = -d\tau^2 + a^2(\tau)\delta_{AB}dy^Ady^B ~~;~~ a(\tau)
  =  \ba(\bt(\tau))\,.
\label{correln23} 
\end{equation} 
Since \eqns{correln13} are \emph{scalar} equations, this
transformation only has the effect of re-expressing all the terms as
functions of the synchronous time $\tau$. Although the transformation 
will change the explicit form of the coordination bivector $\W{a}{j}$,
this change involves only the time coordinate, and in the spatial
averaging limit there is no difference between averages computed
in the VPCs and those computed after the time redefinition.
This again emphasizes the importance of the spatial averaging limit of
spacetime averaging, if we are to succeed operationally in 
explicitly displaying the correlations as corrections to the standard
cosmological equations. The correlation terms in \eqns{correln14} are
therefore still interpreted with respect to the volume preserving
gauge, but are treated as functions of $\tau$. For the scale factor on
the other hand, we have  
\begin{equation}
\ba^3H = \frac{1}{a}\frac{da}{d\tau} \equiv H_{\rm FLRW}
~~;~~ \ba^6\left(\frac{\ddot\ba}{\ba}+3H^2\right) =
\frac{1}{a}\frac{d^2a}{d\tau^2} \,.
\label{correln24}
\end{equation}
Further writing
\begin{equation}
\rho(\tau) = \bar\rho(\bt(\tau))  ~~;~~ p(\tau) = \bar p(\bt(\tau))
\,,  
\label{correln25}
\end{equation}
equations \eqref{correln13} become
\begin{subequations}
\begin{align}
H^2_{\rm FLRW} &= \frac{8\pi G_N}{3}\rho -  \frac{1}{6}\left[
  \Cal{Q}^{(1)} + \Cal{S}^{(1)} \right], 
\label{correln26a} \\&\nonumber\\
\frac{1}{a}\frac{d^2a}{d\tau^2} &= -\frac{4\pi G_N}{3}\left(\rho + 
  3p\right) + \frac{1}{3}\Cal{Q}^{(1)} \,.
\label{correln26b} 
\end{align}
\label{correln26}
\end{subequations}
We emphasize that the quantities $\Cal{Q}^{(1)}$ and $\Cal{S}^{(1)}$,
defined in \eqns{correln14a} and \eqref{correln14b} as correlations in
the \emph{volume preserving} gauge, are to be thought of as functions
of the \emph{synchronous} time $\tau$, where the synchronous time
coordinate itself was defined \emph{after} the spatial averaging. Such
an identification is justified since we are dealing with scalar
combinations of these quantities. Note that $\Cal{Q}^{(1)}$ and
$\Cal{S}^{(1)}$ can be treated independently, apart from the
constraints imposed by \eqn{avgZala63}, which we turn to next. These 
conservation conditions can be decomposed into a scalar part and a
$3$-vector part, given respectively by    
\begin{equation}
\bar v^bC{}^a_{b;a} = 0 ~~;~~ \pi^b_k C{}^a_{b;a} = 0 \,.
\label{correln27}
\end{equation}
In the synchronous gauge \eqref{correln23} for the FLRW metric, the
scalar equation reads
\begin{equation}
\left(\partial_\tau\Cal{Q}^{(1)} + 6H_{\rm FLRW}\Cal{Q}^{(1)}\right) + 
\left(\partial_\tau\Cal{S}^{(1)} + 2H_{\rm FLRW}\Cal{S}^{(1)}\right) = 
0 \,.
\label{correln28}
\end{equation}
We recall that this equation is a consequence of setting the
correlation 3-form and the correlation 4-form to zero, and it relates
the evolution of $\Cal{Q}^{(1)}$ and $\Cal{S}^{(1)}$. The $3$-vector
equation (on imposing the first set of conditions in \eqn{correln15})
simply gives $\partial_\tau C{}^\tau_A = 0$, so that $C{}^\tau_A = 0
=\,$constant, which also implies that $C{}^A_\tau = 0=\,$constant and
hence this equation gives nothing new. (We have used the relations
$C{}^0_0 = C{}^\tau_\tau$, $C{}^0_A = \ba^3 C{}^\tau_A$ and $C{}^A_0 =
(1/\ba^3)C{}^A_\tau$ where $0$ denotes the nonsynchronous time
coordinate $\bt$.) 

The cosmological equations (\ref{correln26}), along with the
constraint equations (\ref{correln22}) and (\ref{correln28}) are the
key results of this section. Subject to the acceptance of the volume
preserving gauge on the underlying manifold ${\cal M}$ they can in
principle be used to study the role of the correction terms resulting
from spatial averaging.

\subsection{Results for an arbitrary gauge choice}
\label{gauge}
\noindent
In this subsection, we will display the results obtained
on assuming that the metric 
\begin{equation}
^{(\Cal{M})}ds^2 = - N^2(t,\rmb{x})dt^2  +
  h_{AB}(t,\rmb{x})dx^Adx^B\,, 
\label{gauge1}
\end{equation}
averages out to the FLRW metric in standard form with a nonsynchronous
time coordinate $t$ in general, to give
\begin{equation}
^{(\Mbar)}ds^2 = -f^2(t)dt^2 + \ba^2(t)\delta_{AB}dx^Adx^B\,.
\label{gauge2}
\end{equation}
In other words, we are assuming that the relations in
\eqn{spatlim-gauge1} hold. Note that the averaging operation is no
longer trivial, although we are still assuming an averaging on domains
corresponding to ``thin'' time slices. We again split the averaged
Einstein equations into scalar equations, and $3$-vector and traceless
$3$-tensor equations. After transforming to the synchronous time
coordinate $\tau$, now defined by 
\begin{equation}
\tau = \int^{t}{f(t^\prime)dt^\prime}\,,
\label{gauge3}
\end{equation}
and again defining $H\equiv(1/\ba)(d\ba/dt)$ and $H_{\rm
  FLRW}\equiv(1/a)(da/d\tau)$ with $a(\tau) = \ba(t(\tau))$, the
modified Friedmann and Raychaudhuri equations read  
\begin{subequations}
\begin{align}
H^2_{\rm FLRW} &= \frac{8\pi G_N}{3}\rho -  \frac{1}{6}\left[
  \tilde{\Cal{P}}^{(1)} + \tilde{\Cal{S}}^{(1)} \right]  \,,
\label{gauge4a}\\ &\nonumber\\
\frac{1}{a}\frac{d^2a}{d\tau^2} &= -\frac{4\pi
  G_N}{3}\left(\rho+3p\right) + \frac{1}{3}\left[
  \tilde{\Cal{P}}^{(1)} + \tilde{\Cal{P}}^{(2)} +
  \tilde{\Cal{S}}^{(2)} \right] \,, 
\label{gauge4b}
\end{align}
\label{gauge4}
\end{subequations}
where the correlation terms are now defined using the relations, 
\begin{subequations}
\begin{align}
\tilde{\Cal{P}}^{(1)} &= \frac{1}{f^2}\left[
  \avg{\bil{\Gamma}{}^A_{0A}\bil{\Gamma}{}^B_{0B}} -
  \avg{\bil{\Gamma}{}^A_{0B}\bil{\Gamma}{}^B_{0A}} - 6H^2\right] \,,   
\label{gauge5a} \\&\nonumber\\
\tilde{\Cal{S}}^{(1)} &=
\avg{\bil{g}^{JK}}\left[\avg{\bil{\Gamma}{}^A_{JB}\bil{\Gamma}{}^B_{KA}}
  - \avg{\bil{\Gamma}{}^A_{JA}\bil{\Gamma}{}^B_{KB}}\right] \,,   
\label{gauge5b} \\&\nonumber\\
\tilde{\Cal{P}}^{(2)} + \tilde{\Cal{P}}^{(1)} &=
-\frac{1}{f^2}\avg{\bil{\Gamma}{}^A_{0A}\bil{\Gamma}{}^0_{00}} -
\avg{\bil{g}^{JK}}\avg{\bil{\Gamma}{}^0_{JA}\bil{\Gamma}{}^A_{0K}} +
\frac{3H}{f^2}\left(\partial_{t}(\ln f) + H\right) \,,
\label{gauge5c} \\&\nonumber\\
\tilde{\Cal{S}}^{(2)} &=
\frac{1}{f^2}\avg{\bil{\Gamma}{}^A_{00}\bil{\Gamma}{}^0_{A0}} +
\avg{\bil{g}{}^{JK}}\avg{\bil{\Gamma}{}^0_{J0}\bil{\Gamma}{}^A_{KA}}
\,.
\label{gauge5d}
\end{align}
\label{gauge5}
\end{subequations}
We emphasize that averaging here refers to spatial averaging. 
Also $\avg{\bil{g}^{JK}} = G^{JK} = (1/\ba^2)\delta^{JK}$, and the 
index $0$ refers to the nonsynchronous time $t$. It is easy to check
using \eqn{correln5}, that $\tilde{\Cal{P}}^{(1)}$ and
$\tilde{\Cal{P}}^{(1)} + \tilde{\Cal{P}}^{(2)}$ correspond to
correlations of (the bilocal extensions of) the extrinsic curvature
with itself and with the time derivative of the lapse
function. $\tilde{\Cal{S}}^{(1)}$ corresponds to correlations between
the Christoffel symbols of the $3$-geometry, 
and $\tilde{\Cal{S}}^{(2)}$ to correlations of the 
spatial derivative of the lapse function with itself and with the
Christoffel symbols of the $3$-geometry. Due to the way we have
defined these correlations, one can also check that when the lapse
function satisfies $N\sqrt{h}=1$ (so that the averaging becomes
trivial), we have $\tilde{\Cal{P}}^{(1)} = {\Cal{Q}}^{(1)}$,
$\tilde{\Cal{S}}^{(1)}={\Cal{S}}^{(1)}$, and $\tilde{\Cal{P}}^{(2)} =
0 = \tilde{\Cal{S}}^{(2)}$, where ${\Cal{Q}}^{(1)}$ and
${\Cal{S}}^{(1)}$ were defined in \eqns{correln14}. The $3$-vector and
traceless $3$-tensor equations become 
\begin{subequations}
\begin{align}
&\frac{1}{f^2}\left[\avg{\bil{\Gamma}{}^0_{0A}\bil{\Gamma}{}^B_{B0}} -
  \avg{\bil{\Gamma}{}^0_{0B}\bil{\Gamma}{}^B_{A0}}\right] 
  +  \avg{\bil{g}^{JK}}\left[
  \avg{\bil{\Gamma}{}^0_{JB}\bil{\Gamma}{}^B_{AK}}  
  - \avg{\bil{\Gamma}{}^0_{JA}\bil{\Gamma}{}^B_{BK}} \right] = 0\,,
\label{gauge6a} \\
&\frac{1}{f^2}\left[\avg{\bil{\Gamma}{}^A_{00}\bil{\Gamma}{}^B_{B0}} -
  \avg{\bil{\Gamma}{}^B_{00}\bil{\Gamma}{}^A_{B0}}\right] 
  +  \avg{\bil{g}^{JK}}\left[
  \avg{\bil{\Gamma}{}^A_{JB}\bil{\Gamma}{}^B_{0K}}  
  - \avg{\bil{\Gamma}{}^A_{J0}\bil{\Gamma}{}^B_{BK}} \right] = 0\,,
\label{gauge6b}\\
&\frac{1}{f^2}\left[\avg{\bil{\Gamma}{}^A_{B0}\bil{\Gamma}{}^m_{0m}} -
 \avg{\bil{\Gamma}{}^A_{m0}\bil{\Gamma}{}^m_{0B}} \right] +
 \avg{\bil{g}^{JK}}
 \left[\avg{\bil{\Gamma}{}^A_{Jm}\bil{\Gamma}{}^m_{KB}} -
 \avg{\bil{\Gamma}{}^A_{JB}\bil{\Gamma}{}^m_{Km}} \right]   
 \nonumber\\
&\ph{\frac{1}{f^2}[\avg{\bil{\Gamma}{}^A_{B0}\bil{\Gamma}{}^m_{0m}} -
 \avg{\bil{\Gamma}{}^A_{m0}\bil{\Gamma}{}^m_{0B}}]} =
 -\frac{1}{3}\delta{}^A_B\left[\tilde{\Cal{P}}^{(2)} +
 \tilde{\Cal{S}}^{(2)} - \tilde{\Cal{S}}^{(1)} - \frac{9H}{f^2}\left(
 H + \frac{1}{3}\partial_{t}(\ln f) \right) \right] \,,
\label{gauge6c}
\end{align}
\label{gauge6}
\end{subequations} 
where the lower case index $m$ in the last equation runs over all 
spacetime indices $0,1,2,3$, with the index $0$ referring to the
nonsynchronous time $t$. It is easy to check that \eqns{gauge6}
reduce to \eqns{correln22} with the choice $N=h^{-1/2}$. The condition
$C{}^a_{b;a}=0$ has the scalar part, 
\begin{equation}
\left(\partial_\tau\tilde{\Cal{P}}^{(1)} + 6H_{\rm
  FLRW}\tilde{\Cal{P}}^{(1)}\right) +
  \left(\partial_\tau\tilde{\Cal{S}}^{(1)} + 2H_{\rm
  FLRW}\tilde{\Cal{S}}^{(1)}\right) + 4H_{\rm
  FLRW}\left(\tilde{\Cal{P}}^{(2)} + \tilde{\Cal{S}}^{(2)} \right) = 0 
  \,,
\label{gauge7}
\end{equation}
while the $3$-vector part, as before, gives nothing new and simply
states $\partial_\tau C{}^\tau_A=0$.

We can now state the main result of this section as
follows : Having assumed that the FLRW spatial sections arise as the
average of some gauge choice with lapse function $N(t,\rmb{x})$,
spatial $3$-metric $h_{AB}(t,\rmb{x})$ and shift vector $N^A$ set to
zero for convenience, we can construct the \emph{scalar} quantities
$C{}^a_b\bar v^b\bar v_a$ and $\pi^b_a C{}^a_b + C{}^a_b\bar
v^b\bar v_a$ which, in coordinates natural to the FLRW metric take
the form,
\begin{equation}
C{}^a_b\bar v^b\bar v_a = \frac{1}{2}\left[\tilde{\Cal{P}}^{(1)} +
  \tilde{\Cal{S}}^{(1)} \right] ~~;~~ \pi^b_a C{}^a_b + C{}^a_b\bar
  v^b\bar v_a = 2\left[\tilde{\Cal{P}}^{(1)} + \tilde{\Cal{P}}^{(2)} +
  \tilde{\Cal{S}}^{(2)}\right] \,,
\label{gauge8}
\end{equation}
with the various quantities being defined in \eqns{gauge5}. These
scalars modify the usual cosmological equations as shown in
\eqns{gauge4}, and are themselves subject to the differential 
conditions \eqref{gauge7}. In addition, for consistency of our
assumptions with the formalism, the underlying inhomogeneous metric is
also subject to the conditions \eqref{gauge6}. 

The combinations on the right hand sides of the relations
\eqref{gauge8} can clearly be treated independently, apart from the
conditions \eqref{gauge7}. Further, since the correlation $2$-form has 
40 independent components $Z^{a\ph{bm}i}_{\ph{a}bm\ph{i}jn}$ after 
imposing all algebraic constraints, and since none of the four
quantities $\tilde{\Cal{P}}^{(1)}$, $\tilde{\Cal{P}}^{(2)}$, 
$\tilde{\Cal{S}}^{(1)}$ and $\tilde{\Cal{S}}^{(2)}$ are trivially
related by these constraints, one can always treat these four
functions independently of each other, subject only to the constraint
in \eqn{gauge7}. In general this constraint may also not be satisfied,
and the correlation tensor may not be independently covariantly
conserved. In chapter 4 we will see the explicit time dependence of
these correlation objects in the perturbative setting in cosmology. 

\section{Comparing the approaches of \B\ and \Z}
\noindent
An important motivation in studying the spatial averaging limit of MG
was to be able to compare its results with those of \B's spatial
averaging. Buchert's averaging is the only approach apart from
Zalaletdinov's MG, which is capable of treating inhomogeneities in a
nonperturbative manner, although it is limited to using only scalar
quantities within a chosen $3+1$ splitting of spacetime. Buchert takes
the trace of the Einstein equations in the \emph{inhomogeneous}
geometry, and averages these inhomogeneous scalar equations. In the
context of Zalaletdinov's MG however, we have used the existence of
the vector field $\bar v^a$ in the FLRW spacetime to construct scalar
equations \emph{after} averaging the full Einstein equations. As far
as observations are concerned, it has been noted by Buchert and
Carfora \cite{buchert2}, that the spatially averaged matter density
$\avgD{\rho}$ defined by Buchert is \emph{not} the appropriate
observationally relevant quantity -- the ``observed'' matter density
(and pressure) is actually defined in a \emph{homogeneous}
space. Since we have done precisely this in \eqn{spatlim-T-ab1}, we
are directly dealing with the appropriate observationally relevant
quantity in the MG framework. 

Another important difference between the two approaches is the
averaging operation itself. Buchert's spatial average, given by
\eqref{avgbuch4}, is different from the averaging operation we
have been using (given by \eqn{spatlim8} using the volume
preserving gauge), which is a limit of a spacetime averaging defined 
using the coordination bivector $\W{a}{j}$. Further, since \B\ only
averages two of the Einstein equations,  a major difference between
the two schemes is the presence of the constraints \eqref{correln22}
on the underlying geometry. These are in general nontrivial and hence
indicate that it is not sufficient to assume that the metric 
of the inhomogeneous manifold averages out to the FLRW form. 

Most importantly though, Buchert's averaging scheme by itself does
not incorporate the concept of an averaged manifold \Mbar\
(although the work of Buchert and Carfora \cite{buchert2} does
deal with $3$-spaces of constant curvature). The question then arises
as to how one should interpret \B's \aD. If one does not wish to
identify \aD\ with the scale factor in FLRW cosmology, one is
compelled to develop a whole new set of ideas in order to try and
compare theory with observation. On the other hand, if one does
(naively) identify \aD\ with the scale factor, comparison with
standard cosmology becomes more convenient, but this introduces a
possible inconsistency since we know from \Z's approach that
additional constraints need to be satisfied. It is not clear how one
should account for these constraints since the non-scalar Einstein
equations are not averaged in \B's approach. 

It is our understanding that the MG approach is a complete and
self-consistent scheme which allows us to meaningfully pose questions
in the averaging paradigm, which are directly interpretable in terms
of standard cosmological ideas. The Buchert approach on the other
hand, is harder to interpret. In the MG approach, there are no
unaveraged shear equations, because the trace of the Einstein
equations has been taken after performing the averaging on the
underlying geometry. Since the averaged geometry is FLRW, its shear is
zero by definition. There is a natural metric on the averaged manifold
by construction, the FLRW metric. The correlations satisfy additional 
constraints, given by \eqns{correln22}. Thus, once a gauge
has been chosen and if one can overcome the computational
complexity of the averaging operation, the cosmological equations
derived by us in the MG approach are complete and ready for
application, without any further caveats.

In spite of these differences, our equations \eqref{correln26} and
\eqref{correln28} for the volume preserving gauge are strikingly
similar to Buchert's effective FLRW equations and their
integrability condition in the dust case; and in the case of
general $N$, the role of Buchert's dynamical backreaction
$\uD{\bar{\Cal{P}}}$ in \eqns{avgbuch17} and \eqref{avgbuch21} is
identical to that of our combination of
$(\tilde{\Cal{P}}^{(2)}+\tilde{\Cal{S}}^{(2)})$ in \eqns{gauge4b}
and \eqref{gauge7}. Concentrating on the volume preserving case,
the structure of the correlation $\Cal{Q}^{(1)}$ is identical to
Buchert's kinematical backreaction $\uD{\Cal{Q}}$ (or
$\uD{\bar{\Cal{Q}}}$ in the general case). The correlation
$\Cal{S}^{(1)}$ appears in place of the averaged $3$-Ricci scalar
$\avgD{\Cal{R}}$ in Buchert's dust equations. This is not
unreasonable since Buchert's $\avgD{\Cal{R}}$ can be thought of as
$\avgD{\Cal{R}} = 6k_\Cal{D}/a_\Cal{D}^2 +\,$corrections, where
$6k_\Cal{D}/a_\Cal{D}^2$ represents the $3$-Ricci scalar on the
averaged manifold which in our case is zero, and hence
$\Cal{S}^{(1)}$ represents the corrections due to averaging.
Further, these similarities are in spite of the fact that our
correlations were defined assuming that a \emph{volume preserving}
gauge averages out to the FLRW $3$-metric in standard form,
whereas Buchert's averaging is most naturally adapted to beginning
with a \emph{synchronous} gauge. This remarkable feature, at least
to our understanding, does not seem to have any deeper meaning --
it simply seems to arise from the structure of the Einstein
equations themselves, together with our assumption
$\rmb{D}_{\bar\Omega}\bZ{a}{b}{i}{j} = 0$. In the absence of this
latter condition, one would have to consider the correlation $3$-
and $4$-forms mentioned earlier, and the structure of the
correlation terms and their ``conservation'' equations would be
far more complicated. In the rest of this thesis we will restrict our
calculations to \Z's approach\footnote{An entirely different outlook
  towards his approach has been emphasized to us by Buchert
  \cite{buchpriv}. According to Buchert, the absence of an averaged
  manifold \Mbar\ is not to be thought of as a `caveat', but as a
  feature deliberately retained `on purpose'. The actual inhomogeneous
  universe is regarded by Buchert as the only fundamental entity, and
  the introduction of an averaged universe is in fact regarded as an
  unphysical and unnecessary approximation. As we mentioned earlier,
  this is probably the most important difference between MG and
  Buchert's approach. In the latter, contact with observations is to
  be made by constructing averaged quantities, such as the scalars
  defined earlier in this section, and by introducing the expansion
  factor $\aD$. The assertion here is that the averaging of
  \emph{geometry}, as discussed in MG or in the Renormalization Group
  approach of Buchert and Carfora \cite{buchert2} is not an
  indispensable step in comparing the inhomogeneous universe with
  actual observations. The need for averaging of geometry is to be
  physically separated from simply looking at effective properties
  (such as the constructed scalars) which can be defined for any
  inhomogeneous metric. Averaging of geometry becomes relevant if (i)
  an observer insists on interpreting the data in a FLRW template
  model, so that (s)he needs a mapping from the actual inhomogeneous
  slice and its average properties to the corresponding properties in
  this template, or (ii) one desires a mock metric, to sort of have a
  thermodynamic effective metric to approximate the real one.}. 

\vskip 0.25in
\noindent
{\large\bf Chapter summary and discussion:}\vskip 0.1in
\noindent
This chapter dealt with \Z's fully covariant framework for averaging
Einstein's equations, named Macroscopic Gravity (MG). Although the
details of this approach are rather involved, its strength lies in
the fact that one can speak in terms of a physically relevant
``averaged metric'' whose behaviour is governed by a set of modified
Einstein equations. We argued that in the context of cosmology, the
standard assumptions regarding the large scale homogeneity and
isotropy of the universe, translate to the requirement of taking
a \emph{spatial averaging limit} of the four-dimensional averaging in
MG. The resulting modified cosmological equations were then compared
with the effective equations derived by Buchert, which we discussed in
the previous chapter. 

We wish to emphasize an issue which is of importance in understanding
the approach we will take in subsequent chapters, which will deal
with \Z's averaging. There is a significant difference between the
original philosophy of the averaging formalism, common to both the
Buchert and Zalaletdinov schemes, and the manner in which we employ
Zalaletdinov's averaging. The original idea as developed by these
authors was to construct a framework which would independently
describe a suitably defined averaged dynamics, with no reference to
the inhomogeneous spacetime whose average leads to this dynamics.
\Z's MG is therefore a new \emph{theory} of gravity describing the
dynamics of an averaged manifold, with no recourse to the underlying
manifold which is described by the usual Einstein equations. The
backreaction $C{}^a_b$ in this approach is actually a new field in the
problem which satisfies its own equations and whose dynamics must be
solved for simultaneously with that of other fields such as the
averaged metric and the averaged energy-momentum tensor for matter. 

Our approach to the backreaction issue is different : We consider it
central to be able to \emph{self-consistently} describe both the
inhomogeneous geometry as well as its averaged counterpart. We find
this necessary since modern cosmology crucially relies on observations
of inhomogeneities around us, and ignoring the evolution of
inhomogeneities when solving for the averaged dynamics does not appear
to be satisfactory. Put another way, when faced with a solution of the
averaged dynamics, we find it essential to answer the question ``which
(if any) inhomogeneous solution could lead to this averaged
homogeneous solution?'' All our subsequent calculations will therefore
focus on solving for the averaged dynamics of specific
inhomogeneities, which we attempt to keep as realistic as
possible. We will start with an application of the spatial averaging
limit of MG, to linear perturbation theory in cosmology, in the
next chapter. The reader is referred to papers by \Z\ and co-workers
in \Cite{zala-results}, for applications of MG as a stand-alone theory
for an averaged manifold. 

\chapter{Backreaction in linear perturbation theory}
In this chapter we will adapt the MG formalism and its spatial
averaging limit to the specific case of linear cosmological
perturbation theory (PT). The motivation behind this excercise is to
verify the self-consistency of cosmological PT in the presence of a
backreaction due to averaging. In other words, we wish to ask whether
cosmological PT is \emph{stable} against the inclusion of dynamical
backreaction terms, or whether a runaway process can render the PT
invalid.

This chapter is organised as follows : In \Sec{app:pertavg} we collect
some useful results from linear cosmological perturbation
theory. \Sec{sec:avgVP} presents 
details of the MG averaging procedure adapted to cosmological PT,
including general expressions for the leading order backreaction
terms, with a discussion of gauge related issues and the definition of
the averaging  operator. The main results are in \Sec{sec:corrscal},
where we derive final expressions for the backreaction, both in real
space and Fourier space, which can be directly utilised in model
calculations. These expressions use a few simplifying
restrictions which can be lifted if necessary in a completely
straightforward manner. \Sec{sec:exs} contains example calculations in
first order PT, which show that the magnitude of the backreaction is,
as expected, negligible compared to the homogeneous energy density of
matter in the radiation dominated era and for a significant part of
the matter dominated era. Throughout the chapter, a prime refers to a
derivative with respect to conformal time unless stated otherwise, and
we will assume that the metric of the universe is a perturbation
around the FLRW metric given by 
\be
ds^2 = a^2(\eta)\left( -d\eta^2 + \gamma_{AB}dx^Adx^B \right)\,.
\label{FLRW-metric}
\ee
Here $a$ is the scale factor and $\eta$ is the conformal time
coordinate related to cosmic time $\tau$ by the differential relation 
\be
d\tau = a(\eta)d\eta
\label{conf-time}
\ee
In \eqn{FLRW-metric} we have allowed the spatial metric to have the
general form $a^2\gamma_{AB}$ where $\gamma_{AB}$ is the metric of a
$3$-space of constant curvature. For the calculations in this chapter
we will assume a flat FLRW background in coordinates such that
$\gamma_{AB}=\delta_{AB}$; however for future reference we shall
present certain expressions in terms of the more general spatial
metric. Defining 
\be
\Cal{H} = \frac{1}{a}\frac{da}{d\eta} \equiv \frac{a^\prime}{a} \,, 
\label{FLRW-H-1}
\ee
and setting $f=a$ in \eqns{gauge5}, the correction terms in the
modified Friedmann and Raychaudhuri equations \eqref{gauge4} now read  
\begin{subequations}
\begin{align}
\Cal{P}^{(1)} &= \frac{1}{a^2}\left[
  \avg{\wti{\Gamma}{}^A_{0A}\wti{\Gamma}{}^B_{0B}} -
  \avg{\wti{\Gamma}{}^A_{0B}\wti{\Gamma}{}^B_{0A}} - 6\Cal{H}^2\right]
  \,, 
\label{MG8a} \\&\nonumber\\
\Cal{S}^{(1)} &=
\avg{\wti{g}^{JK}}\left[\avg{\wti{\Gamma}{}^A_{JB}\wti{\Gamma}{}^B_{KA}}
  - \avg{\wti{\Gamma}{}^A_{JA}\wti{\Gamma}{}^B_{KB}}\right] \,,
\label{MG8b} \\&\nonumber\\
\Cal{P}^{(2)} + \Cal{P}^{(1)} &=
-\frac{1}{a^2}\avg{\wti{\Gamma}{}^A_{0A}\wti{\Gamma}{}^0_{00}} -
\avg{\wti{g}^{JK}}\avg{\wti{\Gamma}{}^0_{JA}\wti{\Gamma}{}^A_{0K}}
+ \frac{6\Cal{H}^2}{a^2} \,, 
\label{MG8c} \\&\nonumber\\
\Cal{S}^{(2)} &=
\frac{1}{a^2}\avg{\wti{\Gamma}{}^A_{00}\wti{\Gamma}{}^0_{A0}} +
\avg{\wti{g}{}^{JK}}\avg{\wti{\Gamma}{}^0_{J0}\wti{\Gamma}{}^A_{KA}}
\,, \label{MG8d}
\end{align}
\label{MG8}
\end{subequations}
where we have dropped the tildes for convenience. The averaging in
\eqns{MG8} is assumed to be a spatial averaging in an unspecified
spatial slicing in the inhomogeneous manifold \Cal{M}; in
\Sec{sec:avgVP} we will specify the averaging procedure more exactly. 
In addition, the ``cross-correlation'' constraints \eqref{gauge6} with
$f=a$, $t=\eta$ and $H=\Cal{H}$ must also be satisfied by the
inhomogeneities. Note that we are not imposing the conservation
condition \eqref{avgZala63}. 

Before we move on to deriving formulae for the correlation terms
\eqref{MG8} in terms of perturbation functions in the metric, there is
one issue which merits discussion. The cosmological perturbation
setting, together with the paradigm of averaging, presents us with a
rather peculiar situation. On the one hand, the time evolution of the
scale factor is needed in order to solve the equations satisfied by
the perturbations. Indeed, the standard practice is to fix the time
evolution of the background once and for all, and to use this in
solving for the evolution of the perturbations. On the other hand, the
evolution of the \emph{perturbations} (i.e. -- the inhomogeneities) is
needed to compute the correlation terms appearing in
\eqns{gauge4}. Until these terms are known, the behaviour with time of
the scale factor cannot be determined; and until we know the scale
factor as a function of time, we cannot solve for the
perturbations. Note that this is a generic feature independent of all
details of the averaging procedure. 

It would appear therefore, that we have reached an impasse. To clear
this hurdle, one can try the following iterative approach :
Symbolically denote the background as $a$, the inhomogeneities as
$\vphi$, and the correlation objects as $C$. Note that $a$, $\vphi$
and $C$ all refer to functions of time. We start with a chosen
background, say a standard flat FLRW background with radiation,
baryons and cold dark matter (CDM), and solve for the perturbations in
the usual way, \emph{without} accounting for the correlation terms
$C$. In other words, for this ``zeroth iteration'', we artificially
set $C$ to zero and obtain $a^{(0)}$ and $\vphi^{(0)}$ using the
standard approach (see e.g. \Cite{dodelson}). Clearly, since the
``true'' background (say $a_\ast$) satisfies \eqns{gauge4} \emph{with
  a nonzero} $C$, we have in general $a^{(0)}\neq a_\ast$. Now, using 
the solution $\vphi^{(0)}$, we can calculate the zeroth iteration
correlation objects $C^{(0)}$ by applying the prescription to be
developed later in this chapter. As a first correction to the solution
$a^{(0)}$, we now solve for a new background $a^{(1)}$, with the
\emph{known functions} $C^{(0)}$ acting as sources in
\eqns{gauge4}. This first iteration will then yield a solution
$\vphi^{(1)}$ for the inhomogeneities, and hence a new set of
correlation terms $C^{(1)}$, and this procedure can be repeatedly
applied. Pictorially,  
\be
a^{(0)} \longrightarrow \vphi^{(0)} \longrightarrow C^{(0)}
\longrightarrow a^{(1)} \longrightarrow \vphi^{(1)} \longrightarrow
\ldots 
\label{MG10}
\ee
As for convergence, if perturbation theory \emph{is} in fact a good
approximation to the real universe, then one can expect that the
correlation terms will tend to be small compared to other background
objects, and will therefore not affect the background significantly at
each iteration, leading to rapid convergence. On the other hand, if
the correlation terms are large, this procedure may not converge and
one might expect a breakdown of the perturbative picture itself. We
will see that in the linear regime of cosmological perturbation
theory, the correlation terms do in fact remain negligibly small.  

\section{Metric perturbations in cosmology}
\label{app:pertavg}
\noindent
For ready reference, in this subsection we present expressions for the
metric, its inverse, and the Christoffel connection in \emph{first
  order} cosmological PT, in an arbitrary, unfixed gauge. The notation
we use is similar to that used in \Cite{bmr07}. We will also give
expressions for the first order gauge transformations of the
perturbation functions (see e.g. \Cite{bruni}). 
The first order perturbed FLRW metric in an arbitrary gauge and in
terms of conformal time $\eta$, can be written as 
\be
ds^2 = a^2(\eta)\left[ -(1+2\vphi)d\eta^2 + 2\om_Adx^Ad\eta +
\left((1-2\psi)\gamma_{AB} +  \chi_{AB}\right)dx^Adx^B \right]\,.
\label{linPT1}
\ee
The functions $\vphi$ and $\psi$ are scalars under spatial coordinate 
transformations. The functions $\om_A$ and $\chi_{AB}$ can be
decomposed as follows
\begin{align}
\om_A = \p_A\om + \com_A ~~;~~
\chi_{AB} = D_{AB}\chi+2\nabla_{(A}\cchi_{B)} + \cchi_{AB}\,,
\label{linPT2}
\end{align}
where the parentheses indicate symmetrization; $D_{AB}$ is the
tracefree second derivative defined by 
\be
D_{AB} \equiv \nabla_A\nabla_B - (1/3)\gamma_{AB}\nabla^2 ~~;~~
\nabla^2 \equiv \gamma^{AB}\nabla_A\nabla_B\,,
\label{linPT3}
\ee
with $\nabla_A$ the covariant spatial derivative compatible with
$\gamma_{AB}$; and $\com_A$, $\cchi_A$ and $\cchi_{AB}$ satisfy
\be
\nabla_A\com^A = 0 = \nabla_A\cchi^A ~~;~~ \nabla_A\cchi{}^A_B = 0 = 
\cchi{}^A_A\,, 
\label{linPT4}
\ee
where spatial indices are raised and lowered using $\gamma_{AB}$ and
its inverse $\gamma^{AB}$. From their definitions it is clear that
$\vphi$, $\psi$, $\om$ and $\chi$ each correspond to one scalar degree
of freedom, the transverse $3$-vectors $\com_A$ and $\cchi_A$ each
correspond to two functional degrees of freedom, and the transverse
tracefree $3$-tensor $\cchi_{AB}$ corresponds also to two functional
degrees of freedom. This totals to $10$ degrees of freedom, of which
$4$ are coordinate degrees of freedom which can be arbitrarily fixed,
which is what one means by a gauge choice. For example, the
\emph{conformal Newtonian} or \emph{longitudinal} or \emph{Poisson
  gauge} \cite{mukhanov,bruni} is defined by the
conditions 
\be
\om = 0 = \chi~~;~~ \cchi^A=0\,.
\label{linPT5}
\ee
For the metric \eqref{linPT1} we have at first order,
\be
\sqrt{-{\rm det}\,g} = a(\eta)^4\left(1+ \vphi - 3\psi\right)\,.
\label{linPT6}
\ee
The inverse of metric \eqref{linPT1}, correct to first order, has the
components 
\begin{align}
&g^{00} = -\frac{1}{a^2}(1-2\vphi) ~~;~~ g^{0A} = \frac{1}{a^2}\om^A
\,, \nonumber\\
&g^{AB} = \frac{1}{a^2}\left((1+2\psi)\gamma^{AB} - \chi^{AB} \right)\,.
\label{linPT7}
\end{align}
With $\Cal{H} = (a^\prime/a)$, the first order accurate Christoffel
symbols are  
\begin{align}
&\G^0_{00} = \Cal{H} + \vphi^\prime ~~;~~ \G^0_{0A} = \p_A\vphi +
  \Cal{H}\om_A ~~;~~ \G^A_{00} = \p^A\vphi + \om^{A\prime}
  + \Cal{H}\om^A\,,  \nonumber\\ 
&\G^0_{AB} = \left(\Cal{H}-\psi^\prime-2\Cal{H}(\vphi+\psi)
  \right)\gamma_{AB} - \nabla_{(A}\om_{B)}
  + \frac{1}{2}\chi^\prime_{AB} + 
  \Cal{H}\chi_{AB}\,, \nonumber\\ 
&\G^A_{0B} = \left(\Cal{H} - \psi^\prime\right)\delta{}^A_B +
  \frac{1}{2}\left(\nabla_B\om^A - \nabla^A\om_B \right)
 + \frac{1}{2}\chi^{A\prime}_B\,, \nonumber\\ 
&\G^A_{BC} = {}^{(3)}\bar\G^A_{BC} - \left(\delta{}^A_B\p_C\psi +
  \delta{}^A_C\p_B\psi - \gamma_{BC}\p^A\psi \right) 
 -\Cal{H}\om^A\gamma_{BC} +
  \frac{1}{2}\left(\nabla_C\chi^A_B + \nabla_B\chi^A_C -
  \nabla^A\chi_{BC} \right)\,,  
\label{linPT8}
\end{align}
where ${}^{(3)}\bar\G^A_{BC}$ denotes the Christoffel connection
associated with the homogeneous $3$-metric $\gamma_{AB}$. 

\subsection{Gauge transformations}
\noindent
While the concept of gauge
transformations can be described in a rather sophisticated language
using pullback operators between manifolds \cite{bruni}, for our
purposes it suffices to implement a gauge transformation using the
simpler notion of an infinitesimal coordinate transformation (also
known as the ``passive'' point of view). Hence, denoting the
coordinates and perturbation functions in the new gauge with a tilde
(i.e. $\ti{x}^a$, $\ti{\vphi}$, $\ti{\com_A}$, and so on), we have  
\be
\ti{x}^a = x^a + \xi^a(x) ~~;~~ x^a = \ti{x}^a - \xi^a \,, 
\label{linPT9}
\ee
where the infinitesimal $4$-vector $\xi^a$ can be decomposed as
\be
\xi^a = \left(\xi^0, \xi^A\right) = \left( \alpha, \p^A\beta + d^A
\right)\,, 
\label{linPT10}
\ee
where $\alpha$ and $\beta$ are scalars and $d^A$ is a transverse
$3$-vector satisfying $\nabla_Ad^A = 0$.

It is then easy to show that if this transformation is assumed to
change the metric \eqref{linPT1} by changing only the perturbation
functions but leaving the background intact (a so-called ``steady''
coordinate transformation), then the old perturbations and the new are 
related by \cite{bruni}
\begin{align}
&\vphi = \ti{\vphi} + \alpha^\prime + \Cal{H}\alpha \,,\nonumber\\
&\psi = \ti{\psi} - \frac{1}{3}\nabla^2\beta - \Cal{H}\alpha
\,,\nonumber\\ 
&\om = \ti{\om} - \alpha + \beta^\prime \,,\nonumber\\
&\com^A = \ti{\com^A} + d^{A\prime} \,,\nonumber\\
&\chi = \ti{\chi} + 2\beta \,,\nonumber\\
&\cchi^A = \ti{\cchi^A} + d^A \,,\nonumber\\
&\cchi_{AB} = \ti{\cchi_{AB}}\,.
\label{linPT11}
\end{align}
The last equality shows that the transverse tracefree tensor
perturbations are gauge invariant. They correspond to gravitational
waves.

\section{The Averaging Operation and Gauge Related Issues}
\label{sec:avgVP}
\noindent
In this section, we will describe the details of the MG (spatial)
averaging procedure adapted to the setting of cosmological PT. 

\subsection{Volume Preserving (VP) Gauges and the Correlation Scalars} 
\label{sec:avgVP-A}
\noindent
It will greatly simplify the discussion if we start with symbolic
calculations which allow us to see the broad structure of the objects
we are after. Since the correlation objects in \eqns{gauge4} depend
only on derivatives of the metric, we will primarily deal with metric 
fluctuations; matter perturbations will only come into play when 
solving for the actual dynamics of the system. Before dealing with the
issue of which gauge to choose in order to set the condition
\eqref{MG-avgcond} with the average connection taken to be the FLRW
one, we will show that irrespective of this choice, the leading order
contribution to the correlations requires knowledge of only
\emph{first order} perturbation functions.  

We will use the following symbolic notation :
\begin{itemize}
\item Inhomogeneous connection: \G 
\item FLRW connection: \GF 
\item Perturbation in the connection : $\dG \equiv \G -
  \GF = \dg + \dG^{(2)}  + \ldots$ \vskip .05in  
\item Coordination bivector : $W \equiv 1 + \dW = 1 + \dw +
  \dW^{(2)} + \ldots$ 
\item Bilocal extension of the connection : \Gb
\item Inhomogeneous part of the bilocal extension of the connection : $\dGb
  \equiv \Gb - \GF = \dgb + \dGb^{(2)} + \ldots$
\item Correlation object : $C$
\end{itemize}
The integer superscripts denote the order of perturbation. The form of
the coordination bivector arises from the fact that in perturbation
theory, \emph{in the spatial averaging limit}, a transformation from
an arbitrary gauge to a VP one can be achieved by an infinitesimal
coordinate transformation. By a VP gauge we mean a gauge in which the
metric determinant is independent of the \emph{spatial coordinates} to
the relevant order in PT, but may be a function of time. It can be
shown that such a function of time (which will typically be some power
of the scale factor), is completely consistent with all definitions
and requirements of MG in the spatial averaging limit. An easy way of
seeing this is to note that in any averaged quantity, the metric
determinant appears in two integrals, one in the numerator and the
other in the denominator (which gives the normalising volume). In the
``thin time slicing'' approximation we are using to define the
averaging, any overall time dependent factor in the metric determinant
therefore cancels out. Also, a fully volume preserving coordinate
system can clearly be obtained from any VP gauge as defined above, by
a suitable rescaling of the time coordinate. It is not hard to show
that in the thin time slicing approximation, this gives the same
coordination bivector \Wxx{a}{b}\ as the VP gauge definition above.

To see that first order perturbations are sufficient to calculate
$C$ to leading order, we only have to note that the
background connection \GF\ satisfies
\be
\avg{\GF} = \GF \,,
\label{avgVP1}
\ee
and that the structure of the correlation is $C= \avg{\Gb^2} -
\avg{\Gb}^2$, which then leads to
\be
C  = \avg{\dGb^2} - \avg{\dGb}^2\,,
\label{avgVP2}
\ee
which is exact. Clearly, the correlation is quadratic in the
perturbation as expected, and hence to leading order, \dGb\ above can
be replaced by \dgb. 

\eqns{avgVP1} and \eqref{avgVP2} treat the averaging operation at a
conceptual level only. To make progress however, we also need to
prescribe how to \emph{practically} impose the averaging assumption 
\be
\avg{\Gb} = \GF ~~\text{i.e.}~~ \avg{\dGb}=0\,,
\label{avgVP3}
\ee
in any given perturbative context. This requires some discussion
since, for example, the bilocal extension of the connection
\Gb\ has the structure 
\be
\Gb = \Wi\G W^2 + \Wi(\p + W\pp)W \,,
\label{avgVP4}
\ee
where \p\ is a derivative at $x$ and \pp\, a derivative at
$x^\prime$ (see \eqn{avgZala35}). The actual MG averaging operation in
general is therefore a rather involved procedure. Additionally, it is
also necessary to address certain gauge related issues. 

To clarify the situation, let us start with a fictitious setting in
which the geometry has \emph{exactly} the flat FLRW form, with no
physical perturbations. Clearly, if we work
in the standard comoving coordinates in which the metric $\gamma_{AB}$
of \eqn{FLRW-metric} is simply $\gamma_{AB}=\delta_{AB}$, then since
these coordinates are volume preserving in the sense described above,
the coordination bivector becomes trivial. The averaging involves a
simple integration over $3$-space, and we can easily see that
\eqn{avgVP1} is \emph{explicitly} recovered. 

Now suppose that we perform an infinitesimal coordinate
transformation, \emph{after} imposing \eqn{avgVP1}. Since the
averaging operation is covariant, then from the point of view of a 
general coordinate transformation, \emph{both} sides of \eqn{avgVP1}
will be affected in the same way. However, suppose that we had
performed the transformation \emph{before} imposing \eqn{avgVP1}.
In the language of cosmological PT, we would then be dealing with some
``pure gauge'' perturbations around the fixed, spatially homogeneous
background. If we did not know that these perturbations were pure
gauge, we might naively construct the nontrivial coordination bivector
for this metric, compute the bilocal extension of the connection
according to \eqn{avgVP4} and try to impose \eqn{avgVP3}. This would
be incorrect since these perturbations were arbitrarily generated and
need not average to zero (for example they could be positive definite
functions). In order to maintain consistency, it is then necessary to
ensure in practice that the averaging condition \eqref{avgVP3} is
applied only to gauge invariant fluctuations (which is rather obvious
in hindsight).   

There is another problem associated with the structure of the
coordination bivector, even when there \emph{are} real, gauge
invariant inhomogeneities present. Note from \eqns{avgZala24} and
\eqref{avgZala26} that the coordination bivector has the structure  
\be
W = \left .\frac{\p x}{\p x_V}\right|_{x^\prime} \left .\frac{\p
  x_V}{\p x}\right|_x \,,
\label{avgVP5}
\ee
where $x$ denotes the coordinates we are working in and $x_V$ a set of
VPCs. In perturbation theory (in the spatial averaging limit) we will
have, at leading order,
\be
x = x_V - \xi~~;~~ x_V = x + \xi\,,
\label{avgVP6}
\ee
where $\xi$ symbolically denotes an infinitesimal $4$-vector defining
the transformation, and hence
\be
(\p x_V)/(\p x) = 1 + \p\xi\,,
\label{avgVP7}
\ee
and so on, which gives us
\be
W = 1 - (\p\xi)|_{x^\prime} + (\p\xi)|_x + \ldots = 1 + \dw + \ldots \,.
\label{avgVP8}
\ee
Now when we compute a quantity such as \avg{\GF\dw}\, which appears in
the expression \eqref{avgVP4} for \avg{\Gb}\,, we will be left with a
fluctuating ($\vec{x}$-dependent) term of the form $\GF(\avg{\p\xi} -
\p\xi)$, where $\vec{x}$ denotes the $3$ spatial coordinates. Hence if
we try to impose \eqn{avgVP3} we will ultimately be left with
equations of the type  
\be
\avg{f}(\vec{x}) - f(\vec{x}) = 0\,,
\label{avgVP9}
\ee
for some functions derived from the inhomogeneities which we have
collectively denoted $f$. In other words, consistency would seem to
demand that the inhomogeneities vanish in this coordinate system,
which is of course not desirable. 

It therefore appears that we are forced to impose \eqn{avgVP3}
\emph{in a volume preserving gauge}, since by definition, only in such
a gauge will we have $W=1$ exactly. We emphasize that this is a
purely practical aspect related to defining the averaging operation,
and is completely decoupled from, e.g. the choice of gauge made when
studying the time evolution of perturbations. We are in no way
breaking the usual notion of gauge invariance by choosing an averaging
operator. The conditions \eqn{avgVP9} now reduce to the form  
\be
\avg{f_{VPC}}(\vec{x}) = 0\,,
\label{avgVP10}
\ee
which are far more natural than \eqn{avgVP9}. The averaging condition
is now unambiguous, \emph{but depends on a choice of the VP gauge
  which defines the averaging operation},
an issue we shall discuss in the next subsection. For now, all we can
assert is that this VP gauge must be such that in the \emph{absence}
of gauge invariant fluctuations, it must reduce to the standard
comoving (volume preserving) coordinates of the background geometry as
in \eqn{FLRW-metric}. This of course is simply the statement that the
VP gauge must be \emph{well defined} and must not contain any residual
degrees of freedom.

The averaging operation now takes on an almost trivial form as we have
seen in chapter 3 -- to leading order, for any quantity
$f(\eta,\vec{x})$ (with or without indices), the spatial average of
$f$ in a VP gauge is given by
\be
\avg{f}(\eta,\vec{x}) = \frac{1}{V_L}\int_{\Cal{V}(\vec{x})}{ d^3y
  f(\eta,y)} \,, 
\label{avgVP11}
\ee
where the integral is over a spatial domain $\Cal{V}(\vec{x})$ with a 
constant volume $V_L$. The spatial coordinates are the comoving
coordinates of the background metric, and at leading order the
boundaries of $\Cal{V}(\vec{x})$ can be specified in a straightforward
manner as, e.g., 
\be
\Cal{V}(\vec{x}) = \{\vec{y}~|~x^A-L/2 < y^A < x^A+L/2, A=1,2,3. \}\,,
\label{avgVP12}
\ee
where $L$ is a comoving scale over which the averaging is performed
(in which case $V_L=L^3$). The averaging definition can be written
more compactly in terms of a window function $W_L(\vec{x},\vec{y})$ as  
\begin{align}
&\avg{f}(\eta,\vec{x}) = \int{d^3y W_L(\vec{x},\vec{y})f(\eta,\vec{y})}
 ~~;~~  \int{d^3y W_L(\vec{x},\vec{y})} = 1\,,
\label{avgVP13}
\end{align}
where $W_L(\vec{x},\vec{y})$ vanishes everywhere except in the region 
$\Cal{V}(\vec{x})$, with the integrals now being over all space. This
expression will come in handy when working in Fourier space, as we
shall do in later sections. 

A couple of comments are in order at this stage. Firstly, we have not
specified the magnitude of the averaging scale $L$. The general
philosophy is that this scale must be large enough that a single
averaging domain encompasses several realisations of the random
inhomogeneous fluctuations, and small enough that the observable
universe contains a large number of averaging domains. However, as we
will show later in \Sec{sec:corrscal}, if one is ultimately interested
in quantities which are formally averaged over an ensemble of
realisations of the universe (as is usually done in interpreting
observations), then the actual value of the averaging scale becomes
irrelevant. 

This brings us to the second issue. The above discussion is valid only
in the situation where there are no fluctuations at arbitrarily large
length scales, since in the presence of such fluctuations the
averaging condition \eqref{avgVP3} loses meaning (in such a
situation it would be impossible to isolate the background from the
perturbation by an averaging operation on any finite length
scale). Indeed, we shall see a manifestation of this restriction in
\Sec{sec:corrscal}, where the correlation scalars will be seen to
diverge in the presence of a nonzero amplitude at arbitrarily large
scales, of the power spectrum of metric fluctuations. 

We will end this subsection by explicitly writing out the averaging
condition in an ``unfixed VP'' gauge, to be defined below, and also
writing the correlation terms appearing in \eqn{gauge4}, in this
gauge.  As we can see from \eqn{linPT6}, the basic condition to be
satisfied by  a VP gauge is
\be
\ti{\vphi} = 3\ti{\psi}\,.
\label{avgVP14}
\ee
Hereafter, all VP gauge quantities will be denoted using a
tilde. This should not be confused with the similar notation that was
used so far for the bilocal extension, which will
not be needed in the rest of the chapter. $\ti{\vphi}$ and $\ti{\psi}$
are the scalar potentials appearing in the perturbed FLRW metric
\eqref{linPT1}. The single condition \eqref{avgVP14} leaves $3$ degrees
of freedom to be fixed, in order to completely specify the VP gauge
one is working with. The MG formalism by itself does not prescribe a
method to choose a particular VPC system; in fact this freedom of
choice of VPCs is an inherent part of the formalism. We shall return to
this issue in the next subsection. For now we define the ``unfixed VP 
(uVP) gauge'' by the single requirement \eqref{avgVP14}, with $3$
unfixed degrees of freedom, and present the expressions for the
averaging condition and the correlation scalars, with this choice. 

It is straightforward to determine the consequences of requiring
\eqn{MG-avgcond} to hold, with the right hand side corresponding to
the FLRW connection in conformal coordinates, and remembering that the
coordination bivector (in the spatial averaging limit) is now simply a
Kronecker delta. Together with some additional reasonable
requirements, namely
\be
\avg{\nabla^2s} = 0 = \avg{\nabla^2\p_As}\,,
\label{avgVP15}
\ee
for any scalar $s(\eta,\vec{x})$, the averaging condition in the
uVP gauge reduces to
\begin{align}
&\avg{\ti{\psi}} = 0 ~~;~~ \avg{\p_A\ti{\psi}} = 0 =
  \avg{\ti{\psi}^\prime}\,, \nonumber\\ 
&\avg{\ti{\om}_A} = 0 = \avg{\ti{\om}^\prime_A} ~~;~~
  \avg{\ti{\chi}^\prime_{AB}} = 0\,, \nonumber\\ 
&\avg{\nabla_C\ti{\chi}^A_B} + \avg{\nabla_B\ti{\chi}^A_C} -
  \avg{\nabla^A\ti{\chi}_{BC}} = 0\,, \nonumber\\ 
&\avg{\nabla_A\ti{\om}_B} = \avg{\nabla_B\ti{\om}_A} =
  \Cal{H}\avg{\ti{\chi}_{AB}} \,,
\label{avgVP-avgcond}
\end{align}
where we have used the expressions in \eqn{linPT8} with the uVP
condition \eqref{avgVP14}. We will also make the additional reasonable
requirement that 
\be
\avg{\ti{\chi}_{AB}} = 0\,,
\label{avgVP16}
\ee
using which it is easy to see that the perturbed FLRW metric
\eqref{linPT1} and its inverse \eqref{linPT7}, in the uVP 
gauge, \emph{both} on averaging reduce to their respective homogeneous
counterparts, namely
\be
\avg{g_{ab}} = g_{ab}^{(FLRW)} ~~;~~ \avg{g^{ab}} = g^{ab}_{(FLRW)}\,.  
\label{avgVP-metricavg}
\ee
Using these results, the expressions \eqref{MG8} simplify to give, in
the uVP gauge,
\begin{subequations}
\begin{align}
\Cal{P}^{(1)} &= \frac{1}{a^2} \left[ 6\avg{(\ti{\psi}^\prime)^2} +
    \avg{\nabla_{[A}\ti{\om}_{B]} \nabla^{[A}\ti{\om}^{B]}} -
    \frac{1}{4}\avg{\ti{\chi}_{AB}^\prime 
      \ti{\chi}^{AB\prime}} \right] \,, \label{avgVP-corrscal-a}\\ 
&\nonumber\\
\Cal{S}^{(1)} &= \frac{1}{a^2} \left[ -10\avg{\p_A\ti{\psi}
      \p^A\ti{\psi}} - 2\avg{\p_A\ti{\psi}\nabla_B\ti{\chi}^{AB}}
  + \frac{1}{4}
    \avg{\nabla^B\ti{\chi}^{AC}\left(2\nabla_A\ti{\chi}_{BC}
    - \nabla_B\ti{\chi}_{AC}\right) }  \right]
  \,,\label{avgVP-corrscal-b}\\ 
&\nonumber\\
\Cal{P}^{(1)} + \Cal{P}^{(2)} &= \frac{1}{a^2} \left[
    6\avg{(\ti{\psi}^\prime)^2} -
    24\Cal{H}\avg{\ti{\psi}^\prime\ti{\psi}} -
    \avg{\ti{\psi}^\prime\nabla^2\ti{\om}} +
    \frac{1}{2}\avg{\ti{\chi}^\prime_{AB}\nabla^A\ti{\om}^B}
    - \frac{1}{4}
    \avg{\ti{\chi}^\prime_{AB}\left(\ti{\chi}^{AB\prime} +
      2\Cal{H}\ti{\chi}^{AB} \right)
    }\right]\,, \label{avgVP-corrscal-c}\\ 
\Cal{S}^{(2)} &= \frac{1}{a^2} \left[
    3\avg{\ti{\om}^{A\prime}\p_A\ti{\psi}} +
    \Cal{H}\avg{\ti{\om}^A\ti{\om}^\prime_A} \right]
  \,, \label{avgVP-corrscal-d} 
\end{align}
\label{avgVP-corrscal}
\end{subequations}
where square brackets denote antisymmetrization.

\subsection{Choice of VP Gauge} 
\label{sec:ginvar}
\noindent
In this subsection we will prescribe a choice for the VP gauge which 
defines the averaging operation. In general, the class of volume
preserving coordinate systems for any spacetime, is very large (see
\Cite{mars} for a detailed characterisation). We have so far managed
to pare it down by requiring that the VP gauge we choose should reduce
to the standard FLRW coordinates in the absence of fluctuations. It
turns out to be somewhat difficult to go beyond this step, since there
does not appear to be any unambiguously clear guiding principle
governing this choice. We will therefore motivate a choice for the VP
gauge based on certain details of cosmological PT which one knows from
the standard treatments of the subject.

In particular, we shall make use of certain nice properties of the
conformal Newtonian or longitudinal or Poisson
gauge, which is defined by the conditions \eqref{linPT5}
\cite{bruni} (henceforth we shall refer to this gauge as the
cN gauge for short). Since this gauge is well defined and has no
residual degrees of freedom, all the nonzero perturbation functions in
the cN gauge, namely $\vphi$, $\psi$, $\com_A$ and $\cchi_{AB}$ in the
notation of Appendix C, are equal to gauge invariant
objects. This is trivially true for $\cchi_{AB}$, as seen in the last
equation in \eqref{linPT11}. For the rest, note that in any arbitrary
unfixed gauge, the following combinations are gauge invariant at first
order 
\begin{align}
&\Phi_B = \vphi + \frac{1}{a}\p_\eta\left[a\left( \om -
  \frac{1}{2}\chi^\prime \right)\right] \,,\nonumber\\
&\Psi_B = \psi - \Cal{H}\left(\om - \frac{1}{2}\chi^\prime \right) +
\frac{1}{6}\nabla^2\chi\,,\nonumber\\
&{\hat V}_A = \com_A - \cchi^\prime_A\,,
\label{ginvar2}
\end{align}
which can be easily checked using \eqns{linPT11}, and in the cN gauge, 
$\om$, $\chi$ and $\cchi_A$ all vanish. Here $\Phi_B$ and $\Psi_B$ are
the Bardeen potentials \cite{bardeen} (upto a sign), and $\Psi_B$ in
particular has the physical interpretation of giving the gauge
invariant \emph{curvature perturbation}, which is the quantity on
which initial conditions are imposed post inflation
\cite{liddle-lyth}.  

Additionally, it is also known that the cN gauge \emph{for the metric}
remains stable even during structure formation, when matter
inhomogeneities have become completely nonlinear\footnote{See
  \Cite{wald} for an intuitive description of why this is so. We will
  also see an explicit demonstration in a toy model of structure
  formation in chapter 5.}. We believe that this is a strong argument
in favour of using the cN gauge to define a VP gauge which will then
define the averaging operation in the perturbative context. This will
ensure that this ``truncated'' averaging operation, defined for first
order PT, will remain valid \emph{at leading order} even during the
nonlinear epochs of structure formation. 

To implement this in practice, consider a transformation from the cN
gauge to the uVP gauge defined by \eqn{avgVP14}. The transformation
equations \eqref{linPT11} reduce to  
\begin{align}
&\alpha^\prime + 4\Cal{H}\alpha + \nabla^2\beta = \vphi - 3\psi
  \,,\nonumber\\ 
&\ti{\psi} = \frac{1}{3}\vphi - \alpha^\prime - \Cal{H}\alpha
  \,,\nonumber\\ 
&\ti{\om} = \alpha - \beta^\prime \,,\nonumber\\
&\ti{\com^A} = \com^A - d^{A\prime} \,,\nonumber\\
&\ti{\chi} = - 2\beta \,,\nonumber\\
&\ti{\cchi^A} = - d^A \,,\nonumber\\
&\ti{\cchi_{AB}} = \cchi_{AB}\,.
\label{ginvar3}
\end{align}
Recall that to completely specify a VP gauge, we need to fix $3$
degrees of freedom in the uVP gauge. Our requirement regarding the
``well defined''-ness of the VP gauge, forces us to set $d^A=0$, and
to choose $\alpha$ and $\beta$ such that they vanish in the case where
$\vphi=0=\psi$. 

This has fixed $2$ degrees of freedom, in addition to the condition
\eqref{avgVP14} which is just the definition of the uVP gauge, and has
hence not yielded a uniquely specified VP gauge. To do this, we shall
make the following additional requirement. Since we are dealing with a
spatial averaging, it seems reasonable to require that the VP gauge
being used to define the averaging, should be ``as close as possible''
to the cN gauge in terms of \emph{time slicing}, and for this reason
we shall set the function $\alpha$ to zero.  
To summarize, the VP gauge chosen is defined in terms of the gauge
transformation functions $\xi^a = (\alpha,\p^A\beta+d^A)$ between the
cN gauge and the VP gauge, by the following relations
\be
\alpha = 0 = d^A \,,
\label{ginvar-VP1}
\ee
and
\begin{subequations}
\begin{align}
&\ti{\vphi} = 3\ti{\psi} = \vphi\,,\label{ginvar-VP2-a}\\ 
&\nabla^2\beta = \vphi - 3\psi\,,\label{ginvar-VP2-b}\\
&\ti{\om} = -\beta^\prime ~~;~~ \ti{\chi} = -2\beta
  \,,\label{ginvar-VP2-c}\\ 
&\ti{\cchi}_A = 0\,,\label{ginvar-VP2-d}\\
&\ti{\com}_A = \com_A ~~;~~ \ti{\cchi}_{AB} =
  \cchi_{AB}\,,\label{ginvar-VP2-e} 
\end{align}
\label{ginvar-VP2}
\end{subequations}
where the function $\beta$ is restricted not to contain any nontrivial
solution of the homogeneous (Laplace) equation $\nabla^2\beta=0$. 

Having made this choice for the VP gauge, we are now assured that all
averaged quantities which we compute are gauge invariant : our choice
ensures that the averaging procedure does not introduce any pure gauge
modes, and the philosophy of ``steady'' coordinate transformations
ensures that all background objects are, by assumption, unaffected by
gauge transformations. In particular, the correlation objects in
\eqns{MG8} are all gauge invariant\footnote{Note that all these
  arguments are valid at first order in PT, which is sufficient for
  our present purposes. A consistent treatment at second order would
  require more work, although as long as one is interested only in the
  leading order effect, these arguments are expected to go
  through.}. This is different from the gauge invariance conditions
derived in \Cite{abramo}, where the background was also taken to 
change under gauge transformations at second order in the
perturbations. It is at present not clear how these results are
related to ours.

\section{The Correlation Scalars}
\label{sec:corrscal}
\noindent
With the VP gauge choice defined by \eqns{ginvar-VP2}, it is
straightforward to rewrite the correlation objects in
\eqns{avgVP-corrscal} (which are in the uVP gauge) in terms 
of the perturbation functions in the cN gauge. We will restrict the
subsequent calculations in this chapter to the case where there are no
transverse vector perturbations, i.e.,
\be
\com_A = 0\,,
\label{corrscal1}
\ee
in the cN gauge. This is a reasonable choice since such vector
perturbations, even if they are excited in the initial conditions,
decay rapidly and do not source the other perturbations at first order
\cite{dodelson}. 
In addition, for simplicity we will
choose to ignore the gauge invariant tensor perturbations as well,
\be
\cchi_{AB} = 0\,.
\label{corrscal2}
\ee
In terms of the scalar perturbations in the cN gauge, for a flat FLRW
background, the correlation objects \eqref{avgVP-corrscal} reduce to
\begin{subequations}
\begin{align}
\Cal{P}^{(1)} &= \frac{1}{a^2} \bigg[ \, 2\avg{(\psi^\prime)^2} + 
    \avg{\left(\vphi^\prime - \psi^\prime\right)^2}  -
    \avg{\left(\nabla_A\nabla_B\beta^\prime\right)
      \left(\nabla^A\nabla^B\beta^\prime\right)} \,  \bigg]
  \,, \label{corrscal-a}\\   
&\nonumber\\
\Cal{S}^{(1)} &= -\frac{1}{a^2} \bigg[ 6\avg{\p_A\psi\p^A\psi} +
    \avg{\p_A(\vphi-\psi)\p^A(\vphi-\psi)}  -
    \avg{(\nabla_A\nabla_B\nabla_C\beta)
      (\nabla^A\nabla^B\nabla^C\beta)}  \bigg]  
  \,,\label{corrscal-b}\\ 
&\nonumber\\
\Cal{P}^{(1)} + \Cal{P}^{(2)} &= \frac{1}{a^2} \bigg[
    \avg{\vphi^\prime(\vphi^\prime-\psi^\prime)}
    - 2\Cal{H}\left\{\,
    \avg{\vphi^\prime\vphi} -  \avg{\psi^\prime\psi}  +
    \avg{\psi^\prime(\vphi-\psi)}  \right.  \nonumber\\   
&\ph{\frac{1}{a^2} \bigg[
    \avg{\vphi^\prime(\vphi^\prime-\psi^\prime)} } \left.  +  
    \avg{\psi(\vphi^\prime-\psi^\prime)}  +
    \avg{(\nabla_A\nabla_B\beta)(\nabla^A\nabla^B\beta^\prime)}\, 
    \right\}  \bigg]\,, \label{corrscal-c}\\  
\Cal{S}^{(2)} &= -\frac{1}{a^2} \bigg[
    \avg{\p^A\beta^{\prime\prime} \left(\p_A\vphi -
      \Cal{H}\p_A\beta^\prime \right)}  \bigg] 
  \,, \label{corrscal-d} 
\end{align}
\label{corrscal}
\end{subequations}
where $\beta$ is defined in \eqn{ginvar-VP2-b}.

Since we are working with a flat FLRW background, it becomes
convenient to transform our expressions in terms of Fourier space
variables. This will also highlight the problem with large scale
fluctuations which was mentioned in \Sec{sec:avgVP}. We will use the
following Fourier transform conventions : For any scalar function
$f(\eta,\vec{x})$, its Fourier transform $f_{\vec{k}}(\eta)$ satisfies 
\begin{align}
f(\eta,\vec{x}) &= \int{\frac{d^3k}{(2\pi)^3}e^{i\vec{k}\cdot\vec{x}}
  f_{\vec{k}}(\eta)} \,,\nonumber\\
f_{\vec{k}}(\eta) &= \int{d^3x
  e^{-i\vec{k}\cdot\vec{x}}f(\eta,\vec{x})} \,.
\label{corrscal3}
\end{align}
Consider an average of a generic quadratic product of two scalars
$f^{(1)}(\vec{x})$ and $f^{(2)}(\vec{x})$ where we have suppressed the
time dependence since it simply goes along for a ride. Using the
definition \eqref{avgVP13}, and keeping in mind that the scalars are
real, it is easy to show that we have  
\be
\avg{f^{(1)}f^{(2)}}(\vec{x}) =
\int{\frac{d^3k_1d^3k_2}{(2\pi)^6}W^\ast_L(\vec{k}_1-\vec{k}_2,\vec{x})
f^{(1)}_{\vec{k}_1}f^{(2)\ast}_{\vec{k}_2}} \,,
\label{corrscal4}
\ee
where $W_L(\vec{k},\vec{x})$ is the Fourier transform of the window
function $W_L(\vec{x},\vec{y})$ on the variable $\vec{y}$, and the
asterisk denotes a complex conjugate.

In the present context, the functions $f^{(1)}$ and $f^{(2)}$ will
typically be derived in terms of the initial random fluctuations in
the metric $\vphi_{\vec{k}i}$ which are assumed to be drawn from a
\emph{statistically homogeneous and isotropic} Gaussian distribution
with some given power spectrum. In order to ultimately make contact
with observations, it seems necessary to 
perform a formal ensemble average over all possible realisations of
this initial distribution of fluctuations. The statistical homogeneity
and isotropy of the initial distribution implies that the functions
$f^{(1)}$ and $f^{(2)}$ will satisfy a relation of the type
\be
\eavg{f^{(1)}_{\vec{k}_1}f^{(2)\ast}_{\vec{k}_2}} =
(2\pi)^3\delta^{(3)}(\vec{k}_1 - \vec{k}_2)P_{f_1f_2}(|\vec{k}_1|)\,,
\label{corrscal5}
\ee
for some function $P_{f_1f_2}(k,\eta)$ which is derivable in terms of the
initial power spectrum of metric fluctuations, and where $\eavg{...}$
denotes an ensemble average and $\delta^{(3)}(\vec{k})$ is the Dirac
delta distribution. 

Applying an ensemble average to \eqn{corrscal4} introduces a Dirac
delta which forces $\vec{k}_1=\vec{k}_2$. Further, the normalisation
condition on the window function in \eqn{avgVP13} implies that we have
\be
W_L(\vec{k}=0,\vec{x}) = 1\,,
\label{corrscal6}
\ee
which means that all dependence on the averaging scale and domain
drops out, and we are left with
\be
\eavg{\avg{f^{(1)}f^{(2)}}} = \int{\frac{d^3k}{(2\pi)^3}
  P_{f_1f_2}(k)}\,.  
\label{corrscal7}
\ee
Note however, that the right hand side of \eqn{corrscal7} is precisely 
what we would have obtained, had we treated the spatial average
$\avg{...}$ to be the ensemble average $\eavg{...}$ to begin
with. Therefore for all practical purposes, we are justified in
replacing all the spatial averages in the expressions for the
correlation scalars \eqref{corrscal}, by ensemble averages. 

It is convenient to define the transfer function $\Phi_k(\eta)$ via
the relation  
\be
\vphi_{\vec{k}}(\eta) = \vphi_{\vec{k}i}\Phi_k(\eta)\,.
\label{corrscal8}
\ee
For the calculations in this chapter, we shall
assume that the cN gauge scalars $\vphi(\eta,\vec{x})$ and
$\psi(\eta,\vec{x})$ are equal 
\be
\vphi(\eta,\vec{x}) = \psi(\eta,\vec{x})\,,
\label{corrscal9}
\ee
a choice which is valid in first order PT when anisotropic stresses
are negligible (see \Cite{dodelson}). This simplifies many of the
expressions we are dealing with. The Fourier transform of $\beta$ can
be written, using \eqns{ginvar-VP2-b} and \eqref{corrscal9}, as
\be
\beta_{\vec{k}}(\eta) = \frac{2}{k^2}\vphi_{\vec{k}}(\eta)\,.
\label{corrscal10}
\ee
Finally, in terms of the transfer function $\Phi_k(\eta)$ and
the initial power spectrum of metric fluctuations defined by
\be
\eavg{\vphi_{\vec{k_1}i}\vphi^\ast_{\vec{k_2} i}} = (2\pi)^3
\delta^{(3)}(\vec{k_1} - \vec{k_2}) P_{\vphi i}(k_1)\,,
\label{corrscal-powspec}
\ee
the correlation scalars \eqref{corrscal} can be written as
\begin{subequations}
\begin{align}
&\nonumber\\
\Cal{P}^{(1)} &= -\frac{2}{a^2} \int{\frac{dk}{2\pi^2}
    k^2P_{\vphi i}(k) \left(\Phi_k^\prime \right)^2 } 
\,, \label{corrscal-kspace-a}\\   
&\nonumber\\
\Cal{S}^{(1)} &= -\frac{2}{a^2} \int{\frac{dk}{2\pi^2}
  k^2P_{\vphi i}(k) \left( k^2 \Phi_k^2\right)}
  \,,\label{corrscal-kspace-b}\\ 
&\nonumber\\
\Cal{P}^{(1)} + \Cal{P}^{(2)} &= -\frac{8\Cal{H}}{a^2}
  \int{\frac{dk}{2\pi^2} k^2P_{\vphi i}(k) \left( \Phi_k\Phi_k^\prime 
    \right) } 
\,, \label{corrscal-kspace-c}\\  
&\nonumber\\
\Cal{S}^{(2)} &= -\frac{2}{a^2}   \int{\frac{dk}{2\pi^2}
  k^2P_{\vphi i}(k) \Phi_k^{\prime\prime}\left( \Phi_k -
  \frac{2\Cal{H}}{k^2}\Phi_k^\prime \right) }  
  \,. \label{corrscal-kspace-d} 
\end{align}
\label{corrscal-kspace}
\end{subequations}
These expressions highlight the problem of having a finite amplitude
for fluctuations at arbitrarily large length scales ($k\to0$), which
was mentioned in \Sec{sec:avgVP}. As a concrete example, consider the
frequently discussed Harrison-Zel'dovich scale invariant spectrum
\cite{scalinvarpowspec} which satisfies the condition
\be
k^3P_{\vphi i}(k) = {\rm constant}\,.
\label{scale-invar}
\ee
\eqns{corrscal-kspace} now show that if the transfer function
$\Phi_k(\eta)$ has a finite time derivative at large scales (as it
does in the standard scenarios -- see the next section), then the
correlation objects $\Cal{P}^{(1)}$, $\Cal{P}^{(2)}$ and
$\Cal{S}^{(2)}$ all diverge due to contributions from the $k\to0$
regime. This demonstrates the importance of having an initial power
spectrum in which the amplitude dies down sufficiently rapidly on
large length scales (which is a known issue, see
\Cite{liddle-lyth}). Keeping this in mind, we shall concentrate on 
initial power spectra which display a long wavelength cutoff. Models
of inflation leading to such power spectra have been discussed in the
literature \cite{cutoff-theory}, and more encouragingly, analyses of
WMAP data seem to indicate that such a cutoff in the initial power
spectrum is in fact realised in the universe \cite{cutoff-obsvns}.   

A final comment before proceeding to explicit calculations : In
addition to picking up nontrivial correlation corrections in the
cosmological equations, the averaging formalism also requires that the
``cross-correlation'' constraints in \eqns{gauge6} be satisfied. It is 
straightforward to show that the statistical homogeneity and isotropy
of the metric fluctuations implies that these constraints are
identically satisfied, for \emph{all} types of perturbations (scalar,
vector and tensor). At the lowest order therefore, these
constraints do not impose any additional conditions on the
perturbation theory, which is reassuring.

\section{Worked out examples}
\label{sec:exs}
\noindent
We will now turn to some explicit calculations of the backreaction,
which will show that the magnitude of the effect remains negligibly
small for most of the evolution duration in which linear PT is
valid. At early times, linear PT is valid at practically all scales
including the smallest scales at which we wish to apply general
relativity. As matter fluctuations grow, the small length scales
progressively approach nonlinearity, and linear PT 
breaks down at these scales. As we will see, however, by the time a
particular length scale becomes nonlinear, its contribution to the
amplitude of the \emph{metric} fluctuations correspondingly becomes 
negligible. In practice therefore, one can extend the linear
calculation well into the matter dominated era, with the expectation
that the order of magnitude of the various integrals will not change 
significantly due to nonlinear effects. 

The model we will use is the standard Cold Dark Matter (sCDM) model
consisting of radiation and CDM \cite{dodelson}. We will neglect
the contribution of baryons, and at the end we shall discuss the
effects this may have on the final results. We shall also discuss,
without explicit calculation, the effects which the introduction of a
cosmological constant is likely to have. In the following,
$\Omega_r$ and $\Omega_m$ denote the density parameters of
radiation and CDM respectively at the present epoch $\tau_0$, with
$\tau$ denoting cosmic time. $\Omega_r$ is assumed to contain
contributions from photons and $3$ species of massless,
out-of-equilibrium neutrinos. At the ``zeroth iteration'' we have   
\be
\left(\frac{1}{a}\frac{da}{d\tau}\right)^2 = H^2(a) = H_0^2\left[ 
  \frac{\Omega_m}{a^3} + \frac{\Omega_r}{a^4} \right]  \,,
\label{exs1}
\ee
where $H_0$ is the standard Hubble constant, the scale factor is
normalised so that $a(\tau_0)=1$, and $\Cal{H}$ and $H$ are related by  
\be
\Cal{H}(a) = aH(a)\,.
\label{exs2}
\ee
The comoving wavenumber corresponding to the scale which enters at the
matter radiation equality epoch, is given by
\be
k_{eq} = a_{eq}H(a_{eq}) =
H_0\left(\frac{2\Omega^2_m}{\Omega_r}\right)^{1/2} \sim H_0\cdot
10^{5/2}\,, 
\label{exs3}
\ee
where we have set (see \Cites{kolb-turner,dodelson} for details)
\begin{align}
\Omega_r &= \Omega_{photon} + 3\Omega_{neutrino}
\nonumber\\ 
&= \Omega_{photon}\left(1 +
3\cdot\frac{7}{8}\left(\frac{4}{11}\right)^{4/3} \right) \nonumber\\
&= 4.15\times10^{-5}h^{-2}\,, 
\label{exs4}
\end{align}
where $h$ is the dimensionless Hubble parameter defined by $H_0 =
100h$ km/s/Mpc. For all calculations we shall set $h=0.72$
\cite{hst-key}. 
\subsection{EdS background and non-evolving potentials}
\label{sec:exs:sub:eds}
\noindent
Before dealing with the full model (which requires a numerical
evolution) let us consider the simpler situation, described by an
Einstein-deSitter (EdS) background, with negligible radiation and a
nonevolving potential $\vphi = \vphi(\vec{x})$ (which is a consistent
solution of the Einstein equations in the sCDM model at least at
subhorizon scales at late times \cite{dodelson}). Although not fully
accurate, this example requires some very simple integrals and will
help to give us a feel for the structure and magnitude of the
backreaction. 

With a constant potential, the only correlation object which survives
is $\Cal{S}^{(1)}$, which evolves like $\sim a^{-2}$, where the scale
factor refers to the ``zeroth iteration''. The constant of
proportionality can be written in terms of the BBKS transfer function
$T_{BBKS}(k/k_{eq})$ \cite{bbks,dodelson}, to give 
\be
\Cal{S}^{(1)} = -\frac{2}{a^2}\int{\frac{dk}{2\pi^2}k^4
  P_{\vphi i}(k)T^2_{BBKS}(k/k_{eq})}\,,
\label{exs-eds1}
\ee
where we have \cite{bbks}
\begin{align}
T_{BBKS}(x) &= \frac{\ln{[1+0.171x]}}{(0.171x)} \bigg[
  1 + 0.284x   + (1.18x)^2 + (0.399x)^3 + (0.490x)^4 \bigg]^{-0.25}\,, 
\label{exs-eds2}
\end{align}
where $x\equiv (k/k_{eq})$.

The integral in \eqn{exs-eds1} is well-behaved even in the presence of 
power at arbitrarily large scales, for a (nearly) scale invariant
spectrum. Since we are only looking for an estimate, we shall evaluate
the integral in the absence of a large scale cutoff, and leave a more
accurate calculation for the next subsection. For the initial spectrum
given by 
\be
\frac{k^3P_{\vphi i}(k)}{2\pi^2} = A (k/H_0)^{n_s - 1}\,,
\label{exs-eds3}
\ee
where the scalar spectral index $n_s$ is close to unity, the integral
in \eqn{exs-eds1} can be easily performed numerically and has the
order of magnitude  
\be
\int{\frac{dk}{2\pi^2}k^4 P_{\vphi i}(k)T^2_{BBKS}(k/k_{eq})} \sim
A\left(k_{eq}\right)^2 \sim AH_0^2\cdot10^5 \,, 
\label{exs-eds4}
\ee
upto a numerical prefactor of order $1$. Since the amplitude
of the power spectrum is $A\sim 10^{-9}$ \cite{liddle-normalisn}, the 
overall contribution of the backreaction is  
\be
\frac{\Cal{S}^{(1)}}{H_0^2} \sim -\frac{1}{a^2}(10^{-4})\,.
\label{exs-eds5}
\ee
Now, as long as the correlation objects give a negligible
backreaction to the usual background quantities, when we proceed with
the \emph{next} iteration, the effect of the backreaction on the
evolution of the \emph{perturbations} will also remain negligible (at
least at the leading order). Hence in practice there will be
essentially no difference between the zeroth iteration and first
iteration perturbation functions. This amounts to saying that when the
backreaction is negligible, convergence to the ``true'' solution for
the scale factor \emph{at the leading order}, is essentially achieved
in a single calculation. We will discuss the issue of convergence in
somewhat more detail at the end of the chapter.

\subsection{Radiation and CDM without baryons} 
\label{sec:exs:sub:nobar}
\noindent
Let us now turn to the full sCDM model (without baryons). An
analytical discussion of this model in various regions of
$(k,\eta)$-space, can be found e.g. in \Cite{dodelson}. Since we are
interested in integrals over $k$ across a range of epochs $\eta$, it
is most convenient to solve this model numerically. It is further
convenient to use $(\ln a)$ in place of $\eta$, as the variable with
which to advance the solution. Also, it is useful to introduce
transfer functions like $\Phi_k(\eta)$ for all the relevant
perturbation functions in exactly the same manner (see
\eqn{corrscal8}), namely by pulling out a factor of
$\vphi_{\vec{k}i}$, since the initial conditions are completely
specified by the initial metric perturbation. For a generic
perturbation function $s_{\vec{k}}(\eta)$ (other than the metric 
fluctuation $\vphi_{\vec{k}}$) the transfer function
corresponding to $s$ will be denoted by a caret, so that 
\be
s_{\vec{k}}(\eta) = \vphi_{\vec{k}i}\hat s_k(\eta)\,.
\label{exs-nobar1}
\ee
The relevant Einstein equations can be brought to the following closed
set of first order ordinary differential equations (adapted from
Eqns. (7.11)-(7.15) of \Cite{dodelson}),  
\begin{subequations}
\begin{align}
&\frac{\p\Phi_k}{\p(\ln a)}= -\left[ \left(1 +
    \frac{K^2}{3E^2}\right)\Phi_k  +
    \frac{1}{2E^2a}\left(\Omega_m\cdelta_k + 
    \frac{4}{a}\Omega_r\ctheta_{0k}  \right)
    \right]\,, \label{exs-nobar2a}\\ 
&\frac{\p\cdelta_k}{\p(\ln a)} = -\frac{K}{E}\cv_k +
  3\frac{\p\Phi_k}{\p(\ln a)}  \,,\label{exs-nobar2b}\\
&\frac{\p\ctheta_{0k}}{\p(\ln a)} = -\frac{K}{E}\ctheta_{1k} +
  \frac{\p\Phi_k}{\p(\ln a)}  \,,\label{exs-nobar2c}\\ 
&\frac{\p\ctheta_{1k}}{\p(\ln a)} = \frac{K}{3E}\left(\ctheta_{0k} + 
  \Phi_k \right)   \,,\label{exs-nobar2d}\\ 
&\frac{\p \cv_k}{\p(\ln a)}= -\cv_k + \frac{K}{E}\Phi_k
  \,.\label{exs-nobar2e} 
\end{align}\label{exs-nobar2}
\end{subequations}
Here we have introduced the dimensionless variables
\be
K\equiv \frac{k}{H_0} ~~;~~ E(a) \equiv \frac{\Cal{H}(a)}{\Cal{H}_0} =
\frac{\Cal{H}(a)}{H_0} \,,
\label{exs-nobar3}
\ee
and the various perturbation functions are defined as follows :
$\delta_k$ is the $k$-space density contrast of CDM, $\Theta_{0k}$ and
$\Theta_{1k}$ are the monopole and dipole moments respectively of the
$k$-space temperature fluctuation of radiation, and $(-iV_k)$ is the
$k$-space peculiar velocity scalar potential of CDM (i.e., the real
space peculiar velocity is $v_A = \p_Av$ where $v$ is the Fourier
transform of $(-iV_k)$). 

Assuming adiabatic perturbations, the initial conditions satisfied by
the transfer functions at $a=a_i$ are (adapted from Ch. 6 of
\Cite{dodelson}) 
\begin{align}
&\Phi_k(a_i) = 1 ~~;~~ \cdelta_k(a_i) = -\frac{3}{2} ~~;~~
\ctheta_{0k}(a_i) = -\frac{1}{2} 
~~;~~ \cv_k(a_i) = 3\ctheta_{1k}(a_i) =
\frac{1}{2}\frac{K}{E(a_i)}\,. 
\label{exs-nobar-init}
\end{align}
We choose $a_i = 10^{-16}$, which corresponds to an initial background
radiation temperature of $T\sim10^3$GeV. While this is not as far back in
the past as the energy scale of inflation (which is closer to
$\sim10^{15}$Gev), it is on the edge of the energy scale where known
physics begins \cite{liddle-lyth}. This makes \eqn{exs4} unrealistic
since we have ignored all of Big Bang nucleosynthesis and also the fact
that neutrinos were in equilibrium with other species at temperatures
higher than about $1$Mev. However the modifications due to these
additional details are not expected to drastically change the final
results, and these assumptions lead to some simplifications in the
code used. The goal here is only to demonstrate an application of the
formalism; more realistic calculations accounting for the effects of
baryons can also be performed (see, e.g. Behrend et
al. \cite{behrend} who incorporate these effects for the
post-recombination era, albeit in the Buchert formalism).  

In order to partially account for the fact that inflationary initial
conditions are actually set much earlier than $a=10^{-16}$, we impose
an absolute \emph{small wavelength} cutoff at the scale which enters
the horizon at the initial epoch which \emph{we} have chosen. In the
above notation this corresponds to setting
$K_{max}=E(a_i)\sim10^{13}$. This makes sense since scales satisfying
$K\gg K_{max}$ have already entered the horizon and decayed
considerably by the epoch $a=10^{-16}$. There is a source of error due
to ignoring scales $K\gtrsim K_{max}$ which have not yet decayed
significantly, but this error rapidly decreases with time as
progressively larger length scales enter the horizon and decay. [In
  fact, in practice to compute the integrals at any given epoch
  $a=a_\ast$, one only needs to have followed the evolution of modes
  with $K<\,\sim5000E(a_\ast)$ : more on this in the next subsection.]
More important is the cutoff at \emph{long} wavelengths, which we set
at $K_{min} = 1$ (corresponding to $k_{min}=H_0$), which is firstly a
natural choice given that $H_0^{-1}$ is the only large scale in the
system, and is secondly also guided by analyses of CMB data which
have detected such a cutoff \cite{cutoff-obsvns}. We will see that
reducing $K_{min}$ even by a few orders of magnitude, does not
affect the final qualitative results significantly. 

\subsubsection{Numerical Results}  
\label{sec:exs:sub:nobar:sub:numres}
\noindent
Equations \eqref{exs-nobar2} with initial conditions
\eqref{exs-nobar-init} were solved with a standard $4$th order
Runge-Kutta integrator with adaptive stepsize control (based on the
algorithm given in \Cite{NumRec}). For the integrals in
\eqns{corrscal-kspace}, only the function $\Phi_k(a)$ needs to be
tracked accurately. Hence, although $\ctheta_{0k}$ and $\ctheta_{1k}$
are difficult to follow accurately beyond the matter radiation
equality $a_{eq} = (\Omega_r/\Omega_m) \simeq 8\times10^{-5}$ due to
rapid oscillations, the integrals can still be reliably computed since
$\ctheta_{0k}$ and $\ctheta_{1k}$ do not significantly affect the
evolution of $\Phi_k$ in the matter dominated era (as seen in
\eqn{exs-nobar2a}). 
\begin{figure}[t]
\subfigure[\small The evolution of two large scale modes. Also shown
  is the Kodama-Sasaki analytical solution in the large scale limit
  $k\eta\ll1$, given by
  \eqn{exs-numres1}.]{\includegraphics[width=.475\textwidth]{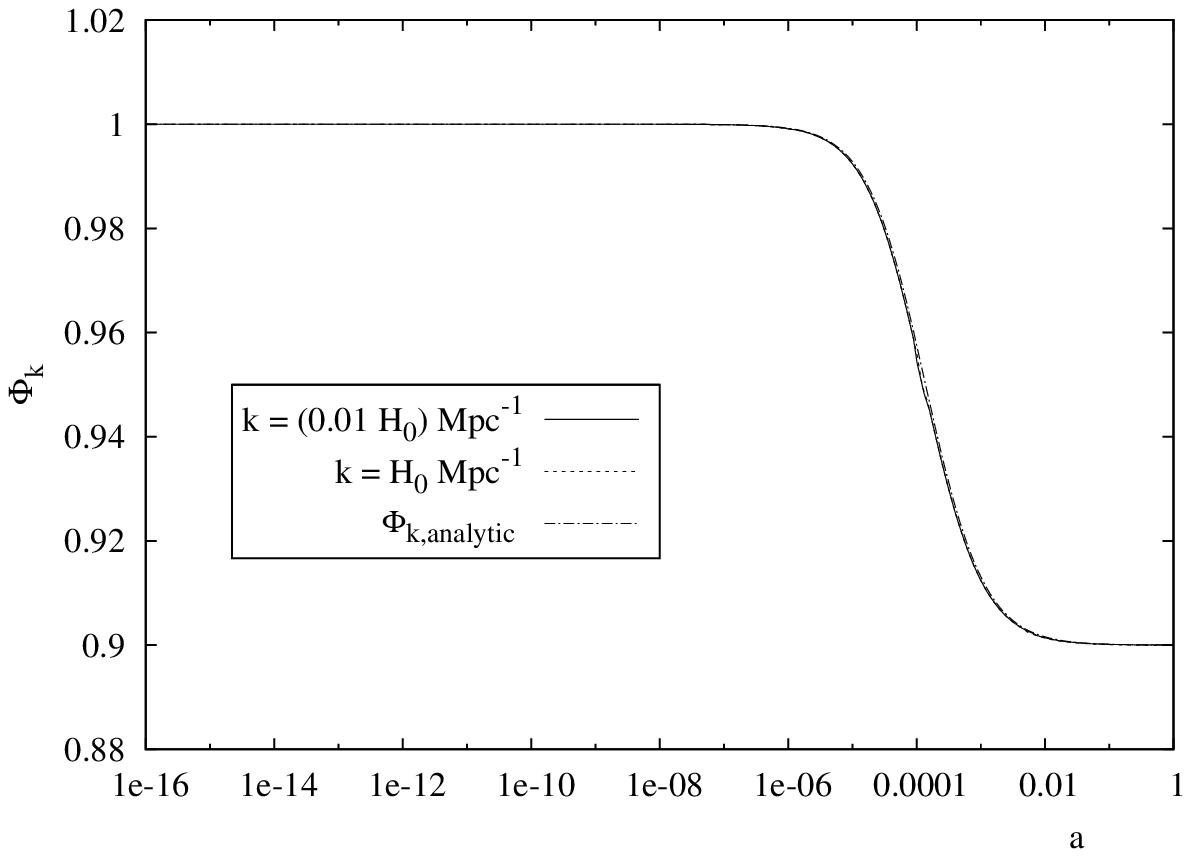} 
\label{fig3}}
\hspace{.05\textwidth} 
\subfigure[\small The transfer function $\Phi_k$ normalised by its
  constant value at large scales, at the epoch $a=500a_{eq}$. The
  dotted line is the BBKS transfer function
  \eqref{exs-eds2}.]{\includegraphics[width=.475\textwidth]{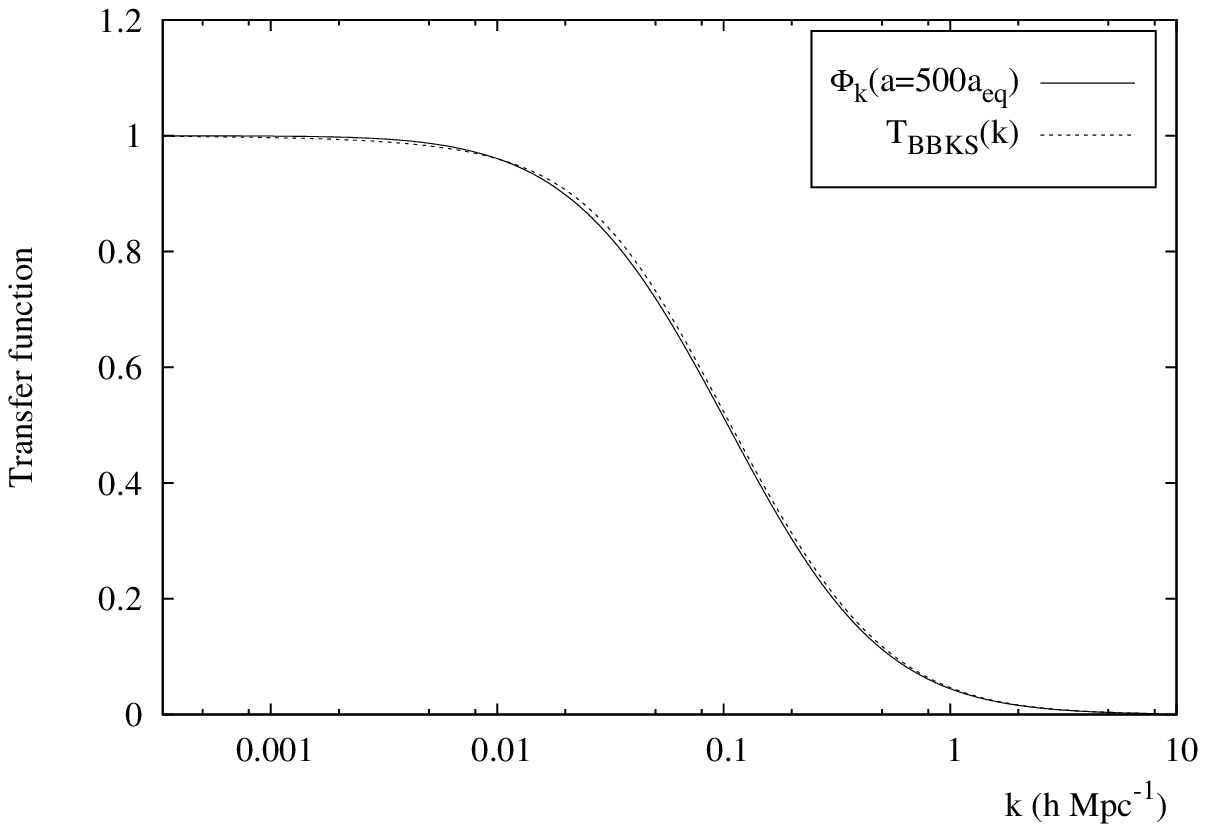}   
\label{fig4}}
\caption{\small Numerical results for the transfer function.} 
\end{figure}

To see that known results are being reproduced by the code, consider
\figs{fig3} and \ref{fig4} as examples. \fig{fig3} shows the
evolution of two scales corresponding to $K=1$ ($k=H_0\,{\rm
  Mpc}^{-1}$) and $K=0.01$. The first enters the horizon at the
present epoch, while the second remains superhorizon for the entire
evolution, satisfying $k\eta\ll1$. In this limit an analytical
solution exists in the sCDM model, due to Kodama and Sasaki
\cite{kodama,dodelson}, given by 
\be
\Phi_k(y) = \frac{1}{10y^3}
\left[16\sqrt{1+y} + 9y^3 + 2y^2 -8y -16 \right] \,,
\label{exs-numres1}
\ee
where $y\equiv a/a_{eq}$, and this function is also shown. Clearly all
the curves in \fig{fig3} are practically identical.
\fig{fig4} shows the function $\Phi_k$ normalised by its (constant)
value at large scales, at the epoch $a=500a_{eq}\simeq 0.04$,
(which is well into the matter dominated era). The dotted line is the
BBKS fitting form given in \eqn{exs-eds2} with $k_{eq}$ given by
\eqn{exs3}. 

To numerically estimate the integrals in \eqns{corrscal-kspace}, the
values of $\Phi_k$ and its first and second derivatives with respect
to $(\ln a)$ are needed across a range of $K$ values. For reference,
note that the following relations hold for a generic function of time
$w(\eta)$, 
\be
\frac{dw}{d\eta} = a\Cal{H}\frac{dw}{da} = \Cal{H}\frac{dw}{d(\ln
  a)} \,.
\label{exs-numres2}
\ee
Based on the earlier discussion, the initial power spectrum
$P_{\vphi i}(k)$ is taken to satisfy
\be
\frac{k^3P_{\vphi i}(k)}{2\pi^2} = A ~,~~ {\rm for\,\,} H_0 < k < 
k_{max}=\Cal{H}(a_i) \,,
\label{exs-numres3}
\ee
and zero otherwise, and we set
\be
A = 1.0\times10^{-9}\,,
\label{exs-numres4}
\ee
which, for the sCDM model follows from the convention (see Eqn.(6.100)
of \Cite{dodelson}) $A = (5\delta_H/3)^2$ with
$\delta_H\approx2\times10^{-5}$ (see, e.g. \Cite{liddle-normalisn}).  
\begin{figure}[t]
\subfigure[\small The dimensionless integrand of $\Cal{S}^{(1)}$,
  namely the function $(k/H_0)^2\Phi_k^2$, at three sample values of
  the scale factor. The function dies down rapidly for large $k$, with
  the value at some $k$ being progressively smaller with increasing
  scale factor. The declining behaviour of the curves for $a=a_{eq}$
  and $a=200a_{eq}$ extrapolates to large
  $k$.]{\includegraphics[width=.475\textwidth]{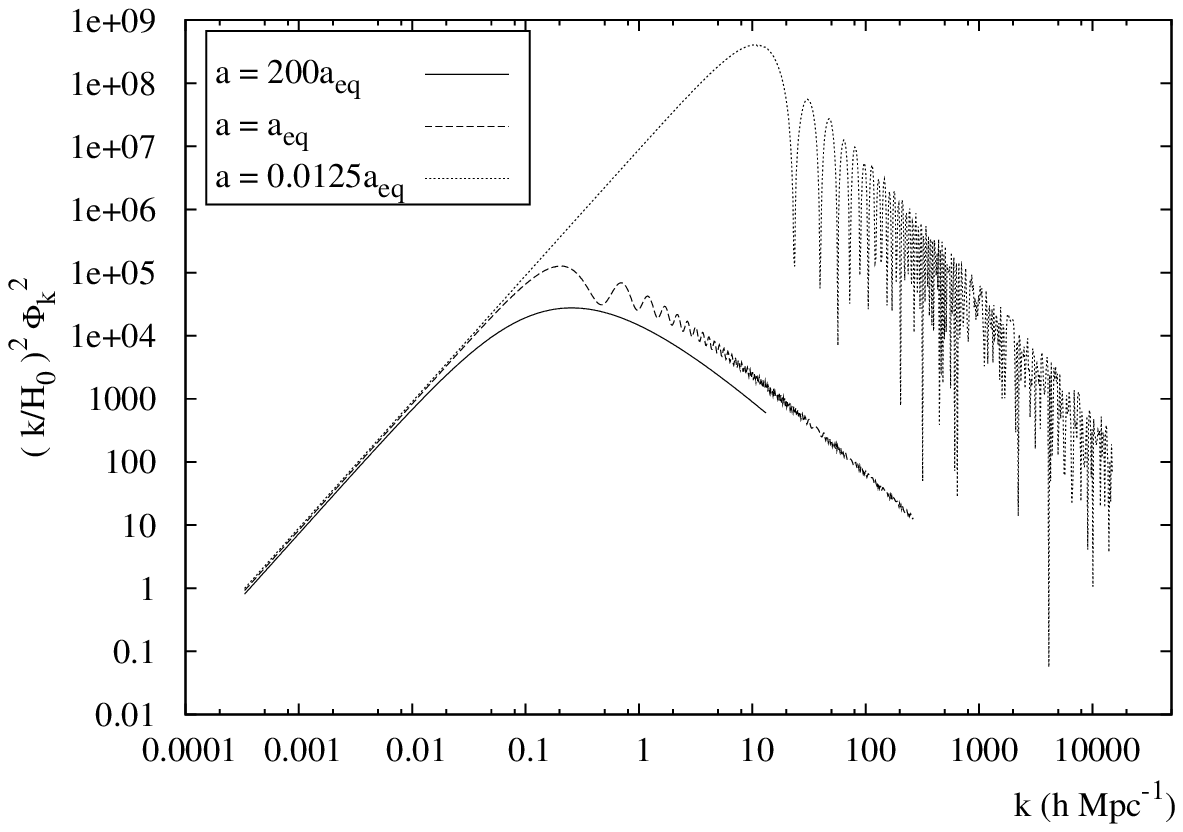}  
\label{fig5}}
\hspace{.05\textwidth} 
\subfigure[\small The dimensionless CDM density contrast. Together
  with \fig{fig5} this shows that nonlinear scales do not impact the
  backreaction integrals
  significantly. ]{\includegraphics[width=.475\textwidth]{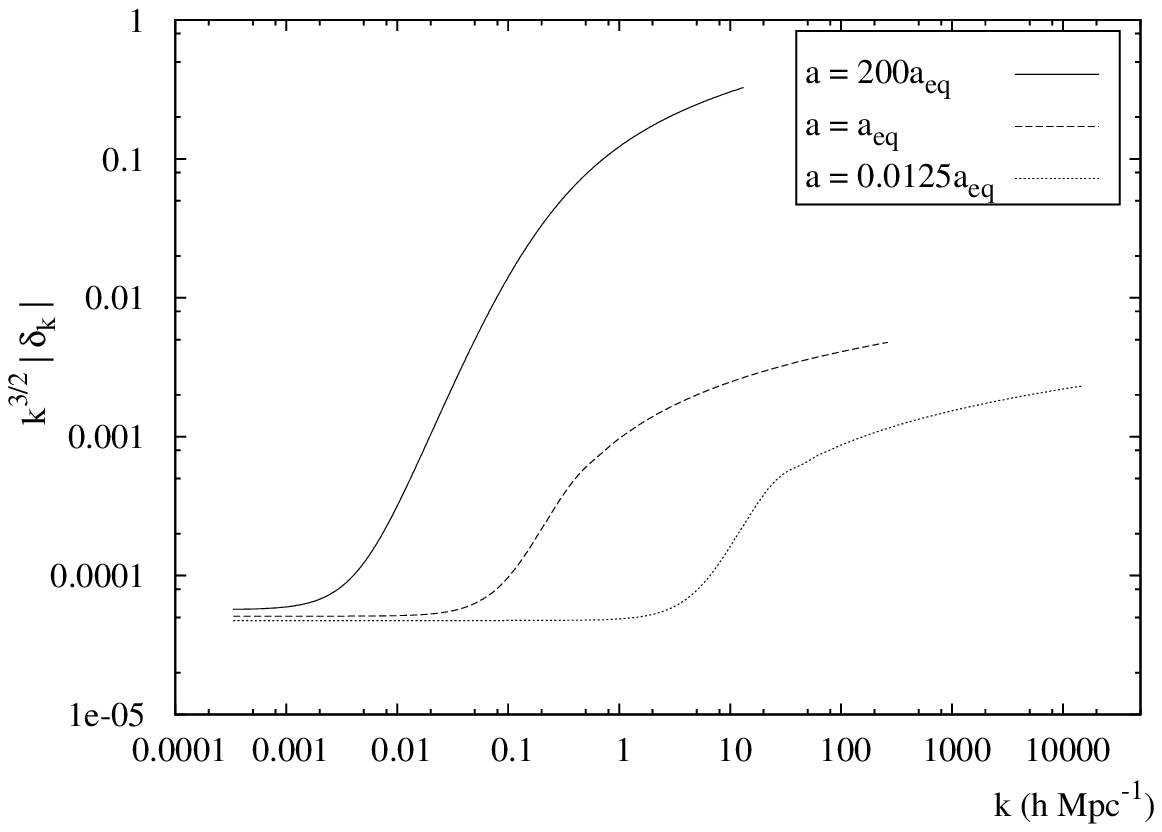}     
\label{fig6}}
\caption{\small Backreaction and nonlinearity.} 
\end{figure}

Consider \figs{fig5} and \ref{fig6}, which highlight two issues
discussed earlier. \fig{fig5} shows the integrand of $\Cal{S}^{(1)}$
at three sample epochs, and we see that the integrand dies down
rapidly at increasingly smaller $k$ values for progressively later
epochs. (The other integrands, not displayed here, also show this
rapid decline for large $k$.) [We have not shown the integrand at the
  later two epochs for all values of $k$ since this was
  computationally expensive, but the declining trend of the curves 
  can be extrapolated to large $k$, which is well understood
  analytically \cite{dodelson}.] This justifies the statement in
the beginning of this section, that at any epoch $a_\ast$ it is
sufficient to have followed the evolution of scales satisfying 
$K<5000E(a_\ast)$ for computing the integrals. Secondly, \fig{fig6}
shows the behaviour of $k^{3/2}|\delta_k| = A^{1/2}|\cdelta_k|$ at the
same three epochs, and comparing with \fig{fig5} we see that at any
epoch, the region of $k$-space where linear PT has broken down, does
not contribute significantly to the integrals. This is in line with
the conjecture in \Cite{paddy} that the effects of the backreaction
should remain small since the mass contained in the nonlinear scales
is subdominant. We will return to this issue in chapter 5.

Due to the structure of the integrals and the chosen initial power
spectrum, it is convenient to compute the integrands in
\eqns{corrscal-kspace} equally spaced in $(\ln K)$, and then perform
the integrals using the extended Simpson's rule \cite{NumRec}. If
$2^N+1$ points are used to evaluate a given integral, resulting in a
value $\Cal{I}_N$ say, then the error can be estimated by computing
the integral with $2^{N-1}+1$ points to get $\Cal{I}_{N-1}$, and
estimating the relative error as $|\Cal{I}_{N-1}/\Cal{I}_N| -1$. With
$N=10$, the estimated errors in all the integrals at all epochs were
typically less than $0.1\%$. A bigger error is incurred in computing
the integrand itself at any given epoch, leading to estimated errors
of order $\sim1\%$ in $\Cal{S}^{(1)}$, $\Cal{P}^{(1)}$ and
$\Cal{P}^{(1)}+\Cal{P}^{(2)}$, with a larger error in $\Cal{S}^{(2)}$
as explained below.

The second derivative $\p^2\Phi_k/\p(\ln a)^2$ proves to be difficult
to track numerically. At early times, when most scales are
superhorizon, the Kodama-Sasaki analytical solution
\eqref{exs-numres1} is a good approximation for most values of
$k$. Using this one can see that at early times the value of the
derivative is numerically very small, and is difficult to reliably
estimate due to roundoff errors. For this reason the integral
$\Cal{S}^{(2)}$ could not be accurately estimated at early
times. However, the structure of the integrand of $\Cal{S}^{(2)}$
\eqref{corrscal-kspace-d} shows that the largest contribution comes
from large (superhorizon) scales (the small scales being subdominant
due to the presence of $\Phi_k$ and $1/k^2$). An analysis using the
Kodama-Sasaki solution then shows in a fairly straightforward manner
that the behaviour of the backreaction term is
$|\Cal{S}^{(2)}/H^2|\sim 10^{-6}(a/a_{eq})(H_0/k_{min})^2$ for our
choices of parameters, where $(a/a_{eq})\ll1$. At intermediate times
around $a\sim a_{eq}$ and later, although it becomes computationally
expensive to obtain convergent values for the second derivative at all
relevant scales\footnote{Convergence was tested by varying a global
  parameter which dynamically controls the stepsize during evolution
  (by stepsize doubling/halving, see \Cite{NumRec}). The integrals
  other than $\Cal{S}^{(2)}$ show convergence at $3$ or more
  significant digits for all epochs, whereas convergence can be
  obtained for $\Cal{S}^{(2)}$ only at epochs sufficiently close to
  matter radiation equality, and there only for $1$-$2$ significant
  digits, by setting stringent conditions on stepsize doubling.},
moderately good accuracy ($1$-$5\%$) can be achieved.   
\begin{figure}[t]
\centering
\includegraphics[height=0.3\textheight]{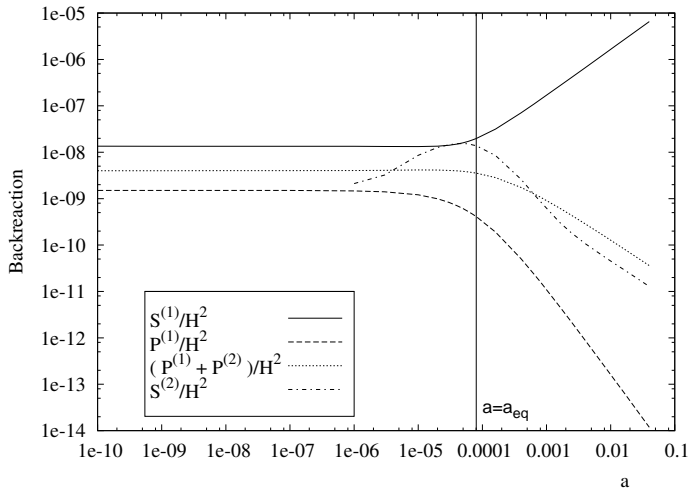}
\caption{\small The correlation scalars (``backreaction'') for the
  sCDM model, normalised by $H^2(a)$. $\Cal{S}^{(1)}$, $\Cal{P}^{(1)}$
  and $\Cal{S}^{(2)}$ are negative definite and their magnitudes have
  been plotted. The vertical line marks the epoch of matter radiation
  equality $a=a_{eq}$.}    
\label{fig7}
\end{figure}

The results are shown in \fig{fig7}, in which the
magnitudes of the correlation integrals of \eqn{corrscal-kspace},
normalised by the Hubble parameter squared $H^2(a) = (\Cal{H}/a)^2$
are plotted as a function of the scale factor in a log-log plot. The
values for $\Cal{S}^{(2)}$ are shown only for epochs later than
$a\simeq0.01a_{eq}\sim10^{-6}$. We see that at
all epochs, the correlation terms remain negligible compared to the
chosen zeroth iteration background. Also, in the radiation dominated
epoch all the correlation scalars (except $\Cal{S}^{(2)}$ whose
evolution couldn't be accurately obtained) track the $\sim a^{-4}$
behaviour of the background radiation density (see also
\Cite{martineau}). The discussion above shows however that the 
magnitude of $\Cal{S}^{(2)}$ is far smaller than the other
backreaction functions at early times, for a cutoff at
$k_{min}=H_0$. On the other hand, in the matter dominated epoch
$\Cal{S}^{(1)}$ dominates the backreaction and settles into a
curvature-like $\sim a^{-2}$ behaviour (note that in the matter  
dominated epoch we have $H^2\sim a^{-3}$). As for the signs of the
correlations, $\Cal{S}^{(1)}$, $\Cal{S}^{(2)}$, and $\Cal{P}^{(1)}$
are negative throughout the evolution while
$\Cal{P}^{(1)}+\Cal{P}^{(2)}$ is positive throughout.   

Finally, a few comments regarding the effects of ignoring baryons,
nonlinear corrections, etc. Including baryons in the problem (with a
background density parameter of $\Omega_b\simeq0.05$) will lead to a
significant suppression of small scale power (by introducing pressure
terms which will tend to wipe out inhomogeneities) and also a small
suppression of large scale power. This effect causes a
(downward) change in the late time transfer function of roughly
$15$-$20\%$ \cite{dodelson}, and therefore cannot increase the
contribution of the backreaction. Quasi-linear corrections can lead to
significant changes in the transfer function, but do not cause shifts
by several orders of magnitude (see \Cite{halo} and references
therein). Hence accounting for changes due to quasi-linear behaviour
will also not increase the magnitude of the backreaction by a large
amount.  As for effects from fully nonlinear
scales, we have seen that these can be expected to remain small
(see also chapter 5).  

Adding a cosmological constant (and retaining a flat background
geometry) will change the qualitative features of the correlation
functions by shifting the scale $k_{eq}$ (due to a reduced $\Omega_m$,
which will also increase the power spectrum amplitude \cite{dodelson},
but again not by orders of magnitude). Also, the late time behaviour
of the correlation scalars will be affected since the potential
$\Phi_k$ will decay at late times instead of remaining
constant (see also below). Regardless, the backreaction is expected to 
remain small even in this case (which is also indicated by the
calculations of Behrend et al. \cite{behrend} in the Buchert
framework).  

\vskip 0.25in
\noindent
{\large\bf Chapter summary and discussion:}\vskip 0.1in
\noindent
This chapter showed how the spatial averaging limit of MG can be
adapted to the case of linear perturbation theory (PT) in cosmology,
to calculate the backreaction in gauge invariant manner. In doing so
we also saw the significance of volume preserving gauges in defining
self-consistent averaging operators. The formalism leaves some freedom
in the choice of the averaging operator (via a choice of a volume
preserving gauge), and therefore we cannot claim that our results are
unique down to all numerical factors. However, since the final
explicit form of the late time backreaction in \fig{fig7} matches
closely with the (essentially order of magnitude) estimate
of \eqn{exs-eds5} which is actually independent of any details of the
averaging procedure, we do expect our results to be qualitatively
robust.

\fig{fig7} also shows that in the absence of a cosmological constant,
the backreaction after a single iteration of the procedure outlined in
section 4.1, tracks the radiation density in the radiation dominated
era, and essentially behaves like a curvature term in the matter
dominated era. Two issues arise from this behaviour. Denote
the corrections to the Friedmann equation \eqref{flrw3a} and the
acceleration equation \eqref{flrw3b} as $\Cal{C}_F$ and
$\Cal{C}_{acc}$ respectively. Then firstly, we find that in the
radiation dominated era, although both $\Cal{C}_F$ and $\Cal{C}_{acc}$
behave like $\sim a^{-4}$, their numerical coefficients do not combine
so as to preserve the conservation criterion \eqref{avgZala63}
separately. Since the backreaction must now necessarily couple to the
background radiation density, this points to a very tiny
gravitationally induced correction in the equation of state for
radiation. This effect can be traced back essentially to the presence
of a small but non-zero correlation 3-form and 4-form, arising from
higher order perturbative effects. In the matter dominated era, at
least in the ``zeroth'' iteration, this effect seems to be highly
suppressed since we now have $\Cal{C}_F\sim a^{-2}$ and
$|\Cal{C}_{acc}|\ll|\Cal{C}_F|$, which is approximately consistent
with \eqref{avgZala63}.

The second issue concerns what happens at higher iterations, and is
important from the point of view of obtaining a convergent answer for
the backreaction. The basic cycle that one needs to keep in mind is
that the backreaction affects $H^2$, which affects the equations for
the density and metric perturbations, which in turn define the
backreaction. Consider the situation in the matter dominated era,
which is easier to handle since firstly only one term
$\Cal{S}^{(1)}$ contributes to the backreaction and secondly the
linear PT solution has a simple analytic form. The estimate
in \eqn{exs-eds5} shows that most of the contribution comes
from (quasi)linear subhorizon scales $k\sim k_{eq}$ for which the
Poisson equation holds, so that if the density contrast behaves like
$\delta_{k}\sim D(a)$ then the metric transfer function behaves like
$\Phi_{k} \sim D/a$. Standard linear PT \cite{dodelson} tells us that
$D(a)$ is the so-called growth function which can be written as
$D \sim E\int{da/(aE)^3}$ upto some numerical coefficient, where
$E\equiv H/H_0$. An analysis similar to the one leading
to \eqn{exs-eds5} then shows that we should expect $\Cal{C}_F\sim
a^{-2}(D/a)^2$ at late times. For a flat universe without a
cosmological constant, $D(a)=a$ and we recover the single iteration
result that we have been discussing so far.  

The crucial thing to note is that since the backreaction affects only
the background equations and \emph{not} the perturbation equations,
$D(a)$ is completely determined by the Hubble parameter
$H(a)=H_0E(a)$, so that at any iteration $i$ we will have
\be
(E^{(i+1)})^2 = \Omega_m^{(i+1)}a^{-3}+\eps_{\rm
bkrxn}^{(i)}a^{-2}(D^{(i)}/a)^2 ~~;~~ D^{(i)}(a)\sim
E^{(i)}\int{da/(aE^{(i)})^3} \,,
\la{converg1}
\ee
where we expect $\eps^{(i)}_{\rm bkrxn}\sim10^{-4}$. This immediately 
suggests that the limit of this series is the solution of the
integral equation
\be
E^2 = \Omega_ma^{-3}+\eps_{\rm
bkrxn}a^{-2}(D/a)^2 ~~;~~ D(a)\sim
E\int{da/(aE)^3} \,.
\la{converg2}
\ee
This equation can in principle be solved perturbatively by exploiting
the smallness of the parameter $\eps_{\rm bkrxn}$, and we expect the
solution to be close to the ``zeroth'' iteration answer $D\sim a$. To
understand why, notice that at the zeroth iteration we found
$\Cal{S}^{(1)}$ to be negative, so that the first iteration Hubble
parameter $E^{(1)}(a)$ is effectively that of an \emph{open}
universe with a small negative curvature. Standard analysis shows that
the growth factor in an open universe is suppressed compared to that
in a flat matter dominated one, and hence the Hubble parameter at
the \emph{second} iteration $E^{(2)}(a)$ will have a slightly smaller
contribution from the backreaction than $E^{(1)}(a)$. This will
correspondingly slightly \emph{enhance} the contribution of the
backreaction to $E^{(3)}(a)$ over the contribution to $E^{(2)}(a)$,
and so on until the solution converges. 

This convergent solution will, like the radiation dominated case,
mildly violate the conservation criterion \eqn{exs-eds5}. Further, the
analysis above generalises to the case when the cosmological constant
is nonzero. In this case the late time growth factor is suppressed
compared to the EdS case even at the zeroth iteration \cite{dodelson},
and the convergent solution will violate the conservation criterion by
an amount comparable to the backreaction itself. What is important
however is that in \emph{all} cases, the backreaction as well as the
violation of matter conservation remain negligibly small,
approximately at the level of one part in $10^4$. This
analysis ignored all contributions from scales which have become fully
nonlinear in the matter density contrast at late times. The reasoning
was that these scales are not expected to contribute significantly to
the backreaction due to a suppression in the transfer function
$\Phi_k$. In the next chapter, we will confirm this expectation in the
context of a toy model of nonlinear structure formation.

\chapter{Nonlinear structure formation and backreaction}
We have seen so far that applying the averaging framework during
epochs when linear perturbation theory (PT) is expected to be valid,
leads to only negligible modifications in the standard cosmological 
equations. This is not unexpected, since the estimate in
e.g. \eqn{exs-eds5} does not depend crucially on the details of the
averaging procedure, and is therefore robust. Most of the interest in
the backreaction issue however, has been from the point of view of
\emph{late time} cosmology when matter fluctuations at least have
become \emph{nonlinear} on small scales. It has been claimed using
some simple models of structure formation \cite{rasanen} that using
the perturbed FLRW ansatz for the metric is no longer a good
approximation in this regime, and that one should expect backreaction
effects to grow large at these times. We will investigate this issue
in this chapter, in the framework of an exact toy model of structure
formation based on the LTB solutions of general relativity (see
Appendix B). 

We have already seen in our linear PT calculation that \emph{if} the
metric at late times continues to be of the perturbed FLRW form
\eqref{linPT1}, then backreaction effects of the nonlinear scales are
in fact likely to be small, which follows from studying the structure
of the integrands of the various backreaction functionals (see the
discussion of \figs{fig5} and \ref{fig6}). Order of magnitude
arguments such as those in \Cite{wald} suggest that the perturbed FLRW
metric does remain a good approximation at late times. In this chapter
we will see an explicit demonstration that a fully relativistic,
highly nonlinear collapsing system can be described by a perturbed
FLRW metric, provided \emph{peculiar velocities} of the matter remain
nonrelativistic. Further, in our model (which will satisfy this
condition) we can then use the formalism developed in chapter 4 to
explicitly compute the backreaction. We will see that the backreaction
in this case does remain small.

\section{Spherical Collapse : Setting up the model}
\noindent
Our model is based on the LTB solution described in Appendix B, and is
completely determined once the initial conditions are specified.

\subsection{Initial conditions}
\noindent
While choosing the initial density, velocity and coordinate scaling
profiles, we make the important assumption that at initial
time, a well-defined \emph{global} background FLRW solution can be
identified, with scale factor $a(t)$, Hubble parameter $H(t)$ and
density $\rho_b(t)$. This is reasonable since the CMB data (combined
with the Copernican principle) assure us that inhomogeneities at the
last scattering epoch were at the level of 10 parts per million. This
assumption plays a crucial role in deciding which regions are
overdense and will eventually collapse, and which regions will keep
expanding.
\begin{itemize}
\item {\bf Initial density profile} $\rho(t_i,r)$ : \\
The initial density is chosen to be
\begin{equation}
\rho(t_i,r)=\rho_{bi}\left\{
\begin{array}{l}
(1+\delta_\ast),~~~~r<r_\ast\\
(1-\delta_v),~~~r_\ast<r<r_v\\
1,~~~~~~~~~~~~r>r_v\,,
\end{array}\right .
\label{nonlin-2eq6}
\end{equation}
where $\rho_{bi}=\rho_b(t_i)$. Initially, the region $r<r_\ast$ is
assumed to contain a tiny overdensity and the region $r_\ast<r<r_v$, 
an underdensity. In other words, 
\begin{equation}
0 < \delta_\ast,\delta_v \ll 1\,.
\label{nonlin-2eq7}
\end{equation}
The discontinuities in the initial density profile can be smoothed out
by replacing the step functions appropriately. We will not do this
here, since the step functions make calculations very simple. This is
not expected to affect the qualitative features of our final results. 
\item {\bf Initial conditions on scaling and velocities} : \\
We match the initial velocity and coordinate scaling to the global
background solution, by requiring 
\begin{align}
R(t_i,r) &= a_i r\,,
\label{nonlin-2eq8} \\
\dot R(t_i,r) &= a_i H_i r \,,
\label{nonlin-2eq9}
\end{align}
with $a_i$ and $H_i$ denoting the initial values of the scale factor
and Hubble parameter respectively of the global background. This
amounts to setting the initial velocities to match the Hubble flow,
ignoring initial peculiar velocities. This is only a convenient choice
and the introduction of initial peculiar velocities is not expected to
modify our final results qualitatively. 
\end{itemize}
For the FLRW background we consider an Einstein-deSitter (EdS)
solution with scale factor and Hubble parameter given by 
\begin{align}
&a(t) = (t/t_0)^{2/3} ~~;~~ t_0 = 2/(3H_0) \,,
\label{nonlin-2eq10} \\
&H(t) \equiv \dot a/a = 2/(3t)\,,
\label{nonlin-2eq11}
\end{align}
with $t_0$ denoting the present epoch. $a_i$ fixes the initial time as 
\begin{equation}
t_i = 2/(3H_0) a_i^{3/2}\,.
\label{nonlin-2eq12}
\end{equation}
We will always use $a_i=10^{-3}$, so that the initial conditions are
being set around the CMB last scattering epoch; in general $a_i$ must
be treated as one of the parameters in the problem. The initial EdS
background density is given in terms of $H_0$ and $a_i$ as
\begin{equation}
\rho_{bi}= \frac{3}{8\pi G}H_0^2a_i^{-3}\,.
\label{nonlin-2eq13}
\end{equation}
\subsection{Mass function $M(r)$ and curvature function $k(r)$}
\noindent
We now have enough information to fix $M(r)$ and $k(r)$. Using
\eqn{LTB2b} at initial time together with the scaling in
\eqn{nonlin-2eq8} gives us
\begin{equation}
GM(r)=\frac{1}{2}H_0^2r^3\left\{ 
\begin{array}{l}
1 + \delta_\ast,~~~~0<r<r_\ast\\
1 + \delta_v\left( \left(r_c/r\right)^3 - 1 \right),
~~~r_\ast<r<r_v\\    
1+(\delta_v/r^3)\left(r_c^3 - r_v^3\right),~~r>r_v\,, 
\end{array}\right .
\label{nonlin-2eq14}
\end{equation}
where we have defined a ``critical'' radius $r_c$ by the equation
\begin{equation}
\left(\frac{r_c}{r_\ast}\right)^3 = 1 + \frac{\delta_\ast}{\delta_v} 
\,. 
\label{nonlin-2eq15}
\end{equation}
The significance of $r_c$ will become apparent shortly.
Using the initial conditions \eqns{nonlin-2eq8} and
\eqref{nonlin-2eq9} in the evolution equation \eqref{LTB2a} at
initial time, gives 
\begin{equation}
k(r) r^2 = \frac{2GM(r)}{a_i r} - a_i^2 H_i^2 r^2\,,
\label{nonlin-2eq16}
\end{equation}
with $H_i^2 = H_0^2a_i^{-3}$, and hence
\begin{equation}
k(r)=\frac{H_0^2}{a_i}\left\{ 
\begin{array}{l}
\delta_\ast,~~~~r<r_\ast\\
\delta_v\left(\left(r_c/r\right)^3-1\right), ~~~ r_\ast<r<r_v\\    
(\delta_v/r^3)\left(r_c^3 - r_v^3\right), ~~ r>r_v\,.  
\end{array}\right .
\label{nonlin-2eq17}
\end{equation}
The significance of $r_c$ is now clarified. Since
$\delta_\ast,\delta_v>0$, we have $r_c>r_\ast$ by definition
(\eqn{nonlin-2eq15}). The following possibilities arise :
\begin{itemize}
\item If $r_c>r_v$, then $k(r)>0$ for all $r$, and every shell will 
ultimately collapse, including the ``void'' region $r_\ast<r<r_v$. 
\item If $r_c<r_v$, then $k(r)>0$ for $r<r_c$ and changes sign at
  $r=r_c$. Hence, the region $r_\ast<r<r_c$ will collapse even
  though it is underdense, while the region $r>r_c$ will expand
  forever.  
\item If $r_c=r_v$, then the ``void'' exactly compensates for the
  overdensity, and the universe is exactly EdS for $r>r_v$. [$GM(r) = 
  (1/2)H_0^2r^3$ and $k(r)=0$.] Also the ``void'' will eventually
  collapse.  
\end{itemize}
Clearly the most interesting case for us is the one with $r_c<r_v$,
and we will hence make this choice for our model. We realize that the
model as it stands is not a very realistic depiction of the (nearly
spherical) voids we see in our Universe \cite{voids-obs}, since these 
voids are seen to be \emph{surrounded} by ``walls'' of
matter. However, our goal is to describe two regions, one of which
collapses while the other expands ever more rapidly, and our model is
capable of doing so while retaining its fully relativistic
character. 

Although we have set up the model for all values of the radial
coordinate $r$, hereon we will concentrate on the region
$0<r<r_v$. One reason is that most of the interesting dynamics takes
place in this region. Another is that the region $r>r_v$ develops
shell-crossing singularities due to the sharp rise in density across
$r=r_v$. A more realistic model would be able to incorporate the
pressures that are expected to build up when a shell-crossing occurs
\cite{bert}, but the LTB
model is limited in this respect due to its pressureless character. We
will therefore ignore the region $r>r_v$.

\subsection{The solution in the region $0<r<r_v$}
\noindent
The region of interest can be split into three parts : region 1
$=\left\{0<r<r_\ast\right\}$, region 2 $=\left\{r_\ast<r<r_c\right\}$
and region 3 $=\left\{r_c<r<r_v\right\}$. The solution in the three
regions is as follows :
\begin{itemize}
\item {\bf region 1 ($0<r<r_\ast$) :}  
\begin{subequations}
\begin{align}
&R = \frac{1}{2}\left(\frac{a_i}{\delta_\ast}\right) r (1+\delta_\ast)
(1-\cos{u}) \,,
\label{nonlin-2eq18a}\\
&u-\sin{u} = \frac{2H_0}{1+\delta_\ast}
\left(\frac{\delta_\ast}{a_i}\right)^{3/2} (t-t_i) + (u_i - \sin{u_i})
\,, 
\label{nonlin-2eq18b}\\
&1 - \cos{u_i} = \frac{2\delta_\ast}{1+\delta_\ast} \,,
\label{nonlin-2eq18c}\\
&R^2R^\prime = \frac{R^3}{r}\,.
\label{nonlin-2eq18d}
\end{align}
\label{nonlin-2eq18}
\end{subequations}
\end{itemize}
For regions $2$ and $3$, it is convenient to define a function
$\varepsilon(r)$ as
\begin{equation}
\varepsilon(r) \equiv \delta_v \left( \left(\frac{r_c}{r}\right)^3 - 1
\right) = \frac{a_i}{H_0^2}k(r) \,, ~~~ r_\ast < r < r_v\,.
\label{nonlin-2eq19}
\end{equation}
\begin{itemize}
\item {\bf region 2 ($r_\ast<r<r_c$) :}  
\begin{subequations}
\begin{align}
&R = \frac{1}{2}\left(\frac{a_i}{\varepsilon}\right) r (1+\varepsilon) 
(1-\cos{\alpha}) \,,
\label{nonlin-2eq20a}\\
&\alpha-\sin{\alpha} = \frac{2H_0}{1+\varepsilon}
\left(\frac{\varepsilon}{a_i}\right)^{3/2} (t-t_i) + (\alpha_i -
\sin{\alpha_i}) \,, 
\label{nonlin-2eq20b}\\
&1 - \cos{\alpha_i(r)} = \frac{2\varepsilon}{1+\varepsilon} \,,
\label{nonlin-2eq20c}\\
&R^2R^\prime = \frac{R^3}{r} \left(1 -
  \frac{r\varepsilon^\prime}{\varepsilon(1+\varepsilon)}\left\{1 -
  \frac{\varepsilon^{3/2}}{(1-\cos{\alpha})^2}
  \left[H_i(t-t_i)\sin{\alpha}
  \left(\frac{3+\varepsilon}{1+\varepsilon}\right) \right. \right. \right. 
\nonumber\\    
&\ph{R^2R^\prime = \frac{R^3}{r}\left[1 -
    \frac{r\varepsilon^\prime}{\varepsilon(1+\varepsilon)} \right]  }
  \left. \left. \left. +  \,
\frac{4\varepsilon^{1/2}}{(1+\varepsilon)^2}
\left(\frac{\sin{\alpha}}{\sin{\alpha_i}}\right) \right]\right\}  
  \right)    \,.
\label{nonlin-2eq20d}
\end{align}
\label{nonlin-2eq20}
\end{subequations}
\item {\bf region 3 ($r_c<r<r_v$) :}  
\begin{subequations}
\begin{align}
&R = \frac{1}{2}\left(\frac{a_i}{|\varepsilon|}\right) r (1+\varepsilon)  
(\cosh{\eta}-1) \,,
\label{nonlin-2eq21a}\\
&\sinh{\eta}-\eta = \frac{2H_0}{1+\varepsilon}
\left(\frac{|\varepsilon|}{a_i}\right)^{3/2} (t-t_i) + (\sinh{\eta_i} -
\eta_i) \,,  
\label{nonlin-2eq21b}\\
&\cosh{\eta_i(r)}-1 = \frac{2|\varepsilon|}{1+\varepsilon} \,,
\label{nonlin-2eq21c}\\
&R^2R^\prime = \frac{R^3}{r} \left(1 -
  \frac{r\varepsilon^\prime}{\varepsilon(1+\varepsilon)}\left\{1 -
  \frac{|\varepsilon|^{3/2}}{(\cosh{\eta}-1)^2}
  \left[H_i(t-t_i)\sinh{\eta}
  \left(\frac{3+\varepsilon}{1+\varepsilon}\right) \right. \right. \right. 
\nonumber\\    
&\ph{R^2R^\prime = \frac{R^3}{r}\left[1 -
    \frac{r\varepsilon^\prime}{\varepsilon(1+\varepsilon)} \right]  }
  \left. \left. \left. +  \,
\frac{4|\varepsilon|^{1/2}}{(1+\varepsilon)^2}
\left(\frac{\sinh{\eta}}{\sinh{\eta_i}}\right) \right]\right\}  
  \right)    \,.
\label{nonlin-2eq21d}
\end{align}
\label{nonlin-2eq21}
\end{subequations}
\end{itemize}
The crossover from region 1 to region 2 is discontinuous in $R^\prime$
(but not in $R$) due to our discontinuous choice of initial
density. Smoothing out the density will also smooth out
$R^\prime$. The crossover from region 2 to region 3 can be shown to be
smooth, by considering the limits $r\to r_c^-$ and $r\to r_c^+$ or
equivalently $\varepsilon\to0^-$ and $\varepsilon\to0^+$. Note that the
results in \eqns{nonlin-2eq18}, \eqref{nonlin-2eq20} and
\eqref{nonlin-2eq21} are exact, and do not involve any perturbative
expansions in $\delta_\ast$ or $\delta_v$, even though these
parameters are small. 

\subsection{Behaviour of the model}
\noindent
Each shell in the inner, homogeneous and overdense region 1 behaves as
a closed FLRW universe, expanding out to a maximum radius $R_{max}(r)$
given by 
\begin{equation}
R_{max}(r) = \frac{a_i}{\delta_\ast}r(1+\delta_\ast)\,.
\label{nonlin-2eq22}
\end{equation}
\begin{table}[t]
\centering
\begin{tabular}{|c|c|}\hline
Parameter name & Parameter value \\ [1ex]\hline \hline
$a_i$ & $0.001$ \\ [0.5ex]\hline
$H_0$ & $~~~1/13.59\, {\rm Gyr}^{-1} ~~ (=72 \,{\rm km/s/Mpc})~~~$  
\\ [0.5ex] \hline  
$t_0$ & $2/(3H_0) = 9.06 {\rm Gyr}$ \\ [0.5ex] \hline
$c$ & $306.6\, {\rm Mpc Gyr}^{-1}$\\ [0.5ex] \hline
$\delta_\ast$ & $1.25 a_i(3\pi/4)^{2/3} = 2.21 \times 10^{-3}$
\\ [0.5ex] \hline 
$\delta_v$ & $0.005$ \\ [0.5ex] \hline 
$r_\ast$ & $0.004 c/H_0 = 16.7\, {\rm Mpc}$ \\ [0.5ex] \hline 
$t_{turn}/t_0$ & $0.72$ \\ [0.5ex] \hline
$r_c$ & $r_\ast\left(1+\delta_\ast/\delta_v\right)^{1/3} = 18.8 \,
   {\rm Mpc}$ \\ [0.5ex] \hline 
$r_v$ & $1.25 r_c = 23.5\, {\rm Mpc}$\\ [0.5ex] \hline 
$R(t_0,r_\ast)$ & $6.8\,{\rm Mpc}$\\ [0.5ex] \hline 
$R(t_0,r_v)$ & $33.3\,{\rm Mpc}$\\ [0.5ex] \hline 
\end{tabular}
\caption{\small Values of various parameters used in generating
  plots.}  
\label{tab1}
\end{table}
All the inner shells reach their maximum radius and turn around at the
same time $t_{turn}$ given by
\begin{equation}
t_{turn} = t_i + \frac{1+\delta_\ast}{2H_0}
\left(\frac{a_i}{\delta_\ast}\right)^{3/2}
\left(\pi-(u_i-\sin{u_i})\right) \approx
t_0\left(\frac{3\pi}{4}\right)
\left(\frac{a_i}{\delta_\ast}\right)^{3/2} \,,
\label{nonlin-2eq23}
\end{equation}
where we have used the smallness of $a_i$ and $\delta_\ast$ to make
the last approximation. By appropriately choosing a value of
$\delta_\ast$, we can arrange for the turnaround of region 1 to occur
either before or after the present epoch. 

\begin{figure}[t]
\centering
\subfigure[]{\includegraphics[width=.4\textwidth]
{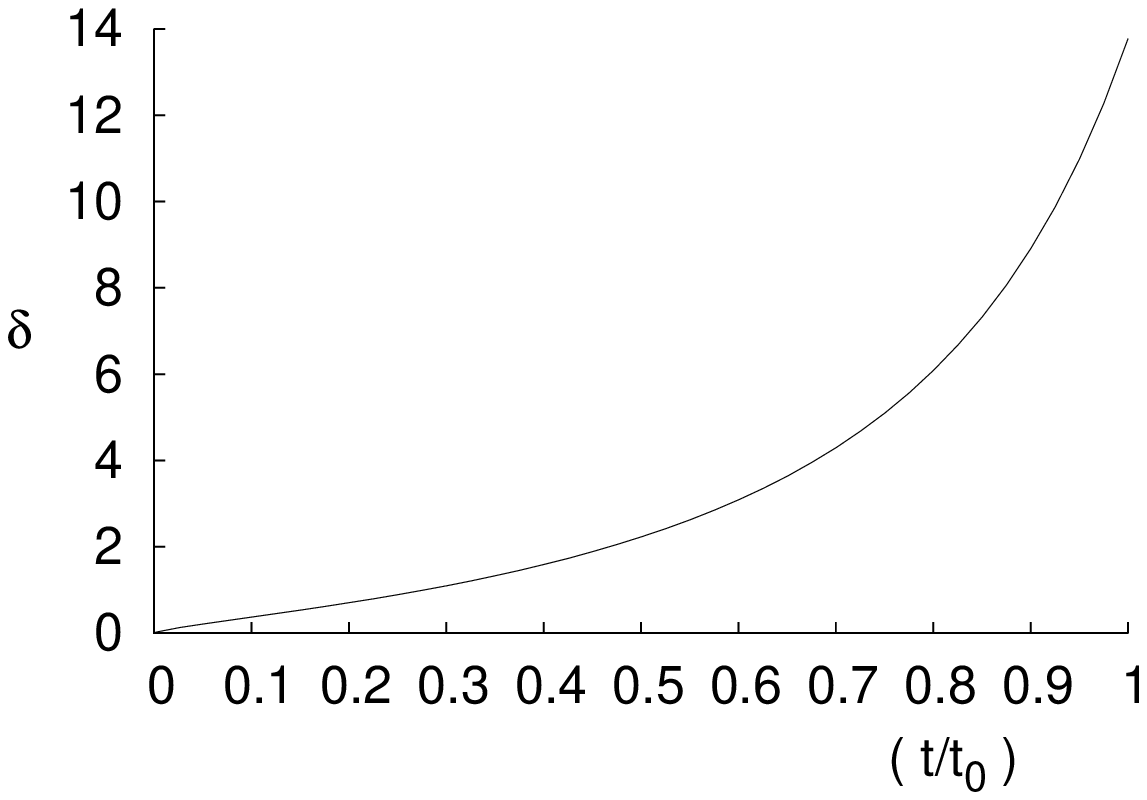}  \label{fig8a}}  
\hspace{.05\textwidth} 
\subfigure[]{\includegraphics[width=.4\textwidth]
{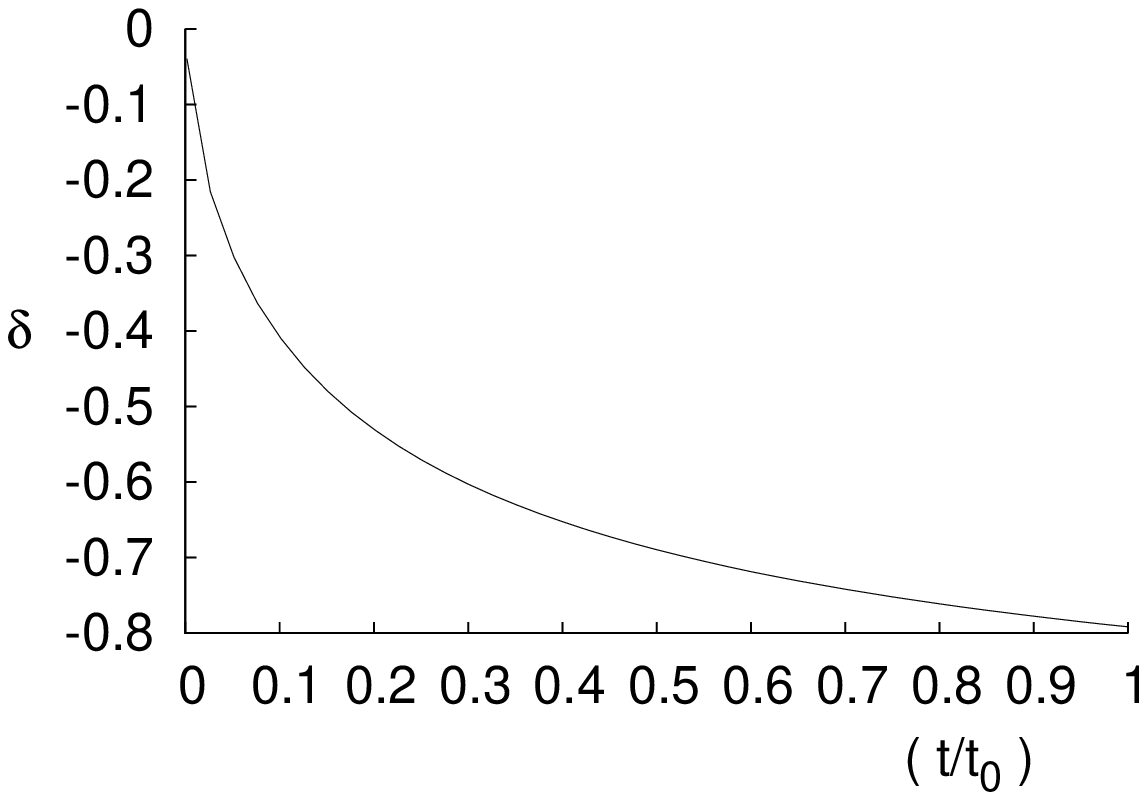}  \label{fig8b}}  
\caption{\small The evolution of the density contrast $\delta(t,r)$,
 using parameter values from \tab{tab1} evaluated at (a) $r=r_\ast/2$
 in region 1 and (b) $r=(r_c + r_v)/2$ in region 3.}  
\label{fig8}
\end{figure}

In \tab{tab1} we have listed the parameter values which we will use
frequently in displaying plots. Along with the parameter set
$\{a_i,H_0,\delta_\ast,\delta_v,r_\ast,r_v\}$, we have also listed the
values of the derived quantities $\{r_c,t_i,t_0,t_{turn}\}$ and speed
of light $c$ in units of ${\rm Mpc Gyr}^{-1}$. We have also shown the
values of the present day physical area radius $R(t_0,r)$ at
$r=r_\ast$ and $r=r_v$. The density contrasts are to be understood to
reflect the inhomogeneities in the dark matter density close to last
scattering, and not the inhomogeneities of the baryons which were much
smaller \cite{dodelson}. 
In \fig{fig8} we have shown the evolution of the density contrast
$\delta(t,r)$ defined in the usual way by 
\begin{equation}
1+\delta(t,r) = \frac{\rho(t,r)}{\rho_b(t)}\,,
\label{nonlin-2eq24}
\end{equation}
for the parameter choices of \tab{tab1}, for which one has
$t_{turn}/t_0 \simeq 0.72$, so that the collapse is well under 
way in region 1 at the present epoch. The two panels show the contrast
for two representative values of $r$, one in region 1 and the other in
region 3.

\subsection{Aside : Acceleration from initial conditions}
It is interesting to note that our model is capable of qualitatively
reproducing results derived by earlier by R\"as\"anen \cite{rasanen}
in the context of a very simple model of structure
formation. R\"as\"anen's model can be summarized as follows : one
considers two disjoint regions, one overdense and the other completely
empty, each evolving according to the FLRW evolution equations. (The 
embedding of these regions in an FLRW background, and the behaviour of 
the region \emph{between} these two regions, is not considered.) The
scale factor in the overdense region therefore behaves as
$a_1\propto(1-\cos{u})$ with $t\propto (u-\sin{u})$, and the scale
factor in the empty region behaves as $a_2\propto t$. It is then
straightforward to show that if one defines a volume averaged scale
factor by $a^3\equiv a_1^3+a_2^3$, then the effective deceleration
parameter given by $q\equiv-(\ddot a a)/{\dot a}^2$ becomes negative
(indicating acceleration) around the time that the overdense region
turns around and starts collapsing.

\begin{figure}[t]
\centering
\subfigure[$t_{turn}= 1.278 t_0$]{\includegraphics[width=.4\textwidth] 
{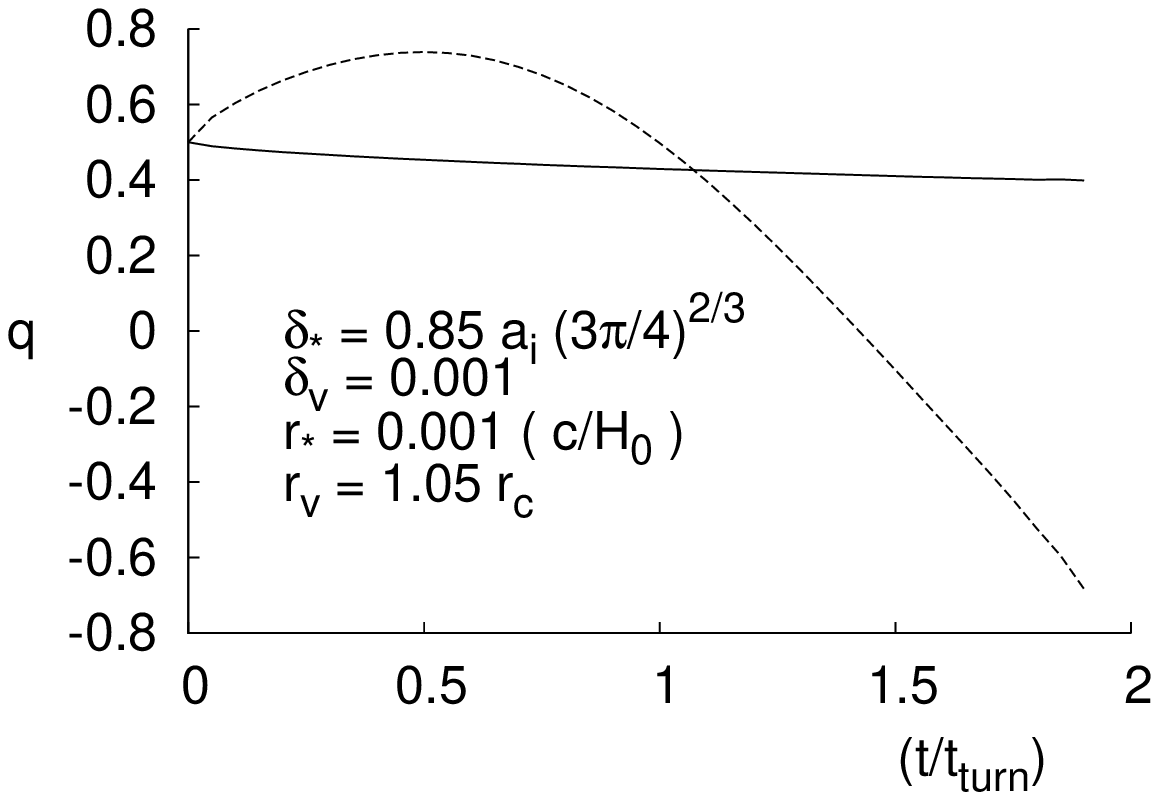}  \label{fig9a}}  
\hspace{.05\textwidth} 
\subfigure[$t_{turn}= 1.0018 t_0$]{\includegraphics[width=.4\textwidth] 
{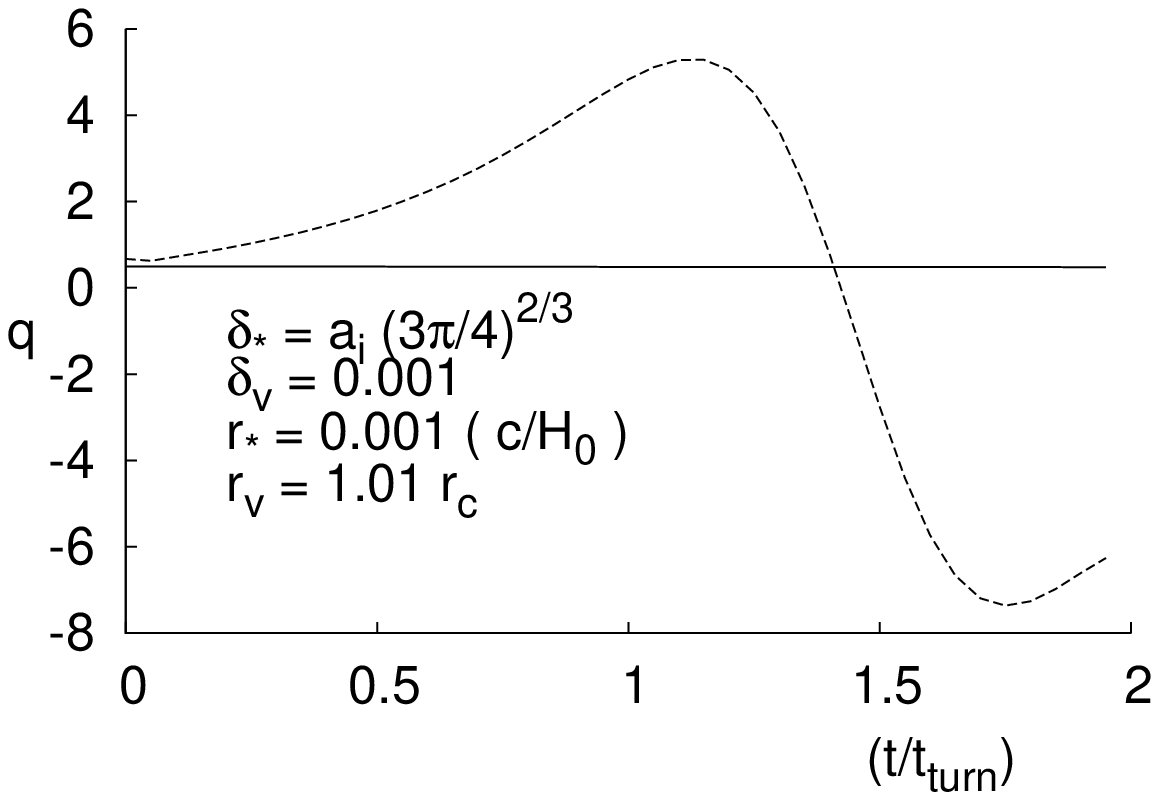}  \label{fig9b}}\\
\subfigure[$t_{turn}= 0.868 t_0$]{\includegraphics[width=.4\textwidth]
{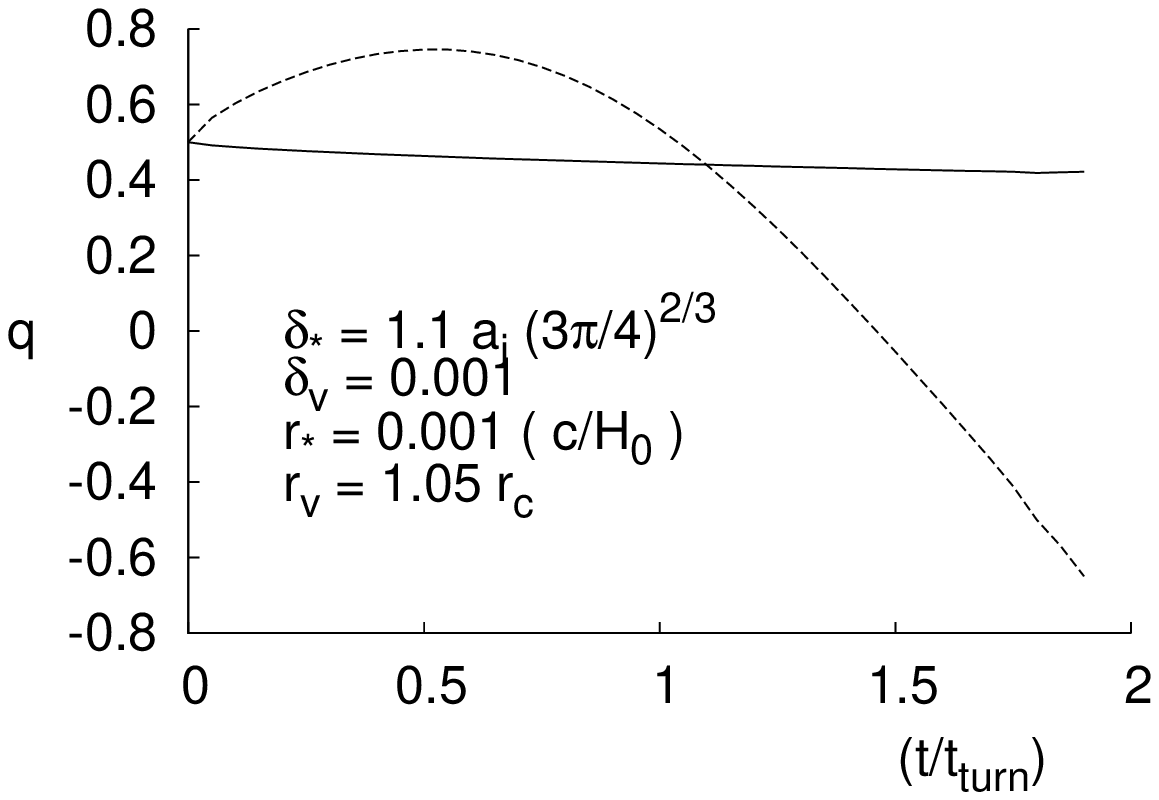}  \label{fig9c}}
\hspace{.05\textwidth} 
\subfigure[$t_{turn}= 0.717 t_0$]{\includegraphics[width=.4\textwidth]  
{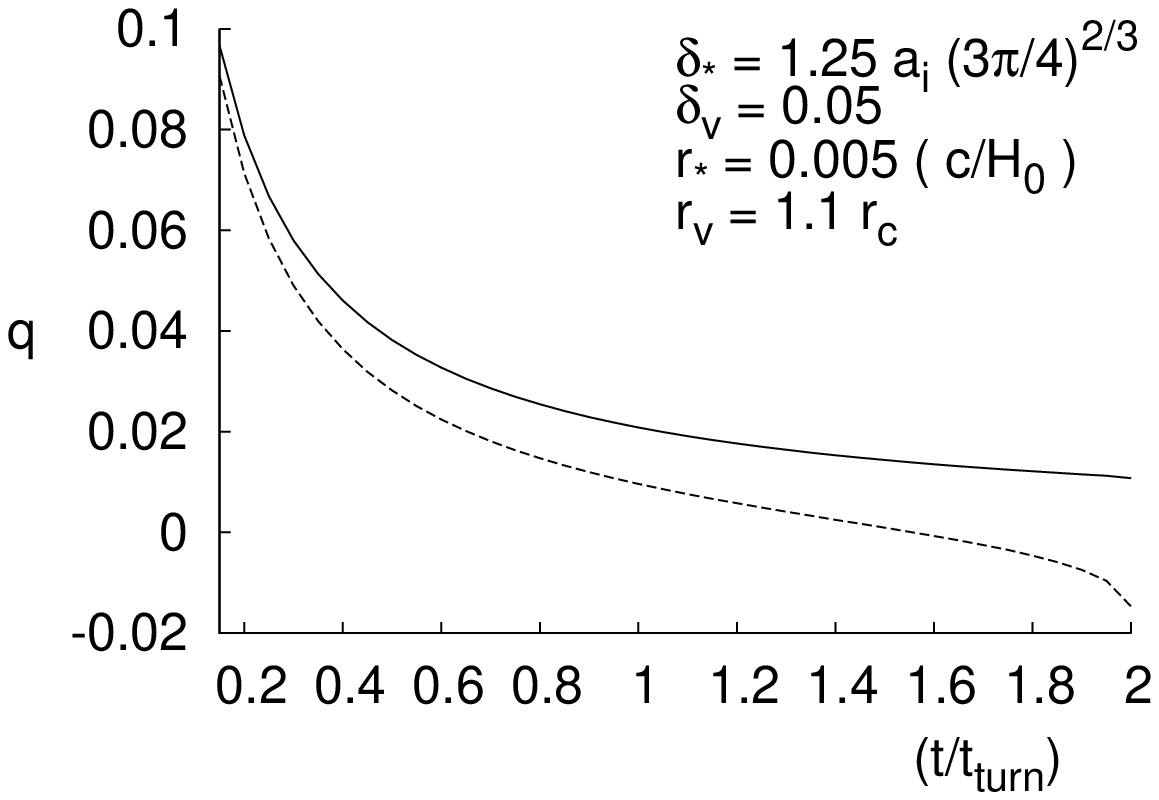}  \label{fig9d}}
\caption{\small The deceleration parameters for a range of parameter
  values. The dashed lines correspond to $q_{mod}$ and the solid lines
  to $q$. The $x$-axis shows $t/t_{turn}$, where $t_{turn}$ is the
  time at which region 1 turns around, and is different for each
  plot. The values for $a_i$, $H_0$ and $c$ are the same as those
  listed in \tab{tab1}. [Both curves in \fig{fig9d} begin at
    $q\sim0.5$ at $t=t_i$.]}    
\label{fig9}
\end{figure}

If we define the volume of each of our three comoving
regions separately, as
\begin{equation}
V_1 \equiv
4\pi\int_0^{r_\ast}{\frac{R^2R^\prime}{\sqrt{1-k(r)r^2}}dr} ~~;~~ V_2
\equiv
4\pi\int_{r_\ast}^{r_c}{\frac{R^2R^\prime}{\sqrt{1-k(r)r^2}}dr} ~~;~~ 
V_3 \equiv
4\pi\int_{r_c}^{r_v}{\frac{R^2R^\prime}{\sqrt{1-k(r)r^2}}dr} \,,
\label{nonlin-2eq25}
\end{equation}
then the total volume of the region can be used to define a
``Buchert-style'' volume averaged scale factor as
\begin{equation}
a(t)\equiv \left(\frac{V(t)}{V(t_0)}\right)^{1/3} ~~;~~ V(t) \equiv 
V_1(t)+V_2(t)+V_3(t) \,,
\label{nonlin-2eq26}
\end{equation}
and hence an effective deceleration parameter $q$ given by
\begin{equation}
q\equiv -\frac{\ddot a a}{{\dot a}^2} = 2 - 3\frac{\ddot V V}{{\dot
    V}^2} \,,
\label{nonlin-2eq27}
\end{equation}
whereas R\"as\"anen's model can be mimicked
more closely by ours, if we simply remove the region 2, by hand. By
doing so we are left with two disjoint regions, each spherically
symmetric, one of which is collapsing and the other expanding ever
rapidly and becoming ever emptier. There is no physical
reason to throw away region 2 in this manner, but for the sake of
comparison we will define a ``modified'' scale factor $a_{mod}$ and
it's corresponding deceleration parameter $q_{mod}$ by
\begin{equation}
a_{mod}(t) \equiv \left( \frac{V_1(t)+V_3(t)}{V_1(t_0)+V_3(t_0)}
\right)^{1/3} ~~;~~  
q_{mod} \equiv -\frac{\ddot a_{mod} a_{mod}}{{\dot a_{mod}}^2}\,.
\label{nonlin-2eq28}
\end{equation}
In \fig{fig9} we plot $q(t)$ and
$q_{mod}(t)$, for several sets of initial conditions which are close
to our ``base set'' listed in \tab{tab1} (except for \fig{fig9d} which
has a large value for $\delta_v$)\footnote{Both curves in \fig{fig9d}
  begin at $q\sim0.5$ at $t=t_i$. To enhance the contrast between the
  curves, we have plotted them for times $t>0.15 t_{turn}$. The
  remaining plots (Figs. \ref{fig9a}--\ref{fig9c}) are plotted
  starting from $t=t_i$.}. The
various initial conditions correspond to turnaround times that are
slightly greater than, or slightly less than, or significantly less
than the present epoch. The results are therefore valid regardless of
whether the collapse has just begun or is well under way at the
present epoch.   
We see that while the modified scale factor does accelerate as in
R\"as\"anen's model, the scale factor $a(t)$ does not show this
effect. The reason for this can be understood as follows. The region 2
is of a rather peculiar nature -- it is underdense initially and
becomes emptier with time, however its evolution is closely linked to
that of the \emph{overdense} region 1. Namely, the whole of region 2
(except its boundary at $r=r_c$), is dragged along with region 1 and
eventually turns around, instead of expanding away to infinity like
its counterpart region 3. Now, if one ignores region 2, then
R\"as\"anen's arguments about the remaining two regions stand -- one
region is contracting and the other is expanding faster than the
global mean, and this stand-off leads to an acceleration of the
effective scale factor $a_{mod}$, as we see in the plots of
\fig{fig9}. But if we account for region 2 as well, then we bring in a
counter-balancing influence of a large \emph{underdense} volume which
is expanding \emph{slower} than average, and this reduces the
accelerating influence of region 3 to the point of making the effect
completely disappear. Note that at late times, the volume of region 1
contributes negligibly to the total volume, and the volumes of regions
2 and 3 are comparable. 

We wish to highlight two points. First, it is very important to note
the role played by the initial conditions in this entire
excercise. The function $k(r)$ is defined in a continuous fashion once
the initial density, velocity and coordinate scaling are given, and
$k(r)$ then decides which shells will eventually collapse and which
will not. The continuity of $k(r)$ assures us that in models such as
ours, with an overdensity surrounded by an underdensity, the
underdense region \emph{will always} contain a subregion in which
$k(r)>0$. We see therefore that the existence of region 2, is a
generic feature not restricted to our specific choice of discontinuous
initial density or vanishing initial peculiar velocities. Further, as
we see in \fig{fig9d}, it is possible to make $q$ deviate even more
significantly from the EdS value than the $\sim10\%$ effect of the
first three figures, but this requires an unnaturally high value of
$\delta_v\gtrsim0.01$ (the figure has $\delta_v=0.05$), which
contradicts CMB data.  
Secondly, one may argue about the ``naturalness'' of choosing one set
of regions over another set, in order to compute volumes. But this
itself places the physicality of the acceleration effect into question
-- if one has to judiciously choose a specific set of averaging
domains in order to obtain acceleration on average, then the effect
would appear to be an artifact of this choice rather than something
which observers would see. 

\section{Transforming to Perturbed FLRW form}
\noindent
We now turn to the main calculation of this section.
We ask whether the LTB metric \eqref{LTB1} for our model can be
brought to the perturbed FLRW form with scalar perturbations, at any
arbitrary stage of the collapse. Namely, we want a coordinate
transformation $(t,r)\to(\tau,\ti{r})$ such that the metric in the
new coordinates is
\begin{equation}
ds^2 = -(1+2\vphi)d\tau^2 + a^2(\tau)(1-2\psi)\left(
d\ti{r}^2 + \ti{r}^2d\Omega^2\right)\,,
\label{nonlin-3eq1}
\end{equation}
with at least the conditions
\begin{equation}
\mid\vphi\mid\ll1 ~~;~~ \mid\psi\mid\ll1\,,
\label{nonlin-3eq2}
\end{equation}
being satisfied. We will ignore conditions on the derivatives of
$\psi$ and $\vphi$ for now (see the end of Section 3.2). The
scale factor is the EdS solution, with $\tau$ as the argument. The
coordinate $\ti{r}$ is comoving with the (fictitious) background
Hubble flow, but not with the matter itself. On physical grounds we
expect that this transformation should be possible as long as the
gravitational field is weak and matter velocity is small. We will see
below that this is exactly what happens. In the new coordinates, all
matter shells labelled by $\ti{r}$ expand with the Hubble flow, with a
superimposed peculiar velocity. 

Since we want $\ti{r}$ to be comoving with the background, the natural
choice for this coordinate would be $\ti{r}\sim R/a$, at least at
early times.  Also, we need to account for the local spatial curvature
induced by the initial conditions. As an ansatz for the coordinate
transformation therefore, we consider the equations 
\begin{subequations}
\begin{equation}
\ti{r} = \frac{R(t,r)}{a(t)}\left(1 + \xi(t,r)\right) \,, 
\label{nonlin-3eq3a}
\end{equation}
\begin{equation}
\tau = t + \xi^0(t,r)\,,
\label{nonlin-3eq3b}
\end{equation}
\label{nonlin-3eq3}
\end{subequations}
where $\xi(t,r)$ and $\xi^0(t,r)$ are expected to satisfy
\begin{equation}
{\mid\xi\mid}\, \ll\, 1~~;~~ {\mid\xi^0H\mid}\, \ll\, 1\,.
\label{nonlin-3eq4}
\end{equation}
This form of the transformation keeps us close to the standard
gauge transformation of cosmological perturbation theory, while still
accounting for the deviations in the evolution from the background
FLRW, caused by structure formation. We will show that a
self-consistent transformation exists, which preserves the conditions
\eqref{nonlin-3eq2} and \eqref{nonlin-3eq4} for most of the
evolution. We will use the metric transformation rule given by 
\begin{equation}
\ti{g}_{ab}(\ti{x})\frac{\partial\ti{x}^a}{\partial x^i}
\frac{\partial\ti{x}^b}{\partial x^j} = g_{ij}(x)\,,
\label{nonlin-3eq5}
\end{equation}
and \emph{expand to leading order} in the small functions
$\xi$, $\xi^0H$, $\vphi$, $\psi$ and also $k(r)r^2$ which, as we
see from \eqn{nonlin-2eq17}, remains small in the entire region of 
interest. The relations in \eqn{nonlin-3eq5} must be analysed for t he
cases $(ij)=\{(tt), (tr), (rr), (\theta\theta) \}$, in each of the
three regions. (The remaining cases can be shown to lead to trivial or 
non-independent relations.) The analysis is similar to the standard
gauge transformation analysis in relativistic perturbation theory
\cite{dodelson}. Since the calculations involved are straightforward
but tedious, we will only present an outline of the calculation and 
highlight certain issues. At the end we will present equations for all
three regions and numerically show that the transformation is
well-behaved in the regime of interest.  

The case $(ij)= (\theta\theta)$ is easily analysed and
leads to
\begin{equation}
\psi = \xi^0H + \xi \,,
\label{nonlin-3eq6}
\end{equation}
The cases $(ij)= (tr)$ and $(tt)$ both require
$\mid\partial_t\ti{r}\mid\ll1$ for consistency (since the
RHS of \eqn{nonlin-3eq5} in these cases has no zero order term to
balance a large $\partial_t\ti{r}$). Note that since $t$ is the proper
time of each matter shell, the quantity $\partial_t\ti{r}$ is simply
the velocity of matter in the $(\tau,\ti{r})$ frame (which is
comoving with the Hubble flow). In other words,   
\begin{equation}
\ti{v} \equiv \frac{\partial\ti{r}}{\partial t}\,,
\label{nonlin-3eq8}
\end{equation}
is the radial comoving peculiar velocity of the matter shells in the 
$(\tau,\ti{r})$ frame. We will soon see that whereas the quantities
$\xi$ and $\xi^0$ behave roughly as $\sim(H_0r)^2$, the peculiar velocity
$a\ti{v}$ behaves roughly as $\sim(H_0r)$. We will therefore treat
$(a\ti{v})^2$ as a small quantity of the same order as $\xi$, etc. The
case $(ij)=(tr)$ then leads to 
\begin{equation}
\xi^{0\prime} = a\ti{v}R^\prime\,,
\label{nonlin-3eq9}
\end{equation}
the case $(ij)=(tt)$ gives
\begin{equation}
\vphi = -\dot\xi^0 + \frac{1}{2}(a\ti{v})^2\,,
\label{nonlin-3eq10}
\end{equation}
and the case $(ij)=(rr)$ gives\footnote{This corrects an error in
Eqn. 35 of \pap{4}. I am grateful to Karel Van Acoleyen for pointing
this out to me.}
\begin{equation}
\xi^\prime = \frac{1}{2}\left(k(r)r^2 +
(a\ti{v})^2\right)\left(\frac{R^\prime}{R}\right) \,.   
\label{nonlin-3eq7}
\end{equation}
The
equations \eqref{nonlin-3eq6}, \eqref{nonlin-3eq7}, \eqref{nonlin-3eq9}, 
and 
\eqref{nonlin-3eq10} are valid in the entire range $0<r<r_v$, provided the
peculiar velocity remains small in magnitude. The comoving peculiar
velocity is given by
\begin{equation}
\ti{v} = \partial_t\left(\frac{R}{a}\right)\,, 
\label{nonlin-3eq11}
\end{equation}
where we have assumed for consistency that $|\partial_t(R/a)|\ll1$ and
have dropped the term $(R/a)\dot\xi$ since it is expected to be of
higher order than $\partial_t(R/a)$. (This can be seen from simple
dimensional considerations -- we have $\partial_t(R/a)\sim HR/a$, and
since, from \eqns{nonlin-3eq7} and \eqref{nonlin-2eq17}, $\xi\sim
(HR)^2$, we also have $(R/a)\dot\xi\sim (HR)^3/a$.) We will see that
these conditions do indeed hold for most of the evolution, throughout
the region of interest. 
\subsection{The transformation in region 1 }
Since region 1 corresponds to a homogeneous solution, the integrals in
\eqns{nonlin-3eq9} and \eqref{nonlin-3eq7} can be analytically
performed. Since $R$ has the structure $R=ry_1(u(t))$, we have
$a\ti{v}=r(\dot y_1-y_1H)$ and \eqn{nonlin-3eq9} then leads to  
\begin{equation}
\xi^0 = \frac{1}{2}a\ti{v}R\,,
\label{nonlin-3eq13}
\end{equation}
after setting an arbitrary function of time to zero,
while \eqn{nonlin-3eq7} gives
\begin{equation}
\xi = \frac{r^2}{4}\left[ \frac{\delta_\ast}{a_i} H_0^2 + (\dot y_1 -
y_1H)^2\right]\,,    
\label{nonlin-3eq12}
\end{equation}
after setting another arbitrary function of time to zero\footnote{Note
  that it might be more meaningful to fix the two arbitrary functions
  of time $\xi(t,0)$ and $\xi^0(t,0)$, by requiring that $\xi(t,r_c)$
  and $\xi^0(t,r_c)$ vanish. This would be in line with the shell
  $r=r_c$ expanding like the flat EdS background. However, this
  complicates some of the expressions we evaluate, and does not change
  the order of magnitude of any of the final results. Hence we will
  continue to assume that the transformation functions $\xi$ and
  $\xi^0$ vanish at $r=0$ rather than at $r=r_c$.}. The peculiar
  velocity can be explicitly calculated to be  
\begin{equation}
a(t)\ti{v}(t,r) = (H_0r)
\left(\frac{\delta_\ast}{a_i}\right)^{1/2}
\left[\frac{\sin{u}}{(1-\cos{u})} - \frac{2}{3}
\frac{1-\cos{u}}{(u-\sin{u}+B)} \right] \,, 
\label{nonlin-3eq14}
\end{equation}
where the various functions are defined in \eqns{nonlin-2eq18}, and we
have defined the constant $B$ by
\begin{equation}
B \equiv \frac{2H_0t_i}{1+\delta_\ast} \left( \frac{\delta_\ast}{a_i}
\right)^{3/2} - \left(u_i-\sin{u_i} \right)\,.
\label{nonlin-3eq15}
\end{equation} 
\begin{figure}[t]
\centering
\fbox{
\includegraphics[height=0.25\textheight]{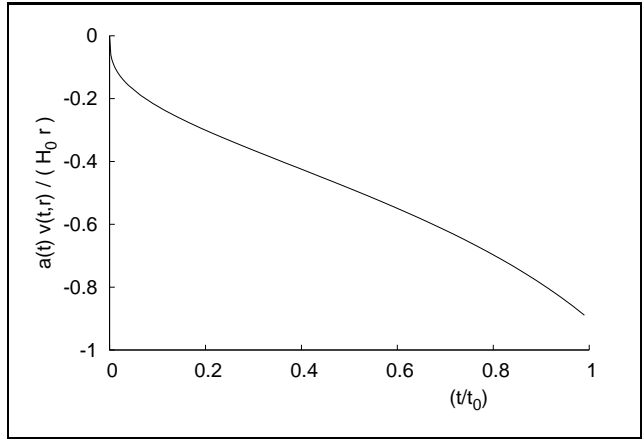}
}
\caption{\small The quantity $a\ti{v}/(H_0r)$ in region 1, plotted
  using parameter values from \tab{tab1}. Since $(H_0r_\ast/c) \sim
  0.001$, the peculiar velocity $a\ti{v}$ remains small.}
\label{fig10}
\end{figure}
In the rest of this section we will use the parameter values listed in 
\tab{tab1}. In \fig{fig10} we have plotted $a\ti{v}/(H_0r)$ in region
1. We see that this dimensionless quantity remains of order $\sim1$
throughout the evolution. For our choice of
$(H_0r_\ast/c)\sim10^{-3}$, which corresponds to an overdensity
spanning a few Mpc today, the peculiar velocity is of order
$\sim10^{-3}$ in region 1.

Direct calculation shows that $\vphi$ and $\psi$ are equal and given
by   
\be
\vphi = \psi = \frac{1}{4}\left(\frac{\delta_\ast}{a_i}\right)
(H_0r)^2 \left[ \frac{2}{1-\cos u} - \frac{4}{9} \frac{(1-\cos
u)^2}{(u-\sin u +B)^2} \right]\,,
\la{nonlin-3eq16}
\ee
It is not hard to see that for our parameter choices, $\vphi$ and
$\psi$ remain of order $\sim(H_0r)^2\sim10^{-6}$ for most of 
the evolution (the $1/(1-\cos u)$ factor will start becoming
significant only at very late times which are larger than $t_0$ for
our parameter choices). 
The fact that $\vphi=\psi$ is in fact a general result which follows
from the absence of anisotropic stresses in the problem. It can be
shown (see e.g. \Cites{karel,dodelson}) that the difference between
$\vphi$ and $\psi$ is governed by the stress-tensor component
$T_{\theta\theta}$ which vanishes for the spherically symmetric dust
we are considering\footnote{This is independent of whether we use
$(t,r)$ or $(\tau,\ti{r})$, since the angular coordinates are not
affected by this transformation.}. This is fortunate, since the form
of $\xi$ in regions 2 and 3 is complicated, and is cumbersome to
evaluate numerically. All we need however is $(a\ti{v})$ which can be
directly evaluated, and $\xi^0$ which can be found after one
integration in \eqref{nonlin-3eq9}. These are sufficient to determine
$\vphi$ and the form for $\psi$ immediately follows, assuming that
$\vphi$ and $\psi$ vanish at the same radius $r$ (which in our case is
$r=0$).

\subsection{The transformation in regions 2 and 3}
For the calculation in regions 2 and 3, the integrals involved cannot
be computed analytically. We will therefore display the expressions we
obtain for $\ti{v}$ and $\xi^0$, and plot the results of numerically 
computing $\vphi=\psi$ from these quantities.  
\begin{itemize}
\item {\bf region 2 ($r_\ast<r<r_c$):} \\
In region 2 we have
\begin{align}
&\ti{v} = \frac{R}{a} \left[ C(t,r) - H 
\right] \,,
\label{nonlin-3eq20}\\
&\xi^0(t,r) = \xi^0(t,r_\ast) + a(t)\int_{r_\ast}^r{ \ti{v}(t,\bar r) 
  R^\prime(t,\bar r)d\bar r}\,,
\label{nonlin-3eq21}
\end{align}
where $\xi^0(t,r_\ast)$ is computed from \eqn{nonlin-3eq13} at
$r=r_\ast$, and we have defined
\be
C(t,r) \equiv \frac{H_i\sin{\alpha}}{(1-\cos{\alpha})^2}
\frac{2\varepsilon^{3/2}}{1+\varepsilon}\,.
\label{nonlin-3eq18c}
\ee
$\vphi$ must now be computed using
\eqref{nonlin-3eq10}.

\begin{figure}[t]
\centering
\fbox{
\includegraphics[height=0.25\textheight]{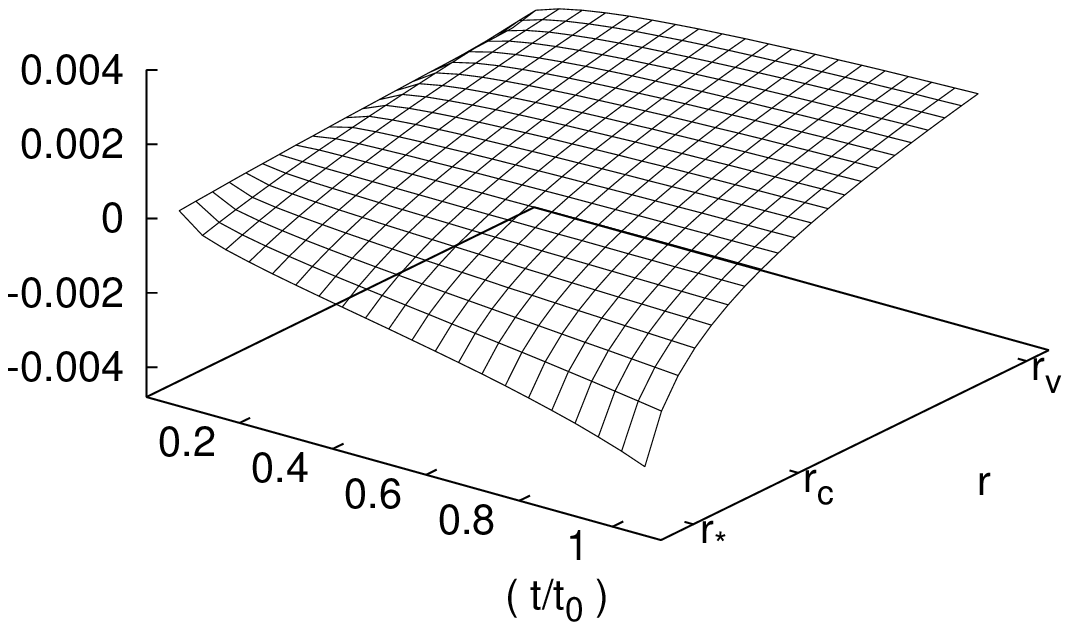}
}
\caption{\small The peculiar velocity $a\ti{v}/c$ in
  regions 2 and 3 using parameter values from \tab{tab1}.}
\label{fig11}
\end{figure}
\item {\bf region 3 ($r_c<r<r_v$) :}\\
The analysis is very similar to that in region 2. We find
\begin{align}
&\ti{v} = \frac{R}{a} \left[ D(t,r) - H 
\right] \,,
\label{nonlin-3eq24}\\
&\xi^0(t,r) = \xi^0(t,r_c) + a(t)\int_{r_c}^r{ \ti{v}(t,\bar r) 
  R^\prime(t,\bar r)d\bar r}\,,
\label{nonlin-3eq25}
\end{align}
where $\xi^0(t,r_c)$ is obtained
from \eqref{nonlin-3eq21}, evaluated in the limit $r\to r_c^-$, and we
have defined
\be
D(t,r)  \equiv \frac{H_i\sinh{\eta}}{(\cosh{\eta}-1)^2}
\frac{2|\varepsilon|^{3/2}}{1+\varepsilon}\,,
\label{nonlin-3eq22c}
\ee
\end{itemize}

\begin{figure}[t]
\centering
\fbox{
  \includegraphics[width=0.4\textwidth]{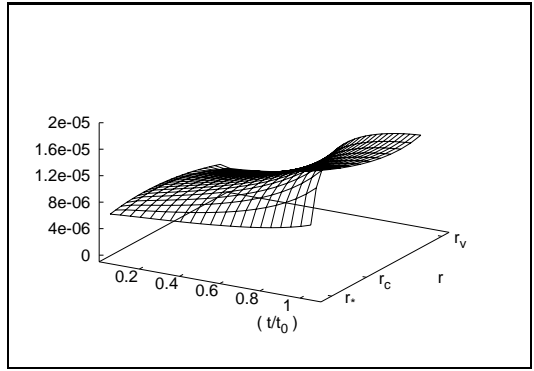}}
\caption{\small The metric function
  $\vphi(t,r)$ in regions 2 and 3 using parameter values from
  \tab{tab1}. The time axis begins at $t=50t_i$.}   
\label{fig12}
\end{figure}

In \fig{fig11}, we have plotted the velocity $a\ti{v}/c$ in regions 2 
and 3 for a range of time. It can be shown that at the order of
approximation we are working at, $a\ti{v}$ changes sign at
$r=r_c$\footnote{Recall $\varepsilon(r_c)=0$ and hence this shell
expands exactly like the EdS background. The metric in the   
$(\tau,\ti{r})$ coordinates will not be exactly EdS at $r=r_c$, due to
our unusual choice of normalisation for $\xi$ and $\xi^0$ at
$r=0$. This does not pose any problem for our conclusions.}. 
In \fig{fig12} we plot $\vphi$. We see that this function is well
behaved and remains small for the 
entire region of interest (in space and time). Hence the perturbed
FLRW picture is indeed valid for this system, even though each region
by itself appears to be very different from FLRW in the synchronous
coordinates comoving with the matter. Due to numerical difficulties
close to the initial time $t=t_i$, we have plotted the time axis
starting from $t=50t_i$. 
Note that the magnitude of $\vphi$ is sensitive to
the overall size of the region, determined by the value of
$R(t,r_v)$. For our parameter choices given in \tab{tab1}, the size of
the region at the present epoch is $\sim33 {\rm Mpc}$, which is a
typical size for observed voids. The dependence is roughly $(HR)^2$,
and hence a void which is about 10 times larger in length scale than
the above value, would have metric functions about 100 times larger.

We end this subsection by noting the following. 
It is known that simply having a metric of the
form \eqref{nonlin-3eq1} with only the \emph{magnitude} of the
perturbations being small, is not enough to guarantee consistency with
Einstein's equations written as a perturbation series; additional
constraints on the \emph{derivatives} of these functions must be
satisfied. These constraints, given in e.g. \Cite{wald}, take the form
(for the metric \eqref{nonlin-3eq1} with 
$\ti{\psi}=-\vphi$),
\begin{equation}
\left| \frac{\partial\vphi}{\partial t} \right|^2 \ll
\frac{1}{a^2} \nabla^\alpha\vphi \nabla_\alpha\vphi \,,~~~
\left(\nabla^\alpha\vphi \nabla_\alpha\vphi \right)^2 \ll
\left(\nabla^\alpha\nabla^\beta\vphi \right)
\nabla_\alpha\nabla_\beta\vphi\,, 
\label{nonlin-3eq26} 
\end{equation}
where $\alpha,\beta=1,2,3$, and $\nabla_\alpha$ is the spatial
covariant derivative associated with the flat 3-space metric. On
dimensional grounds, treating $\vphi\sim (HR)^2 \ll1$, $\partial_t
\sim H$ and $\nabla\sim aR^{-1}$, it is easy to see that these
constraints will be satisfied by our solution. This should also be
expected since we started from an exact solution of the Einstein
equations and performed a self-consistent coordinate transformation.

\subsection{The magnitude of the backreaction}
One can now legitimately ask the question, ``How large is the effect
of the small metric inhomogeneities?'' Naively, one would argue that
small inhomogeneities must lead to small effects. Indeed, the question
of the magnitude of the backreaction in the Newtonianly perturbed FLRW
setting has been investigated by Behrend, \emph{et al.} \cite{behrend}
in the linear and quasilinear regimes, and they find that corrections
to the FLRW equations remain at the level of one part in
$10^5$. However, what we are dealing with is a 
situation in which the \emph{matter} perturbations are completely
nonlinear, and it is not \emph{a priori} clear that the same arguments
would carry through. Indeed, we saw in section 2 that the deceleration
parameter $q$ deviated from its EdS value by about $\sim10\%$. Here we
give an argument based on dimensional considerations supplemented with
realistic numbers, which will show that this effect is scale
dependent, and is not expected to be present if a sufficiently large
averaging scale is chosen.

In the following we will work at the present epoch $t_0$. Consider a
model situation similar to the one we have been considering so far,
such that at present epoch the physical extent of the overdense region
is $R_\ast$, and that of the underdense is $R_v$. For order of
magnitude estimates, we assume that in the perturbed FLRW metric
\eqref{nonlin-3eq1} (which is valid for this system provided
$H_0R_v \ll1$), $\vphi \sim -\ti{\psi}$. Also assume that the density
contrast in the overdense region is $\delta_{\ast0}$ and that in the
underdense region is $\delta_{v0}$, where we take $\delta_{\ast0}$ and
$\delta_{v0}$ to be constant in space, which is fine for an order of
magnitude estimate. The backreaction in the Buchert approach contains,
among other terms, the spatial average of the quantity $\nabla^2\vphi$ 
which appears in the spatial curvature \cite{buchert1a,behrend}, where 
$\nabla^2$ is the Laplacian operator for the flat $3$-space
metric. The spatial curvature has the structure 
\begin{equation}
\mathcal{R} \sim \frac{1}{a^2}\left[(\#_1) \nabla^2\vphi +
  (\#_2) \vphi\nabla^2\vphi + (\#_3) (\nabla\vphi)^2 \right]\,,
\label{nonlin-3eq27}
\end{equation}
where $\#_1,\#_2,\#_3$ are constants whose values are irrelevant for
this order of magnitude argument. Due to the Einstein equations in the
small scale Newtonian approximation, the leading order effect in the
\emph{nonlinear} regime, comes from $\nabla^2\vphi$ which satisfies
\begin{equation}
\nabla^2\vphi \sim\,\, \left\{
\begin{array}{l}
H_0^2\delta_{\ast0} \,,~~~~~ \text{overdense region}\,,\\
H_0^2\delta_{v0} \,,~~~~~ \text{underdense region}\,.
\end{array} \right .
\label{nonlin-3eq28}
\end{equation}
Consider the situation when, at present epoch,
$R_\ast \sim 6$Mpc, $R_v \sim 30$Mpc, $\delta_{\ast0} \sim 10^2$
and $\delta_{v0} \sim -0.9$. These are typical numbers for clusters
of galaxies and voids. It is straightforward to now show that the
spatial average of $\nabla^2\vphi$ over a domain comprising the
overdense and underdense region, works out to be 
\begin{align}
 \langle \nabla^2\vphi \rangle &\sim \frac{H_0^2}{R_\ast^3 + R_v^3}
 \left[ R_\ast^3 \delta_{\ast0}  + R_v^3\delta_{v0} \right]\,,\nonumber \\
&\simeq -0.1 H_0^2\,.
\label{nonlin-3eq29}
\end{align}
It would appear  therefore, that this spatial average of
$\nabla^2\vphi$ (which is usually neglected) thus turns out to be a
significant contributor to the backreaction. (In fact it is the most
significant contributor, since the other terms are clearly of at least
one higher order in the small quantity $(H_0R_v)^2$, for such a
model.) 

As we now argue, however, the above effect can be deceptive, and is
really scale dependent. Let the initial density contrasts in the
overdense and underdense regions be $\delta_{\ast i}$ and
$\delta_{vi}$ respectively, so that $\delta_{\ast
  i},|\delta_{vi}|\ll1$. If $M_{\ast i}$, $M_{vi}$, $M_{\ast}$ and 
$M_v$ are the masses at initial time and today, in the overdense and
underdense region respectively, and $\rho_i$ and $\rho_0$ are the
values of the background density at initial time and today, then at
initial time 
\begin{equation}
M_{\ast i}  \approx \rho_i (a_i R_\ast)^3 = \rho_0 R_\ast^3\,,~~
M_{vi} \approx \rho_0 R_v^3 
\,, 
\label{nonlin-3eq30}
\end{equation}
and at present time,
\begin{equation}
M_\ast = \rho_0 (1 + \delta_\ast) R_\ast^3  >  M_{\ast i}\,, ~~ M_v =
\rho_0 (1 + \delta_v) R_v^3  <  M_{vi}\,.
\label{nonlin-3eq31}
\end{equation}
We now make the crucial observation that \emph{if the averaging scale
  is large enough}, and we are counting several such ``pairs'' of
overdense and underdense regions, then the mass ejected from the
underdense region must have all gone into the overdense region. It is
then easy to show, that
\begin{equation}
\delta_\ast R_\ast^3 \approx - \delta_v R_v^3\,,
\label{nonlin-3eq32}
\end{equation}
which means that, just like in the linear theory, the average of
$\nabla^2\vphi$ is expected to be negligible on such a scale. In the
real universe, we do expect that the averaging scale must be at least
of the order of the homogeneity scale, and on such a scale we will be
sampling several pairs of overdense and underdense regions. The only
cumulative effects that may arise with such a choice of scale are
from terms such as $(\nabla\vphi)^2$, which as we mentioned earlier,
are of one higher order in the perturbation and will give effects of
the size $\sim H_0^2 (H_0R_v)^2 \ll H_0^2$. (For a demonstration of
the scale dependence of the effect, see e.g. the work of Li and
Schwarz, the first paper in \Cite{buchert-obs}.) 


\section{Backreaction during nonlinear growth of structure}
\noindent
We can do better than the estimates for the magnitude of the
backreaction during late stages of structure formation. In chapter 4
we have already developed a formalism in place to calculate the
backreaction whenever the metric has the perturbed FLRW form
(irrespective of matter inhomogeneities). We can use this procedure on
our LTB model in the $(\tau,\ti{r})$ coordinates to explicitly
evaluate the backreaction functions.

The expressions for the backreaction in \eqns{corrscal} were derived 
under the requirement that the averaging operation be free of gauge
related ambiguities, in \emph{linear} perturbation theory. However,
the actual conditions used to derive \eqns{corrscal} only depended on
the fact that one is working with leading order effects in
the \emph{metric} perturbations. In particular, a key step was the
transformation \eqref{ginvar-VP2} between the metric in the conformal
Newtonian gauge  (in Cartesian spatial coordinates) and the
corresponding volume preserving form. In the 
present context, the same transformation remains valid \emph{at the
  leading order}, and hence the expressions \eqref{corrscal} for the
backreaction are physically relevant here as well. We emphasize that
this truncated averaging operation remains valid even at late times
since the weak field approximation for gravity works well during the
nonlinear phase of structure formation.

Since our numerical results are in terms of the LTB variables $(t,r)$,
where $r$ ``comoves'' with the matter but not with the FLRW
background, we need to reexpress the averaging
operation \eqref{avgVP11} in terms of these variables. It is easy to
show that, at the leading order, the average of a scalar $s(t,r)$
defined in \eqn{avgVP11} can be written as 
\be
\avg{s} = \frac{3}{(a(t)L)^3} \int_0^{r_L(t)}{sR^2R^\prime dr} \,,
\label{4eq2}
\ee
where the function $r_L(t)$ solves the equation
\be
R(t,r_L(t)) = a(t)L \,.
\label{4eq3}
\ee
\eqn{4eq2} gives the average of $s$ over a
single domain centered at the origin, which is what we will restrict
ourselves to in this section. There are two reasons behind this choice
: firstly this is the most natural choice given the symmetry of the
system, and secondly since our model is constructed as a ``typical
representative'' of nonlinear inhomogeneities, it makes sense to use
averages over the single central region as representative of more
general averages. As discussed in section 5.1, we are
constrained to consider values $r<r_v$, due to unphysical shell
crossing singularities in the region beyond. For this reason the
largest value of $L$ which we can choose is $L=r_v$, which then
ensures $r_L(t)<r_v$ since $r_L(t)$ is a decreasing function for this
choice. This gives us an averaging scale of $L=23.5$ Mpc (comoving with
the FLRW background), which is smaller than the more realistic
expected value of $\sim100h^{-1}$Mpc. One consequence
is that our model does not strictly satisfy the condition that the
potentials $\vphi$ and $\psi$ and their spatial and time derivatives
should average to zero, as was assumed in chapter 4. One can check
that the actual average values of the form $\avg{(\p_A\vphi)}^2$ are
small ($\lesssim10\%$) compared to terms like $\avg{(\p_A\vphi)^2}$
which are needed in the backreaction calculations, for all times,
although it turns out that the \emph{time} derivatives satisfy
$\avg{\dot\vphi}^2\sim\avg{(\dot\vphi)^2}$, throughout the
evolution. However, since the averaging scale chosen here is large
enough to encompass all the inhomogeneity of \emph{this} system, we
expect that our estimates for the backreaction functions in
\eqns{corrscal} are fairly representative. 

Consider now the function $\beta$. Since $\beta$ satisfies the Poisson
equation $\nabla^2\beta=\vphi-3\psi$ on a flat $3$-space background,
with no nontrivial solutions of the corresponding homogeneous equation 
allowed, we can directly write the solution for $\beta$ in terms of
the background radial coordinate $\ti{r}$ as,
\begin{align}
\nonumber
\beta(\tau,\ti{r}) &= -\frac{1}{4\pi} \int{
  \frac{q(\tau,\vec{y})}{|\vec{\ti{r}}-\vec{y}|} d^3y }
\nonumber\\ 
& = -\frac{1}{\ti{r}}\int_0^{\ti{r}}{q\, y^2dy} -
\int_{\ti{r}}^\infty{q\, ydy}\,,
\label{5eq1}
\end{align}
where $q \equiv -2\vphi$ (we have set $\vphi=\psi$ at leading order)
and the integration is over the spatial coordinates comoving with the
\emph{background}. The following relations turn out to be useful in
the calculations,  
\begin{subequations}
\begin{align}
\p_{\ti{r}}\beta &= \frac{1}{\ti{r}^2} \int_0^{\ti{r}}{
  q\,y^2dy}\,, \label{5eq3a}\\ 
\p_{\ti{r}}^2\beta &= q -
\frac{2}{\ti{r}}\p_{\ti{r}}\beta\,, \label{5eq3b} \\
\p_{\ti{r}}^3\beta &= \frac{6}{\ti{r}^2}\p_{\ti{r}}\beta -
\frac{2}{\ti{r}}q + \p_{\ti{r}}q\,. \label{5eq3c}
\end{align}
\label{5eq3}
\end{subequations}
We will need the quantities $\p_{\ti{r}}\beta$, $\p_{\ti{r}}^2\beta$
and $\p_{\ti{r}}^3\beta$ as functions of the LTB variables $(t,r)$,
which can be done by replacing $\tau$ and $\ti{r}$ at leading order by
$t$ and $R(t,r)/a(t)$ respectively. This gives us (treating $q$ as a
function of $t$ and $r$),
\begin{subequations}
\begin{align}
(\p_{\ti{r}}\beta)(t,r) &= \frac{1}{aR^2} \int_0^{r}{
  q\,R^2R^\prime dr}\,, \label{5eq4a}\\ 
(\p_{\ti{r}}^2\beta)(t,r) &= q -
\frac{2}{R^3}\int_0^{r}{ q\,R^2R^\prime dr}\,, \label{5eq4b} \\ 
(\p_{\ti{r}}^3\beta)(t,r) &= \frac{6a}{R^4}\int_0^{r}{q\,R^2R^\prime
  dr} - \frac{2a}{R}q + \frac{a}{R^\prime}q^\prime\,, \label{5eq4c}
\end{align}
\label{5eq4}
\end{subequations}
where in the last equation we have used the fact that, at leading
order, $\p_{\ti{r}}q = (a/R^\prime)q^\prime$.

Also, noting that the time derivatives in \eqns{corrscal} are taken
keeping the coordinate $\ti{r}$ fixed, we have at the leading order,
and in terms of $t\approx\tau$ 
\begin{subequations}
\begin{align}
(\p_{\ti{r}}\dot\beta)(t,r) &= \frac{1}{aR^2} \int_0^{r}{
  \dot q\,R^2R^\prime dr}\,, \label{5eq5a}\\ 
(\p_{\ti{r}}{\ddot\beta})(t,r) &= \frac{1}{aR^2} \int_0^{r}{
  \ddot q\,R^2R^\prime dr}\,, \label{5eq5b} \\ 
(\p_{\ti{r}}^2\dot\beta)(t,r) &= \dot q - \frac{2}{R^3}\int_0^{r}{ 
  \dot q\,R^2R^\prime dr}
  \,, \label{5eq5c} 
\end{align}
\label{5eq5}
\end{subequations}
which follow from \eqns{5eq3}.
The expressions in \eqns{corrscal}, rewritten in terms of the
LTB proper time $t$ and valid at leading order in the various small
quantities, reduce to
\begin{align}
&\Cal{P}^{(1)} =  \bigg[ \, 2\avg{(\dot\vphi)^2} 
-    \avg{(\p_{\ti{r}}^2\dot\beta)^2} -
    2\avg{(1/\ti{r}^2)(\p_{\ti{r}}\dot\beta)^2} \,  \bigg] 
  \,, \nonumber\\
&\Cal{S}^{(1)} = -\frac{1}{a^2} \bigg[ 6\avg{(\p_{\ti{r}}\vphi)^2} 
  -  \avg{(\p_{\ti{r}}^3\beta)^2} 
- 6\avg{(\p_{\ti{r}}\beta
      - \ti{r}\p_{\ti{r}}^2\beta)^2/\ti{r}^4}  \bigg]  
  \,,\nonumber\\ 
&\Cal{P}^{(1)} + \Cal{P}^{(2)} = - 2H\bigg[ 
    \avg{(\p_{\ti{r}}^2\beta)(\p_{\ti{r}}^2\dot\beta)} +
    2\avg{(1/\ti{r}^2)(\p_{\ti{r}}\beta)(\p_{\ti{r}}\dot\beta)} \,  
 \bigg]\,, \nonumber\\  
&\Cal{S}^{(2)} = \avg{(\p_{\ti{r}}{\ddot\beta} +
    H\p_{\ti{r}}{\dot\beta})(a^2H\p_{\ti{r}}{\dot\beta} -
    \p_{\ti{r}}\vphi)}  
  \,, \label{5eq6}
\end{align}
where the angular brackets are now defined by \eqn{4eq2} and the
various integrands can be read off using \eqns{5eq4}, \eqref{5eq5} and
the results $\p_{\ti{r}}\vphi \approx (a/R^\prime)\vphi^\prime$ and
$\ti{r}\approx (R/a)$, at leading order.
\begin{figure}[t]
\centering
\includegraphics[height=0.35\textheight]{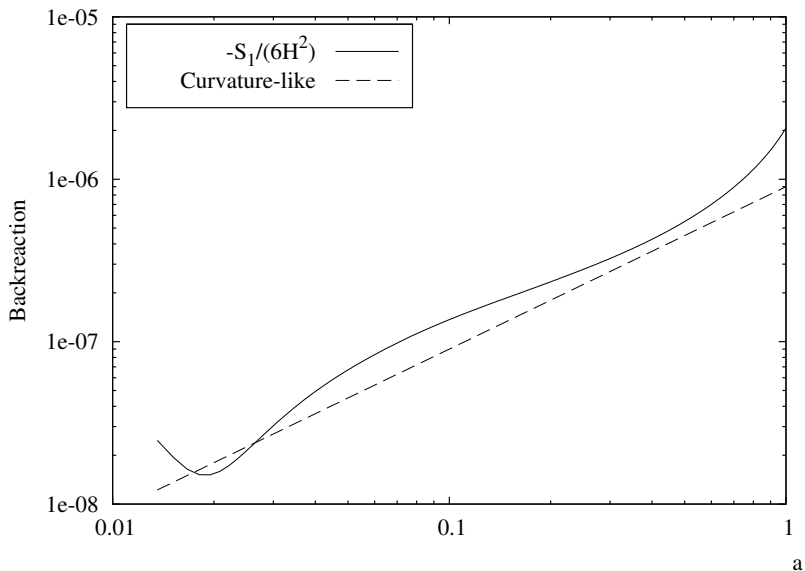}
\caption{\small The evolution of $|\Cal{S}^{(1)}|$, normalised by
  $6H^2$. Also shown is a 
  hypothetical curvature-like correction, evolving like $\sim
  a^{-2}$.}   
\label{fig13}
\end{figure}
\figs{fig13} and \ref{fig14} show results of numerical calculations
performed with {\it Mathematica}. \fig{fig13} shows the evolution of 
$-\Cal{S}^{(1)}/(6H^2)$, the dominant correction, as a function
of the scale factor. The dotted line shows a hypothetical curvature
like correction. Clearly the evolution of the actual 
backreaction is more complicated, due to significant evolution of
$\vphi$. Note that the largest value of
$|\Cal{S}^{(1)}/H^2|$ computed here is $\sim10^{-6}$, whereas estimates
using \emph{linear} theory in chapter 4 suggested that this value
should be around $\sim10^{-4}$. This discrepancy highlights an issue
we noticed earlier in chapter 4, namely that nonlinear inhomogeneities
on small scales \emph{do not contribute significantly} to the
backreaction. Our model has no large scale inhomogeneities and
underestimates the backreaction. Reassuringly, accounting for the
deficit only requires a calculation in \emph{linear} theory, such as
the one in chapter 4. \fig{fig14} shows the evolution of the remaining 
integrals, also normalised by $6H^2$. An initial rapid decay of
$\Cal{P}^{(1)}/H^2$ starting from values of $\sim10^{-8}$ has not been
shown, in order to enhance the contrast in the late time behaviour of
the three functions. The other functions remain subdominant compared
to $\Cal{P}^{(1)}$ at the early times not shown. 

\begin{figure}[t]
\centering
\includegraphics[height=0.35\textheight]{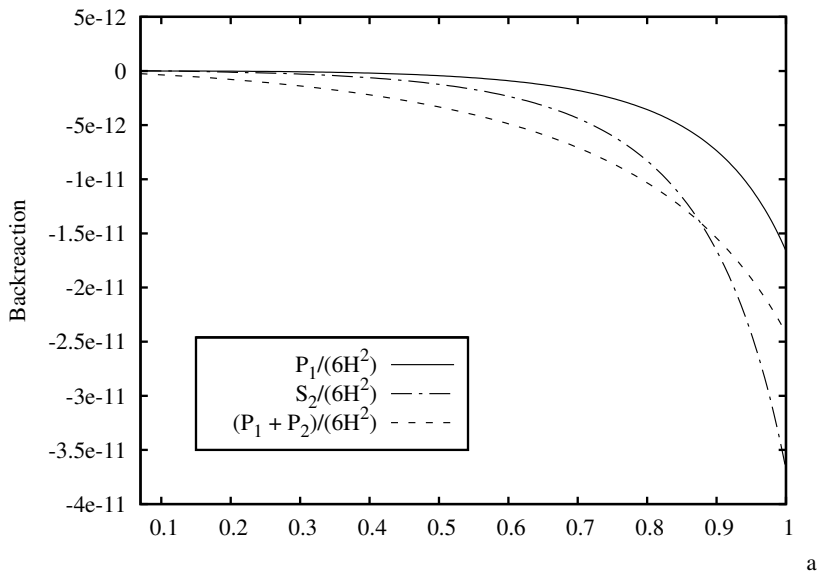}
\caption{\small The normalised evolution of the backreaction functions
  other than $\Cal{S}^{(1)}$. 
To enhance contrast, a strongly decaying
  early time mode for $\Cal{P}^{(1)}/H^2$ 
has not been shown. }
\label{fig14}
\end{figure}

This completes the picture of the effects of backreaction in the
cosmic expansion history.
Our covariant and self-consistent calculation of the backreaction in
this spherical collapse model establishes that inhomogeneities have an
insignificant impact on the average cosmological dynamics. In
particular, the observed cosmic acceleration cannot be explained by
the averaging of inhomogeneities. Our nonlinear dust model can be
regarded as representing a realistic situation, because it has a
overdensity-void structure, and departure from sphericity, tidal
interactions, and second order corrections are not expected to
introduce any significant change in the results. What appears true in
general is that as long as peculiar velocities remain small, as seems
to be the case in the real universe, a description as a perturbed FLRW
model is valid, and this keeps the backreaction small. 

\vskip 0.25in
\noindent
{\large\bf Chapter summary and discussion:}\vskip 0.1in
\noindent
In this chapter we addressed the question of whether or not the
perturbed FLRW form for the metric remains a valid approximation at
late times in the cosmological history. Heuristic arguments such as
those presented by Ishibashi and Wald \cite{wald} indicate that this
should in fact be the case. We have studied this issue in the context
of an exactly solvable, fully relativistic toy model which allows us
to track the behaviour of both the metric and matter perturbations
unambiguously, well into the regime where the matter perturbations
have become completely nonlinear.

In these late stages of structure formation, a perturbation theory in
the density contrast $\delta$ is no longer valid. As expected from
standard Newtonian analyses however \cite{peebles}, we found that a
perturbative expansion in \emph{peculiar velocities} remains valid
until fairly late times. Our model parameters were chosen to reflect
typical inhomogeneities on scales of tens of Mpc. While the
perturbative description of our \emph{model} eventually breaks down
(see e.g. the $(1-\cos u)^{-1}$ factor in the
expression \eqref{nonlin-3eq16} for $\vphi$), the model itself is not
expected to reflect realistic conditions at very late times. For
example, real clusters of galaxies eventually achieve virial
equilibrium due to random motions of the galaxies, rather than
collapsing to a singularity. Our pressureless model cannot accomodate
such a behaviour and must be abandoned beyond the point where
virialisation is expected to occur. Heuristic reasoning further
suggests that in the virial phase, the peculiar velocity with respect
to the Hubble flow will essentially be $|a\ti{v}|\sim H R_{\rm vir}$
where $R_{\rm vir}$ is the virial radius, and the backreaction which
is controlled by $(a\ti{v})^2$ should therefore \emph{decrease} with
time in this phase. Our model parameters were chosen so that the
physical radius of the overdensity at $t=t_0$ is close to typical
virial radii for corresponding realistic clusters, and it is therefore
meaningful to stop the model evolution at $t=t_0$.

Additionally we also computed the backreaction in our model using the
formalism developed in chapter 4. We emphasize that this formalism is
fully covariant, and is guaranteed to yield results which are
coordinate invariant at the leading order. The metric perturbation
function $\vphi$ does not satisfy the equations of linear perturbation
theory at late times, and neither is it expected to. This is reflected
in the complicated behaviour of $\Cal{S}^{(1)}$ seen
in \fig{fig13}. However it is still true that the metric can be
brought to the perturbed FLRW form, which is all that is required for a
reliable calculation of the backreaction. And finally, since the
nonlinear inhomogeneities do not contribute as much to the
backreaction as those on scales $k\sim k_{eq}$, our entire discussion
regarding convergence of the iterative calculation (see chapter 4
summary and discussion) is expected to remain approximately valid even
after accounting for nonlinearity. This is important since it
establishes that cosmological perturbation theory is stable against
including the effects of the backreaction from averaging
inhomogeneities. 

\chapter{Conclusions}
Backreaction as an explanation for the late time cosmic acceleration
would have truly been the most conservative solution to the dark
energy problem. Not only would it have resolved the discrepancy
between observed data and what is generally considered to be
``ordinary physics'', but more importantly 
it would have obviated the need for statements such as ``We do not
understand what 70\% of the universe is made of''\footnote{Including 
  dark matter would take
  this number up to approximately 95\%. If we further take
  into account the fact that the only component which we directly
  measure with great precision is the CMB radiation, then one might
  say that we truly understand only $\sim10^{-4}$ of the
  universe!}. Needless to say, 
this approach has captured the imagination of many cosmologists, and
the (possibly incomplete) list of references cited in the Introduction
is testimony to this fact. Due to the technically challenging nature
of the problem however, it is very important to proceed systematically
and rigorously while determining the size and nature of the effects of
backreaction. This is especially true since order of magnitude
estimates on the one hand indicate that the effect can never be large
\cite{wald}, while simple toy models indicate exactly the opposite
\cite{rasanen}. It has been our goal in this thesis to provide a
reliable and self-consistent calculation to estimate the nature of the
backreaction. 

We have used a fully covariant averaging formalism, adapted to the
specific needs of cosmology, and developed it further to allow
estimates of the effects of perturbative metric inhomogeneities in a
gauge invariant manner. We have used this formalism in the linear
regime of cosmological perturbations and have shown that linear PT is
stable against the inclusion of backreaction effects. We have also 
demonstrated using simple but realistic toy models of nonlinear
structure formation, that (a) a description of the universe in terms
of the perturbed FLRW metric is indeed valid at late times as order of
magnitude estimates suggest, and (b) the backreaction due to these
perturbative metric inhomogeneities remains small, irrespective of
whether the \emph{matter} inhomogeneities are in the linear or
nonlinear regime. One might argue that more precise nonlinear
calculations, e.g. in numerical simulations, might enhance the
effect. While this might be true, this enhancement is not expected to
be by orders of magnitude, since the transfer functions routinely
found in nonlinear calculations do not deviate drastically enough from
the linear theory ones \cite{nonlinearPT}, to lead to an order unity 
backreaction. The cosmological expansion history therefore appears to
be largely insensitive to the presence of the backreaction terms.  

Of course, this means that backreaction from averaging cannot solve
the dark energy problem. There have been claims in the literature
which suggest otherwise; we will comment on a few of these in what
follows. Kolb et al. \cite{marra-phd} study a model with very
large nonlinear inhomogeneities (a $\sim1\,$Gpc sized void) to claim
that in such a case the perturbed FLRW form for the metric cannot be
recovered at all times, and that backreaction effects can be
large. Our results do not contradict this, since we have seen that the
quantity $HR$ controls the late time perturbative expansion, and a
$\sim1\,$Gpc size inhomogeneity with a density contrast of order unity
today will imply a breakdown in the perturbation theory. What is
important to bear in mind however, is whether such large
inhomogeneities are \emph{generic}. If the universe is dominated by
Gpc sized voids then our analysis will indeed break down; but this
does not appear to be the case observationally.

Wiltshire \cite{wiltshire} models the scale dependence of
inhomogeneities in a nonperturbative manner using Buchert's
formalism. The assumption is that the clocks of observers in voids run
at very different rates from those of observers in the ``walls''
surrounding voids. The argument then is that since the universe today
is dominated by voids (of sizes ranging from $30$-$50h^{-1}\,$Mpc
\cite{voids-obs}), ``wall'' observers such as ourselves are
atypical. The difference in clock rates between average observers and 
wall-observers is then fitted to data and can account for several sets
of observations. This was a rather simplistic picture of Wiltshire's
model, which is actually far more involved in its
construction. However, Wiltshire's basic final result, the clock rate 
difference mentioned above, at least superficially does not agree with
our results, since in our toy model we explicitly see that clock rate
differences for any pair of observers are governed by the metric
potential $\vphi$ which remains perturbatively small. It is not clear
where this discrepancy arises from, and this matter is further
complicated by the fact that Wiltshire's model does not have a
concrete setup in which to follow the evolution of inhomogeneities.

To date, perhaps the most physically clear attempts to explain
e.g. supernova data without invoking dark energy, have involved the
``non-Copernican'' models which place us at a special location in the
universe \cite{voids-DE}. At the risk of some repetition, we will
briefly discuss a few issues that arise in this context. As we
discussed in the Introduction, a lot of attention has been focused on
studying light propagation in the inhomogeneous spacetime of a void
modelled by the LTB metric. These calculations involve standard GR,
without any averaging and the associated complications. As far as
supernova data are concerned, the calculations involve light
propagation in a known metric, and the two functional degrees of
freedom in LTB models can be suitably employed to construct a void
geometry (which will typically span several hundred $h^{-1}\,$Mpc)
which fits the supernova data without any dark energy/cosmological
constant. This apparently straightforward result leads to a host of
questions. Do such ``super''-voids exist in the universe? Galaxy
surveys at least do not show any evidence of this, although some
analyses of CMB data, specifically using the integrated Sachs-Wolfe
effect, have thrown up some interesting results \cite{szapudi}. Even
if such voids exist, is there any independent evidence that we reside
close to the center of one? (Were we not close to the center, the CMB
dipole would have been much higher.) Current data is not sufficient to 
answer this question, although future surveys are likely to make some
headway in this issue \cite{copernican-test}. On the theoretical
front, how does one analyse the growth of structure in a geometry
which contains a nonperturbative void? This question has begun to
receive some attention only recently (see the papers by Zibin, Moss
and Scott, and Clifton, Ferreira and Zuntz, in \Cite{voids-DE}). The
jury is therefore still out on whether or not ``dark energy'' is
simply a consequence of our being in a special location. The most
encouraging aspect of the ``void instead of dark energy'' scenario is
that it is amenable to observational testing, which is very likely to
occur in the foreseeable future.

To conclude, backreaction from inhomogeneities cannot solve the dark
energy problem. Void-like inhomogeneities, while
having the potential to solve this problem, await further
observational evidence. And for now, we
still do not understand what (at least) 70\% of the universe is made
of. The future continues to hold significant challenges for cosmology
and theoretical physics.

\appendix
\chapter{Basics of FLRW cosmology}
The unique line element for a spacetime which admits homogeneous and
isotropic spatial sections (i.e. $3$-surfaces of constant
curvature) is given by \cite{weinberg}
\begin{equation}
ds^2=-d\tau^2+a^2(\tau)\left(\frac{dr^2}{1-kr^2} +
r^2d\theta^2+r^2\sin^2\theta d\phi^2\right) \,.
\label{flrw1}
\end{equation}
Here $k$ is a parameter related to the spatial curvature of $3$-space
(${}^{(3)}\Cal{R}=6k/a^2$) and $a(\tau)$ is the single dynamical
function - the \emph{scale factor} -  which describes the evolution of
the universe. This thesis will mostly deal with the case $k=0$
corresponding to spatially flat sections. The line element is written
in coordinates which are \emph{comoving} with those observers who see
a homogeneous and isotropic $3$-space. The coordinate $\tau$ is called
cosmic time. The subscript $0$ will refer to the present epoch
$\tau_0$ and the scale factor is always normalized so that
$a(\tau_0)\equiv a_0=1$. 

The energy-momentum tensor of matter is taken to describe a perfect
fluid, which is homogeneous and isotropic in its rest frame (which
therefore coincides with the comoving reference frame
of \eqn{flrw1}). This has the form  
\begin{equation}
T^a_{\ph{a}b} \equiv {\rm
diag}\left(-\rho(\tau),p(\tau),p(\tau),p(\tau)\right)\,, 
\label{flrw2}
\end{equation}
with $\rho(\tau)$ the energy density and $p(\tau)$ the pressure as
measured in the fluid rest frame. 

The Einstein equations $E_{ab}=8\pi GT_{ab}$, with 
$E_{ab}$ the Einstein tensor constructed using \eqref{flrw1} and 
$T_{ab}$ as given above, reduce to\footnote{Eqn.\eqref{flrw3b} is
actually obtained after substituting \eqref{flrw3a} in the Einstein
equation for $\addot$.}
\begin{subequations}
\begin{equation}
\left(\frac{\adot}{a}\right)^2+\frac{k}{a^2}=\frac{8\pi G}{3}\rho \,, 
\label{flrw3a}
\end{equation}
\begin{equation}
\frac{\addot}{a}=-\frac{4\pi G}{3}\left(\rho +3p\right)\,,
\label{flrw3b}
\end{equation}
\label{flrw3}
\end{subequations}
with the overdot referring to a time derivative
$\dot{}\equiv\partial_\tau$. We will refer to \eqns{flrw3a}
and \eqref{flrw3b} as the Friedmann equation and the acceleration
equation respectively. Together with an equation of state $p=p(\rho)$,
these equations can be solved to get an  expression for $\rho(a)$ and
$a(\tau)$. (One could also use the Friedmann equation \eqref{flrw3a}
in conjunction with the continuity equation $d(\rho a^3)+pd(a^3)=0$
which follows from $T^a_{b;a}=0$, to solve for $\rho(a)$ and
$a(\tau)$.) For example, with the simple assumption $p=w\rho$ for
constant $w$, it can be shown that $\rho\propto
a^{-3\left(1+w\right)}$, and further if $k=0$ then   
\begin{align}
a(\tau)&\propto \tau^{\frac{2}{3\left(1+w\right)}}~\text{,}~w\neq 
    -1\nonumber\\     &\propto {\rm exp}(H \tau)~\text{,}~w=-1 \,,
\label{flrw4}
\end{align}
where $H$ is constant. Some commonly occurring values of the equation
of state parameter $w$ are $w=0$ (pressureless matter or dust) and
$w=1/3$ (radiation). The case $w=-1$ corresponds to the 
cosmological constant $\Lambda$, which gives rise to
a  constant energy density $\rho_\Lambda=\Lambda/(8\pi G)$, and
consequently a value of $H=\sqrt{\Lambda/3}$ in \eqn{flrw4} above. 

In general we define $H\equiv(\adot/a)$, called the Hubble
parameter. Its value at the present epoch ($H_0$) is called the
Hubble constant, and is usually parametrized as $H_0=100h\,{\rm km\,
s^{-1}Mpc^{-1}}$, with the current consensus on its numerical value
being $0.5\lesssim h\lesssim0.8$. Define the \emph{critical density} 
$\rho_c$ at the present epoch as $\rho_c\equiv(3H_0^2)/(8\pi G)$, and
also the quantities $\Omega_i\equiv\rho_{i0}/\rho_c$ where $\rho_{i0}$
is the value of the density of the $i^{th}$  matter component, at the
present epoch. Here $i$ labels the components radiation ($R$), baryons
($b$), dark matter ($DM$) and an additional ``dark energy''
represented by the cosmological constant ($\Lambda$). The total energy
density is simply $\rho(a)=\sum_i{\rho_i(a)}$ and we can write
$\Omega(a)\equiv\rho(a)/\rho_c$, with  the value of $\Omega(a)$ at the
present epoch being $\Omega\equiv\Omega(a=1)=\sum_i{\Omega_i}$. 
The Friedmann equation \eqref{flrw3a} can be written as 
\begin{align}
H^2(a)=H_0^2&\left[\left(1-\Omega\right)a^{-2} +  
\Omega_Ra^{-4} + \left(\Omega_b+\Omega_{DM}\right) a^{-3}
+\Omega_{\Lambda}\right]\,, 
\label{flrw5}
\end{align} 
where the cases $k=0,k>0,k<0$ correspond respectively to $\Omega=1$, 
$\Omega>1$, $\Omega<1$, and in general one would have
$\Omega_\Lambda=\Omega_\Lambda(a)$. The cosmological redshift $z$ of a
source which emits light of wavelength $\lambda_{\rm em}$ at time
$\tau_{\rm em}$ and is observed ``here-and-now'' at time $\tau_0$ with
wavelength $\lambda_0$, is
\begin{equation}
1+z\equiv \frac{\lambda_0}{\lambda_{\rm em}} = \frac{1}{a(\tau_{\rm
em})}\,. 
\label{flrw6}
\end{equation} 
Using this in \eqref{flrw5} gives an expression for $H(z)$. Often it
is useful to work in terms of conformal time $\eta$ defined by
\be
\eta = \int^\tau{\frac{d\ti{\tau}}{a(\ti{\tau})}}\,,
\la{flrw7}
\ee
with the corresponding Hubble parameter
\be
\Cal{H} \equiv \frac{a^\prime}{a} \equiv \frac{1}{a}\frac{da}{d\eta} =
aH(a) \,.
\la{flrw8}
\ee

\chapter{The Lema\^itre-Tolman-Bondi solution}
In this appendix we describe the \emph{spherically symmetric}
Lema\^itre-Tolman-Bondi (LTB) solution of the Einstein equations for
matter comprising a pressureless ``dust'' \cite{LTB}. For an arbitrary
dust configuration, the metric can always be expressed in coordinates
which are ``synchronous'' ($g_{00}=-1$; $g_{0A}=0$) and ``comoving''
(world lines of the fluid elements are orthogonal to
$3$-space) \cite{landau-tof}. Specifically, the LTB metric is given in  
the synchronous and comoving gauge, by  
\begin{equation}
ds^2=\,-dt^2+\frac{R^{\prime 2}(t,r)}{1-k(r)r^2}dr^2+
R^2(t,r)\left(d\theta^2 + \sin^2\theta d\phi^2\right)\,.
\label{LTB1}
\end{equation}
Throughout this appendix, a prime and a dot will refer to partial
derivatives with respect to $r$ and $t$ respectively. The Einstein
equations simplify to 
\begin{subequations}
\begin{equation}
\dot R^2(t,r)=\frac{2GM(r)}{R(t,r)}-k(r)r^2\,,
\label{LTB2a}
\end{equation}
\begin{equation}
4\pi\rho(t,r)=\frac{M^\prime(r)}{R^\prime(t,r)R^2(t,r)}\,.
\label{LTB2b}
\end{equation}
\end{subequations}
Surfaces of constant $r$ are $2-$spheres having area $4\pi R^2(t,r)$. 
$\rho(t,r)$ is the energy density of dust, while $k(r)$ and $M(r)$ are
arbitrary functions that arise  on integrating the dynamical
equations. Solutions can be found for three  cases $k(r)<0$, $k(r)=0$
and $k(r)>0$. The solution for $k(r)=0$  (the
marginally bound case) has the particularly simple form 
\begin{equation}
R(t,r)=\bigg(\frac{9GM(r)}{2}\bigg)^{1/3}
\left(t-t_0(r)\right)^{2/3},~~~~\text{for}~k(r)=0\,.
\label{LTB3}
\end{equation}
Here $t_0(r)$ is another arbitrary function arising from integration.
The solution describes an expanding region, with the initial time
$t_{in}$ chosen such that $t>t_{in}\geq t_0(r)$ for all $r$. For the
other two cases, the solutions can be written in parametric form 
\begin{subequations}
\begin{equation}
R=\frac{GM(r)}{-k(r)r^2}\left(\cosh\eta -1\right)~~~
;~~~t-t_0(r)=\frac{GM(r)}{\left(-k(r)r^2\right)^{3/2}}\left(\sinh\eta 
- \eta\right)~,~0\leq\eta<\infty~,~~~\text{for}~k(r)<0\,.
\label{LTB4a}
\end{equation}
\begin{equation}
R=\frac{GM(r)}{k(r)r^2}\left(1-\cos\eta\right)~~~;~~~
t-t_0(r)=\frac{GM(r)}{\left(k(r)r^2\right)^{3/2}}
\left(\eta -\sin\eta\right)~,~0\leq\eta\leq 2\pi,~~~
\text{for}~k(r)>0\,.
\label{LTB4b} 
\end{equation}
\end{subequations}
In the unbound case ($k(r)<0$), $R(t,r)$ increases monotonically with
$t$,  for every shell with label $r$. In the bound case ($k(r)>0$),
$R(t,r)$ increases to a maximum value $R_{max}(r)$ for each shell $r$
and then decreases back to $0$ in a finite time. 

In all cases, there are two physically different free functions,
although  three arbitrary functions $k$, $M$ and $t_0$ appear. One of
the three  represents the freedom to rescale the coordinate $r$. We
use this freedom to set\footnote{In the main text we also use a
slightly modified rescaling when convenient.} $R(t_{in},r)\equiv
R_{in}(r)=r$. To completely specify the solution, we specify the 
initial density $\rho_{in}(r)$ and the function $k(r)$ (which can be
related to the initial velocity profile $\dot R_{in}(r)$
using \eqn{LTB2a} evaluated at initial time). This specifies
$M(r)=4\pi\int_0^r{\rho_{in}(\tilde r)\tilde r^2d\tilde r}$ (which in
the marginally bound case is interpreted as the mass contained  in a
comoving shell), and $t_0(r)$ can be solved for using
Eqns. \eqref{LTB3}, \eqref{LTB4a} or \eqref{LTB4b} as the case may be,
at time $t=t_{in}$. The FLRW solution is a special case and is
recovered by setting $\rho_{in}=\,$constant, $k=\,$constant.  

\section{Regularity conditions}
It is useful to keep in mind certain regularity conditions when
constructing LTB models. In any LTB model, the functions $M(r)$ and
$k(r)$ are to be specified by initial conditions at $t=t_{in}$, and
the choice of scaling $R(r,t_{in})=r$ fixes $t_0(r)$ as 
\be
t_0(r)=t_{in}-\frac{GM(r)}{\left(|k(r)r^2|\right)^{3/2}} S_{in}(r)
~~~;~~~C_{in}(r)=\frac{|k(r)r^3|}{GM(r)}\,,
\la{reg1}
\ee
where $S_{in}(r) \equiv \left(\sinh\eta_{in}(r)-\eta_{in}(r)\right)$
and $C_{in}(r) \equiv \cosh\eta_{in}(r)-1$ for $k(r)<0$; $S_{in}(r)
\equiv \left(\eta_{in}(r)-\sin\eta_{in}(r)\right)$  and $C_{in}(r) 
\equiv 1 - \cos\eta_{in}(r)$ for $k(r)>0$.
\vskip 0.1in
The regularity conditions imposed on this model, and their
consequences, are as follows  
\begin{itemize}
\item \underline{{\it No evolution at the symmetry centre}}:\\ 
This is required in order to maintain spherical symmetry about the
same point at all times,  and translates as $\dot R(t,0)=0$ for all
$t$. The right hand side of \eqn{LTB2a} must therefore vanish in
the limit $r\to 0$.  
Since the functions involved are non-negative, we assume that we can
write  
\begin{equation}
|k(r\to 0)|\sim r^\mu~,~\mu >-2~~~;~~~
M(r\to 0)\sim r^\alpha~~;~~
R(t,r\to 0)\sim r^\beta f(t)~,~\alpha>\beta\geq 0\,.
\label{reg2}
\end{equation}
Consistency requires $\beta$ to be constant, and our scaling choice
further requires $\beta=1$. We do not require the exponents $\mu$
and $\alpha$ to necessarily be integers.\\ 
\item \underline{{\it No shell-crossing singularities}}:\\
Physically, we demand that an outer shell (labelled by a larger value
of $r$) have a larger area  radius $R$ than an inner shell, at any
time $t$. Unphysical shell-crossing singularities arise when this
condition is not met. Mathematically, this requires 
\begin{equation}
R^\prime(t,r)>0~~~\text{for all $r$, for all $t$.}
\label{reg3}
\end{equation}

\item \underline{{\it Regularity of energy density}}:\\
We demand that the energy density $\rho(t,r)$ remain finite and
strictly positive for all values of $r$ and $t$. Combining this with
\eqns{LTB2b} and \eqref{reg3} gives (assuming that $R^\prime$
is finite for all $r$ and since $\beta=1$)
\begin{equation}
\lim_{r\to
0}\rho(t,r) = \,\text{finite}\Rightarrow\,\alpha-1-2\beta=0\Rightarrow\, 
\alpha=3\,.   
\label{reg4}
\end{equation}

\item \underline{{\it No trapped shells}}:\\
In order for an expanding shell to not be trapped
initially, it must satisfy the condition $r>2GM(r)$. Near the regular
center, this condition is automatically satisfied independent of the
exact form of $M(r)$, since there $M\sim r^3$.\\
\end{itemize}  

\chapter{Cosmology in MG}\label{mainapp}
\noindent 
In this appendix we give proofs of several results that were used in
chapter 2.

\section{Analysis of $\rmb{D}_{\bar\Omega}\bar g^{ab}=0$}
\label{app-gbar}
\noindent
We start with the metric
\begin{equation}
^{(\Cal{M})}ds^2 = g_{00}(t,\rmb{x})dt^2 + g_{AB}(t,\rmb{x})dx^Adx^B
  \,, 
\label{appA1}
\end{equation}
on \Cal{M}\ and assume that it averages out to the FLRW form
(\eqn{spatlim-gauge1}): 
\begin{equation}
G_{00} = \avg{\bil{g}_{00}} = -f^2(t) ~~;~~ G_{0A} = \avg{\bil{g}_{0A}}
= 0 ~~;~~ G_{AB} = \avg{\bil{g}_{AB}} = a^2(t)\delta_{AB} \,.
\label{appA2}
\end{equation} 
We will analyze the second relation of \eqn{avgZala53} and show that it
leads to the result $U^{ij}\equiv\bar g^{ij}-G^{ij}=0$, where
$\barOm{a}{b}$ refers to the connection $1$-forms associated with
$G_{ij}$ given by
\begin{align}
\barOm{0}{0} = \partial_t(\ln f)\ext t ~~&;~~ \barOm{A}{0} =
H\delta{}^A_B\dx{B} \,,\nonumber\\  
\barOm{0}{A} = \frac{a^2}{f^2}H\delta_{AB}\dx{B} ~~&;~~ \barOm{A}{B} = 
H\delta{}^A_B\ext t \,, 
\label{appA3}
\end{align}
where $H=(1/a)(da/dt)$ for this section. We have
\begin{equation}
\ext\bar g^{ab} + \barOm{a}{j}\bar g^{jb} + \barOm{b}{j}\bar g^{aj} =
0\,. 
\label{appA4}
\end{equation} 
Consider the three cases ($a=b=0$), ($a=0,b=B$) and ($a=A,b=B$) in
turn. The first case gives
\begin{equation}
\ext\bar g^{00} + 2\barOm{0}{0}\bar g^{00} + 2\barOm{0}{A}\bar g^{0A}
= 0\,,
\label{appA5}
\end{equation}
which reduces to
\begin{equation}
\left[\partial_t\bar g^{00} + 2\partial_t(\ln f)\bar g^{00}
  \right]\ext t + \left[\partial_A\bar g^{00} +
  2\frac{a^2}{f^2}H\delta_{AB}\bar g^{0B}\right] \dx{A} = 0\,. 
\label{appA6}
\end{equation}
We can conclude that
\begin{subequations}
\begin{equation}
\bar g^{00}(t,\rmb{x}) = -\frac{k(\rmb{x})}{f^2(t)} \,, 
\label{appA7a}
\end{equation}
\begin{equation}
\partial_A k(\rmb{x}) = 2a^2H\delta_{AB}\bar g^{0B} \,.
\label{appA7b}
\end{equation}
\label{appA7}
\end{subequations}
where $k(\rmb{x})$ is a positive definite function (so that the metric
signature is preserved) which arises as an integration constant and is
constrained by \eqn{appA7b}. The second case ($a=0,b=B$) leads to 
\begin{equation}
\left[\partial_t\bar g^{0B} + \partial_t\ln(af)\bar g^{0B} \right]
\ext t + \left[\partial_J\bar g^{0B} + \frac{a^2}{f^2}H\delta_{AJ}\bar
  g^{AB}  + \frac{k(\rmb{x})}{f^2}H\delta{}^B_J\right] \dx{J} = 0 \,, 
\label{appA8}
\end{equation}
which gives us
\begin{subequations}
\begin{equation}
\bar g^{0B} = \frac{m^B(\rmb{x})}{a(t)f(t)}\,,
\label{appA9a}
\end{equation}
\begin{equation}
\frac{1}{af}\partial_Jm^B(\rmb{x}) + \frac{a^2}{f^2}H\delta_{AJ}\bar
g^{AB} - \frac{k(\rmb{x})}{f^2}H\delta{}^B_J = 0\,.
\label{appA9b}
\end{equation}
\label{appA9}
\end{subequations}
where $m^B(\rmb{x})$ is a $3$-vector that arises as a constant of
integration like $k(\rmb{x})$, and is constrained by
\eqn{appA9b}. Finally, the last case ($a=A,b=B$) leads to
\begin{equation}
\left[\partial_t\bar g^{AB} + 2H\bar g^{AB}\right] \ext t +
\left[\partial_J\bar g^{AB} +
  \frac{1}{af}H\left(\delta{}^A_Jm^B(\rmb{x}) +
  \delta{}^B_Jm^A(\rmb{x}) \right) \right] \dx{J} = 0\,,   
\label{appA10}
\end{equation}
which gives us
\begin{subequations}
\begin{equation}
\bar g^{AB} = \frac{1}{a^2(t)}s^{AB}(\rmb{x}) \,,
\label{appA11a}
\end{equation}
\begin{equation}
\frac{1}{a^2}\partial_Js^{AB}(\rmb{x}) +
\frac{1}{af}H\left(\delta{}^A_Jm^B(\rmb{x}) + \delta{}^B_Jm^A(\rmb{x})
\right) = 0\,.
\label{appA11b}
\end{equation}
\label{appA11}
\end{subequations}
Here $s^{AB}(\rmb{x})$ is another constant of integration, a symmetric
$3$-tensor. Now, since the left hand side of \eqn{appA7b} is
independent of time, either the time dependence of the right hand side
must cancel, or both sides must vanish. For the time dependence to
cancel, we need $f\propto(da/dt)$ which is not expected \emph{a
  priori}. Therefore both sides of \eqn{appA7b} must
vanish, which immediately tells us that the vector $m^B(\rmb{x})$ must
vanish, and the function $k(\rmb{x})$ must be a constant,
\begin{equation}
k(\rmb{x}) = k = {\rm constant} ~~;~~ m^B(\rmb{x}) = 0\,.
\label{appA12}
\end{equation}
Equations \eqref{appA9b} and \eqref{appA11b} then give us
\begin{equation}
s^{AB}(\rmb{x}) = k \delta^{AB}\,,
\label{appA13}
\end{equation}
with the same constant $k$ as in \eqn{appA12}. Finally, putting
everything together we find
\begin{align}
\bar g^{00} = -\frac{k}{f^2} ~~;~~ \bar g^{0A} = 0 ~~&;~~ \bar g^{AB} =
\frac{k}{a^2}\delta^{AB} \,,\nonumber\\
\Rightarrow \bar g^{ij} = kG^{ij}\,.
\label{appA14}
\end{align}
The constant $k$ is not constrained by any of the equations and
appears to be a free parameter in the theory. The modified Einstein
equations \eqref{avgZala57} show that $k$ can be absorbed into the
averaged energy momentum tensor. We will for simplicity assume $k$ to
be unity thereby obtaining, as required  
\begin{equation}
U^{ij} \equiv \bar g^{ij} - G^{ij} = 0\,.
\label{appA15}
\end{equation}

\section{Analysis of the condition $\avg{\Gamma{}^a_{bc}} =
\,^{(\rm FLRW)}\Gamma{}^a_{bc}$}
\label{app-gamma}
\noindent
Here we will assume that the line element on \Cal{M}\ is in the volume
preserving gauge
\begin{equation}
^{(\Cal{M})}ds^2=-\frac{d\bt^2}{h(\bt,\rmb{x})} +
  h_{AB}(\bt,\rmb{x})dx^Adx^B\,,
\label{appA16}
\end{equation}
so that the averaging is trivial, and the metric and averages out to
the FLRW line element on \Mbar\ given by 
\begin{equation}
^{(\Mbar)}ds^2 = -\frac{d\bt^2}{\avg{h}\!(\bt)} +
  \ba^2(\bt)\delta_{AB}dx^Adx^B \,,
\label{appA17}
\end{equation}
where we used the condition $\avg{1/h} = 1/\avg{h}$ that follows from
$\bar g^{00} = G^{00}$. The conditions $\avg{\Gamma{}^a_{bc}} =
\,^{(\rm FLRW)} \Gamma{}^a_{bc}$ then result in the following : 
\begin{subequations}
\begin{align}
&\Gamma{}^0_{00} ~~:~~ \avg{\partial_{\bt}(\ln\sqrt{h})} =
\partial_{\bt}(\ln\sqrt{\avg{h}})\,, 
\label{appA18a} \\
&\Gamma{}^0_{0A} ~~:~~ \avg{\partial_A(\ln\sqrt{h})} = 0\,,
\label{appA18b} \\
&\Gamma{}^A_{00} ~~:~~ \avg{\frac{h^{AB}}{h}\partial_B(\ln\sqrt{h})} = 
0\,, 
\label{appA18c} \\
&\Gamma{}^A_{0B} ~~:~~ \avg{\frac{1}{\sqrt h}\Theta{}^A_B} =
H\delta{}^A_B \,,
\label{appA18d} \\
&\Gamma{}^0_{AB} ~~:~~ \avg{\sqrt{h}\Theta_{AB}} =
\avg{h}\ba^2H\delta_{AB} \,,
\label{appA18e} \\
&\Gamma{}^A_{BC} ~~:~~ \avg{\,^{(3)}\Gamma{}^A_{BC}} = 0 \,. 
\label{appA18f}
\end{align}
\label{appA18} 
\end{subequations}
\eqns{appA18b} and \eqref{appA18f} are consistent with each other since
$\,^{(3)}\Gamma{}^A_{BA} = \partial_B(\ln\sqrt{h})$, and \eqn{appA18c}
is consistent with the assumption \eqn{avgZala52a}. The trace of
\eqn{appA18d} gives $\avg{(1/\sqrt h)\Theta} = 3H$. However
we have $(1/\sqrt h)\Theta = N\Theta =
\partial_{\bt}(\ln\sqrt{h})$, and combined with \eqn{appA18a} this
gives 
\begin{equation}
\frac{1}{2}\partial_{\bt}(\ln\avg{h}) = 3\partial_{\bt}(\ln\ba)
\Rightarrow \avg{h} = \ba^6\,,
\label{appA19}
\end{equation}
where we have set an arbitrary proportionality constant (representing
rescaling of the time coordinate by a constant) to unity. This
establishes the last equality in \eqn{spatlim10}. 

Finally, consider the trace
$(\avg{h^{AB}}/\avg{h})\avg{\sqrt{h}\Theta_{AB}}$ : using the
condition $\bar g^{AB} = G^{AB}$, \eqn{appA18e} and the trace of 
\eqn{appA18d}, this gives us 
\begin{equation}
\frac{\avg{h^{AB}}}{\avg{h}}\avg{\sqrt{h}\Theta_{AB}} =
\frac{1}{\avg{h}}\frac{\delta^{AB}}{\ba^2} \avg{\sqrt{h}\Theta_{AB}} =
3H = \avg{\frac{1}{\sqrt h}\Theta} =
\avg{\frac{h^{AB}}{h}(\sqrt{h}\Theta_{AB})}\,.  
\label{appA20}
\end{equation}
On using the condition \eqref{avgZala52a} this leads to
\begin{equation}
\left(\frac{\avg{h^{AB}}}{\avg{h}} -
\avg{\frac{h^{AB}}{h}}\right)\avg{\Gamma{}^0_{AB}} = 0\,,
\label{appA21}
\end{equation}
which is consistent with the assumption
\begin{equation}
\frac{\avg{h^{AB}}}{\avg{h}} = \avg{\frac{h^{AB}}{h}}\,.
\label{appA22}
\end{equation}

\label{appendix}

\end{onehalfspacing}


\newpage
\begin{thebibliography}{10}
\bibitem{hubble-1929} 
E P Hubble, {\it Proc. Natl. Acad. Sci. USA} {\bf 15}, 168 (1929);\\
R P Kirshner {\it Proc. Natl. Acad. Sci. USA} {\bf 101}, 8 (2004).
\bibitem{history}
For a history of the Big Bang model, see J-P Luminet,
arXiv:0704.3579 (2007). 
\bibitem{weinberg}
S Weinberg, {\it Cosmology}, Oxford Univ. Press (2008).
\bibitem{COBE} 
G F Smoot et al., {\it Astrophys. J.} {\bf 396}, L1 (1992).
\bibitem{wmap}
D N Spergel et al., {\it Astrophys. J. Suppl.} {\bf 170}, 377
(2007) [arXiv:astro-ph/0603449].
\bibitem{2df}
J A Peacock et al., {\it Nature} {\bf 410}, 169 (2001)
[arXiv:astro-ph/0103143]. 
\bibitem{sdss}
C Hikage et al. (SDSS collaboration), {\it Publ. Astron. Soc. Jap.}
  {\bf 55}, 911 (2003) [arXiv:astro-ph/0304455];\\   
J Yadav et al., {\it Mon. Not. Roy. Astron. Soc.} {\bf 364}, 601
(2005) [arXiv:astro-ph/0504315]
\bibitem{homog}
D W Hogg et al., {\it Astrophys. J.} {\bf 642}, 54 (2005)
[arXiv:astro-ph/0411197];\\
M Joyce et al., {\it Astron. \& Astrophys.} {\bf 443}, 11 (2005)
[arXiv:astro-ph/0501583];\\ 
M Kerscher, J Schmalzing, T Buchert and H Wagner, {\it Astron. \&
  Astrophys.} {\bf 333}, 1 (1998) [arXiv:astro-ph/9704028];\\
M Kerscher et al., {\it Astron. \& Astrophys.} {\bf 373}, 1 (2001)
[arXiv:astro-ph/0101238]. 
\bibitem{sylos}
F Sylos Labini, N L Vasilyev and Y V Baryshev, {\it Europhys. Lett.}
{\bf 85}, 29002 (2009) [arXiv:0812.3260];\\
F Sylos Labini et al., arXiv:0805.1132 (2008).
\bibitem{lss-rev}
V J Martinez, arXiv:0804.1536, to appear in {\it ``Data Analysis
  in Cosmology'', Lecture Notes in Physics} (2008), eds. V J Martinez,
et al., Springer-Verlag.
\bibitem{peebles}
P J E Peebles, {\it Principles of Physical Cosmology}, Princeton
Univ. Press, New Jersey (1993). 
\bibitem{dodelson}
S Dodelson, {\it Modern Cosmology}, Academic Press, San Diego
(2003). 
\bibitem{mukhanov}
V F Mukhanov, H A Feldman and R H Brandenberger, {\it Phys. Rept.}
{\bf 215}, 203 (1992).
\bibitem{press-schechter}
W H Press and P Schechter, {\it Astrophys. J.} {\bf 187}, 425
(1974);\\ 
R K Sheth and G Tormen, {\it Mon. Not. Roy. Astron. Soc.} {\bf 329},
61 (2002) [arXiv:astro-ph/0105113].
\bibitem{nonlinearPT}
F Bernardeau, S Colombi, E Gaztanaga and R Scoccimarro, {\it
Phys. Rept.} {\bf 367}, 1 (2002) [arXiv:astro-ph/0112551].
\bibitem{simulations}
M Trenti and P Hut, arXiv:0806.3950 (2008), invited refereed review
for the Scholarpedia Encyclopedia of Astrophysics, available online at \\
http://www.scholarpedia.org/article/N-body\_simulations\_(gravitational)   
\bibitem{ellis}
G F R Ellis, in {\it General Relativity and Gravitation} (D.
Reidel Publ. Co., Dordrecht),  Eds. B. Bertotti et al., (1984). 
\bibitem{accln-evid}
A Riess et al., {\it Astrophys. J.} {\bf 607}, 664 (2004)
[arXiv:astro-ph/0402512]; \\  
M Seikel and D J Schwarz, {\it JCAP} {\bf 02}(2008)007
[arXiv:0711.3180]; \\ 
E M\"ortsell and C Clarkson, arXiv:0811.0981 (2008), JCAP in press.
\bibitem{sami}
E J Copeland, M Sami and S Tsujikawa, {\it Int. J. Mod. Phys.}, {\bf
D15}, 1753, (2006) [arXiv:hep-th/0603057].
\bibitem{isaacson}
R A Isaacson, {\it Phys. Rev.} {\bf 166}, 1272 (1968).
\bibitem{brill-hartle}
D R Brill and J B Hartle, {\it Phys. Rev.} {\bf 135}, B271 (1964).
\bibitem{noonan}
T W Noonan, {\it Gen. Rel. Grav.} {\bf 16}, 1103 (1984).
\bibitem{futamase}
T Futamase, {\it Phys. Rev. Lett.} {\bf 61}, 2175 (1988); \\
T Futamase, {\it Phys. Rev.} {\bf D53}, 681 (1996).
\bibitem{kasai-avg}
M Kasai, {\it Phys. Rev. Lett.} {\bf 69}, 2330 (1992).
\bibitem{boersma}
J P Boersma, {\it Phys. Rev.} {\bf D57}, 798 (1998)
[arXiv:gr-qc/9711057]. 
\bibitem{avg}
W R Stoeger, G F R Ellis and C Hellaby, {\it
Mon. Not. Roy. Astron. Soc.} {\bf 226}, 373 (1987);\\
M Carfora and K Piotrkowska, {\it Phys. Rev.} {\bf D52}, 4393 (1995)
[arXiv:gr-qc/9502021]; \\
N Mustapha, B A Bassett, C Hellaby and G F R  Ellis, {\it
Class. Quant. Grav.} {\bf 15}, 2363 (1998) [arXiv:gr-qc/9708043];\\ 
A Krasi\'{n}ski, {\it Inhomogeneous Cosmological Models}, Cambridge
Univ. Press (1997);\\
M Reiris, {\it Class. Quant. Grav.} {\bf 25}, 085001 (2008)
[arXiv:0709.0770];\\
C Anastopoulos, arXiv:0902.0159 (2009).
\bibitem{wald}
A Ishibashi and R M Wald, {\it Class. Quant. Grav.} {\bf 23}, 235
(2006) [arXiv:gr-qc/0509108].
\bibitem{martineau}
P Martineau and R H Brandenberger, arXiv:gr-qc/0509108 (2005).
\bibitem{abramo}
L R Abramo, R H Brandenberger and V F Mukhanov, {\it Phys. Rev.} {\bf
D56}, 3248 (1997) [arXiv:gr-qc/9704037].
\bibitem{barausse}
E Barausse, S Matarrese and A Riotto, {\it Phys.Rev.} {\bf D71},
063537 (2005) [arXiv:astro-ph/0501152].
\bibitem{kolb-2005}
E W Kolb, S Matarrese, A Notari and A Riotto, arXiv:hep-th/0503117
(2005). 
\bibitem{nosuperhorznbackrxn}
C M Hirata and U Seljak, {\it Phys. Rev.} {\bf D72}, 083501 (2005) 
[arXiv:astro-ph/0503582]; \\ 
G Geshnizjani, D J H Chung and N Afshordi, {\it Phys. Rev.} {\bf D72}
023517 (2005) [arXiv:astro-ph/0503553]; \\
E E Flanagan, {\it Phys. Rev.} {\bf D71}, 103521 (2005)
[arXiv:hep-th/0503202]; \\
S R\"as\"anen, {\it Class. Quant. Grav.} {\bf 23}, 1823 (2006)
[arXiv:astro-ph/0504005];\\
M F Parry, {\it JCAP} {\bf 0606}:016 (2006)
[arXiv:astro-ph/0605159];\\ 
N Kumar and E E Flanagan, {\it Phys. Rev.} {\bf D78}, 063537 (2008)
[arXiv:0808.1043].
\bibitem{misc-backrxn}
W R Stoeger, A Helmi, D F Torres, {\it Int. J. Mod. Phys.} {\bf D16},
1001 (2007) [arXiv:gr-qc/9904020];\\
E A Calzetta, B L Hu and F D Mazzitelli, {\it Phys. Rept.} {\bf 352},
459 (2001) [arXiv:hep-th/0102199];\\
C Wetterich, {\it Phys. Rev.} {\bf D67}, 043513 (2003)
[arXiv:astro-ph/0111166];\\
E R Siegel and J N Fry, {\it Astrophys. J.} {\bf 628}, L1 (2005)
[arXiv:astro-ph/0504421];\\
A Gruzinov, M Kleban, M Porrati and M Redi, {\it JCAP} {\bf 0612}:001
(2006) [arXiv:astro-ph/0609553];\\
M Gasperini, G Marozzi and G Veneziano, arXiv:0901.1303 (2009).
\bibitem{notari}
A Notari, {\it Mod. Phys. Lett.} {\bf A21}, 2997 (2006)
[arXiv:astro-ph/0503715]
\bibitem{karel}
K Van Acoleyen, {\it JCAP} {\bf 0810}:028 (2008) [arXiv:0808.3554]. 
\bibitem{schwarz-1}
N Li and D J Schwarz, {\it Phys. Rev.} {\bf D76}, 083011 (2007)
[arXiv:gr-qc/0702043].
\bibitem{rasanen}
S R\"as\"anen, {\it JCAP} {\bf 0611}:003 (2006)
[arXiv:astro-ph/0607626].   
\bibitem{wiltshire}
D L Wiltshire, {\it New J. Phys.} {\bf 9}, 377 (2007)
[arXiv:gr-qc/0702082]; {\it Int. J. Mod. Phys.} {\bf D17}, 641 (2008)
[arXiv:0712.3982]; {\it Phys. Rev. Lett.} {\bf 99}, 251101 (2007)
[arXiv:0709.0732];\\
B M Leith, S C Cindy Ng, D L Wiltshire, {\it Astrophys. J.} {\bf 672},
L91 (2008) [arXiv:0709.2535].
\bibitem{kwan}
J Kwan, M J Francis and G F Lewis, arXiv:0902.4249 (2009).
\bibitem{buchert1a}
T Buchert, {\it Gen. Rel. Grav.} {\bf 32}, 105 (2000)
[arXiv:gr-qc/9906015]. 
\bibitem{buchert1b}
T Buchert, {\it Gen. Rel. Grav.} {\bf 33}, 1381 (2000) 
[arXiv:gr-qc/0102049].
\bibitem{buchert2}
T Buchert and M Carfora, {\it Class. Quant. Grav.} {\bf 19}, 6109
(2002) [arXiv:gr-qc/0210037]; {\it Phys. Rev. Lett.} {\bf 90}, 031101
(2003) [arXiv:gr-qc/0210045].
\bibitem{zala1}
R M Zalaletdinov, {\it Gen. Rel. Grav.} {\bf 24}, 1015 (1992); {\it
Gen. Rel. Grav.} {\bf 25}, 673 (1993). 
\bibitem{mars}
M Mars and R M Zalaletdinov, {\it J. Math. Phys.}, {\bf 38}, 4741,
(1997) [dg-ga/9703002].
\bibitem{zala2}
R M Zalaletdinov, {\it Bull. Astron. Soc. India}, {\bf 25}, 401, (1997)
[gr-qc/9703016].
\bibitem{buchert-ehlers}
T Buchert and J Ehlers, {\it Astron. \& Astrophys.} {\bf 320}, 1
(1997) [arXiv:astro-ph/9510056].
\bibitem{kerscher}
T Buchert, M Kerscher and C Sicka, {\it Phys. Rev.} {\bf D62}, 043525
(2000) [arXiv:astro-ph/9912347].
\bibitem{buchertavg-misc}
D Palle, {\it Nuovo Cim.} {\bf 117B}, 687 (2002)
[arXiv:astro-ph/0205462];\\
S R\"as\"anen, {\it JCAP} {\bf 0402}:003 (2004)
[arXiv:astro-ph/0311257]; {\it JCAP} {\bf 0411}:010 (2004)
[arXiv:gr-qc/0408097]; {\it JCAP} {\bf 0804}:026 (2008)
[arXiv:0801.2692]; {\it JCAP} {\bf 0902}:011 (2009)
[arXiv:0812.2872];\\ 
G F R  Ellis and T Buchert, {\it Phys. Lett.} {\bf A347}, 38 (2005)
[arXiv:gr-qc/0506106];\\
E W Kolb, S Matarrese and A Riotto, {\it New J. Phys.} {\bf 8}, 322
(2006) [arXiv:astro-ph/0506534];\\
Y Nambu and M Tanimoto, arXiv:gr-qc/0507057 (2005);\\
J D Barrow and C G Tsagas, {\it Class. Quant. Grav.} {\bf 24}, 1023
(2007) [arXiv:gr-qc/0609078];\\
A E Romano, {\it Phys. Rev.} {\bf D75}, 043509 (2007)
[arXiv:astro-ph/0612002]; \\
M-N Celerier, {\it New Adv. Phys.} {\bf 1}, 29 (2007)
[arXiv:astro-ph/0702416];\\
T Buchert, {\it Gen. Rel. Grav.} {\bf 40}, 467 (2008)
[arXiv:0707.2153];\\ 
T Buchert and M Carfora,  {\it Class. Quant. Grav.} {\bf 25},
195001 (2008) [arXiv:0803.1401];\\
V F Cardone and G Esposito, arXiv:0805.1203 (2008);\\
R A Sussman, arXiv:0807.1145 (2008);\\
K Bolejko and L Andersson, {\it JCAP} {\bf 0810}:003 (2008)
[arXiv:0807.3577];\\ 
K Bolejko, arXiv:0808.0376 (2008);\\
J Larena, arXiv:0902.3159 (2009).
\bibitem{behrend}
J Behrend, I A Brown and G Robbers, {\it JCAP} {\bf 0801}:013 (2008)
[arXiv:0710.4964].
\bibitem{buch-lar-alimi}
T Buchert, J Larena and J-M Alimi, {\it Class. Quant. Grav.} {\bf 23},
6379 (2006) [arXiv:gr-qc/0606020];\\
\bibitem{marra-phd}
V Marra, arXiv:0803.3152 (2008), PhD Thesis; \\
E W Kolb, V Marra and S Matarrese, {\it Phys. Rev.} {\bf D78}, 103002
(2008) [arXiv:0807.0401].
\bibitem{buchert-obs}
N Li and D J Schwarz, {\it Phys. Rev.} {\bf D78}, 083531 (2008)
[arXiv:0710.5073];\\
J Larena, J-M Alimi, T Buchert, M Kunz and P-S Corasaniti,
arXiv:0808.1161 (2008);\\
E Rosenthal and E E Flanagan, arXiv:0809.2107 (2008);\\
K Bolejko, A Kurek and M Szydlowski, arXiv:0811.4487 (2008).
\bibitem{zala-results}
A A Coley, N Pelavas and R M Zalaletdinov, {\it Phys. Rev. Lett.},
{\bf 95}, 151102, (2005) [gr-qc/0504115];\\
A A Coley and N Pelavas, {\it Phys. Rev.} {\bf D74}, 087301, (2006) 
[astro-ph/0606535]; {\it Phys. Rev.} {\bf D75}, 043506, (2007)
[gr-qc/0607079];\\
R J Van Den Hoogen, {\it Gen. Rel. Grav.} {\bf 40}, 2213 (2008)
[arXiv:0710.1823];\\
A A Coley, arXiv:0812.4565 (2008).
\bibitem{backrxn-DE}
J W Moffat, {\it JCAP} {\bf 0605}:001 (2006)
[arXiv:astro-ph/0505326];\\
T Mattsson and M Ronkainen, {\it JCAP} {\bf 0802}:004 (2008)
[arXiv:0708.3673];\\
G M Hossain, arXiv:0709.3490 (2007);\\
T Mattsson, arXiv:0711.4264 (2007). 
\bibitem{inhom-lightprop-early}
J Kristian and R K Sachs, {\it Astrophys. J} {\bf 143}, 379 (1966);\\ 
C C Dyer and R C Roeder, {\it Astrophys. J} {\bf 174}, L115 (1972);
{\it Astrophys. J} {\bf 189}, 167 (1974).
\bibitem{voids-obs}
F Hoyle and M S Vogeley, {\it Astrophys. J.} {\bf 566}, 641 (2002)
[arXiv:astro-ph/0109357]; {\it Astrophys. J.} {\bf 607}, 751 (2004)
[arXiv:astro-ph/0312533];\\
M S Vogeley et al., in {\it Proc. IAU Colloquium
  No. 195} (2004), A Diaferio, ed. \\
S G Patiri, et al., {\it Mon. Not. R. Astr. Soc.} {\bf 369},
335 (2006) [arXiv:astro-ph/0506668];\\
A V Tikhonov and I D Karachentsev, {\it Astrophys. J.} {\bf 653}, 969
(2006) [astro-ph/0609109]. 
\bibitem{voids-DE}
M Sasaki, {\it Mon. Not. Roy. Astron. Soc.} {\bf 228}, 653 (1987);\\ 
N Sugiura, K Nakao and T Harada,  {\it Phys. Rev.} {\bf D60}, 103508
(1999); \\ 
M-N Celerier, {\it Astron. \& Astrophys.} {\bf 353}, 63 (2000)
[arXiv:astro-ph/9907206];\\ 
K Tomita, {\it Mon. Not. Roy. Astron. Soc.} {\bf 326}, 287 (2001)
[arXiv:astro-ph/0011484]; \\ 
H Iguchi, T Nakamura and K Nakao,  {\it Prog. of
Theo. Phys.} {\bf 108}, 809 (2002); \\
W Godlowski, J Stelmach and M Szydlowski, {\it Class. Quant. Grav.}
{\bf 21}, 3953 (2004) [arXiv:astro-ph/0403534];\\
H Alnes, M Amazguioui and O Gron, {\it Phys. Rev.} {\bf D73}, 083519
(2006) [arXiv:astro-ph/0512006];\\
R A Vanderveld, E E Flanagan and I Wasserman, {\it Phys. Rev.} {\bf
D74}, 023506 (2006) [arXiv:astro-ph/0602476]; {\it Phys. Rev.} {\bf
D76}, 083504 (2007) [arXiv:0706.1931]; \\ 
C H Chuang, J A Gu and W Y P Hwang, {\it Class. Quant. Grav.} {\bf
25}, 175001 (2008) [arXiv:astro-ph/0512651];\\ 
K Bolejko, {\it PMC Phys.} {\bf A2}, 1 (2008)
[arXiv:astro-ph/0512103]; \\
P Apostolopoulos, N Brouzakis, N Tetradis and E Tzavara,
{\it JCAP} {\bf 0606}:009 (2006) [arXiv:astro-ph/0603234]; \\ 
D Garfinkle, {\it Class. Quant. Grav.} {\bf 23}, 4811 (2006)
[arXiv:gr-qc/0605088];\\ 
T Biswas, R Mansouri and A Notari, {\it
JCAP} {\bf 0712}:017 (2007) [arXiv:astro-ph/0606703];\\
H Alnes and M Amazguioui, {\it Phys. Rev.} {\bf D74}, 103520 (2006)
[arXiv:astro-ph/0607334]; \\
K Enqvist and T Mattsson, {\it JCAP} {\bf 0702}:019 (2007)
[arXiv:astro-ph/0609120]; \\ 
T Biswas and A Notari, {\it JCAP} {\bf 0806}:021 (2008)
[arXiv:astro-ph/0702555]; \\
V Marra, E W Kolb, S Matarrese and A Riotto, {\it Phys. Rev.} {\bf
D76}, 123004 (2007) [arXiv:0708.3622];\\
V Marra, E W Kolb and S Matarrese, {\it Phys. Rev.} {\bf D77}, 023003
(2008) [arXiv:0710.5505];\\
J Garcia-Bellido and T Haugboelle, {\it JCAP} {\bf 0804}:003 (2008)
[arXiv:0802.1523];\\ 
I Jakacka and J Stelmach, {\it Class. Quantum Grav.} {\bf 18}, 2643
(2001) [arXiv:0802.2284]; \\
C Yoo, T Kai and K Nakao, {\it Prog. Theor. Phys.} {\bf 120}, 937
(2008) [arXiv:0807.0932];\\
P Hunt and S Sarkar, arXiv:0807.4508 (2008);\\
R A Vanderveld, E E Flanagan and I Wasserman, {\it Phys. Rev.} {\bf
D78}, 083511 (2008) [arXiv:0808.1080];\\
J P Zibin, A Moss and D Scott, {\it Phys. Rev. Lett.} {\bf 101},
251303 (2008) [arXiv:0809.3761];\\
T Clifton, P G Ferreira and J Zuntz, arXiv:0902.1313 (2009);\\
T Clifton and J Zuntz, arXiv:0902.0726 (2009);\\
W Valkenburg, arXiv:0902.4698 (2009).
\bibitem{obs-cosmo}
M E Araujo, W R Stoeger, R C Arcuri and M L Bedran, {\it Phys. Rev.}
{\bf D78}, 063513 (2008) [arXiv:0807.4193].
\bibitem{future-obs}
J Frieman, M Turner and D Huterer, arXiv:0803.0982 (2008), invited
review for {\it Ann. Rev. Astron. and Astrophys.}
\bibitem{copernican-test}
R R Caldwell and A Stebbins, {\it Phys. Rev. Lett.} {\bf 100}, 191302
(2008) [arXiv:0711.3459];\\
J-P Uzan, C Clarkson and G F R Ellis, {\it Phys. Rev. Lett.} {\bf
  100}, 191303 (2008) [arXiv:0801.0068];\\
K Bolejko and J S B Wyithe, {\it JCAP} {\bf 0902}:020 (2009)
[arXiv:0807.2891]; \\
C Quercellini, M Quartin and L Amendola, arXiv:0809.3675 (2008).
\bibitem{LTB} 
Lema\^\i tre, G. {\it Ann. Soc. Sci. Brux.}, A
  {\bf 53}, 51, (1933) (in French)\\ Lema\^\i tre, G. {\it
  Gen. Rel. and Grav.}, {\bf 29}, 5, (1997) (reprint).\\ Tolman,
  R. C., {\it Proc. Nat. Acad. Sci}, {\bf 20}, 169, (1934).\\ Bondi, H.,
  {\it   Mon. Not. R. Astron. Soc.}, {\bf 107}, 410, (1947). 
\bibitem{landau-tof}
L D Landau and E M Lifshitz, {\it The Classical Theory of Fields}, 4th
rev. English ed., Butterworth-Heinemann, Oxford (1975).
\bibitem{bmr07}
N Bartolo, S Matarrese and A Riotto, Lectures given at Les Houches
Summer School - Session 86: Particle Physics and Cosmology: The Fabric
of Spacetime, Les Houches, France, 31 Jul - 25 Aug 2006,
[arXiv:astro-ph/0703496]. 
\bibitem{bruni}
M Bruni, S Matarrese, S Mollerach and S Sonego, {\it
  Class. Quant. Grav.} {\bf 14}, 2585 (1997) [arXiv:gr-qc/9609040]. 
\bibitem{tps} 
T P Singh and P S Joshi, {\it Class. Quant. Grav.} {\bf 13}, 559
(1996) [arXiv:gr-qc/9409062].
\bibitem{mtw}
C W Misner, K S Thorne and J A Wheeler, {\it Gravitation}, W H Freeman
and Co., New York, (1970).
\bibitem{coley-lightcone}
A A Coley, [arXiv:0905.2442 [gr-qc]] (2009).
\bibitem{buchpriv} 
T Buchert, private communication.
\bibitem{bardeen}
J M Bardeen, {\it Phys. Rev.} {\bf D22}, 1882 (1980).
\bibitem{liddle-lyth}
A R Liddle and D H Lyth, {\it Cosmological Inflation and Large Scale
  Structure}, Cambridge Univ. Press (2000).  
\bibitem{scalinvarpowspec} 
E R Harrison, {\it Phys. Rev.} {\bf D1}, 2726 (1970);\\
Y B Zel'dovich, {\it Mon. Not. Roy. Astron. Soc.} {\bf 160}, 1P
(1972). 
\bibitem{cutoff-theory}
See, e.g.\\
A Vilenkin and L H Ford, {\it Phys. Rev.} {\bf D26}, 1231 (1982);\\
J Silk and M S Turner, {\it Phys. Rev.} {\bf D35}, 419 (1987);\\
L A Kofman, A D Linde, {\it Nucl. Phys.} {\bf B282}, 555 (1987).
\bibitem{cutoff-obsvns}
A Shafieloo and T Souradeep, {\it Phys. Rev.} {\bf D70}, 043523
(2004) [arXiv:astro-ph/0312174];\\
D Tocchini-Valentini, Y Hoffman and J Silk, {\it
  Mon. Not. Roy. Astron. Soc.} {\bf 367}, 1095 (2006)
[arXiv:astro-ph/0509478];\\ 
See, however,\\
L Verde and H V Peiris [arXiv:0802.1219 [astro-ph]] (2008).
\bibitem{kolb-turner}
E W Kolb and M S Turner, {\it The Early Universe}, Addison-Wesley
(1990). 
\bibitem{hst-key}
W L Freedman et al., {\it Astrophys. J.} {\bf 553}, 47 (2001)
[arXiv:astro-ph/0012376]. 
\bibitem{bbks}
J M Bardeen, J R Bond, N Kaiser and A S Szalay, {\it Astrophys. J.}
{\bf 304}, 15 (1986).
\bibitem{liddle-normalisn}
A R Liddle,D Parkinson, S M Leach and P Mukherjee, {\it Phys. Rev.}
{\bf D74}, 083512 (2006) [arXiv:astro-ph/0607275].
\bibitem{NumRec}
W H Press, S A Teukolsky, W T Vetterling and B P Flannery, {\it
  Numerical Recipes in C}, Cambridge Univ. Press (1988).
\bibitem{kodama}
H Kodama and M Sasaki, {\it Prog. Theor. Phys. Suppl.} {\bf 78}, 1
(1984). 
\bibitem{paddy}
T Padmanabhan, {\it Gen. Rel. Grav.} {\bf 40}, 529 (2008)
[arXiv:0705.2533 [gr-qc]] \,.
\bibitem{halo}
R E Smith et al. (The Virgo Consortium) {\it
  Mon. Not. Roy. Astron. Soc.} {\bf 341}, 1311 (2003)
[arXiv:astro-ph/0207664]. 
\bibitem{bert}
For a Newtonian treatment of shell crossings in spherical collapse,
see e.g. -- \\
E Bertschinger, {\it Astrophys. J. S.} {\bf 58}, 1, (1985); {\it
  Astrophys. J. S.} {\bf 58}, 39 (1985); \\
G L Hoffman, D W Olson and E E Salpeter, {\it Astrophys. J.} {\bf
  242}, 861 (1980). \\
M A Sanchez-Conde, J Betancort-Rijo and F Prada, {\it
  Mon. Not. R. Astron. Soc.} {\bf 378}, 339 (2007)
[arXiv:astro-ph/0609479]. \\
See also\\
K Bolejko, A Krasinski and C Hellaby, {\it Mon. Not. R. Astron. Soc.}
{\bf 362}, 213 (2005) [arXiv:gr-qc/0411126];\\
R K Sheth and R van de Weygaert, {\it Mon. Not. R. Astron. Soc.} {\bf
  350}, 517 (2004) [arXiv:astro-ph/0311260].
\bibitem{szapudi}
B R Granett, M C Neyrinck and I Szapudi, {\it Astrophys. J.} {\bf
683}, L99 (2008) [arXiv:0805.3695 [astro-ph]].

\end{thebibliography}
\end{document}